\pdfoutput=1
\documentclass[a4paper]{article}
\usepackage[utf8]{inputenc}

\usepackage{hyperref}
\usepackage{amsmath,amsfonts,amssymb}
\usepackage{bbm}
\usepackage{amsthm}
\usepackage{color}
\usepackage{graphicx}
\usepackage{tikz-cd}
\usepackage{tikz-3dplot}
\usetikzlibrary {decorations.pathmorphing,shadows,fadings}
\usepackage{pgfplots}
\usepackage{cite}

\def\ds{\displaystyle}
\def\bea{\begin{array}{c}}
\def\ea{\end{array}}
\def\be{\begin{equation}\bea\ds}
\def\ee{\ea\end{equation}}
\def\bee{\begin{equation}\begin{array}{rcl}\ds}
\def\eee{\end{array}\end{equation}}

\def\Lc{{\mathcal{L}}}
\def\Hc{{\mathcal{H}}}

\def\Nc{{\mathcal{N}}}
\def\Dc{{\mathcal{D}}}

\def\Kc{{\mathcal{K}}}

\def\tr{{\rm tr}\,}
\def\Tr{{\rm Tr}\,}
\def\trm{\Tr\!_{\rm M}}
\def\Svn{S_{\rm vN}}

\title{Tutorial on Knots and Quantum Mechanics}
\author{Dmitry Melnikov}
\date{}

\begin{document}

\maketitle

\begin{center}
\textit{\small International Institute of Physics, Federal University of 
Rio Grande do Norte, \\ Campus Universit\'ario, Lagoa Nova, Natal-RN  
59078-970, Brazil}

\vspace{2cm}

\end{center}

\vspace{-2cm}

\begin{abstract}
These notes review a description of quantum mechanics in terms of the topology of spaces, basing on the axioms of Topological Quantum Field Theory and path integral formalism. In this description quantum states and operators are encoded by the topology of spaces that are used as modules to build the quantum mechanical model, while expectation values and probabilities are given by topological invariants of spaces, knots and links. The notes focus on the specific way the topology encodes quantum mechanical features, or, equivalently, on how these features can be controlled through the topology. A topological classification of entanglement is discussed, as well as properties of entanglement entropy and basic quantum protocols. The primary aim is to build a less conventional diagrammatic intuition about quantum mechanics,  expanding the paradigm of ``Quantum Picturalism".
\end{abstract}

\tableofcontents

\section{Introduction}

Physics contributed to the birth of knot theory in the 19th century, when Lord Kelvin hypothesized that atoms are knotted vortices of aether~\cite{Thomson:1869vor}. Although the idea was doomed to fail,\footnote{The beauty of the concept continued to enchant physicists and led to a new reincarnation of the idea at the end of the 20th century. In the new formulation, atomic nuclei were proposed to be viewed as knots of pion fields, extending the Skyrme model~\cite{Faddeev:1996zj}.} it is said to inspire the first knot classification by Peter Guthrie Tait. Tait's knot table had significant impact on developing knot theory as a branch of pure mathematics, a subfield of topology. Until late 20th century advances in knot theory were predominantly limited to math.

In 1980s new connections between knot theory and physics started to emerge. Vaughn Jones discovers a new type of topological invariant of knots~\cite{Jones:1985dw}, now known as Jones polynomial, by studying operator algebras relevant for quantum mechanical systems (von Neumann algebras). Connections between gauge theories and various types of topological invariants are being unearthed, e.g.~\cite{Belavin:1975fg,Schwarz:1978cn,Donaldson:1983wm,Floer:1988ins}, including the one between the oldest link invariant of Gauss and Chern-Simons theory~\cite{Polyakov:1988md}. General methods of constructing knot invariants are emerging in integrable statistical physics models~\cite{Kauffman:1987sta,Turaev:1988eb} and two-dimensional conformal field theories (CFT)~\cite{Tsuchiya:1987rv,Verlinde:1988sn,Moore:1988uz,Segal:1988con}. A benchmark achievement is the work of Edward Witten~\cite{Witten:1988hf}, which brings together all the different types of evidence, tying CFT, Chern-Simons theories and the Jones polynomial, and opening gates for a plethora of generalizations. (See the following reviews of modern topics in knot theory~\cite{Manturov:2018book,Przytycki:2024book} and its connections to physics~\cite{Kauffman:2001book}.)

A notable feature of these manifestations of knot theory is their quantumness. At the technical level this is explained by the observation that the relevant topological invariants, as defined in mathematics, can be cast as path integrals (partition functions) of some physical theories. At a more abstract level, topology is intrinsically discrete, as is the quantum theory. (See, for example, the discussion in~\cite{Atiyah:1989vu}.) Physical theories in the path integral should satisfy certain conditions. To be topological, they should be insensitive to smooth local variations of the geometry of the space they live in. In other words, they should be metric independent. One way to achieve this is to have their Lagrangians constructed without any explicit or implicit use of the metric tensor~\cite{Witten:1988ze,Schwarz:2000ct}.

On the physics side, path integrals provide a convenient formal description of quantum mechanics. They compute matrix elements of the evolution operator and, by construction, possess a natural concatenation, or composition, property, which expresses a big matrix element as a product of infinitesimal ones. Path integrals are, essentially, matrices, or tensors, though commonly represented in an uncountably infinite-dimensional space.

Yet, on the mathematical side, path integrals are ill-defined, what motivated Michael Atiyah to give an alternative, axiomatic, definition of the \emph{topological quantum field theories} (TQFTs), to somewhat convince mathematicians about the legitimacy and merits of the physical approach~\cite{Atiyah:1989vu}. Atiyah's axioms replace the path integral by topological spaces with boundaries (the cobordisms, a sort of fat Feynman diagrams) that have a similar concatenation property, as well as other properties compatible with path integrals. By construction, these topological spaces should also be viewed as relevant quantum mechanical matrices, such as the evolution operator. In other words, Atiyah's TQFTs construct quantum mechanics from topological spaces, bypassing the definition of the path integral.

The purpose of this review is to use the Atiyah's definition of TQFT and its relation to knot theory in order to build some general intuition about quantum mechanics, and in particular, about quantum correlations. The basis for this intuition will be general manipulations we can apply to topological spaces, such as twisting, braiding, knotting etc., besides the mentioned concatenation. We will use these manipulations to ``visualize'' quantum states and quantum correlations, explain quantum entanglement and some of its properties, and demonstrate how topological manipulations can be used in construction of quantum algorithms. Hopefully, this will help provide a useful complementary perspective on quantum phenomena and quantum physics. 

As such the review has overlaps with two particular subjects in quantum theory. One is Topological Quantum Computation, a large modern and actively developing field, which started from the idea that topological phases of matter provide an appealing setup for realization of the quantum computer. Unlike bosons and fermions, anyons -- the quasiparticles of quantum Hall systems -- may possess an unusual \emph{nonabelian} statistics, which makes them simple low-dimensional quantum-mechanical systems with respect to a permutation, a mesoscopic operation of exchanging two particle positions. Together with the robustness of quantum Hall systems to local decoherence, the properties of anyons led Alexei Kitaev to the idea to propose such systems as fault-tolerant quantum computers~\cite{Kitaev:1997wr}. In this proposal world lines of permuted anyons form braids, and their knotlike closures mathematically compute matrix elements of the evolution operator, as we will also see in this review.   

Mathematical foundations of anyon-based quantum computing stem from knot theory. As the name of the field implies it should pursue the following two main goals: search for physical systems in which suitable topological phases and quantum operations can be realized and design of protocols and algorithms for given systems and sets of operations (gates). In some cases this approach becomes too ``hardware-oriented'', making the role of knots and topology more technical and obscure. The present review takes a more abstract approach trying to explore the tools available in the topological features themselves, though less concerning with important practical details. The reader may compare the approaches by looking at standard references on Topological Quantum Computations, such as the classic review~\cite{Nayak:2008zza} and a recent book~\cite{Simon:2023top}.

The second overlap is with a recently-introduced subfield of mathematics referred as Categorical Quantum Mechanics (CQM). The name means that in this field quantum mechanics is formulated in terms of category theory -- the axiomatic approach of Atiyah is another example of a categorical formulation. CQM started in the works of Samson Abramsky and Bob Coecke~\cite{Abramsky:2004doh}, based on the mathematical studies of category theory in~\cite{Kelly:1980coh,Joyal:1991geo,Carboni:1987car,Lack:2004com}. One of the central tools of CQM is the diagrammatic presentation of quantum states and processes. Unlike Atiyah's TQFTs the diagrams do not depict topological spaces but more abstract objects (categories). This makes the description more universal, but at the same time less intuitive. One of the goals of this review is thus to advocate for the topological version of the categorical approach as for more intuitive one. Hopefully this also contributes to the proposals spelled in~\cite{Coecke:2006kin,Coecke:2010swv,Dundar-Coecke:2023xyf} to use the categorical approach, or \emph{Quantum Picturalism}, as an early introduction of quantum mechanics. The following books are recommended as general references for CQM: \cite{Coecke:2017pic,Heunen:2019cat}.

The purpose of this review can also be seen as filling some gaps between the Topological Quantum Computation and CQM. It is based on some recent work, which will be introduced along the way. Some earlier work that served as motivation for the description built here might not receive deserved attention, so the reader is referred to~\cite{Kauffman:2013bh} for additional information anf literature, such as works of Louis Kauffman, Sam Lomonaco and others on the application of knot calculus in quantum mechanics and in the study of quantum entanglement, e.g.~\cite{Kauffman:2002qua,Kauffman:2003ent,Kauffman:2004qua,Kauffman:2004bra,Kauffman:2007qde,Lomonaco:2008qua}. 

These notes are organized as follows. Section~\ref{sec:knots} gives a short introduction into knot theory with a minimal amount of necessary techniques to calculate topological invariants to be used in the remaining parts. Section~\ref{sec:Jonespolynomial} introduces the Jones polynomial (and its ``bracket'' version due to Kauffman) and section~\ref{sec:matrix} -- a useful matrix formalism for its evaluation.

In section~\ref{sec:topobits} we will explain how braids and more general topological spaces can be understood as quantum mechanical states. We use a minimal knowledge of category theory to review Atiyah's definition of TQFT in section~\ref{sec:tqft}. In section~\ref{sec:TQM} we describe a one dimensional TQFT, to which we refer as topological quantum mechanics (TQM). This TQM is in fact more naturally formulated in three dimensions, as a Chern-Simons theory, as we review in section~\ref{sec:CS}. 

Section~\ref{sec:entanglement} is dedicated to a study of quantum entanglement using the topological viewpoint. We explain the general relation of space connectivity and topology with the entanglement in section~\ref{sec:tangling}. In section~\ref{sec:SLOCC} we discuss a classification of entanglement using stochastic local operations and classical communication and how this classification emerges in topological spaces describing bipartite entanglement. In section~\ref{sec:entropy} we introduce the von Neumann entropy and explain its topological meaning. Some of the properties of the entropy are discussed, including famous inequalities in section~\ref{sec:inequalities}.

Section~\ref{sec:algorithms} reviews two quantum algorithms in the topological formulation: quantum teleportation in section~\ref{sec:teleport} and dense coding in section~\ref{sec:densecoding}. The purpose is to illustrate the sense, in which entanglement is a resource for quantum tasks.

We give concluding remarks in section~\ref{sec:outlook}, where we also review some open questions and possible applications, focusing on multipartite entanglement and quantum gravity.

%%%%%%%%%%%%%%%%%%%%%%%%%%%%%%%%%%%%%%%%%%%%%%%%%%%%%%%%%%%%%
\section{Basics of knot theory}
\label{sec:knots}

In this section we give a minimal introduction to knots and their topological invariants, necessary to make quantum mechanical computations. Introductory material into knot theory can be found in these books~\cite{Kauffman:2001book, Adams:2004book,Sossinsky:2023book}.

We understand a knot $\Kc$ as a map of a circle $\mathbb{S}^1$ into an ambient three-dimensional space $\mathbb{R}^3$. For example:\footnote{The 3D pictures of knots were obtained from Wikimedia Commons.}
\be
 {\cal K}:\ \begin{array}{c}
         \includegraphics[height=0.08\linewidth]{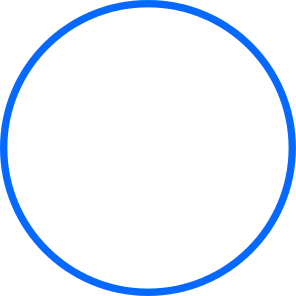}  
     \end{array}
     \quad \longrightarrow \quad
     \begin{array}{c}
    \includegraphics[height=0.1\linewidth]{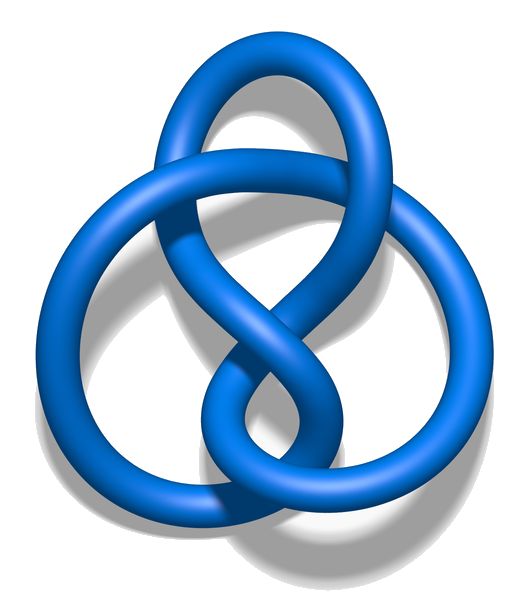}
    \end{array}.
\ee
The trivial map takes $\mathbb{S}^1$ into a circle in $\mathbb{R}^3$, which is called the \emph{unknot}.

Similarly, we define a link through a collection of circles, $\Lc:~\mathbb{S}^1\times \mathbb{S}^1\times \cdots\times \mathbb{S}^1\to \mathbb{R}^3$:
\be
\label{Borromean}
  {\cal L}:\ \begin{array}{c}
         \includegraphics[height=0.08\linewidth]{./figs/circle.png}  
     \end{array}
     \begin{array}{c}
         \includegraphics[height=0.08\linewidth]{./figs/circle.png}  
     \end{array}
     \begin{array}{c}
         \includegraphics[height=0.08\linewidth]{./figs/circle.png}  
     \end{array}
     \quad \longrightarrow \quad \begin{array}{c}
          \includegraphics[height=0.1\linewidth]{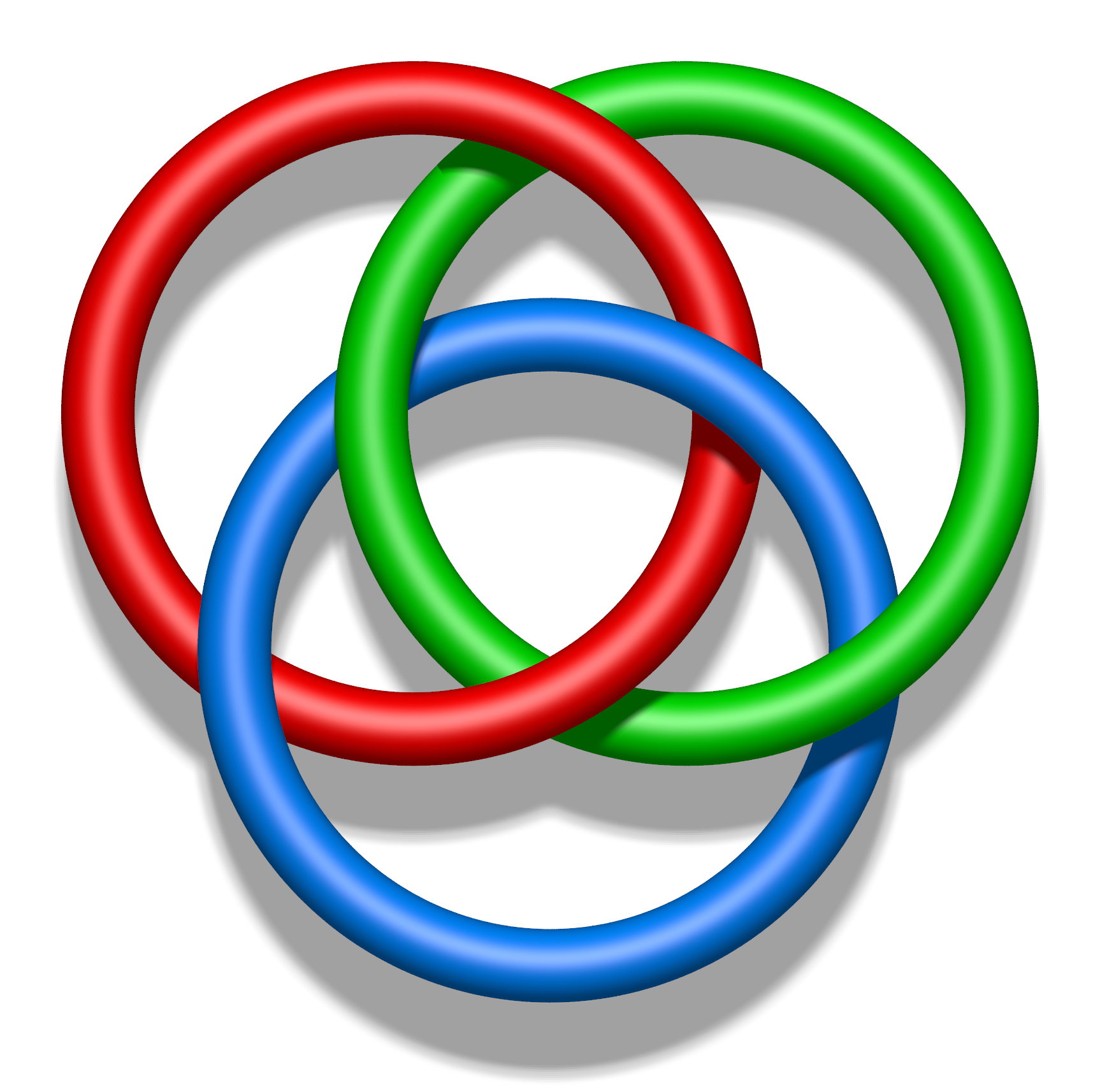} 
     \end{array}.
\ee

We will use the notion of a knot (or link) diagram, which is a projection of a three-dimensional knot (link) on any two dimensional plane, as the following diagram intends to illustrate:
\be
 \begin{array}{c}
          \includegraphics[height=0.1\linewidth]{./figs/figure8.png} 
     \end{array}
     \quad \longrightarrow \quad
     \begin{array}{c}
    \includegraphics[height=0.1\linewidth]{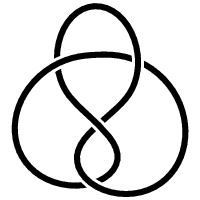}
    \end{array}.
\ee
In most cases the projection will produce a number of ambiguous points corresponding to apparent intersections of lines. In order to maintain the full information about the topology the intersections of the diagram are \emph{resolved} into one of the two possibilities,
\be
\begin{array}{c}
\begin{tikzpicture}[line width=1.5]
\draw (0,0) -- (0.7,0.7);
\draw (0,0.7) -- (0.3,0.4);
\draw (0.4,0.3) -- (0.7,0);
\end{tikzpicture}
\end{array}
\qquad 
\text{or}\,,
\qquad
\begin{array}{c}
\begin{tikzpicture}[line width=1.5]
\draw (0,0.7) -- (0.7,0.0);
\draw (0,0.0) -- (0.3,0.3);
\draw (0.4,0.4) -- (0.7,0.7);
\end{tikzpicture}
\end{array},
\ee
keeping track of which of the lines appears on top of the other in the projection.

The same knot or link can give origin to a pair of apparently distinct knot diagrams, so one could be interested in introducing an equivalence relation on the diagrams, namely any two diagrams should be equivalent if they come from projections of two homeomorphic (3D-equivalent) knots or links.

A theorem proven by James Alexander and Garland Briggs (and independently by Kurt Reidemeister) states that a pair of diagrams correspond to the same knot or link if they can be transformed into each other through a sequence of the following three \emph{Reidemeister moves}:
\be
\begin{array}{ccccc}
   \begin{array}{c}
   \begin{tikzpicture}[line width=2]
     \draw[rounded corners=3] (0,0) -- (0,0.3) -- (0.4,0.7) -- (0.6,0.7) -- (0.6,0.3) -- (0.4,0.3) -- (0.3,0.4);
     \draw[rounded corners=3] (0.1,0.6) -- (0,0.7) -- (0,1);
   \end{tikzpicture}
   \end{array} \leftrightarrow \begin{array}{c}
   \begin{tikzpicture}[line width=2]
     \draw[rounded corners=3] (0.0,0.0) -- (0,1);
   \end{tikzpicture}
   \end{array}
     & & \begin{array}{c}
   \begin{tikzpicture}[line width=2]
     \draw[rounded corners=3] (0,0) -- (0,0.1) -- (0.4,0.4) -- (0.4,0.6) -- (0.0,0.9) -- (0.0,1);
     \draw[rounded corners=3] (0.4,0) -- (0.4,0.1) -- (0.3,0.2);
     \draw[rounded corners=3] (0.1,0.4) -- (0.0,0.5) -- (0.1,0.6);
     \draw[rounded corners=3] (0.3,0.8) -- (0.4,0.9) -- (0.4,1);
   \end{tikzpicture}
   \end{array} \leftrightarrow \begin{array}{c}
   \begin{tikzpicture}[line width=2]
     \draw[rounded corners=3] (0.0,0.0) -- (0,1);
     \draw[rounded corners=3] (0.4,0.0) -- (0.4,1);
   \end{tikzpicture}
   \end{array} 
   & & \begin{array}{c}
   \begin{tikzpicture}[line width=2]
     \draw[rounded corners=3] (0,0) -- (0,0.3) -- (0.4,0.7) -- (0.4,1);
     \draw[rounded corners=3] (0.4,0.0) -- (0.4,0.3) -- (0.3,0.4);
     \draw[rounded corners=3] (0.1,0.6) -- (0,0.7) -- (0,1);
     \draw[rounded corners=3] (-0.25,0.2) -- (-0.1,0.2);
     \draw[rounded corners=3] (0.1,0.2) -- (0.3,0.2);
     \draw[rounded corners=3] (0.5,0.2) -- (0.65,0.2);
   \end{tikzpicture}
   \end{array} \leftrightarrow \begin{array}{c}
   \begin{tikzpicture}[line width=2]
     \draw[rounded corners=3] (0,0) -- (0,0.3) -- (0.4,0.7) -- (0.4,1);
     \draw[rounded corners=3] (0.4,0.0) -- (0.4,0.3) -- (0.3,0.4);
     \draw[rounded corners=3] (0.1,0.6) -- (0,0.7) -- (0,1);
     \draw[rounded corners=3] (-0.25,0.8) -- (-0.1,0.8);
     \draw[rounded corners=3] (0.1,0.8) -- (0.3,0.8);
     \draw[rounded corners=3] (0.5,0.8) -- (0.65,0.8);
   \end{tikzpicture}
   \end{array}\\
    \text{I move} & &  \text{II move} & & \text{III move}
\end{array}.
\ee

Then the central problem of knot theory is to understand how many different knots exist and to classify them. In particular, if one draws a pair of random diagrams one might wonder if the diagrams correspond to the same knot or link, or, similarly, whether any of the diagrams represent an unknot, the trivial knot. Anticipating quantum mechanical applications of knot theory, one may formulate the classification problem as finding all possible ways of tangling of the unknot.

Distinguishing a pair of knots requires a discriminator, or a \emph{hash function}, that, when evaluated on a pair of knots, would produce the same \emph{key} (the same value) if two knots are the same, and, ideally, different keys (values) if the knots are different. Such a function is called topological invariant. It should not be sensitive to smooth deformations of the knot that do not change its topology. 

Preferably, the output of a topological invariant is a simple, computable object. A limited number of such functions is known. One of them is the linking number introduced by Carl Friedrich Gauss in the first half of the 19th century. This invariant can be derived in physics by substituting the Biot-Savart law,
\be
{\mathbf{B}}(\mathbf{r_2}) \ = \ \frac{\mu_0 I}{4\pi}\int_{\gamma_1} \frac{d\mathbf{r_1}\times (\mathbf{r}_2-\mathbf{r}_1)}{|\mathbf{r}_2-\mathbf{r}_1|^3}\,,
\ee
for a magnetic field created by a current $I$ flowing in a circuit $\gamma_1$, parameterized by vector $\mathbf{r}_1$, into the Ampere's law,
\be
\int_{\gamma_2} \mathbf{B}(\mathbf{r}_2)\cdot d\mathbf{r}_2 \ = \ \mu_0\, \ell I\,,
\ee
where $\ell$ is computing the algebraic sum of the number of times current $I$ crosses a surface bounded by a closed circuit $\gamma_2$. All physics cancels from the result and one obtains the Gauss' formula:
\be
\ell \ = \ \frac{1}{4\pi} \int_{\gamma_1}\int_{\gamma_2} \frac{(d\mathbf{r_1}\times (\mathbf{r}_2-\mathbf{r}_1))\cdot d\mathbf{r}_2}{|\mathbf{r}_2-\mathbf{r}_1|^3}\,. 
\ee
Although it appears as a complicated double integral, the result is a simple counting of the number of times circuit $\gamma_1$ passes through circuit $\gamma_2$, taking into account the orientations of both circuits. The result does not depend on smooth deformations of $\gamma_1$ and $\gamma_2$, that is deformations that do not make the two circuits cross each other.

As is clear from its definition, Gauss' linking number $\ell$ defines a topological invariant of a two-component link, which means that it is not applicable to the proper knots and to links with more than two components. Moreover, it is not hard to find an example of two topologically distinct links with the same linking number. One example is provided by the Whitehead link
\be
\label{whitehead}
\begin{array}{c}
          \includegraphics[height=0.09\linewidth]{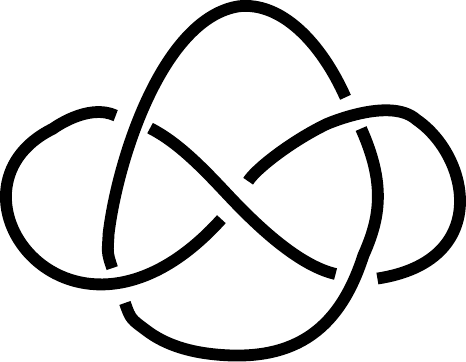} 
     \end{array},
\ee
which has zero linking number.

More powerful invariants were discovered in the 20th century. In 1920s Alexander used homology methods to introduce an invariant of knots and links that evaluates to a polynomial of a single parameter~\cite{Alexander:1928top}. Alexander's polynomial has a much higher resolution power, but still gives same values for many nonequivalent knots. A stronger polynomial invariant was discovered by Vaughan Jones in 1980s by studying von Neumann algebras~\cite{Jones:1985dw,Jones:1987dy}. Even stronger generalizations appeared shortly after. The two-variable colored HOMFLY-PT polynomial, named after Jim Hoste, Adrian Ocneanu, Kenneth Millet, Peter Freyd, William Lickorish, David Yetter as well as after Jozef Przytycki and Pawel Traczyk is believed to be powerful enough to distinguish any pair of knots or links~\cite{Freyd:1985dx,Przytycki:2016inv}.

In the remaining part of this section we will introduce the Jones polynomial and review several methods of computing it.

%%%%%%%%%%%%%%%%%%%%%%%%%%%%%%%%%%%%%%%%%%%%%%%%%%%%%%%%%%%%%%%%%%%%
\subsection{Topological invariants from matrix traces}

It will be useful to review a few important steps explaining the origin of the Jones polynomial, as they will also be helpful for later applications.

An important object is the (Artin) braid group. This group is a generalization of the group of permutations of $n$ elements and can be defined in terms of $n-1$ generators:
\be
    B_n\ := \ \{ b_k,\, k=1,\ldots,n-1|\ b_ib_{i+1}b_i = b_{i+1}b_ib_{i+1},\, i=1,\ldots,n-2;\ b_ib_j=b_jb_i,\, |i-j|>1\}\,.
\ee
In other words, group $B_n$ consists of all words constructed of $n-1$ letters $b_k$, satisfying the above relations. The relations are the same as for the generators of the permutation group, with the exception of the requirement that the generators square to identity. Relaxing this condition makes the braid group infinite, unlike the permutation group.

Braid group has a diagrammatic presentation as an evolution of $n$ points on a line. For example, the generators of the group are represented by
\be
b_1 \ = \ 
\begin{array}{c}
       \begin{tikzpicture}[thick]
           \draw[gray,line width=1.5,opacity=0.6] (0,0) -- (2.2,0);
           \draw[gray,line width=1.5,opacity=0.6] (0,1) -- (2.2,1);
           \draw[rounded corners=3] (0.1,0) -- (0.1,0.3) -- (0.5,0.7) -- (0.5,1);
         \draw[rounded corners=3] (0.5,0) -- (0.5,0.3) -- (0.4,0.4);
         \draw[rounded corners=3] (0.2,0.6) -- (0.1,0.7) -- (0.1,1);
         \draw[rounded corners=3] (0.9,0.0) -- (0.9,1);
         \draw[rounded corners=3] (1.3,0.0) -- (1.3,1);
         \draw[rounded corners=3] (2.1,0.0) -- (2.1,1);
           \foreach \x in {0.1,0.5,...,1.5}
           \fill[black] (\x,0) circle (0.05);
           \foreach \x in {0.1,0.5,...,1.5}
           \fill[black] (\x,1) circle (0.05);
           \fill[black] (2.1,1) circle (0.05);
           \fill[black] (2.1,0) circle (0.05);
           \draw (1.7,0.5) node {$\cdots$};
       \end{tikzpicture}
\end{array},\quad
b_2 \ = \ 
\begin{array}{c}
       \newcommand{\x}{0.4}
       \begin{tikzpicture}[thick]
           \draw[gray,line width=1.5,opacity=0.6] (0,0) -- (2.2,0);
           \draw[gray,line width=1.5,opacity=0.6] (0,1) -- (2.2,1);
           \draw[rounded corners=3] (0.1+\x,0) -- (0.1+\x,0.3) -- (0.5+\x,0.7) -- (0.5+\x,1);
         \draw[rounded corners=3] (0.5+\x,0) -- (0.5+\x,0.3) -- (0.4+\x,0.4);
         \draw[rounded corners=3] (0.2+\x,0.6) -- (0.1+\x,0.7) -- (0.1+\x,1);
         \draw[rounded corners=3] (0.1,0.0) -- (0.1,1);
         \draw[rounded corners=3] (1.3,0.0) -- (1.3,1);
         \draw[rounded corners=3] (2.1,0.0) -- (2.1,1);
           \foreach \x in {0.1,0.5,...,1.5}
           \fill[black] (\x,0) circle (0.05);
           \foreach \x in {0.1,0.5,...,1.5}
           \fill[black] (\x,1) circle (0.05);
           \fill[black] (2.1,1) circle (0.05);
           \fill[black] (2.1,0) circle (0.05);
           \draw (1.7,0.5) node {$\cdots$};
       \end{tikzpicture}
\end{array}, \quad
b_3 \ = \ 
\begin{array}{c}
       \newcommand{\x}{0.8}
       \begin{tikzpicture}[thick]
           \draw[gray,line width=1.5,opacity=0.6] (0,0) -- (2.2,0);
           \draw[gray,line width=1.5,opacity=0.6] (0,1) -- (2.2,1);
           \draw[rounded corners=3] (0.1+\x,0) -- (0.1+\x,0.3) -- (0.5+\x,0.7) -- (0.5+\x,1);
         \draw[rounded corners=3] (0.5+\x,0) -- (0.5+\x,0.3) -- (0.4+\x,0.4);
         \draw[rounded corners=3] (0.2+\x,0.6) -- (0.1+\x,0.7) -- (0.1+\x,1);
         \draw[rounded corners=3] (0.1,0.0) -- (0.1,1);
         \draw[rounded corners=3] (0.5,0.0) -- (0.5,1);
         \draw[rounded corners=3] (2.1,0.0) -- (2.1,1);
           \foreach \x in {0.1,0.5,...,1.5}
           \fill[black] (\x,0) circle (0.05);
           \foreach \x in {0.1,0.5,...,1.5}
           \fill[black] (\x,1) circle (0.05);
           \fill[black] (2.1,1) circle (0.05);
           \fill[black] (2.1,0) circle (0.05);
           \draw (1.7,0.5) node {$\cdots$};
       \end{tikzpicture}
\end{array}\quad \ldots
\ee
The inverse generators have the opposite order of line crossing (left twists, as opposed to right twists):
\be
b_1^{-1} \ = \ 
\begin{array}{c}
       \newcommand{\x}{0.0}
       \begin{tikzpicture}[thick]
           \draw[gray,line width=1.5,opacity=0.6] (0,0) -- (2.2,0);
           \draw[gray,line width=1.5,opacity=0.6] (0,1) -- (2.2,1);
           \draw[rounded corners=3] (0.5+\x,0) -- (0.5+\x,0.3) -- (0.1+\x,0.7) -- (0.1+\x,1);
         \draw[rounded corners=3] (0.1+\x,0) -- (0.1+\x,0.3) -- (0.2+\x,0.4);
         \draw[rounded corners=3] (0.4+\x,0.6) -- (0.5+\x,0.7) -- (0.5+\x,1);
         \draw[rounded corners=3] (0.9,0.0) -- (0.9,1);
         \draw[rounded corners=3] (1.3,0.0) -- (1.3,1);
         \draw[rounded corners=3] (2.1,0.0) -- (2.1,1);
           \foreach \x in {0.1,0.5,...,1.5}
           \fill[black] (\x,0) circle (0.05);
           \foreach \x in {0.1,0.5,...,1.5}
           \fill[black] (\x,1) circle (0.05);
           \fill[black] (2.1,1) circle (0.05);
           \fill[black] (2.1,0) circle (0.05);
           \draw (1.7,0.5) node {$\cdots$};
       \end{tikzpicture}
\end{array}, \quad
b_2^{-1} \ = \ 
\begin{array}{c}
       \newcommand{\x}{0.4}
       \begin{tikzpicture}[thick]
           \draw[gray,line width=1.5,opacity=0.6] (0,0) -- (2.2,0);
           \draw[gray,line width=1.5,opacity=0.6] (0,1) -- (2.2,1);
           \draw[rounded corners=3] (0.5+\x,0) -- (0.5+\x,0.3) -- (0.1+\x,0.7) -- (0.1+\x,1);
         \draw[rounded corners=3] (0.1+\x,0) -- (0.1+\x,0.3) -- (0.2+\x,0.4);
         \draw[rounded corners=3] (0.4+\x,0.6) -- (0.5+\x,0.7) -- (0.5+\x,1);
         \draw[rounded corners=3] (0.1,0.0) -- (0.1,1);
         \draw[rounded corners=3] (1.3,0.0) -- (1.3,1);
         \draw[rounded corners=3] (2.1,0.0) -- (2.1,1);
           \foreach \x in {0.1,0.5,...,1.5}
           \fill[black] (\x,0) circle (0.05);
           \foreach \x in {0.1,0.5,...,1.5}
           \fill[black] (\x,1) circle (0.05);
           \fill[black] (2.1,1) circle (0.05);
           \fill[black] (2.1,0) circle (0.05);
           \draw (1.7,0.5) node {$\cdots$};
       \end{tikzpicture}
\end{array}, \quad \ldots \quad 
b_{n-1}^{-1} \ = \ 
\begin{array}{c}
       \newcommand{\x}{1.6}
       \begin{tikzpicture}[thick]
           \draw[gray,line width=1.5,opacity=0.6] (0,0) -- (2.2,0);
           \draw[gray,line width=1.5,opacity=0.6] (0,1) -- (2.2,1);
           \draw[rounded corners=3] (0.5+\x,0) -- (0.5+\x,0.3) -- (0.1+\x,0.7) -- (0.1+\x,1);
         \draw[rounded corners=3] (0.1+\x,0) -- (0.1+\x,0.3) -- (0.2+\x,0.4);
         \draw[rounded corners=3] (0.4+\x,0.6) -- (0.5+\x,0.7) -- (0.5+\x,1);
         \draw[rounded corners=3] (0.1,0.0) -- (0.1,1);
         \draw[rounded corners=3] (0.5,0.0) -- (0.5,1);
         \draw[rounded corners=3] (0.9,0.0) -- (0.9,1);
           \foreach \x in {0.1,0.5,...,1.0}
           \fill[black] (\x,0) circle (0.05);
           \foreach \x in {0.1,0.5,...,1.0}
           \fill[black] (\x,1) circle (0.05);
           \fill[black] (2.1,1) circle (0.05);
           \fill[black] (2.1,0) circle (0.05);
           \fill[black] (1.7,1) circle (0.05);
           \fill[black] (1.7,0) circle (0.05);
           \draw (1.3,0.5) node {$\cdots$};
       \end{tikzpicture}
\end{array}.
\ee
Group multiplication is organized by stacking the diagrams of the group elements on top of each other:
\be
b_2b_1 \ = \ \begin{array}{c}
       \newcommand{\x}{0.4}
       \newcommand{\y}{1}
       \begin{tikzpicture}[thick]
           \draw[gray,line width=1.5,opacity=0.1] (0,0+\y) -- (2.2,0+\y);
           \draw[gray,line width=1.5,opacity=0.6] (0,1+\y) -- (2.2,1+\y);
           \draw[rounded corners=3] (0.1+\x,0+\y) -- (0.1+\x,0.3+\y) -- (0.5+\x,0.7+\y) -- (0.5+\x,1+\y);
         \draw[rounded corners=3] (0.5+\x,0+\y) -- (0.5+\x,0.3+\y) -- (0.4+\x,0.4+\y);
         \draw[rounded corners=3] (0.2+\x,0.6+\y) -- (0.1+\x,0.7+\y) -- (0.1+\x,1+\y);
         \draw[rounded corners=3] (0.1,0.0+\y) -- (0.1,1+\y);
         \draw[rounded corners=3] (1.3,0.0+\y) -- (1.3,1+\y);
         \draw[rounded corners=3] (2.1,0.0+\y) -- (2.1,1+\y);
           \foreach \x in {0.1,0.5,...,1.5}
           \fill[black] (\x,0+\y) circle (0.05);
           \foreach \x in {0.1,0.5,...,1.5}
           \fill[black] (\x,1+\y) circle (0.05);
           \fill[black] (2.1,1+\y) circle (0.05);
           \fill[black] (2.1,0+\y) circle (0.05);
           \draw (1.7,0.5+\y) node {$\cdots$};
           \renewcommand{\x}{0}
           \renewcommand{\y}{0}
           \draw[gray,line width=1.5,opacity=0.6] (0,0+\y) -- (2.2,0+\y);
           \draw[gray,line width=1.5,opacity=0.1] (0,1+\y) -- (2.2,1+\y);
           \draw[rounded corners=3] (0.1+\x,0+\y) -- (0.1+\x,0.3+\y) -- (0.5+\x,0.7+\y) -- (0.5+\x,1+\y);
         \draw[rounded corners=3] (0.5+\x,0+\y) -- (0.5+\x,0.3+\y) -- (0.4+\x,0.4+\y);
         \draw[rounded corners=3] (0.2+\x,0.6+\y) -- (0.1+\x,0.7+\y) -- (0.1+\x,1+\y);
         \draw[rounded corners=3] (0.9,0.0+\y) -- (0.9,1+\y);
         \draw[rounded corners=3] (1.3,0.0+\y) -- (1.3,1+\y);
         \draw[rounded corners=3] (2.1,0.0+\y) -- (2.1,1+\y);
           \foreach \x in {0.1,0.5,...,1.5}
           \fill[black] (\x,0+\y) circle (0.05);
           \foreach \x in {0.1,0.5,...,1.5}
           \fill[black] (\x,1+\y) circle (0.05);
           \fill[black] (2.1,1+\y) circle (0.05);
           \fill[black] (2.1,0+\y) circle (0.05);
           \draw (1.7,0.5+\y) node {$\cdots$};
       \end{tikzpicture}
       \end{array}, \qquad
       b_2b_2^{-1} \ = \ \begin{array}{c}
       \newcommand{\x}{0.4}
       \newcommand{\y}{1}
       \begin{tikzpicture}[thick]
           \draw[gray,line width=1.5,opacity=0.1] (0,0+\y) -- (2.2,0+\y);
           \draw[gray,line width=1.5,opacity=0.6] (0,1+\y) -- (2.2,1+\y);
           \draw[rounded corners=3] (0.1+\x,0+\y) -- (0.1+\x,0.3+\y) -- (0.5+\x,0.7+\y) -- (0.5+\x,1+\y);
         \draw[rounded corners=3] (0.5+\x,0+\y) -- (0.5+\x,0.3+\y) -- (0.4+\x,0.4+\y);
         \draw[rounded corners=3] (0.2+\x,0.6+\y) -- (0.1+\x,0.7+\y) -- (0.1+\x,1+\y);
         \draw[rounded corners=3] (0.1,0.0+\y) -- (0.1,1+\y);
         \draw[rounded corners=3] (1.3,0.0+\y) -- (1.3,1+\y);
         \draw[rounded corners=3] (2.1,0.0+\y) -- (2.1,1+\y);
           \foreach \x in {0.1,0.5,...,1.5}
           \fill[black] (\x,0+\y) circle (0.05);
           \foreach \x in {0.1,0.5,...,1.5}
           \fill[black] (\x,1+\y) circle (0.05);
           \fill[black] (2.1,1+\y) circle (0.05);
           \fill[black] (2.1,0+\y) circle (0.05);
           \draw (1.7,0.5+\y) node {$\cdots$};
           \renewcommand{\x}{0.4}
           \renewcommand{\y}{0}
           \draw[gray,line width=1.5,opacity=0.6] (0,0+\y) -- (2.2,0+\y);
           \draw[gray,line width=1.5,opacity=0.1] (0,1+\y) -- (2.2,1+\y);
           \draw[rounded corners=3] (0.5+\x,0+\y) -- (0.5+\x,0.3+\y) -- (0.1+\x,0.7+\y) -- (0.1+\x,1+\y);
         \draw[rounded corners=3] (0.1+\x,0+\y) -- (0.1+\x,0.3+\y) -- (0.2+\x,0.4+\y);
         \draw[rounded corners=3] (0.4+\x,0.6+\y) -- (0.5+\x,0.7+\y) -- (0.5+\x,1+\y);
         \draw[rounded corners=3] (0.1,0.0+\y) -- (0.1,1+\y);
         \draw[rounded corners=3] (1.3,0.0+\y) -- (1.3,1+\y);
         \draw[rounded corners=3] (2.1,0.0+\y) -- (2.1,1+\y);
           \foreach \x in {0.1,0.5,...,1.5}
           \fill[black] (\x,0+\y) circle (0.05);
           \foreach \x in {0.1,0.5,...,1.5}
           \fill[black] (\x,1+\y) circle (0.05);
           \fill[black] (2.1,1+\y) circle (0.05);
           \fill[black] (2.1,0+\y) circle (0.05);
           \draw (1.7,0.5+\y) node {$\cdots$};
       \end{tikzpicture}
\end{array}.
\ee
The diagrams give a faithful representation of the group elements and have an obvious relation with knot diagrams. In particular, it is easy to see that the II and the III Reidemeister moves of knot diagrams guarantee the braid group relations.

A theorem, proven by Alexander, states that any knot or link diagram can be obtained as a tracelike closure of some braid. By a tracelike closure one understands the pairwise connection of all top and bottom points of the same order with nonintersecting lines. Here are two examples, how the simplest nontrivial trefoil knot can be obtained from two tracelike closures:
\be
\label{trefoilclosure}
\begin{array}{c}
       \newcommand{\x}{0.0}
       \newcommand{\y}{0}
       \begin{tikzpicture}[thick]
           \draw[gray,line width=1.5,opacity=0.6] (0+\y,-0) -- (0+\y,-0.6);
           \draw[rounded corners=3] (0+\y,-0.5-\x) -- (0.3+\y,-0.5-\x) -- (0.7+\y,-0.1-\x) -- (1+\y,-0.1-\x);
         \draw[rounded corners=3] (0+\y,-0.1-\x) -- (0.3+\y,-0.1-\x) -- (0.4+\y,-0.2-\x);
         \draw[rounded corners=3] (0.6+\y,-0.4-\x) -- (0.7+\y,-0.5-\x) -- (1+\y,-0.5-\x);
         \draw[rounded corners=3] (0.0+\y,-0.9) -- (1+\y,-0.9);
         \draw[rounded corners=3] (0.0+\y,-1.3) -- (1+\y,-1.3);
         \draw[rounded corners=3] (-0+\y,-0.5-\x) -- (-0.2+\y,-0.5-\x) -- (-0.2+\y,-0.9) -- (0.0+\y,-0.9);
         \draw[rounded corners=3] (-0+\y,-0.1-\x) -- (-0.4+\y,-0.1-\x) -- (-0.4+\y,-1.3) -- (0.0+\y,-1.3);
           \foreach \x in {0.1,0.5,...,0.6}
           \fill[black] (0+\y,-\x) circle (0.05);
           \renewcommand{\x}{0.0}
            \renewcommand{\y}{1}
           \draw[rounded corners=3] (0+\y,-0.5-\x) -- (0.3+\y,-0.5-\x) -- (0.7+\y,-0.1-\x) -- (1+\y,-0.1-\x);
         \draw[rounded corners=3] (0+\y,-0.1-\x) -- (0.3+\y,-0.1-\x) -- (0.4+\y,-0.2-\x);
         \draw[rounded corners=3] (0.6+\y,-0.4-\x) -- (0.7+\y,-0.5-\x) -- (1+\y,-0.5-\x);
         \draw[rounded corners=3] (0.0+\y,-0.9) -- (1+\y,-0.9);
         \draw[rounded corners=3] (0.0+\y,-1.3) -- (1+\y,-1.3);
            \renewcommand{\x}{0.0}
            \renewcommand{\y}{2}
            \draw[gray,line width=1.5,opacity=0.6] (1+\y,-0) -- (1+\y,-0.6);
           \draw[rounded corners=3] (0+\y,-0.5-\x) -- (0.3+\y,-0.5-\x) -- (0.7+\y,-0.1-\x) -- (1+\y,-0.1-\x);
         \draw[rounded corners=3] (0+\y,-0.1-\x) -- (0.3+\y,-0.1-\x) -- (0.4+\y,-0.2-\x);
         \draw[rounded corners=3] (0.6+\y,-0.4-\x) -- (0.7+\y,-0.5-\x) -- (1+\y,-0.5-\x);
         \draw[rounded corners=3] (0.0+\y,-0.9) -- (1+\y,-0.9);
         \draw[rounded corners=3] (0.0+\y,-1.3) -- (1+\y,-1.3);
          \draw[rounded corners=3] (1+\y,-0.5-\x) -- (1.2+\y,-0.5-\x) -- (1.2+\y,-0.9) -- (1+\y,-0.9);
         \draw[rounded corners=3] (1+\y,-0.1-\x) -- (1.4+\y,-0.1-\x) -- (1.4+\y,-1.3) -- (1+\y,-1.3);
          \foreach \x in {0.1,0.5,...,0.6}
           \fill[black] (1+\y,-\x) circle (0.05);
       \end{tikzpicture}
\end{array} \quad \sim \quad
\begin{array}{c}
    \includegraphics[height=0.1\linewidth]{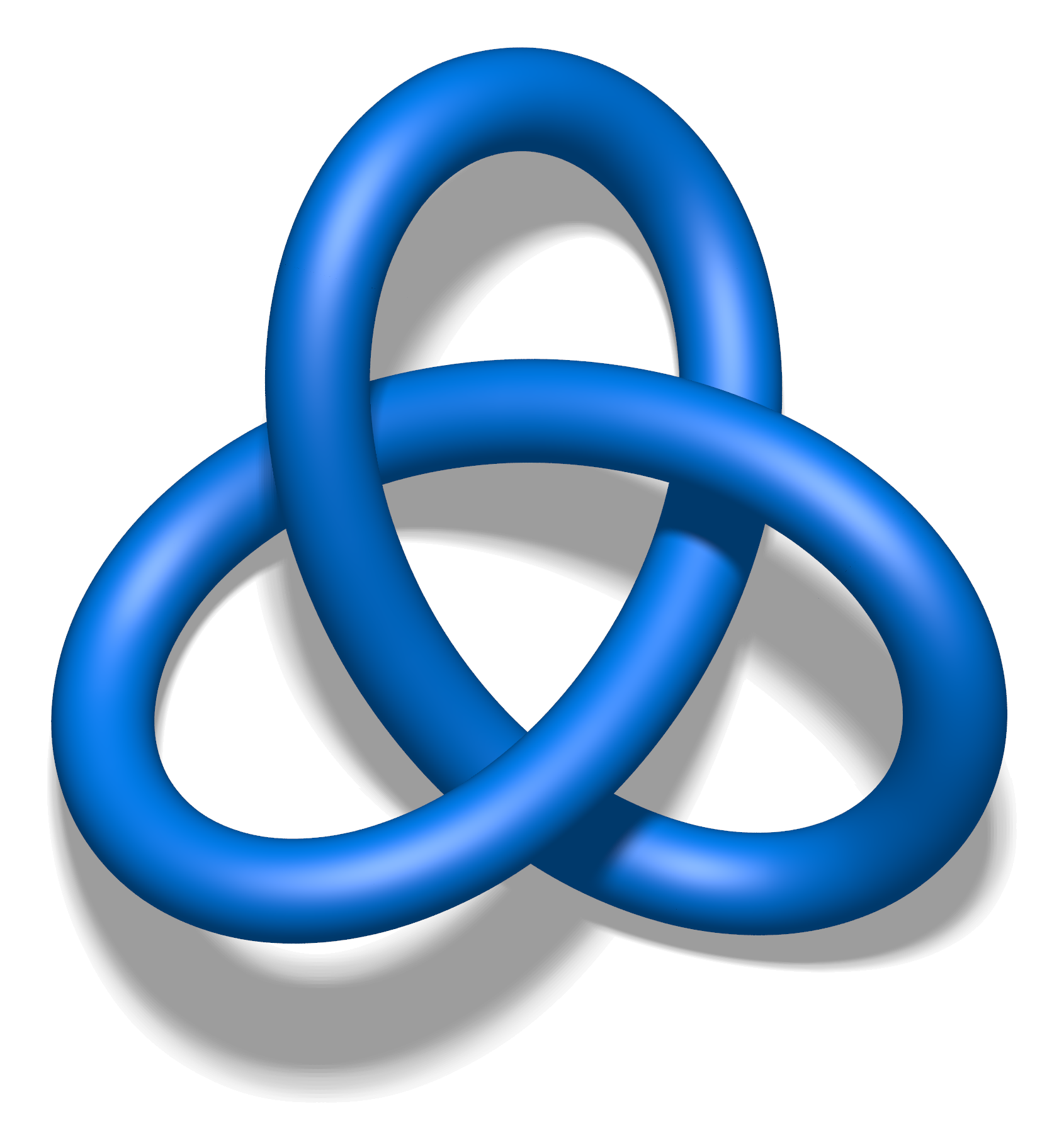}
\end{array} \quad \sim \quad
\begin{array}{c}
       \newcommand{\x}{0.4}
       \newcommand{\y}{0}
       \begin{tikzpicture}[thick]
           \draw[gray,line width=1.5,opacity=0.6] (0+\y,-0) -- (0+\y,-1.0);
           \draw[rounded corners=3] (0+\y,-0.5-\x) -- (0.3+\y,-0.5-\x) -- (0.7+\y,-0.1-\x) -- (1+\y,-0.1-\x);
         \draw[rounded corners=3] (0+\y,-0.1-\x) -- (0.3+\y,-0.1-\x) -- (0.4+\y,-0.2-\x);
         \draw[rounded corners=3] (0.6+\y,-0.4-\x) -- (0.7+\y,-0.5-\x) -- (1+\y,-0.5-\x);
         \draw[rounded corners=3] (0.0+\y,-0.1) -- (1+\y,-0.1);
         \draw[rounded corners=3] (0.0+\y,-1.3) -- (1+\y,-1.3);
         \draw[rounded corners=3] (0.0+\y,-1.7) -- (1+\y,-1.7);
         \draw[rounded corners=3] (0.0+\y,-2.1) -- (1+\y,-2.1);
         \draw[rounded corners=3] (-0+\y,-0.9) -- (-0.2+\y,-0.9) -- (-0.2+\y,-1.3) -- (0.0+\y,-1.3);
         \draw[rounded corners=3] (-0+\y,-0.5) -- (-0.4+\y,-0.5) -- (-0.4+\y,-1.7) -- (0.0+\y,-1.7);
         \draw[rounded corners=3] (-0+\y,-0.1) -- (-0.6+\y,-0.1) -- (-0.6+\y,-2.1) -- (0.0+\y,-2.1);
           \foreach \x in {0.1,0.5,...,1.0}
           \fill[black] (0+\y,-\x) circle (0.05);
            \renewcommand{\x}{0}
       \renewcommand{\y}{1}
           \draw[rounded corners=3] (0+\y,-0.5-\x) -- (0.3+\y,-0.5-\x) -- (0.7+\y,-0.1-\x) -- (1+\y,-0.1-\x);
         \draw[rounded corners=3] (0+\y,-0.1-\x) -- (0.3+\y,-0.1-\x) -- (0.4+\y,-0.2-\x);
         \draw[rounded corners=3] (0.6+\y,-0.4-\x) -- (0.7+\y,-0.5-\x) -- (1+\y,-0.5-\x);
         \draw[rounded corners=3] (0.0+\y,-0.9) -- (1+\y,-0.9);
         \draw[rounded corners=3] (0.0+\y,-1.3) -- (1+\y,-1.3);
         \draw[rounded corners=3] (0.0+\y,-1.7) -- (1+\y,-1.7);
         \draw[rounded corners=3] (0.0+\y,-2.1) -- (1+\y,-2.1);
         \renewcommand{\x}{0.4}
       \renewcommand{\y}{2}
       \draw[rounded corners=3] (0+\y,-0.5-\x) -- (0.3+\y,-0.5-\x) -- (0.7+\y,-0.1-\x) -- (1+\y,-0.1-\x);
         \draw[rounded corners=3] (0+\y,-0.1-\x) -- (0.3+\y,-0.1-\x) -- (0.4+\y,-0.2-\x);
         \draw[rounded corners=3] (0.6+\y,-0.4-\x) -- (0.7+\y,-0.5-\x) -- (1+\y,-0.5-\x);
         \draw[rounded corners=3] (0.0+\y,-0.1) -- (1+\y,-0.1);
         \draw[rounded corners=3] (0.0+\y,-1.3) -- (1+\y,-1.3);
         \draw[rounded corners=3] (0.0+\y,-1.7) -- (1+\y,-1.7);
         \draw[rounded corners=3] (0.0+\y,-2.1) -- (1+\y,-2.1);
         \renewcommand{\x}{0}
       \renewcommand{\y}{3}
       \draw[gray,line width=1.5,opacity=0.6] (1+\y,-0) -- (1+\y,-1.0);
           \draw[rounded corners=3] (0+\y,-0.5-\x) -- (0.3+\y,-0.5-\x) -- (0.7+\y,-0.1-\x) -- (1+\y,-0.1-\x);
         \draw[rounded corners=3] (0+\y,-0.1-\x) -- (0.3+\y,-0.1-\x) -- (0.4+\y,-0.2-\x);
         \draw[rounded corners=3] (0.6+\y,-0.4-\x) -- (0.7+\y,-0.5-\x) -- (1+\y,-0.5-\x);
         \draw[rounded corners=3] (0.0+\y,-0.9) -- (1+\y,-0.9);
         \draw[rounded corners=3] (0.0+\y,-1.3) -- (1+\y,-1.3);
         \draw[rounded corners=3] (0.0+\y,-1.7) -- (1+\y,-1.7);
         \draw[rounded corners=3] (0.0+\y,-2.1) -- (1+\y,-2.1);
         \draw[rounded corners=3] (1+\y,-0.9) -- (1.2+\y,-0.9) -- (1.2+\y,-1.3) -- (1.0+\y,-1.3);
         \draw[rounded corners=3] (1+\y,-0.5) -- (1.4+\y,-0.5) -- (1.4+\y,-1.7) -- (1.0+\y,-1.7);
         \draw[rounded corners=3] (1+\y,-0.1) -- (1.6+\y,-0.1) -- (1.6+\y,-2.1) -- (1.0+\y,-2.1);
           \foreach \x in {0.1,0.5,...,1.0}
           \fill[black] (1+\y,-\x) circle (0.05);
       \end{tikzpicture}
\end{array}.
\ee
In the above examples the braids were oriented horizontally to optimize space.

Since braid diagrams respect two of the three Reidemeister moves and the tracelike closures of braids form knot diagrams, one may wonder if traces of matrix representation of braids can yield topological invariants. In fact, naive trace will not do the job. Andrei Markov proved a theorem explaining what kind of traces gives an invariant. The so-called Markov trace $\Tr\!_{\rm M}$ should satisfy the following two properties:
\begin{itemize}
    \item (cyclicity) for two braids $\alpha$ and $\beta$
    \be
    \trm(\alpha\cdot\beta) \ = \ \trm(\beta\cdot\alpha)
    \ee
    \item (Markov property) for braid $\beta_n\in B_{n}$ and its embedding $\beta_n\cdot b_n$ in $B_{n+1}$,
    \be
    \trm(\beta_n) \ = \ \trm(\beta_n\cdot b_n^{\pm1})
    \ee
\end{itemize}

Markov property can be expressed diagrammatically as
\be
\begin{array}{c}
       \newcommand{\x}{0.0}
       \newcommand{\y}{0}
       \begin{tikzpicture}[thick]
           \draw[gray,line width=1.5,opacity=0.6] (0+\y,-0) -- (0+\y,-0.6);
           \draw[rounded corners=3] (0+\y,-0.5-\x) -- (1+\y,-0.5-\x);
         \draw[rounded corners=3] (0+\y,-0.1-\x) -- (1+\y,-0.1-\x);
         \draw[rounded corners=3] (0.0+\y,-0.9) -- (1+\y,-0.9);
         \draw[rounded corners=3] (0.0+\y,-1.3) -- (1+\y,-1.3);
         \draw[rounded corners=3] (-0+\y,-0.5-\x) -- (-0.2+\y,-0.5-\x) -- (-0.2+\y,-0.9) -- (0.0+\y,-0.9);
         \draw[rounded corners=3] (-0+\y,-0.1-\x) -- (-0.4+\y,-0.1-\x) -- (-0.4+\y,-1.3) -- (0.0+\y,-1.3);
           \foreach \x in {0.1,0.5,...,0.6}
           \fill[black] (0+\y,-\x) circle (0.05);
           \renewcommand{\x}{0.0}
            \renewcommand{\y}{1}
           \draw[rounded corners=3] (0+\y,-0.5-\x) -- (0.3+\y,-0.5-\x) -- (0.7+\y,-0.1-\x) -- (1+\y,-0.1-\x);
         \draw[rounded corners=3] (0+\y,-0.1-\x) -- (0.3+\y,-0.1-\x) -- (0.4+\y,-0.2-\x);
         \draw[rounded corners=3] (0.6+\y,-0.4-\x) -- (0.7+\y,-0.5-\x) -- (1+\y,-0.5-\x);
         \draw[rounded corners=3] (0.0+\y,-0.9) -- (1+\y,-0.9);
         \draw[rounded corners=3] (0.0+\y,-1.3) -- (1+\y,-1.3);
         \fill[gray] (0+\y,-0.7-\x) rectangle (1+\y,0.1-\x);
         \fill[white] (0.1+\y,-0.6-\x) rectangle (0.9+\y,-0.-\x);
         \draw (0.5+\y,-0.3-\x) node {$\beta_n$};
            \renewcommand{\x}{0.0}
            \renewcommand{\y}{2}
            \draw[gray,line width=1.5,opacity=0.6] (1+\y,-0) -- (1+\y,-0.6);
           \draw[rounded corners=3] (0+\y,-0.5-\x) -- (1+\y,-0.5-\x);
         \draw[rounded corners=3] (0+\y,-0.1-\x) -- (1+\y,-0.1-\x);
         \draw[rounded corners=3] (0.0+\y,-0.9) -- (1+\y,-0.9);
         \draw[rounded corners=3] (0.0+\y,-1.3) -- (1+\y,-1.3);
          \draw[rounded corners=3] (1+\y,-0.5-\x) -- (1.2+\y,-0.5-\x) -- (1.2+\y,-0.9) -- (1+\y,-0.9);
         \draw[rounded corners=3] (1+\y,-0.1-\x) -- (1.4+\y,-0.1-\x) -- (1.4+\y,-1.3) -- (1+\y,-1.3);
          \foreach \x in {0.1,0.5,...,0.6}
           \fill[black] (1+\y,-\x) circle (0.05);
       \end{tikzpicture}
\end{array}
\quad \sim \quad
\begin{array}{c}
       \newcommand{\x}{0.4}
       \newcommand{\y}{0}
       \begin{tikzpicture}[thick]
           \draw[gray,line width=1.5,opacity=0.6] (0+\y,-0) -- (0+\y,-1.0);
           \draw[rounded corners=3] (0+\y,-0.5-\x) -- (1+\y,-0.5-\x);
         \draw[rounded corners=3] (0+\y,-0.1-\x) -- (1+\y,-0.1-\x);
         \draw[rounded corners=3] (0.0+\y,-0.1) -- (1+\y,-0.1);
         \draw[rounded corners=3] (0.0+\y,-1.3) -- (1+\y,-1.3);
         \draw[rounded corners=3] (0.0+\y,-1.7) -- (1+\y,-1.7);
         \draw[rounded corners=3] (0.0+\y,-2.1) -- (1+\y,-2.1);
         \draw[rounded corners=3] (-0+\y,-0.9) -- (-0.2+\y,-0.9) -- (-0.2+\y,-1.3) -- (0.0+\y,-1.3);
         \draw[rounded corners=3] (-0+\y,-0.5) -- (-0.4+\y,-0.5) -- (-0.4+\y,-1.7) -- (0.0+\y,-1.7);
         \draw[rounded corners=3] (-0+\y,-0.1) -- (-0.6+\y,-0.1) -- (-0.6+\y,-2.1) -- (0.0+\y,-2.1);
           \foreach \x in {0.1,0.5,...,1.0}
           \fill[black] (0+\y,-\x) circle (0.05);
            \renewcommand{\x}{0}
       \renewcommand{\y}{1}
           \draw[rounded corners=3] (0+\y,-0.5-\x) -- (0.3+\y,-0.5-\x) -- (0.7+\y,-0.1-\x) -- (1+\y,-0.1-\x);
         \draw[rounded corners=3] (0+\y,-0.1-\x) -- (0.3+\y,-0.1-\x) -- (0.4+\y,-0.2-\x);
         \draw[rounded corners=3] (0.6+\y,-0.4-\x) -- (0.7+\y,-0.5-\x) -- (1+\y,-0.5-\x);
         \draw[rounded corners=3] (0.0+\y,-0.9) -- (1+\y,-0.9);
         \draw[rounded corners=3] (0.0+\y,-1.3) -- (1+\y,-1.3);
         \draw[rounded corners=3] (0.0+\y,-1.7) -- (1+\y,-1.7);
         \draw[rounded corners=3] (0.0+\y,-2.1) -- (1+\y,-2.1);
         \fill[gray] (0+\y,-0.7-\x) rectangle (1+\y,0.1-\x);
         \fill[white] (0.1+\y,-0.6-\x) rectangle (0.9+\y,-0.-\x);
         \draw (0.5+\y,-0.3-\x) node {$\beta_n$};
         \renewcommand{\x}{0.4}
       \renewcommand{\y}{2}
       \draw[rounded corners=3] (0+\y,-0.5-\x) -- (0.3+\y,-0.5-\x) -- (0.7+\y,-0.1-\x) -- (1+\y,-0.1-\x);
         \draw[rounded corners=3] (0+\y,-0.1-\x) -- (0.3+\y,-0.1-\x) -- (0.4+\y,-0.2-\x);
         \draw[rounded corners=3] (0.6+\y,-0.4-\x) -- (0.7+\y,-0.5-\x) -- (1+\y,-0.5-\x);
         \draw[rounded corners=3] (0.0+\y,-0.1) -- (1+\y,-0.1);
         \draw[rounded corners=3] (0.0+\y,-1.3) -- (1+\y,-1.3);
         \draw[rounded corners=3] (0.0+\y,-1.7) -- (1+\y,-1.7);
         \draw[rounded corners=3] (0.0+\y,-2.1) -- (1+\y,-2.1);
         \renewcommand{\x}{0}
       \renewcommand{\y}{3}
       \draw[gray,line width=1.5,opacity=0.6] (1+\y,-0) -- (1+\y,-1.0);
           \draw[rounded corners=3] (0+\y,-0.5-\x) -- (1+\y,-0.5-\x);
         \draw[rounded corners=3] (0+\y,-0.1-\x) -- (1+\y,-0.1-\x);
         \draw[rounded corners=3] (0.0+\y,-0.9) -- (1+\y,-0.9);
         \draw[rounded corners=3] (0.0+\y,-1.3) -- (1+\y,-1.3);
         \draw[rounded corners=3] (0.0+\y,-1.7) -- (1+\y,-1.7);
         \draw[rounded corners=3] (0.0+\y,-2.1) -- (1+\y,-2.1);
         \draw[rounded corners=3] (1+\y,-0.9) -- (1.2+\y,-0.9) -- (1.2+\y,-1.3) -- (1.0+\y,-1.3);
         \draw[rounded corners=3] (1+\y,-0.5) -- (1.4+\y,-0.5) -- (1.4+\y,-1.7) -- (1.0+\y,-1.7);
         \draw[rounded corners=3] (1+\y,-0.1) -- (1.6+\y,-0.1) -- (1.6+\y,-2.1) -- (1.0+\y,-2.1);
           \foreach \x in {0.1,0.5,...,1.0}
           \fill[black] (1+\y,-\x) circle (0.05);
       \end{tikzpicture}
\end{array},
\ee
which is a particular expression of the I Reidemeister move. Therefore, Markov traces of braids should be invariant under all the Reidemester moves, or, in other words, they should compute topological invariants of knots and links.

%%%%%%%%%%%%%%%%%%%%%%%%%%%%%%%%%%%%%%%%%%%%%%%%%%%%%%%%%%%%
\subsection{Skein relation}

To construct an explicit example of a Markov trace it is convenient to introduce another structure, the Temperley-Lieb algebra $TL_n$~\cite{Temperley:1971iq}. Similarly to the braid group, it can be introduced via generators as follows:
\begin{multline}
TL_n\ := \ \{ u_k,\, k=1,\ldots,n-1|\ u_i^2=d\cdot u_i,\ i=1,\ldots\, n-1;\ u_iu_{i+1}u_i=u_i,\ i=1,\ldots,n-2;\\ u_{i+1}u_iu_{i+1}=u_{i+1},\ i=1,\ldots,n-2;\ u_iu_j=u_ju_i,\, |i-j|>1\}\,,
\end{multline}
where $d$ is a $\mathbb{C}$-number. Since the above relations define an algebra, all its elements belong to a vector space. The first relation makes this space finite-dimensional. The dimension of $TL_n$ is given by the $n$-th Catalan number,
\be
\label{Catalan}
C_n \ = \ \frac{(2n)!}{n!(n+1)!}\,.
\ee
Note that this algebra also contains the identity element.

As in the case of the braid group it is convenient to think of the Temperley-Lieb algebra in terms of diagrams, which make explicit the defining relations. In this case one has
\be
u_1 \ =  \begin{array}{c}
       \begin{tikzpicture}[thick]
           \draw[gray,line width=1.5,opacity=0.6] (0,0) -- (2.2,0);
           \draw[gray,line width=1.5,opacity=0.6] (0,1) -- (2.2,1);
           \draw[rounded corners=3] (0.1,0) -- (0.1,0.4) -- (0.5,0.4) -- (0.5,0);
         \draw[rounded corners=3] (0.5,1) -- (0.5,0.6) -- (0.1,0.6) -- (0.1,1);
         \draw[rounded corners=3] (0.9,0.0) -- (0.9,1);
         \draw[rounded corners=3] (1.3,0.0) -- (1.3,1);
         \draw[rounded corners=3] (2.1,0.0) -- (2.1,1);
           \foreach \x in {0.1,0.5,...,1.5}
           \fill[black] (\x,0) circle (0.05);
           \foreach \x in {0.1,0.5,...,1.5}
           \fill[black] (\x,1) circle (0.05);
           \fill[black] (2.1,1) circle (0.05);
           \fill[black] (2.1,0) circle (0.05);
           \draw (1.7,0.5) node {$\cdots$};
       \end{tikzpicture}
\end{array},
\quad 
u_2 \ =  \begin{array}{c}
       \begin{tikzpicture}[thick]
           \draw[gray,line width=1.5,opacity=0.6] (0,0) -- (2.2,0);
           \draw[gray,line width=1.5,opacity=0.6] (0,1) -- (2.2,1);
           \draw[rounded corners=3] (0.1,0) -- (0.1,1);
         \draw[rounded corners=3] (0.5,1) -- (0.5,0.6) -- (0.9,0.6) -- (0.9,1);
         \draw[rounded corners=3] (0.9,0.0) -- (0.9,0.4) -- (0.5,0.4) -- (0.5,0);
         \draw[rounded corners=3] (1.3,0.0) -- (1.3,1);
         \draw[rounded corners=3] (2.1,0.0) -- (2.1,1);
           \foreach \x in {0.1,0.5,...,1.5}
           \fill[black] (\x,0) circle (0.05);
           \foreach \x in {0.1,0.5,...,1.5}
           \fill[black] (\x,1) circle (0.05);
           \fill[black] (2.1,1) circle (0.05);
           \fill[black] (2.1,0) circle (0.05);
           \draw (1.7,0.5) node {$\cdots$};
       \end{tikzpicture}
\end{array},
\quad 
u_3 \ =  \begin{array}{c}
       \begin{tikzpicture}[thick]
           \draw[gray,line width=1.5,opacity=0.6] (0,0) -- (2.2,0);
           \draw[gray,line width=1.5,opacity=0.6] (0,1) -- (2.2,1);
           \draw[rounded corners=3] (0.1,0) -- (0.1,1);
         \draw[rounded corners=3] (0.5,1) -- (0.5,0);
         \draw[rounded corners=3] (0.9,0.0) -- (0.9,0.4) -- (1.3,0.4) -- (1.3,0);
         \draw[rounded corners=3] (1.3,1) -- (1.3,0.6) -- (0.9,0.6) -- (0.9,1);
         \draw[rounded corners=3] (2.1,0.0) -- (2.1,1);
           \foreach \x in {0.1,0.5,...,1.5}
           \fill[black] (\x,0) circle (0.05);
           \foreach \x in {0.1,0.5,...,1.5}
           \fill[black] (\x,1) circle (0.05);
           \fill[black] (2.1,1) circle (0.05);
           \fill[black] (2.1,0) circle (0.05);
           \draw (1.7,0.5) node {$\cdots$};
       \end{tikzpicture}
\end{array},
\quad \cdots
\ee
Identity is added to the set of generators and is represented by the diagram with the trivial evolution of points:
\be
I \ = \ \begin{array}{c}
       \begin{tikzpicture}[thick]
           \draw[gray,line width=1.5,opacity=0.6] (0,0) -- (2.2,0);
           \draw[gray,line width=1.5,opacity=0.6] (0,1) -- (2.2,1);
           \draw[rounded corners=3] (0.1,0) -- (0.1,1);
         \draw[rounded corners=3] (0.5,1) -- (0.5,0);
         \draw[rounded corners=3] (0.9,0.0) -- (0.9,1);
         \draw[rounded corners=3] (1.3,1) -- (1.3,0);
         \draw[rounded corners=3] (2.1,0.0) -- (2.1,1);
           \foreach \x in {0.1,0.5,...,1.5}
           \fill[black] (\x,0) circle (0.05);
           \foreach \x in {0.1,0.5,...,1.5}
           \fill[black] (\x,1) circle (0.05);
           \fill[black] (2.1,1) circle (0.05);
           \fill[black] (2.1,0) circle (0.05);
           \draw (1.7,0.5) node {$\cdots$};
       \end{tikzpicture}
\end{array}.
\ee
The multiplication is organized in the same way as for braids. The first defining relation can be cast as
\be
u_k^2 \ = \ \begin{array}{c}
       \begin{tikzpicture}[thick]
           \draw[gray,line width=1.5,opacity=0.6] (0,0) -- (2,0);
           \draw[gray,line width=1.5,opacity=0.6] (0,1) -- (2,1);
           \draw (0.1,0) -- (0.1,1);
           \draw (0.7,0) -- (0.7,1);
           \draw (1.6,0) -- (1.6,1);
           \draw[rounded corners=2] (1.0,0) -- (1.0,0.2) -- (1.3,0.2) -- (1.3,0);
           \draw[rounded corners=2] (1.0,1) -- (1.,0.8) -- (1.3,0.8) -- (1.3,1);
           \draw[rounded corners=2] (1.0,0.5) -- (1.,0.7) -- (1.3,0.7) -- (1.3,0.3) -- (1.0,0.3) -- (1.0,0.5);
           \foreach \x in {0.1,0.4,...,2.1}
           \fill[black] (\x,0) circle (0.05);
           \foreach \x in {0.1,0.4,...,2.1}
           \fill[black] (\x,1) circle (0.05);
           \draw (0.4,0.5) node {$\cdots$};
           \draw (1.9,0.5) node {$\cdots$};
       \end{tikzpicture}
\end{array}  = \ d\ \cdot\! \begin{array}{c}
       \begin{tikzpicture}[thick]
           \draw[gray,line width=1.5,opacity=0.6] (0,0) -- (2,0);
           \draw[gray,line width=1.5,opacity=0.6] (0,1) -- (2,1);
           \draw (0.1,0) -- (0.1,1);
           \draw (0.7,0) -- (0.7,1);
           \draw (1.6,0) -- (1.6,1);
           \draw[rounded corners=2] (1.0,0) -- (1.0,0.4) -- (1.3,0.4) -- (1.3,0);
           \draw[rounded corners=2] (1.0,1) -- (1.,0.6) -- (1.3,0.6) -- (1.3,1);
           \foreach \x in {0.1,0.4,...,2.1}
           \fill[black] (\x,0) circle (0.05);
           \foreach \x in {0.1,0.4,...,2.1}
           \fill[black] (\x,1) circle (0.05);
           \draw (0.4,0.5) node {$\cdots$};
           \draw (1.9,0.5) node {$\cdots$};
       \end{tikzpicture}
\end{array}
\ee
Note that in the multiplication a closed loop is formed. We will always think of a closed loop as of a $\mathbb{C}$-number, while lines with open ends correspond to matrices or tensors. A heuristic explanation for this property is that a disconnected circle can be freely moved around the diagram (in particular, removed) unlike the lines, which are fixed at the endpoints.

It is straightforward to check that the diagrams respect the remaining defining relations. The diagrams also explain the finiteness of the number of the generator words and reduce the counting to a combinatorial problem. Any word should correspond to $n$ pairs of points connected in a way without line intersection. There is only a finite number of such ways, computed by the Catalan numbers.

Given a generator of the Temperley-Lieb algebra, one can construct an element of the braid group. The following set of algebra elements,
\be
\label{protoskein}
b_k \ = \ A I + A^{-1}u_k\,,
\ee
for some $\mathbb{C}$-number $A$, satisfy the braid group relations if
\be
d \ = \ - A^2 - A^{-2}\,.
\ee

Note that relation~(\ref{protoskein}) can be represented diagrammatically as 
\be
\label{skein}
\begin{array}{c}
     \begin{tikzpicture}[thick]
     \draw (0,0.5) -- (0.3,0);
     \draw[line width=3.5,white] (0,0) -- (0.3,0.5);
     \draw (0,0) -- (0.3,0.5);
     \end{tikzpicture}
\end{array}
\ = \ A 
\begin{array}{c}
     \begin{tikzpicture}[thick]
     \draw (0,0) -- (0,0.5);
     \draw (0.3,0) -- (0.3,0.5);
     \end{tikzpicture}
\end{array}
+ A^{-1}
\begin{array}{c}
     \begin{tikzpicture}[thick]
     \draw (0,0) -- (0,0.05) arc (180:0:0.15) -- (0.3,0);
     \draw (0,0.5) -- (0,0.45) arc (180:360:0.15) -- (0.3,0.5);
     \end{tikzpicture}
\end{array}.
\ee
We will refer to this diagrammatic identity as the \emph{skein relation}. 

Let us make the following faithfulness assumption. Assume that the skein relation holds even if we simultaneously rotate all the diagrams by 90 degrees. We will get
\be
\label{iskein}
\begin{array}{c}
     \begin{tikzpicture}[thick]
     \draw (0.3,0.5) -- (0.0,0);
     \draw[line width=3.5,white] (0.3,0) -- (0.0,0.5);
     \draw (0.3,0) -- (0.0,0.5);
     \end{tikzpicture}
\end{array}
\ = \ A 
\begin{array}{c}
     \begin{tikzpicture}[thick]
     \draw (0,0) -- (0,0.05) arc (180:0:0.15) -- (0.3,0);
     \draw (0,0.5) -- (0,0.45) arc (180:360:0.15) -- (0.3,0.5);
     \end{tikzpicture}
\end{array}
+ A^{-1}
\begin{array}{c}
     \begin{tikzpicture}[thick]
     \draw (0,0) -- (0,0.5);
     \draw (0.3,0) -- (0.3,0.5);
     \end{tikzpicture}
\end{array}.
\ee
Note that this rotation takes the braid generator to its inverse and exchanges the identity and the Temperley-Lieb generator. Our assumption imposes an additional relation on the braid group. For example, eliminating the Temperley-Lieb generator from relations~(\ref{skein}) and~(\ref{iskein}) we get a purely braid group relation
\be
\label{skein2}
A\begin{array}{c}
     \begin{tikzpicture}[thick]
     \draw (0,0.5) -- (0.3,0);
     \draw[line width=3.5,white] (0,0) -- (0.3,0.5);
     \draw (0,0) -- (0.3,0.5);
     \end{tikzpicture}
\end{array} 
- A^{-1} \begin{array}{c}
     \begin{tikzpicture}[thick]
     \draw (0.3,0.5) -- (0.0,0);
     \draw[line width=3.5,white] (0.3,0) -- (0.0,0.5);
     \draw (0.3,0) -- (0.0,0.5);
     \end{tikzpicture}
\end{array}
\ = \ (A^2-A^{-2})\begin{array}{c}
     \begin{tikzpicture}[thick]
     \draw (0,0) -- (0,0.5);
     \draw (0.3,0) -- (0.3,0.5);
     \end{tikzpicture}
\end{array},
\ee
which can also be cast as quadratic relation on the group generators:
\be
(b_k)^2 \ = \ (A-A^{-3})b_k + A^{-2}I\,.
\ee
With this additional relation $b_k$ are said to generate the \emph{Hecke algebra}.

%%%%%%%%%%%%%%%%%%%%%%%%%%%%%%%%%%%%%%%%%%%%%%%%%%%%%%%%%%%%%%%%%%%%
\subsection{Bracket and Jones polynomials}
\label{sec:Jonespolynomial}

Skein relation and Hecke algebras allow to compute Markov traces diagrammatically, without explicitly defining matrix representations~\cite{Jones:1987dy}. Recall that we agreed to substitute any disconnected circle by the number $d=-A^2-A^{-2}$. First, one can check that the diagrammatic closure of a braid has the Markov property. Applying the skein relation one finds that
\begin{multline}
\begin{array}{c}
       \newcommand{\x}{0.4}
       \newcommand{\y}{0}
       \begin{tikzpicture}[thick]
         \draw[rounded corners=3] (0.0+\y,-1.7) -- (1+\y,-1.7);
         \draw[rounded corners=3] (0.0+\y,-2.1) -- (1+\y,-2.1);
         \draw[rounded corners=3] (-0+\y,-0.9) -- (-0.2+\y,-0.9) -- (-0.2+\y,-1.3) -- (0.0+\y,-1.3);
         \draw[rounded corners=3] (-0+\y,-0.5) -- (-0.4+\y,-0.5) -- (-0.4+\y,-1.7) -- (0.0+\y,-1.7);
         \draw[rounded corners=3] (-0+\y,-0.1) -- (-0.6+\y,-0.1) -- (-0.6+\y,-2.1) -- (0.0+\y,-2.1);
            \renewcommand{\x}{0}
       \renewcommand{\y}{0}
           \draw[rounded corners=3] (0+\y,-0.5-\x) -- (0.3+\y,-0.5-\x) -- (0.7+\y,-0.1-\x) -- (1+\y,-0.1-\x);
         \draw[rounded corners=3] (0+\y,-0.1-\x) -- (0.3+\y,-0.1-\x) -- (0.4+\y,-0.2-\x);
         \draw[rounded corners=3] (0.6+\y,-0.4-\x) -- (0.7+\y,-0.5-\x) -- (1+\y,-0.5-\x);
         \draw[rounded corners=3] (0.0+\y,-0.9) -- (1+\y,-0.9);
         \draw[rounded corners=3] (0.0+\y,-1.3) -- (1+\y,-1.3);
         \draw[rounded corners=3] (0.0+\y,-1.7) -- (1+\y,-1.7);
         \draw[rounded corners=3] (0.0+\y,-2.1) -- (1+\y,-2.1);
         \fill[gray] (0+\y,-0.7-\x) rectangle (1+\y,0.1-\x);
         \fill[white] (0.1+\y,-0.6-\x) rectangle (0.9+\y,-0.-\x);
         \draw (0.5+\y,-0.3-\x) node {$\beta_n$};
         \renewcommand{\x}{0.4}
       \renewcommand{\y}{1}
       \draw[rounded corners=3] (0+\y,-0.5-\x) -- (0.3+\y,-0.5-\x) -- (0.7+\y,-0.1-\x) -- (1+\y,-0.1-\x);
         \draw[rounded corners=3] (0+\y,-0.1-\x) -- (0.3+\y,-0.1-\x) -- (0.4+\y,-0.2-\x);
         \draw[rounded corners=3] (0.6+\y,-0.4-\x) -- (0.7+\y,-0.5-\x) -- (1+\y,-0.5-\x);
         \draw[rounded corners=3] (0.0+\y,-0.1) -- (1+\y,-0.1);
         \draw[rounded corners=3] (0.0+\y,-1.3) -- (1+\y,-1.3);
         \draw[rounded corners=3] (0.0+\y,-1.7) -- (1+\y,-1.7);
         \draw[rounded corners=3] (0.0+\y,-2.1) -- (1+\y,-2.1);
         \renewcommand{\x}{0}
         \draw[rounded corners=3] (0.0+\y,-1.7) -- (1+\y,-1.7);
         \draw[rounded corners=3] (0.0+\y,-2.1) -- (1+\y,-2.1);
         \draw[rounded corners=3] (1+\y,-0.9) -- (1.2+\y,-0.9) -- (1.2+\y,-1.3) -- (1.0+\y,-1.3);
         \draw[rounded corners=3] (1+\y,-0.5) -- (1.4+\y,-0.5) -- (1.4+\y,-1.7) -- (1.0+\y,-1.7);
         \draw[rounded corners=3] (1+\y,-0.1) -- (1.6+\y,-0.1) -- (1.6+\y,-2.1) -- (1.0+\y,-2.1);
       \end{tikzpicture}
\end{array}
\ = \
A\begin{array}{c}
       \newcommand{\x}{0.4}
       \newcommand{\y}{0}
       \begin{tikzpicture}[thick]
         \draw[rounded corners=3] (0.0+\y,-1.7) -- (1+\y,-1.7);
         \draw[rounded corners=3] (0.0+\y,-2.1) -- (1+\y,-2.1);
         \draw[rounded corners=3] (-0+\y,-0.9) -- (-0.2+\y,-0.9) -- (-0.2+\y,-1.3) -- (0.0+\y,-1.3);
         \draw[rounded corners=3] (-0+\y,-0.5) -- (-0.4+\y,-0.5) -- (-0.4+\y,-1.7) -- (0.0+\y,-1.7);
         \draw[rounded corners=3] (-0+\y,-0.1) -- (-0.6+\y,-0.1) -- (-0.6+\y,-2.1) -- (0.0+\y,-2.1);
            \renewcommand{\x}{0}
       \renewcommand{\y}{0}
           \draw[rounded corners=3] (0+\y,-0.5-\x) -- (0.3+\y,-0.5-\x) -- (0.7+\y,-0.1-\x) -- (1+\y,-0.1-\x);
         \draw[rounded corners=3] (0+\y,-0.1-\x) -- (0.3+\y,-0.1-\x) -- (0.4+\y,-0.2-\x);
         \draw[rounded corners=3] (0.6+\y,-0.4-\x) -- (0.7+\y,-0.5-\x) -- (1+\y,-0.5-\x);
         \draw[rounded corners=3] (0.0+\y,-0.9) -- (1+\y,-0.9);
         \draw[rounded corners=3] (0.0+\y,-1.3) -- (1+\y,-1.3);
         \draw[rounded corners=3] (0.0+\y,-1.7) -- (1+\y,-1.7);
         \draw[rounded corners=3] (0.0+\y,-2.1) -- (1+\y,-2.1);
         \fill[gray] (0+\y,-0.7-\x) rectangle (1+\y,0.1-\x);
         \fill[white] (0.1+\y,-0.6-\x) rectangle (0.9+\y,-0.-\x);
         \draw (0.5+\y,-0.3-\x) node {$\beta_n$};
         \renewcommand{\x}{0.4}
       \renewcommand{\y}{1}
       \draw[rounded corners=3] (0+\y,-0.5-\x) -- (0.3+\y,-0.5-\x) -- (0.3+\y,-0.1-\x) -- (0+\y,-0.1-\x);
         \draw[rounded corners=3] (1+\y,-0.1-\x) -- (0.7+\y,-0.1-\x) -- (0.7+\y,-0.5-\x) -- (1+\y,-0.5-\x);
         \draw[rounded corners=3] (0.0+\y,-0.1) -- (1+\y,-0.1);
         \draw[rounded corners=3] (0.0+\y,-1.3) -- (1+\y,-1.3);
         \draw[rounded corners=3] (0.0+\y,-1.7) -- (1+\y,-1.7);
         \draw[rounded corners=3] (0.0+\y,-2.1) -- (1+\y,-2.1);
         \renewcommand{\x}{0}
         \draw[rounded corners=3] (0.0+\y,-1.7) -- (1+\y,-1.7);
         \draw[rounded corners=3] (0.0+\y,-2.1) -- (1+\y,-2.1);
         \draw[rounded corners=3] (1+\y,-0.9) -- (1.2+\y,-0.9) -- (1.2+\y,-1.3) -- (1.0+\y,-1.3);
         \draw[rounded corners=3] (1+\y,-0.5) -- (1.4+\y,-0.5) -- (1.4+\y,-1.7) -- (1.0+\y,-1.7);
         \draw[rounded corners=3] (1+\y,-0.1) -- (1.6+\y,-0.1) -- (1.6+\y,-2.1) -- (1.0+\y,-2.1);
       \end{tikzpicture}
\end{array}
+A^{-1}\begin{array}{c}
       \newcommand{\x}{0.4}
       \newcommand{\y}{0}
       \begin{tikzpicture}[thick]
         \draw[rounded corners=3] (0.0+\y,-1.7) -- (1+\y,-1.7);
         \draw[rounded corners=3] (0.0+\y,-2.1) -- (1+\y,-2.1);
         \draw[rounded corners=3] (-0+\y,-0.9) -- (-0.2+\y,-0.9) -- (-0.2+\y,-1.3) -- (0.0+\y,-1.3);
         \draw[rounded corners=3] (-0+\y,-0.5) -- (-0.4+\y,-0.5) -- (-0.4+\y,-1.7) -- (0.0+\y,-1.7);
         \draw[rounded corners=3] (-0+\y,-0.1) -- (-0.6+\y,-0.1) -- (-0.6+\y,-2.1) -- (0.0+\y,-2.1);
            \renewcommand{\x}{0}
       \renewcommand{\y}{0}
           \draw[rounded corners=3] (0+\y,-0.5-\x) -- (1+\y,-0.5-\x);
         \draw[rounded corners=3] (0+\y,-0.1-\x) -- (1+\y,-0.1-\x);
         \draw[rounded corners=3] (0.0+\y,-0.9) -- (1+\y,-0.9);
         \draw[rounded corners=3] (0.0+\y,-1.3) -- (1+\y,-1.3);
         \draw[rounded corners=3] (0.0+\y,-1.7) -- (1+\y,-1.7);
         \draw[rounded corners=3] (0.0+\y,-2.1) -- (1+\y,-2.1);
         \fill[gray] (0+\y,-0.7-\x) rectangle (1+\y,0.1-\x);
         \fill[white] (0.1+\y,-0.6-\x) rectangle (0.9+\y,-0.-\x);
         \draw (0.5+\y,-0.3-\x) node {$\beta_n$};
         \renewcommand{\x}{0.4}
       \renewcommand{\y}{1}
       \draw[rounded corners=3] (0+\y,-0.5-\x) -- (1+\y,-0.5-\x);
         \draw[rounded corners=3] (0+\y,-0.1-\x) -- (1+\y,-0.1-\x);
         \draw[rounded corners=3] (0.6+\y,-0.4-\x) -- (0.7+\y,-0.5-\x) -- (1+\y,-0.5-\x);
         \draw[rounded corners=3] (0.0+\y,-0.1) -- (1+\y,-0.1);
         \draw[rounded corners=3] (0.0+\y,-1.3) -- (1+\y,-1.3);
         \draw[rounded corners=3] (0.0+\y,-1.7) -- (1+\y,-1.7);
         \draw[rounded corners=3] (0.0+\y,-2.1) -- (1+\y,-2.1);
         \renewcommand{\x}{0}
         \draw[rounded corners=3] (0.0+\y,-1.7) -- (1+\y,-1.7);
         \draw[rounded corners=3] (0.0+\y,-2.1) -- (1+\y,-2.1);
         \draw[rounded corners=3] (1+\y,-0.9) -- (1.2+\y,-0.9) -- (1.2+\y,-1.3) -- (1.0+\y,-1.3);
         \draw[rounded corners=3] (1+\y,-0.5) -- (1.4+\y,-0.5) -- (1.4+\y,-1.7) -- (1.0+\y,-1.7);
         \draw[rounded corners=3] (1+\y,-0.1) -- (1.6+\y,-0.1) -- (1.6+\y,-2.1) -- (1.0+\y,-2.1);
       \end{tikzpicture}
\end{array}
\\ = \ 
-A^{-3}\begin{array}{c}
\newcommand{\x}{0.0}
       \newcommand{\y}{0}
       \begin{tikzpicture}[thick]
           \draw[gray,line width=1.5,opacity=0.6] (0+\y,-0) -- (0+\y,-0.6);
           \draw[rounded corners=3] (0+\y,-0.5-\x) -- (1+\y,-0.5-\x);
         \draw[rounded corners=3] (0+\y,-0.1-\x) -- (1+\y,-0.1-\x);
         \draw[rounded corners=3] (0.0+\y,-0.9) -- (1+\y,-0.9);
         \draw[rounded corners=3] (0.0+\y,-1.3) -- (1+\y,-1.3);
         \draw[rounded corners=3] (-0+\y,-0.5-\x) -- (-0.2+\y,-0.5-\x) -- (-0.2+\y,-0.9) -- (0.0+\y,-0.9);
         \draw[rounded corners=3] (-0+\y,-0.1-\x) -- (-0.4+\y,-0.1-\x) -- (-0.4+\y,-1.3) -- (0.0+\y,-1.3);
           \foreach \x in {0.1,0.5,...,0.6}
           \fill[black] (0+\y,-\x) circle (0.05);
           \renewcommand{\x}{0.0}
            \renewcommand{\y}{1}
           \draw[rounded corners=3] (0+\y,-0.5-\x) -- (0.3+\y,-0.5-\x) -- (0.7+\y,-0.1-\x) -- (1+\y,-0.1-\x);
         \draw[rounded corners=3] (0+\y,-0.1-\x) -- (0.3+\y,-0.1-\x) -- (0.4+\y,-0.2-\x);
         \draw[rounded corners=3] (0.6+\y,-0.4-\x) -- (0.7+\y,-0.5-\x) -- (1+\y,-0.5-\x);
         \draw[rounded corners=3] (0.0+\y,-0.9) -- (1+\y,-0.9);
         \draw[rounded corners=3] (0.0+\y,-1.3) -- (1+\y,-1.3);
         \fill[gray] (0+\y,-0.7-\x) rectangle (1+\y,0.1-\x);
         \fill[white] (0.1+\y,-0.6-\x) rectangle (0.9+\y,-0.-\x);
         \draw (0.5+\y,-0.3-\x) node {$\beta_n$};
            \renewcommand{\x}{0.0}
            \renewcommand{\y}{2}
            \draw[gray,line width=1.5,opacity=0.6] (1+\y,-0) -- (1+\y,-0.6);
           \draw[rounded corners=3] (0+\y,-0.5-\x) -- (1+\y,-0.5-\x);
         \draw[rounded corners=3] (0+\y,-0.1-\x) -- (1+\y,-0.1-\x);
         \draw[rounded corners=3] (0.0+\y,-0.9) -- (1+\y,-0.9);
         \draw[rounded corners=3] (0.0+\y,-1.3) -- (1+\y,-1.3);
          \draw[rounded corners=3] (1+\y,-0.5-\x) -- (1.2+\y,-0.5-\x) -- (1.2+\y,-0.9) -- (1+\y,-0.9);
         \draw[rounded corners=3] (1+\y,-0.1-\x) -- (1.4+\y,-0.1-\x) -- (1.4+\y,-1.3) -- (1+\y,-1.3);
          \foreach \x in {0.1,0.5,...,0.6}
           \fill[black] (1+\y,-\x) circle (0.05);
       \end{tikzpicture}
\end{array}.
\label{MarkovSkein}
\end{multline}
Note that we are not simply using the I Reidemeister move here, but rather the algebraic properties of braids. It is therefore not so trivial to conclude that the second diagram in the first line is equivalent to the diagram in the second line. It might be intuitively clear to those familiar with tensor algebras and Feynman diagrams.

Note also that the initial and final diagrams differ by an overall numerical factor, which means that the Markov property is only satisfied up to this factor. This factor has a special meaning and the next few paragraphs will be dedicated to explaining it, although, skein relations can be rescaled to eliminate it from the trace.

The discrepancy is related to the intrinsic nonplanarity of the knot and to its attribute called framing. In general, framing is a rule to introduce an orthogonal vector to a curve at each point. The endpoint of this vector is transported along the curve and defines a satellite curve that may arbitrarily wind around its parent curve. In a planar projection this winding (which is nothing but the Gauss' linking number of the satellite and the parent curve) is split into the sum of two components, \emph{twist} and \emph{writhe}, according to the theorem introduced by Gheorghe Calugareanu (Calugareanu-Fuller-White theorem)~\cite{Cualuguareanu:1959int,Cualuguareanu:1961cla,White:1969sel,Fuller:1971wri}.  

While the twist is a single naive winding of one curve around the other -- it appears as a pair of braidings -- the writhe is the number of loops, like the one in the first diagram of~(\ref{MarkovSkein}), that the parent loop has. Calugareanu's theorem states that, for a chosen framing, the sum of the twist and the writhe is a topological invariant. Unmaking the loop in the I Reidemeister move removes a unit of writhe, therefore, assuming that the first and the last diagrams above have the same twist number, the prefactor introduces a compensating factor indicating how much the framing changes in passing from one diagram to another. In all the computations the framing factor will appear as $(-A^3)^w$, where $w$ is the writhe number. Altogether, one says that what is being computed is an invariant of framed knots.

Skein relations can be used to ``resolve'' all crossings in the braid diagram, so that its closure is reduced to a weighted sum of collections of embedded nonintersecting circles, where weights are polynomial functions of $A$. At the final step of the calculation every circle can be replaced by $d$, according to our prescription, giving a Laurent polynomial of $A$ as a final result. This is the bracket polynomial introduced by Louis Kauffman, which is an invariant of framed knots~\cite{Kauffman:1987sta}.

\begin{figure}
    \centering
    \includegraphics[width=0.5\linewidth]{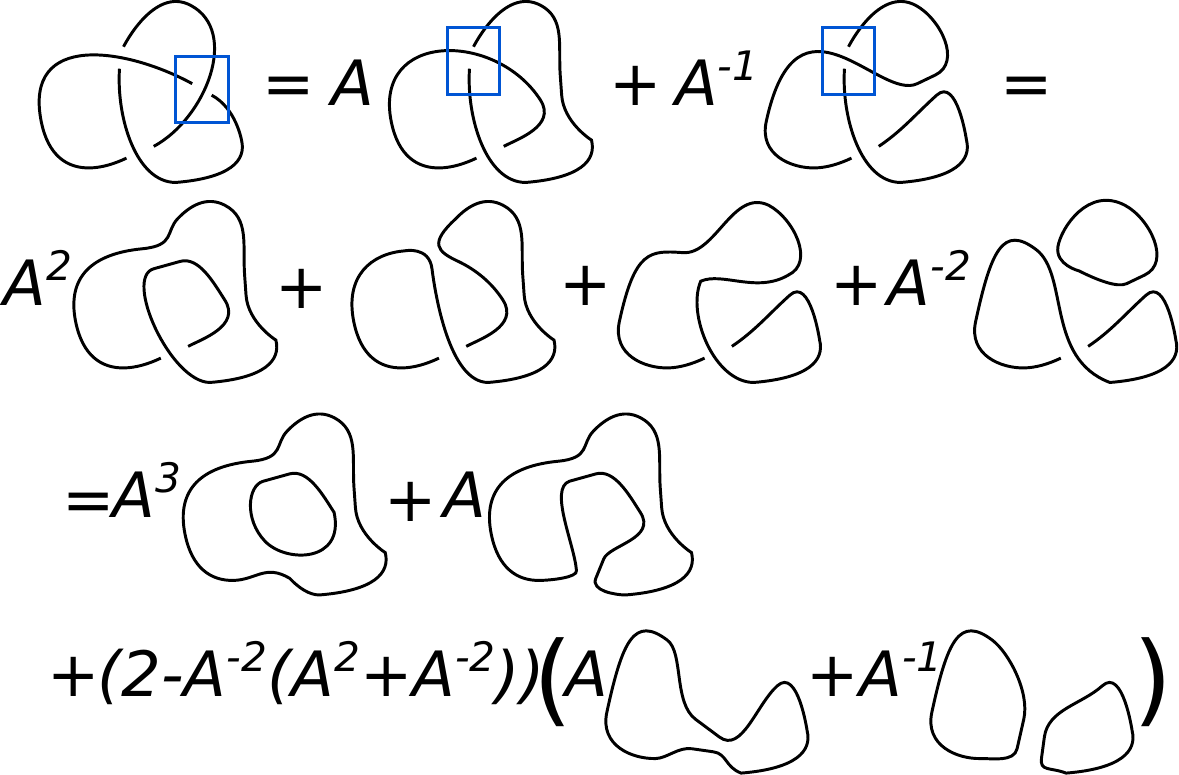} 
    \caption{Calculation of the bracket polynomial of the trefoil knot $3_1$ using skein relations~(\ref{skein}).}
    \label{fig:skein4trefoil}
\end{figure}

Our faithfulness assumption allows not only application of the skein relation to braids, but directly to arbitrary diagrams. Figure~\ref{fig:skein4trefoil} shows a sequence of skein relations applied to the trefoil knot and computing their invariants without casting the knots as braid closures. For the trefoil knot one gets the following bracket polynomial:
\be
\label{bracket3foil}
\left\langle\begin{array}{c}
   \scalebox{-1}[1]{\includegraphics[height=0.1\linewidth]{./figs/trefoil.png}}
\end{array} \right\rangle \ = \  (-A^3)^3(-A^2-A^{-2})(A^{-4}+A^{-12}-A^{-16}).
\ee
One can see in this example the appearance of three units of the writhe factor and another overall factor of $d$. This is the general structure obtained by applying the skein relations. To get the original definition of Kauffman one divides the result by $d$. In such a normalization the invariant of the unknot is unity up to the writhe. The remaining term has a simple relation to the polynomial obtained originally by Jones:
\begin{itemize}
    \item Remove the writhe and the $d$ factors
    \item Substitute $A^4\to q^{-1}$
\end{itemize}
One then gets the following Jones polynomial of the trefoil knot:
\be
V\left(\begin{array}{c}
   \scalebox{-1}[1]{\includegraphics[height=0.08\linewidth]{./figs/trefoil.png}}
\end{array}\right) \ = \ q + q^{3}-q^{4}\,.
\ee

Note that there are two trefoil knots differing by mirror reflection. If one applies the skein procedure to the mirror-reflected knot one will get
\be
V\left(\begin{array}{c}
   \scalebox{1}[1]{\includegraphics[height=0.08\linewidth]{./figs/trefoil.png}}
\end{array}\right) \ = \ q^{-1} + q^{-3}-q^{-4}\,.
\ee
In other words, mirror reflection acts as $q\to q^{-1}$ ($A\to A^{-1}$) on Jones (bracket) polynomials.

In the rest of this review we will work with the Kauffman bracket version of the Jones polynomial. The whole purpose of section~\ref{sec:knots} is to explain what knot is and how the bracket polynomial can be computed, for example, using the skein relation. We will need this calculus for quantum mechanical applications. Let us summarize the main steps:
\begin{itemize}
    \item Draw the knot (link) diagram;
    \item Choose a crossing, apply the skein relation by replacing the diagram by a pair of new diagrams with the crossing modified according to skein relation~(\ref{skein});
    \item Repeat the procedure for the next crossing for each of the generated diagrams until the diagrams are nothing but a collection of nested nonintersecting circles;
    \item replace each circle by factor $d=-A^2-A^{-2}$.
\end{itemize}

%%%%%%%%%%%%%%%%%%%%%%%%%%%%%%%%%%%%%%%%%%%%%%%%%%%%%%%%%%%%%%%%%%%%
\subsection{Matrix representation}
\label{sec:matrix}

To complete the review of knot calculus let us give an alternative way of computing the bracket polynomial using an explicit matrix representation of the braid group and Temperley-Lieb algebra. This method is abstractly described by Nicolai Reshetikhin and Vladimir Turaev in~\cite{Reshetikhin:1991tc} and can be also inferred from papers of Louis Kauffman and others~\cite{Zhang:2005afb,Kauffman:2019,Padmanabhan:2020obt}. See~\cite{Melnikov:2024ddj} for a recent short compilation of the idea.

Let us start from matrix\footnote{In the original arXiv version, the expression for the matrix used conventions incompatible with other formulas below. I am grateful to Marcia Tenser for spotting this inconsistency.}
\be
\label{R}
R \ = \ \left(
\begin{array}{cccc}
    A & & &   \\
     & 0 & A^{-1} & \\
     & A^{-1} & A-A^{-3} & \\
     &  &  & A
\end{array}
\right). 
\ee
This matrix is mentioned in~\cite{Turaev:1988eb} as R matrix of $A_1$ Toda system~\cite{Jimbo:1985ua}. A more general class of such matrices is then described in the ``categorification'' approach to knot invariants~\cite{Reshetikhin:1991tc}. (See explicit formulas in~\cite{Morozov:2010kv}.) The matrix can be thought of as acting on a tensor product of two two-dimensional vector spaces, $R:V\otimes V\to V\otimes V$. Using this matrix one can construct two matrices acting on $V\otimes V\otimes V\equiv V^{\otimes3}$
\be
R_1 = R\otimes I_2\qquad \text{and} \qquad R_2 = I_2\otimes R\,,
\ee
where $I_2$ is the identity matrix on $V$. One can check that $R_1$ and $R_2$ satisfy the (Yang-Baxter) relation of the braid group
\be
R_1R_2R_1\ =\ R_2 R_1 R_2\,.
\ee
We can associate $R_1$ and $R_2$ to the generators of braid group $B_3$:
\be
b_1\ := \  R\otimes I_2 \ =  
\begin{array}{c}
       \begin{tikzpicture}[thick]
           \draw[gray,line width=1.5,opacity=0.6] (0,0) -- (1.0,0);
           \draw[gray,line width=1.5,opacity=0.6] (0,1) -- (1.0,1);
           \draw[rounded corners=3] (0.1,0) -- (0.1,0.3) -- (0.5,0.7) -- (0.5,1);
         \draw[rounded corners=3] (0.5,0) -- (0.5,0.3) -- (0.4,0.4);
         \draw[rounded corners=3] (0.2,0.6) -- (0.1,0.7) -- (0.1,1);
         \draw[rounded corners=3] (0.9,0.0) -- (0.9,1);
           \foreach \x in {0.1,0.5,...,1.0}
           \fill[black] (\x,0) circle (0.05);
           \foreach \x in {0.1,0.5,...,1.0}
           \fill[black] (\x,1) circle (0.05);
       \end{tikzpicture}
\end{array},\qquad
b_2 \ := \ I_2\otimes R \ =
\begin{array}{c}
       \newcommand{\x}{0.4}
       \begin{tikzpicture}[thick]
           \draw[gray,line width=1.5,opacity=0.6] (0,0) -- (1.,0);
           \draw[gray,line width=1.5,opacity=0.6] (0,1) -- (1.,1);
           \draw[rounded corners=3] (0.1+\x,0) -- (0.1+\x,0.3) -- (0.5+\x,0.7) -- (0.5+\x,1);
         \draw[rounded corners=3] (0.5+\x,0) -- (0.5+\x,0.3) -- (0.4+\x,0.4);
         \draw[rounded corners=3] (0.2+\x,0.6) -- (0.1+\x,0.7) -- (0.1+\x,1);
         \draw[rounded corners=3] (0.1,0.0) -- (0.1,1);
           \foreach \x in {0.1,0.5,...,1.0}
           \fill[black] (\x,0) circle (0.05);
           \foreach \x in {0.1,0.5,...,1.0}
           \fill[black] (\x,1) circle (0.05);
       \end{tikzpicture}
\end{array}.
\ee
Note that in the diagrams we can associate a line to each factor of $V$ and tensoring $R$ with $I$ simply means adding a strand to the braid. It is then straightforward to construct the set of generators of any $B_n$:
\be
\label{braidgens}
\begin{array}{ccc}
    b_1 & = & R\otimes {\mathbb{I}}_2\otimes {\mathbb{I}}_2\otimes \cdots\\
    b_2 & = & {\mathbb{I}}_2\otimes R\otimes {\mathbb{I}}_2\otimes\cdots \\
    b_3 & = & {\mathbb{I}}_2\otimes {\mathbb{I}}_2\otimes R\otimes \cdots \\
    & \cdots &
\end{array}
\ee

Next, consider
\be
U \ = \ \left(
\begin{array}{cccc}
    0 & & &   \\
     & -A^{2} & 1 & \\
     & 1 & - A^{-2} & \\
     &  &  & 0
\end{array}
\right).
\ee
According to~(\ref{protoskein}), we expect this to be a generator of the Temperley-Lieb algebra. Indeed,
\be
U^2 \ = \ d U\,.
\ee
Defining
\be
U_1 = U\otimes I_2\qquad \text{and} \qquad U_2 = I_2\otimes U\,,
\ee
it is not hard to check that
\be
U_1U_2U_1 \ = \ U_1 \qquad \text{and} \qquad U_2U_1U_2 \ = \ U_2\,.
\ee
Consequently a full set of generators of $TL_n$ can be constructed in the same way as for $B_n$, (\ref{braidgens}).

We expect that we can define a kind of trace for the braid matrices that would compute the bracket polynomials. It is instructive to check that the regular trace does not work. Instead, let us define the Markov trace as a sort of statistical expectation value. For a single strand (that is for a single $V$ factor) we will define
\be
\trm\, O \ = \ \Tr \rho\cdot O\,, \qquad \rho \ =  \left(
\begin{array}{cc}
 -A^2 &   \\
     & -A^{-2}  \\
\end{array}
\right) = \ e^{2\sigma^3\log iA}\,.
\ee
Consequently, for $V^{\otimes n}$ and braid $\beta_n$,
\be
\trm\, \beta_n \ = \ \Tr \rho^{\otimes n}\beta_n \ = \ \Tr \beta_n\exp\left(\sum\limits_{i=1}^n 2\sigma_i^3\log iA\right).
\ee

As an example, one can compute traces of braids appearing in~(\ref{trefoilclosure}). For the two-strand braid one gets
\be
\begin{array}{c}
       \newcommand{\x}{0.0}
       \newcommand{\y}{0}
       \begin{tikzpicture}[thick]
           \draw[gray,line width=1.5,opacity=0.6] (0+\y,-0) -- (0+\y,-0.6);
           \draw[rounded corners=3] (0+\y,-0.5-\x) -- (0.3+\y,-0.5-\x) -- (0.7+\y,-0.1-\x) -- (1+\y,-0.1-\x);
         \draw[rounded corners=3] (0+\y,-0.1-\x) -- (0.3+\y,-0.1-\x) -- (0.4+\y,-0.2-\x);
         \draw[rounded corners=3] (0.6+\y,-0.4-\x) -- (0.7+\y,-0.5-\x) -- (1+\y,-0.5-\x);
         \draw[rounded corners=3] (0.0+\y,-0.9) -- (1+\y,-0.9);
         \draw[rounded corners=3] (0.0+\y,-1.3) -- (1+\y,-1.3);
         \draw[rounded corners=3] (-0+\y,-0.5-\x) -- (-0.2+\y,-0.5-\x) -- (-0.2+\y,-0.9) -- (0.0+\y,-0.9);
         \draw[rounded corners=3] (-0+\y,-0.1-\x) -- (-0.4+\y,-0.1-\x) -- (-0.4+\y,-1.3) -- (0.0+\y,-1.3);
           \foreach \x in {0.1,0.5,...,0.6}
           \fill[black] (0+\y,-\x) circle (0.05);
           \renewcommand{\x}{0.0}
            \renewcommand{\y}{1}
           \draw[rounded corners=3] (0+\y,-0.5-\x) -- (0.3+\y,-0.5-\x) -- (0.7+\y,-0.1-\x) -- (1+\y,-0.1-\x);
         \draw[rounded corners=3] (0+\y,-0.1-\x) -- (0.3+\y,-0.1-\x) -- (0.4+\y,-0.2-\x);
         \draw[rounded corners=3] (0.6+\y,-0.4-\x) -- (0.7+\y,-0.5-\x) -- (1+\y,-0.5-\x);
         \draw[rounded corners=3] (0.0+\y,-0.9) -- (1+\y,-0.9);
         \draw[rounded corners=3] (0.0+\y,-1.3) -- (1+\y,-1.3);
            \renewcommand{\x}{0.0}
            \renewcommand{\y}{2}
            \draw[gray,line width=1.5,opacity=0.6] (1+\y,-0) -- (1+\y,-0.6);
           \draw[rounded corners=3] (0+\y,-0.5-\x) -- (0.3+\y,-0.5-\x) -- (0.7+\y,-0.1-\x) -- (1+\y,-0.1-\x);
         \draw[rounded corners=3] (0+\y,-0.1-\x) -- (0.3+\y,-0.1-\x) -- (0.4+\y,-0.2-\x);
         \draw[rounded corners=3] (0.6+\y,-0.4-\x) -- (0.7+\y,-0.5-\x) -- (1+\y,-0.5-\x);
         \draw[rounded corners=3] (0.0+\y,-0.9) -- (1+\y,-0.9);
         \draw[rounded corners=3] (0.0+\y,-1.3) -- (1+\y,-1.3);
          \draw[rounded corners=3] (1+\y,-0.5-\x) -- (1.2+\y,-0.5-\x) -- (1.2+\y,-0.9) -- (1+\y,-0.9);
         \draw[rounded corners=3] (1+\y,-0.1-\x) -- (1.4+\y,-0.1-\x) -- (1.4+\y,-1.3) -- (1+\y,-1.3);
          \foreach \x in {0.1,0.5,...,0.6}
           \fill[black] (1+\y,-\x) circle (0.05);
       \end{tikzpicture}
\end{array} \ = \  \Tr \rho^{\otimes 2}R^{-3} \ = \  (-A^{3})^{-3}(-A^2-A^{-2})(A^4+A^{12}-A^{16})\,,
\ee
in accordance with~(\ref{bracket3foil}). As advertised, in the mirror reflected diagrams $A$ is replaced by its inverse. For the three-strand braid,
\be
\begin{array}{c}
       \newcommand{\x}{0.4}
       \newcommand{\y}{0}
       \begin{tikzpicture}[thick]
           \draw[gray,line width=1.5,opacity=0.6] (0+\y,-0) -- (0+\y,-1.0);
           \draw[rounded corners=3] (0+\y,-0.5-\x) -- (0.3+\y,-0.5-\x) -- (0.7+\y,-0.1-\x) -- (1+\y,-0.1-\x);
         \draw[rounded corners=3] (0+\y,-0.1-\x) -- (0.3+\y,-0.1-\x) -- (0.4+\y,-0.2-\x);
         \draw[rounded corners=3] (0.6+\y,-0.4-\x) -- (0.7+\y,-0.5-\x) -- (1+\y,-0.5-\x);
         \draw[rounded corners=3] (0.0+\y,-0.1) -- (1+\y,-0.1);
         \draw[rounded corners=3] (0.0+\y,-1.3) -- (1+\y,-1.3);
         \draw[rounded corners=3] (0.0+\y,-1.7) -- (1+\y,-1.7);
         \draw[rounded corners=3] (0.0+\y,-2.1) -- (1+\y,-2.1);
         \draw[rounded corners=3] (-0+\y,-0.9) -- (-0.2+\y,-0.9) -- (-0.2+\y,-1.3) -- (0.0+\y,-1.3);
         \draw[rounded corners=3] (-0+\y,-0.5) -- (-0.4+\y,-0.5) -- (-0.4+\y,-1.7) -- (0.0+\y,-1.7);
         \draw[rounded corners=3] (-0+\y,-0.1) -- (-0.6+\y,-0.1) -- (-0.6+\y,-2.1) -- (0.0+\y,-2.1);
           \foreach \x in {0.1,0.5,...,1.0}
           \fill[black] (0+\y,-\x) circle (0.05);
            \renewcommand{\x}{0}
       \renewcommand{\y}{1}
           \draw[rounded corners=3] (0+\y,-0.5-\x) -- (0.3+\y,-0.5-\x) -- (0.7+\y,-0.1-\x) -- (1+\y,-0.1-\x);
         \draw[rounded corners=3] (0+\y,-0.1-\x) -- (0.3+\y,-0.1-\x) -- (0.4+\y,-0.2-\x);
         \draw[rounded corners=3] (0.6+\y,-0.4-\x) -- (0.7+\y,-0.5-\x) -- (1+\y,-0.5-\x);
         \draw[rounded corners=3] (0.0+\y,-0.9) -- (1+\y,-0.9);
         \draw[rounded corners=3] (0.0+\y,-1.3) -- (1+\y,-1.3);
         \draw[rounded corners=3] (0.0+\y,-1.7) -- (1+\y,-1.7);
         \draw[rounded corners=3] (0.0+\y,-2.1) -- (1+\y,-2.1);
         \renewcommand{\x}{0.4}
       \renewcommand{\y}{2}
       \draw[rounded corners=3] (0+\y,-0.5-\x) -- (0.3+\y,-0.5-\x) -- (0.7+\y,-0.1-\x) -- (1+\y,-0.1-\x);
         \draw[rounded corners=3] (0+\y,-0.1-\x) -- (0.3+\y,-0.1-\x) -- (0.4+\y,-0.2-\x);
         \draw[rounded corners=3] (0.6+\y,-0.4-\x) -- (0.7+\y,-0.5-\x) -- (1+\y,-0.5-\x);
         \draw[rounded corners=3] (0.0+\y,-0.1) -- (1+\y,-0.1);
         \draw[rounded corners=3] (0.0+\y,-1.3) -- (1+\y,-1.3);
         \draw[rounded corners=3] (0.0+\y,-1.7) -- (1+\y,-1.7);
         \draw[rounded corners=3] (0.0+\y,-2.1) -- (1+\y,-2.1);
         \renewcommand{\x}{0}
       \renewcommand{\y}{3}
       \draw[gray,line width=1.5,opacity=0.6] (1+\y,-0) -- (1+\y,-1.0);
           \draw[rounded corners=3] (0+\y,-0.5-\x) -- (0.3+\y,-0.5-\x) -- (0.7+\y,-0.1-\x) -- (1+\y,-0.1-\x);
         \draw[rounded corners=3] (0+\y,-0.1-\x) -- (0.3+\y,-0.1-\x) -- (0.4+\y,-0.2-\x);
         \draw[rounded corners=3] (0.6+\y,-0.4-\x) -- (0.7+\y,-0.5-\x) -- (1+\y,-0.5-\x);
         \draw[rounded corners=3] (0.0+\y,-0.9) -- (1+\y,-0.9);
         \draw[rounded corners=3] (0.0+\y,-1.3) -- (1+\y,-1.3);
         \draw[rounded corners=3] (0.0+\y,-1.7) -- (1+\y,-1.7);
         \draw[rounded corners=3] (0.0+\y,-2.1) -- (1+\y,-2.1);
         \draw[rounded corners=3] (1+\y,-0.9) -- (1.2+\y,-0.9) -- (1.2+\y,-1.3) -- (1.0+\y,-1.3);
         \draw[rounded corners=3] (1+\y,-0.5) -- (1.4+\y,-0.5) -- (1.4+\y,-1.7) -- (1.0+\y,-1.7);
         \draw[rounded corners=3] (1+\y,-0.1) -- (1.6+\y,-0.1) -- (1.6+\y,-2.1) -- (1.0+\y,-2.1);
           \foreach \x in {0.1,0.5,...,1.0}
           \fill[black] (1+\y,-\x) circle (0.05);
       \end{tikzpicture} 
\end{array}
\ = \ \Tr \rho^{\otimes 3}(R_1^{-1}R_2^{-1})^2 \ = \  (-A^{3})^{-4}(-A^2-A^{-2})(A^4+A^{12}-A^{16})\,.
\ee
This diagram indeed leads to the same Jones polynomial, although the bracket has a framing different by a unit of writhe as compared to the above. All this features are neatly reproduced by the matrix representation.

In fact, one can go beyond computing traces and evaluate the invariants as matrix elements. Since computing a trace corresponds to closing the lines of a braid, that is gluing the beginning and the end of the braid, computing a matrix element should correspond to ``gluing'' states to the beginning and to the end. For example, the naive identification
\be
\label{knotasoverlap}
\begin{array}{c}\begin{tikzpicture}[baseline=0]
\draw[thick,rounded corners=2] (0.5,-0.1) -- (0.3,-0.1) -- (0.15,0);
\draw[thick,rounded corners=2] (-0.15,0.2) -- (-0.3,0.3) -- (-0.5,0.3);
\draw[thick,rounded corners=2] (-0.5,-0.1) -- (-0.3,-0.1) -- (0.3,0.3) -- (0.5,0.3);
\newcommand{\x}{2}
\draw[thick,rounded corners=2] (0.5,0.3) -- (1.5,0.3);
\draw[thick,rounded corners=2] (0.5+\x,-0.1) -- (0.3+\x,-0.1) -- (0.15+\x,0);
\draw[thick,rounded corners=2] (-0.15+\x,0.2) -- (-0.3+\x,0.3) -- (-0.5+\x,0.3);
\draw[thick,rounded corners=2] (-0.5+\x,-0.1) -- (-0.3+\x,-0.1) -- (0.3+\x,0.3) -- (0.5+\x,0.3);
\newcommand{\y}{-0.8}
\draw[thick,rounded corners=2] (0.5,-0.1+\y) -- (0.3,-0.1+\y) -- (0.15,0+\y);
\draw[thick,rounded corners=2] (-0.15,0.2+\y) -- (-0.3,0.3+\y) -- (-0.5,0.3+\y);
\draw[thick,rounded corners=2] (-0.5,-0.1+\y) -- (-0.3,-0.1+\y) -- (0.3,0.3+\y) -- (0.5,0.3+\y);
\draw[thick,rounded corners=2] (0.5,-0.1+\y) -- (1.5,-0.1+\y);
\draw[thick,rounded corners=2] (0.5+\x,-0.1+\y) -- (0.3+\x,-0.1+\y) -- (0.15+\x,0+\y);
\draw[thick,rounded corners=2] (-0.15+\x,0.2+\y) -- (-0.3+\x,0.3+\y) -- (-0.5+\x,0.3+\y);
\draw[thick,rounded corners=2] (-0.5+\x,-0.1+\y) -- (-0.3+\x,-0.1+\y) -- (0.3+\x,0.3+\y) -- (0.5+\x,0.3+\y);
\renewcommand{\x}{1}
\renewcommand{\y}{-0.4}
\draw[thick,rounded corners=2] (0.5+\x,0.3+\y) -- (0.3+\x,0.3+\y) -- (0.15+\x,0.2+\y);
\draw[thick,rounded corners=2] (-0.15+\x,0.+\y) -- (-0.3+\x,-0.1+\y) -- (-0.5+\x,-0.1+\y);
\draw[thick,rounded corners=2] (-0.5+\x,0.3+\y) -- (-0.3+\x,0.3+\y) -- (0.3+\x,-0.1+\y) -- (0.5+\x,-0.1+\y);
\renewcommand{\x}{-1}
\draw[thick,rounded corners=2] (0.5+\x,-0.1+\y) -- (0.3+\x,-0.1+\y) arc (-90:-270:0.2) -- (0.5+\x,0.3+\y);
\renewcommand{\x}{3}
\draw[thick,rounded corners=2] (-0.5+\x,-0.1+\y) -- (-0.3+\x,-0.1+\y) arc (-90:90:0.2) -- (-0.5+\x,0.3+\y);
\renewcommand{\x}{3}
\renewcommand{\x}{-1}
\draw[thick,rounded corners=2] (-0.5+\x,-0.1+\y) -- (-0.3+\x,-0.1+\y) arc (-90:90:0.2) -- (-0.5+\x,0.3+\y);
\renewcommand{\x}{-2}
\renewcommand{\y}{0}
\draw[thick,rounded corners=2] (0.5+\x,-0.1+\y) -- (0.3+\x,-0.1+\y) arc (-90:-270:0.2) -- (0.5+\x,0.3+\y);
\renewcommand{\y}{-0.8}
\draw[thick,rounded corners=2] (0.5+\x,-0.1+\y) -- (0.3+\x,-0.1+\y) arc (-90:-270:0.2) -- (0.5+\x,0.3+\y);
\renewcommand{\x}{4}
\renewcommand{\y}{0}
\renewcommand{\x}{-2}
\draw[thick,rounded corners=2] (0.5+\x,0.3) -- (1.5+\x,0.3);
\draw[line width=2,gray,dashed,opacity=0.5] (0.5+\x,-1.1) -- (0.5+\x,0.5);
\renewcommand{\x}{+2}
\draw[thick,rounded corners=2] (0.5+\x,0.3) -- (0.7+\x,0.3) arc (90:-90:0.6) -- (0.5+\x,-0.9);
\draw[line width=2,gray,dashed,opacity=0.5] (0.5+\x,-1.1) -- (0.5+\x,0.5);
\renewcommand{\y}{-0.8}
\renewcommand{\x}{-2}
\draw[thick,rounded corners=2] (0.5+\x,-0.1+\y) -- (1.5+\x,-0.1+\y);
\end{tikzpicture}
\end{array}
\ = \ \langle\Psi| B|\Phi\rangle\,,
\ee
suggests that we should be able to construct appropriate states $|\Psi\rangle$ and $|\Phi\rangle$ as some sort of bent braids (or tangles).

We will once again resort to the assumption of faithfulness, namely, that the class of matrices we use represent faithfully the diagrams. For diagrams we can easily convert a matrix to a vector: 
\be
U \ \equiv \ \begin{array}{c}
     \begin{tikzpicture}[thick]
     \draw (0,0) -- (0.05,0) arc (-90:90:0.15) -- (0,0.3);
     \draw (0.5,0) -- (0.45,0) arc (-90:-270:0.15) -- (0.5,0.3);
     \end{tikzpicture}
\end{array}\ = \ |u\rangle\langle u|
\qquad 
\longleftrightarrow
\qquad
|\Psi\rangle \ \equiv \ |u\rangle\otimes |u\rangle \ = \ 
\begin{array}{c}
\begin{tikzpicture}[thick]
    \draw (0,0) -- (0.05,0) arc (-90:90:0.15) -- (0,0.3);
    \draw (0,0.6) -- (0.05,0.6) arc (-90:90:0.15) -- (0,0.9);
\end{tikzpicture}
\end{array} .
\ee
To transform the matrix into a vector one just flattens it:
\be
|\Psi\rangle  \ = \  \left(0,0,0,0,0,-A^{2},1,0,0,1,-A^{-2},0,0,0,0,0\right).
\ee
A small subtlety here is related with the fact that the representation of matrix $R$ is not unitary, so one should use simple transposition instead of Hermitian conjugation.

One can be a little bit more precise about the last point and note that the representation can be made pseudounitary if one assumes that $A=\exp(i\theta)$ is a pure phase, $A^\ast=A^{-1}$. There exists matrix 
\be
\Sigma \ = \ 
\left(
\begin{array}{cc}
     & 1 \\
    1 & 
\end{array}
\right),
\ee
such that
\be
R^{-1} \ = \ \Sigma^{\otimes 2} R^\dagger(\Sigma^{\otimes 2})^\dagger.
\ee
Here one should take an appropriate tensor power of $\Sigma$ for objects in $V^{\otimes n}$. $\Sigma$ plays the role of the metric, or $\gamma^0$ in the vector or spinor representation of the Lorentz group. Therefore, the proper conjugation of the vectors is
\be
|\Psi\rangle \ \to \ \left(|\Psi\rangle\right)^\dagger\cdot \Sigma\,,
\ee
which is just equivalent to a simple transposition of $|\Psi\rangle$.

To construct other diagrammatic states, one simply acts on $|\Psi\rangle$ by braids. Thus, state $|\Phi\rangle$ in~(\ref{knotasoverlap}) can be obtained as
\be
|\Phi\rangle \ = \ R_3R_2|\Psi\rangle \ = \ U_2 |\Psi\rangle \ =  \  \left(0,0,0,-A^{2},0,1,0,0,0,0,1,0,-A^{-2},0,0,0\right).
\ee

Collecting all pieces together, we find that
\be
\langle \Psi|U_2R_1^{-1}R_3^{-1}R_2R_3^{-1}R_1^{-1}|\Phi\rangle \ = \ (-{A^3})^{-1}(-A^2-A^{-2})\left(A^{14}-2A^{10} + A^6 - 2 A^2 + A^{-2} - A^{-6}\right)
\ee
is indeed the bracket polynomial of the Whitehead link shown in equation~(\ref{whitehead}). Note that the bracket polynomial has only one unit of writhe, compatible with vanishing Gauss' linking number.

%%%%%%%%%%%%%%%%%%%%%%%%%%%%%%%%%%%%%%%%%%%%%%%%%%%%%%%%%%%%%%%%%%%%
\section{Topological bits}
\label{sec:topobits}

In  this section we will explain how knots can be used in the description of quantum mechanics. In fact, the main idea was already anticipated in the last section, when we introduce a matrix description of knots and their invariants. In this section we will put this idea on a more formal ground using methods of category theory, following the definition given by Michael Atiyah~\cite{Atiyah:1989vu}. Only very basic notions of categories will be used, and they will be explained. The main message that will be conveyed is that basic objects of quantum mechanics, such as states operators and Hilbert spaces can be represented by topological spaces and some basic manipulations on them. The tool that will help establish this connection is the topological quantum field theory (TQFT).

%%%%%%%%%%%%%%%%%%%%%%%%%%%%%%%%%%%%%%%%%%%%%%%%%%%%%%%%%%%%%%%%%%%%
\subsection{Topological Quantum Field Theory}
\label{sec:tqft}

Theories whose description does not explicitly or implicitly use the spacetime metric are called topological~\cite{Witten:1988ze,Witten:1988hf}. Metric independence means that observables of such theories do not depend on distances. This imposes powerful constraints on the theory, one of which is vanishing Hamiltonian.

Recall that the stress-energy tensor can be obtained by the variation of the action of a nongravitational theory with respect to the background metric:
\be
\label{energymomentum}
T^{\mu\nu} \ = \ -\frac{2}{\sqrt{-g}}\frac{\delta S}{\delta g_{\mu\nu}}\,.
\ee
If the theory is metric independent, all the components of the stress-energy tensor are null. Curiously, pure gravity appears to be a type of topological theory that has vanishing of the stress-energy tensor as its dynamical equations. Quantum gravity is also metric-independent in another sense: gravitational path integral is a sum over all possible metrics.

Distance independence implies a lack of the notion of locality and lack of local degrees of freedom. Locality is an important feature of quantum field theory, which distinguishes it from quantum mechanics. Therefore, topological theories are essentially instances of quantum mechanics, which sometimes disguise themselves as quantum field theories.

One famous example of a TQFT is the Chern-Simons theory, which will be formally introduced in a later section. It is an example of a gauge theory, whose local degrees of freedom are pure gauge. Gauge transformations, however, do not affect some global characteristics of fields, which allow for the theory to contain some nontrivial information. Early studies of Chern-Simons theories (for example in \cite{Schwarz:1978cn,Polyakov:1988md,Witten:1988ze}) connected them with knots. Motivated by those studies, Atiyah gave a very formal definition of TQFTs, known as Atiyah's axioms~\cite{Atiyah:1989vu}. 

Atiyah's definition is very useful to motivate the discussion of quantum mechanics that will follow, but to introduce it we will need to first review the basic definitions of category theory. Those interested in learning more about categories and their applications may consult the book~\cite{Simmons:2011book}.

\begin{figure}
    \centering
    \includegraphics[width=0.5\linewidth]{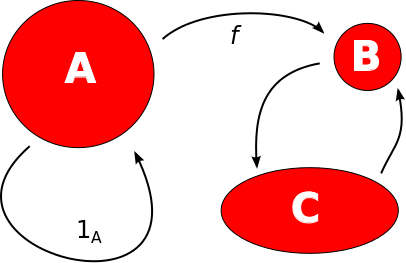}
    \caption{A schematic depiction of a category (objects and morphisms).}
    \label{fig:category}
\end{figure}

A category (figure~\ref{fig:category}) consists of objects ($A$, $B$, $C$ etc.) and morphisms ($f$, $g$, $h$ etc.). Objects can be anything, for example: sets, spaces, algebraic structures etc. Morphisms are maps between objects satisfying the following axioms:
\begin{itemize}
    \item for morphisms $f:A\to B$ and $g:B\to C$ there exists a \emph{composition} $g\circ f: A\to C$. The composition is associative;
    \item for any object $A$ there exists the identity morphism $1_A:A\to A$ such that $f\circ 1_A = f$ or $ 1_A\circ g=g$. This axiom implies that $1_A$ is unique.
\end{itemize}

Linear spaces and linear operators provide a fundamental example of a category -- \textbf{Vect}. This category is the main subject of the study of linear algebra. Note that in this category the objects are not necessarily of the same dimension and morphisms are not necessarily square matrices. Another example is the category of groups -- \textbf{Grp}. Morphisms of the latter are group homomorphisms. 

Group representations and homomorphisms are behind the idea of the \emph{functor} -- a map between categories. A functor $F$ maps objects into objects and morphisms into morhphisms, preserving the composition structure. $F$ satisfies the following axioms:
\begin{itemize}
    \item If $A$ and ${\cal A}$ are objects in two categories, such that ${\cal A} =F(A)$ then the identity morphism is preserved:
    \be
    F(1_A) \ = \ 1_{\cal A}\,;
    \ee
    \item If $f:A\to B$ is a morphism in the first category then $F$ maps it into a morphism between images of $A$ and $B$ in the second category:
    \be
    F(f): \ F(A) \ \to F(B);
    \ee
    \item $F$ preserves the composition of any two morphisms $f$ and $g$:
    \be
    F(f\circ g) \ = \ F(f)\circ F(g)\,.
    \ee
\end{itemize}

To define the topological quantum field theory we will need to introduce one more category -- the category of cobordisms \textbf{Cob}. \emph{Cobordisms} are compact smooth manifolds with boundaries. If the boundaries are compact closed manifolds, one can view the cobordism as connecting different boundaries, that is cobordisms are certain maps between boundaries. Category \textbf{Cob} considers the boundaries as the objects and the cobordisms as morphisms.

Now, TQFT is a functor $Z:\textbf{Cob}\to\textbf{Vect}$, which associates a complex vector space $Z(\Sigma)$ to each $d$-dimensional smooth compact oriented space $\Sigma$ and a vector $Z(M)\in Z(\Sigma)$ to each $d+1$-dimensional compact oriented space $M$ whose boundary is $\Sigma$. If one accepts that the following heuristic $\Sigma$ is a compact closed manifold, which is being mapped on a vector space $V$:
\be
Z: \ \begin{array}{c}
\includegraphics[height=0.08\linewidth]{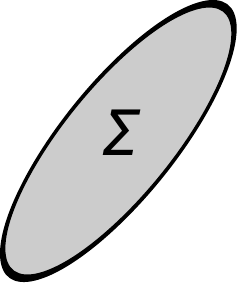} 
\end{array}
\quad \longrightarrow \quad Z(\Sigma) = V\,,
\ee
then vectors in $V$ are obtained from all possible higher dimensional manifolds that can be glued to $\Sigma$, for example,
\be
\label{tqftvector}
Z: \  \begin{array}{c}
\includegraphics[height=0.08\linewidth]{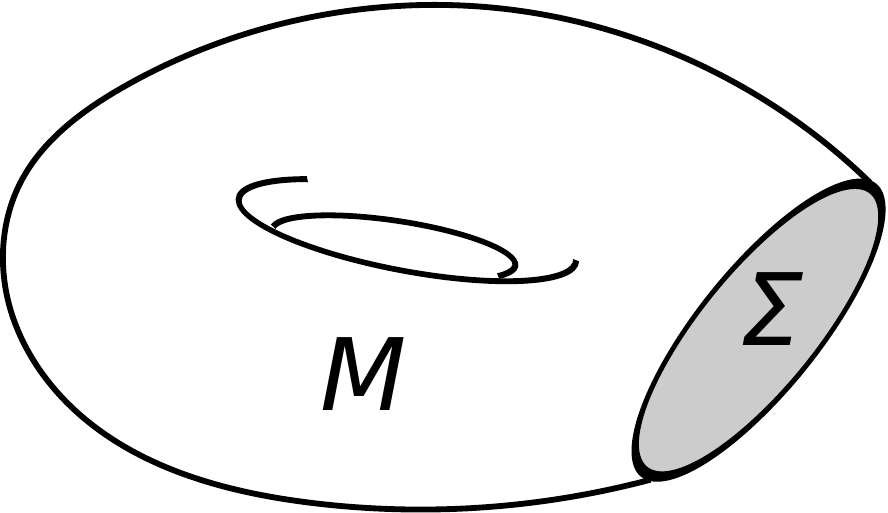} 
\end{array}
\quad \longrightarrow \quad Z(M) = v\, \in \, V\,.
\ee
Note that states can be viewed as cobordisms from a trivial(empty) boundary to $\Sigma$. The existence of infinitely many $M$ explains why $\Sigma$ is a sort of a vector space.

TQFT functor $Z$ satisfies the following axioms:
\begin{itemize}
    \item For a given $\Sigma$, manifold $\Sigma^\ast$, which corresponds to the same manifold, but with the opposite choice of the orientation, is mapped to the dual space of $Z(\Sigma)$:
    \be
    Z(\Sigma^\ast) \ = \ Z(\Sigma)^\ast\,;
    \ee
    \item a disjoint union of boundaries $\Sigma_1\sqcup\Sigma_2$ is mapped to a tensor product of vector spaces:
    \be
    Z(\Sigma_1\sqcup\Sigma_2) \ = \ Z(\Sigma_1)\otimes Z(\Sigma_2)\,;
    \ee
    \item A composition of cobordisms, that is a pair of manifolds $M_1$ ($\partial M_1=\Sigma_1\sqcup\Sigma_2$) and $M_2$ ($\partial M_2=\Sigma_2\sqcup\Sigma_3$) with an identical boundary component $\Sigma_2$, is mapped to the product of linear maps:
    \be
    Z(M_2\circ_{\Sigma_2}M_1) \ = \ Z(M_2)\cdot Z(M_1)\,;
    \ee
    (Note that for this to be consistent, $\Sigma_2$ must be taken with opposite orientations in $M_1$ and $M_2$. This axiom completes the general axiom of a functor by specifying the map $Z(\circ)$.)
    \item If $\Sigma$ is trivial than its image are complex numbers $\mathbb{C}$:
    \be
    Z(\varnothing) \ = \ \mathbb{C}\,;
    \ee
    \item If $I$ is an interval $[0,1]$ then the image of the manifold $\Sigma\times I$ is the identity map on $Z(\Sigma)$:
    \be
    Z(\Sigma\times I) \ = \ 1_{Z(\Sigma)}\,.
    \ee
\end{itemize}

Equipped with Atiyah's axioms we can now start thinking of elements of linear algebra as topological spaces. We have already drawn a vector~(\ref{tqftvector}) as a manifold glued to a given boundary. It is not hard to imagine that an operator $O$ acting on a specific vector in the same vector space should be something of the following kind:
\be
O:\quad v \ = \ \begin{array}{c}
\includegraphics[scale=0.12]{./figs/state2.pdf}
\end{array} \quad \longrightarrow \quad   \begin{array}{c}
\includegraphics[scale=0.2]{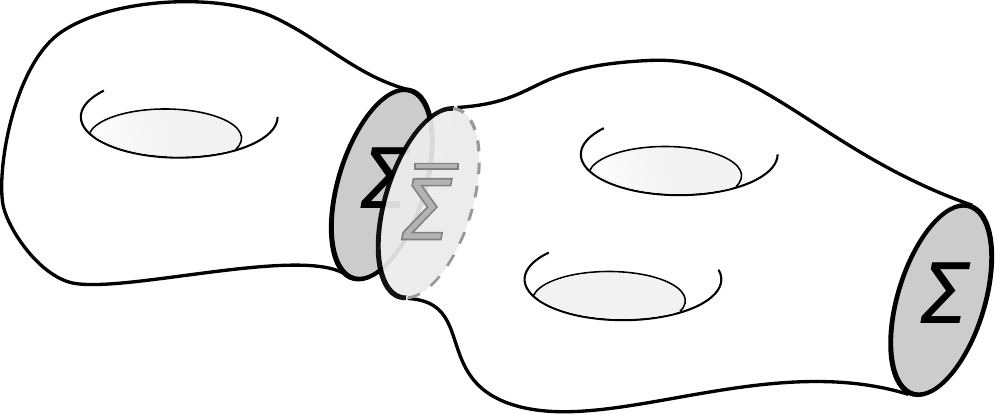}
\end{array} = \  O\cdot v
\ee
This has an apparent analogy with tensor networks and classical and quantum circuits, where application of an operator (gate) is expressed by a diagram
\be
O:\quad \psi \quad \longrightarrow \quad
\begin{array}{c}
\begin{tikzpicture}[line width=1.0]
            \draw (-1,-0.3) -- (1.5,-0.3);
             \fill[gray] (0,-0.7) rectangle (1,0.1);
         \fill[white] (0.1,-0.6) rectangle (0.9,-0.);
         \draw (0.5,-0.3) node {$O$};
         \draw (-1,-0.) node {$\psi$};
         \fill (-1,-0.3) circle (0.05);
\end{tikzpicture}
\end{array}
\,.
\ee
In the TQFT presentation the lines of the evolution are thickened to become spaces, whose nontrivial topological features are the gates. Moreover, states gain their own space presentation that holds information about their structure.

The axioms also instruct us about the inner product: If by gluing a pair of manifolds along a common appropriately oriented boundary we get a manifold with no boundary, the result should correspond to a $\mathbb{C}$-number. Indeed, in this case, the boundary of a closed manifold is an empty space, so the functor maps it to a number. The following diagram illustrates a situation, in which this happens:
\be
\begin{array}{c}
 \includegraphics[scale=0.2]{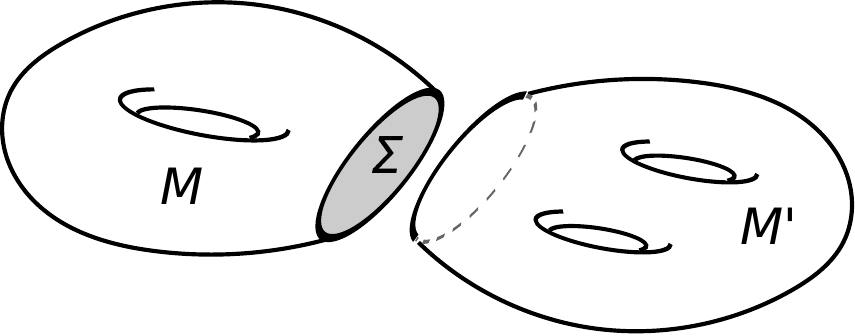}
 \end{array}
 \ \longrightarrow \quad   
 \begin{array}{c}
 \includegraphics[scale=0.2]{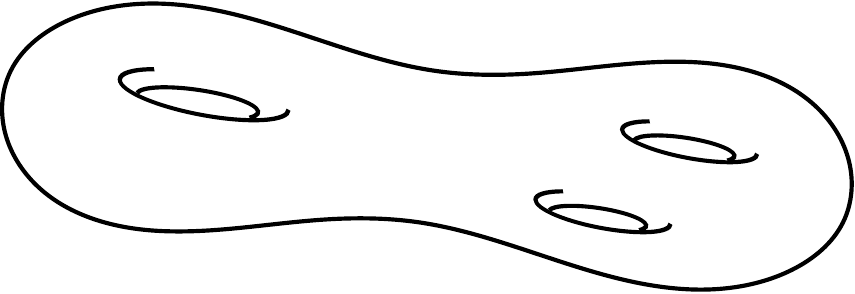}
 \end{array}
  \equiv \ \langle\psi|\psi'\rangle\,.
\ee
The closed manifold represents the inner product of a pair of states.

Another useful trick is the computation of the trace of an operator. For an endomorphism, that is for a linear operators that maps a space to itself, the trace can be computed by gluing together two of its boundaries (closing the space on itself). The following example,
\be
\Tr\left(\begin{array}{c}\includegraphics[width=0.12\linewidth,clip=true,trim=100pt 0pt 300pt 0pt]{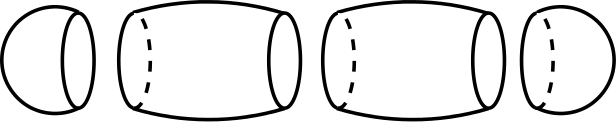} \end{array}\right) = \begin{array}{c}\includegraphics[scale=0.2]{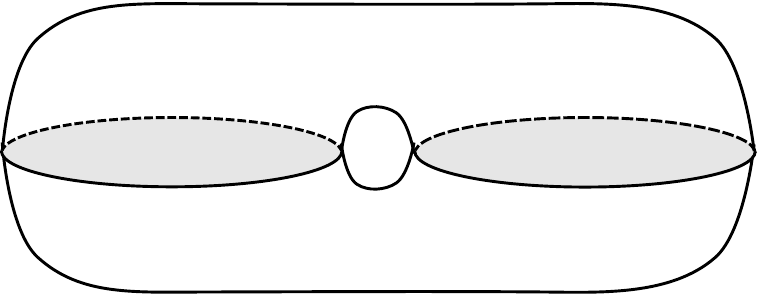}\end{array},
\ee
can be thought as computing the trace of an identity operator, which gives the dimension of the space associated with the boundary. Note that all the above pictures are rather heuristic as the boundaries themselves appear to have boundaries, which is not a true representation of the situation we assume in the TQFT. However, if, in the last example, the TQFT is two-dimensional, and the boundary is a circle, then the trace is indeed a torus.

In the following sections we will give explicit examples of TQFT, which will allow to compute all the data of the linear algebra exactly, preserving the general topological interpretation.

%%%%%%%%%%%%%%%%%%%%%%%%%%%%%%%%%%%%%%%%%%%%%%%%%%%%%%%%%%%%%%%%%%%%

\subsection{Topological Quantum Mechanics}
\label{sec:TQM}

Let us imagine the most simple situation: a one-dimensional TQFT. We will follow a physics-level exposition in~\cite{Melnikov:2022qyt}. See~\cite{Lurie:2009keu,Freed:2012hx} for a more formal mathematical construction.

A $0+1$ dimensional quantum field theory is called quantum mechanics, so we will consider a topological quantum mechanics. In this case, the role of (0-dimensional) $\Sigma$ will be played by a collection of points. For our convenience, we will organize the points along a line, as for the braids, or circuit diagrams, since it will be natural to think of lines attached to the points as their evolution.

Following the TQFT axioms, the states in the Hilbert space associated with a collection of points must be lines connecting those points. For example,
\be
\label{stateXX}
|\psi\rangle \ =  \begin{array}{c} 
\begin{tikzpicture}[thick]
    \draw[gray,opacity=0.5] (0,0) -- (0,1.3);
    \draw[rounded corners=2] (0,0.2) -- (1.9,0.2) -- (1.9,0.5) -- (1.6,0.5) -- (1.3,0.8) -- (1.1,0.8) -- (0.8,0.5) -- (0.6,0.5) -- (0.5,0.6);
    \draw[white,double distance=2,rounded corners=2] (0,1.1) -- (1.9,1.1) -- (1.9,0.8) -- (1.6,0.8) -- (1.3,0.5);
    \draw[rounded corners=2] (0,1.1) -- (1.9,1.1) -- (1.9,0.8) -- (1.6,0.8) -- (1.3,0.5) -- (1.1,0.5) -- (1.0,0.6);
    \draw[rounded corners=2] (0.9,0.7) -- (0.8,0.8) -- (0.6,0.8) -- (0.3,0.5) -- (0,0.5);
    \draw[rounded corners=2] (0.4,0.7) -- (0.3,0.8) -- (0,0.8);
    \foreach \y in {0.2,0.5,...,1.1}
           \fill[black] (0,\y) circle (0.04cm);
\end{tikzpicture}
\end{array}.
\ee
We would like the Hilbert space to be spanned by all possible ways a given set of points is connected. Moreover, the above example suggests that we would like also to distinguish connections beyond specifying which point is connected to which. In particular, we will distinguish the above state $|\psi\rangle$ and state
\be
\label{stateX}
|\chi\rangle \ =  \begin{array}{c} 
\begin{tikzpicture}[thick]
    \draw[gray,opacity=0.5] (0,0) -- (0,1.3);
    \draw[rounded corners=2] (0,0.2) -- (0.9,0.2) -- (0.9,0.5) -- (0.6,0.5) -- (0.3,0.8) -- (0,0.8);
    \draw[white,double distance=2,rounded corners=2] (0,1.1) -- (0.9,1.1) -- (0.9,0.8) -- (0.6,0.8) -- (0.3,0.5) -- (0,0.5);
    \draw[rounded corners=2] (0,1.1) -- (0.9,1.1) -- (0.9,0.8) -- (0.6,0.8) -- (0.3,0.5) -- (0,0.5);
    \foreach \y in {0.2,0.5,...,1.1}
           \fill[black] (0,\y) circle (0.04cm);
\end{tikzpicture}
\end{array},
\ee
which has the same pairs of points connected. In other words, the one-dimensional theory will be implicitly embedded in three-dimensional space. In terms of the evolution of 0-dimensional particles, one can think that such particles can pairwise scatter by a right or a left twist, or they can annihilate or be created in pairs from nothing. Since we want to think of the ket-vectors as initial states, in the ket-states the particles will be created and the evolution will flow from right to left. 

To obtain a bra-state from a ket-state we will reverse the evolution. For state~(\ref{stateX}) this will take the form 
\be
\langle \chi| \ =  \begin{array}{c} 
\begin{tikzpicture}[thick]
    \draw[gray,opacity=0.5] (0,0) -- (0,1.3);
    \draw[rounded corners=2] (0,0.2) -- (-0.9,0.2) -- (-0.9,0.5) -- (-0.6,0.5) -- (-0.3,0.8) -- (0,0.8);
    \draw[white,double distance=2,rounded corners=2] (0,1.1) -- (-0.9,1.1) -- (-0.9,0.8) -- (-0.6,0.8) -- (-0.3,0.5) -- (0,0.5);
    \draw[rounded corners=2] (0,1.1) -- (-0.9,1.1) -- (-0.9,0.8) -- (-0.6,0.8) -- (-0.3,0.5) -- (0,0.5);
    \foreach \y in {0.2,0.5,...,1.1}
           \fill[black] (0,\y) circle (0.04cm);
\end{tikzpicture}
\end{array}.
\ee
The reversed evolution replaces the right twists by the left ones and vice versa. As we will see this is necessary if the evolution is unitary.

It is straightforward to compute the scalar products in this model. For states~(\ref{stateXX}) and~(\ref{stateX}) the scalar product must produce
\be
\langle\chi|\psi\rangle \ =  \begin{array}{c} 
\begin{tikzpicture}[thick]
    \draw[rounded corners=2] (0,0.2) -- (1.9,0.2) -- (1.9,0.5) -- (1.6,0.5) -- (1.3,0.8) -- (1.1,0.8) -- (0.8,0.5) -- (0.6,0.5) -- (0.5,0.6);
    \draw[white,double distance=2,rounded corners=2] (0,1.1) -- (1.9,1.1) -- (1.9,0.8) -- (1.6,0.8) -- (1.3,0.5);
    \draw[rounded corners=2] (0,1.1) -- (1.9,1.1) -- (1.9,0.8) -- (1.6,0.8) -- (1.3,0.5) -- (1.1,0.5) -- (1.0,0.6);
    \draw[rounded corners=2] (0.9,0.7) -- (0.8,0.8) -- (0.6,0.8) -- (0.3,0.5) -- (0,0.5);
    \draw[rounded corners=2] (0.4,0.7) -- (0.3,0.8) -- (0,0.8);
    \draw[rounded corners=2] (0,0.2) -- (-0.9,0.2) -- (-0.9,0.5) -- (-0.6,0.5) -- (-0.3,0.8) -- (0,0.8);
    \draw[white,double distance=2,rounded corners=2] (0,1.1) -- (-0.9,1.1) -- (-0.9,0.8) -- (-0.6,0.8) -- (-0.3,0.5) -- (0,0.5);
    \draw[rounded corners=2] (0,1.1) -- (-0.9,1.1) -- (-0.9,0.8) -- (-0.6,0.8) -- (-0.3,0.5) -- (0,0.5);
    \foreach \y in {0.2,0.5,...,1.1}
           \fill[black] (0,\y) circle (0.04cm);
\end{tikzpicture}
\end{array}.
\ee

To complete the definition of this quantum mechanics model we need to specify the dimension of the Hilbert space and an explicit algebra behind the calculation of scalar products. It must be obvious by now that the purpose of the construction is to relate quantum states and overlaps with knots, so we can take advantage of the calculus developed in the previous sections. In particular, we will use the representation of the braids connected to Jones polynomials, which is based on the skein relation~(\ref{skein}).

Let us assume that the above diagrams representing states satisfy skein relations. This means, for example, that
\be
|\chi\rangle \ =  \begin{array}{c} 
\begin{tikzpicture}[thick]
    \draw[gray,opacity=0.5] (0,0) -- (0,1.3);
    \draw[rounded corners=2] (0,0.2) -- (0.9,0.2) -- (0.9,0.5) -- (0.6,0.5) -- (0.3,0.8) -- (0,0.8);
    \draw[white,double distance=2,rounded corners=2] (0,1.1) -- (0.9,1.1) -- (0.9,0.8) -- (0.6,0.8) -- (0.3,0.5) -- (0,0.5);
    \draw[rounded corners=2] (0,1.1) -- (0.9,1.1) -- (0.9,0.8) -- (0.6,0.8) -- (0.3,0.5) -- (0,0.5);
    \foreach \y in {0.2,0.5,...,1.1}
           \fill[black] (0,\y) circle (0.04cm);
\end{tikzpicture}
\end{array} 
= A \begin{array}{c} 
\begin{tikzpicture}[thick]
    \draw[gray,opacity=0.5] (0,0) -- (0,1.3);
    \draw[rounded corners=2] (0,0.2) -- (0.7,0.2) -- (0.7,1.1) -- (0,1.1);
    \draw[rounded corners=2] (0,0.5) -- (0.4,0.5) -- (0.4,0.8) -- (0,0.8);
    \foreach \y in {0.2,0.5,...,1.1}
           \fill[black] (0,\y) circle (0.04cm);
\end{tikzpicture}
\end{array}
+ A^{-1}
\begin{array}{c} 
\begin{tikzpicture}[thick]
    \draw[gray,opacity=0.5] (0,0) -- (0,1.3);
    \draw[rounded corners=2] (0,0.2) -- (0.4,0.2) -- (0.4,0.5) -- (0,0.5);
    \draw[rounded corners=2] (0,0.8) -- (0.4,0.8) -- (0.4,1.1) -- (0,1.1);
    \foreach \y in {0.2,0.5,...,1.1}
           \fill[black] (0,\y) circle (0.04cm);
\end{tikzpicture}
\end{array}.
\ee
Let us assume that the basis in the Hilbert space is generated only by diagrams with no crossings. Then, for the 4-point Hilbert space used in the examples above, there are only two basis diagrams:
\be
\label{4pbasis}
|e_0\rangle = \begin{array}{c}\scalebox{0.6}{
\begin{tikzpicture}[very thick]
\draw[gray,opacity=0.5] (0,-0.2) -- (0,1.1);
\fill[black] (0,0.0) circle (0.05cm);
\fill[black] (0,0.3) circle (0.05cm);
\fill[black] (0,0.6) circle (0.05cm);
\fill[black] (0,0.9) circle (0.05cm);
\draw (0,0) -- (0.4,0) arc (-90:90:0.15cm) -- (0,0.3);
\draw (0,0.6) -- (0.4,0.6) arc (-90:90:0.15cm) -- (0,0.9);
\end{tikzpicture}} 
\end{array}\,,\qquad |e_1\rangle = \begin{array}{c}\scalebox{0.6}{
\begin{tikzpicture}[very thick]
\draw[gray,opacity=0.5] (0,-0.2) -- (0,1.1);
\fill[black] (0,0.0) circle (0.05cm);
\fill[black] (0,0.3) circle (0.05cm);
\fill[black] (0,0.6) circle (0.05cm);
\fill[black] (0,0.9) circle (0.05cm);
\draw (0,0) -- (0.1,0) arc (-90:90:0.45cm) -- (0,0.9);
\draw (0,0.6) -- (0.1,0.6) arc (90:-90:0.15cm) -- (0,0.3);
\end{tikzpicture}} 
\end{array}.
\ee
Clearly all other states should be reducible to linear combinations of these two states through the skein relations.

Generalizing this idea to any even number of points one can notice a connection between the basis diagrams for the $2n$-point Hilbert space and the elements of the Temperley-Lieb algebra $TL_n$. The basis diagrams are elements of $TL_{n}$ bent in a way to align the points at the top with the points at the bottom:
\be
\begin{array}{c}
       \begin{tikzpicture}[thick]
           \draw[gray,line width=1.5,opacity=0.6] (0,0) -- (1.0,0);
           \draw[gray,line width=1.5,opacity=0.6] (0,1) -- (1.0,1);
           \draw[rounded corners=3] (0.1,0) -- (0.1,1);
         \draw[rounded corners=3] (0.5,0) -- (0.5,1);
         \draw[rounded corners=3] (0.9,0.0) -- (0.9,1);
           \foreach \x in {0.1,0.5,...,1.0}
           \fill[black] (\x,0) circle (0.05);
           \foreach \x in {0.1,0.5,...,1.0}
           \fill[black] (\x,1) circle (0.05);
       \end{tikzpicture}
\end{array}
\quad \longrightarrow \quad
\begin{array}{c}
\begin{tikzpicture}[thick]
\draw[gray,opacity=0.5] (0,-0.2) -- (0,1.5);
\draw (0,0.2) -- (0.1,0.2) arc (-90:90:0.45cm) -- (0,1.1);
\draw (0,0.8) -- (0.1,0.8) arc (90:-90:0.15cm) -- (0,0.5);
\draw (0,1.4) -- (0.1,1.4) arc (90:-90:0.75cm) -- (0,-0.1);
\foreach \y in {-0.1,0.2,...,1.5}
           \fill[black] (0,\y) circle (0.05);
\end{tikzpicture}
\end{array},
\qquad
\begin{array}{c}
       \begin{tikzpicture}[thick]
           \draw[gray,line width=1.5,opacity=0.6] (0,0) -- (1.0,0);
           \draw[gray,line width=1.5,opacity=0.6] (0,1) -- (1.0,1);
           \draw[rounded corners=3] (0.1,0) -- (0.1,0.4) -- (0.5,0.4) -- (0.5,0);
         \draw[rounded corners=3] (0.5,1) -- (0.5,0.6) -- (0.1,0.6) -- (0.1,1);
         \draw[rounded corners=3] (0.9,0.0) -- (0.9,1);
           \foreach \x in {0.1,0.5,...,1.0}
           \fill[black] (\x,0) circle (0.05);
           \foreach \x in {0.1,0.5,...,1.0}
           \fill[black] (\x,1) circle (0.05);
       \end{tikzpicture}
\end{array}
\quad \longrightarrow \quad
\begin{array}{c}
\begin{tikzpicture}[thick]
\draw[gray,opacity=0.5] (0,-0.2) -- (0,1.5);
\draw (0,0.2) -- (0.1,0.2) arc (-90:90:0.15cm) -- (0,0.5);
\draw (0,1.1) -- (0.1,1.1) arc (90:-90:0.15cm) -- (0,0.8);
\draw (0,1.4) -- (0.1,1.4) arc (90:-90:0.75cm) -- (0,-0.1);
\foreach \y in {-0.1,0.2,...,1.5}
           \fill[black] (0,\y) circle (0.05);
\end{tikzpicture}
\end{array},
\qquad
\cdots
\ee
Therefore, there are $C_n$ different basis diagrams for the $2n$-point Hilbert space, with $C_n$ being the Catalan numbers~(\ref{Catalan}). 

Hermitian conjugation is consistent with the diagrammatic presentation of states if $A^\ast=A^{-1}$, that is $A$ is a complex phase. This is the same condition that we encountered in section~\ref{sec:matrix} discussing a matrix representation of braids. This condition will ensure that the scalar product is positive definite. In order to compute an overlap of any pair of diagrammatic states we can use the rules of computing the bracket polynomial summarized at the end of section~\ref{sec:Jonespolynomial}. Besides the skein relation the rules include substituting any disconnected trivial circle by factor $d=-A^{2}-A^{-2}$.

Note that if we compute the Gram matrix of basis vectors~(\ref{4pbasis}) we will find that it is not diagonal. Specifically,
\be
\langle e_0|e_0\rangle \ = \ \begin{array}{c}\scalebox{0.6}{
\begin{tikzpicture}[very thick]
\fill[black] (0,0.0) circle (0.05cm);
\fill[black] (0,0.3) circle (0.05cm);
\fill[black] (0,0.6) circle (0.05cm);
\fill[black] (0,0.9) circle (0.05cm);
\draw (0,0) -- (0.4,0) arc (-90:90:0.15cm) -- (0,0.3);
\draw (0,0.6) -- (0.4,0.6) arc (-90:90:0.15cm) -- (0,0.9);
\draw (0,0) -- (-0.4,0) arc (-90:-270:0.15cm) -- (0,0.3);
\draw (0,0.6) -- (-0.4,0.6) arc (-90:-270:0.15cm) -- (0,0.9);
\end{tikzpicture}} 
\end{array}=\ d^2\,, \qquad \langle e_0|e_1\rangle \ = \ \begin{array}{c}\scalebox{0.6}{
\begin{tikzpicture}[very thick]
\fill[black] (0,0.0) circle (0.05cm);
\fill[black] (0,0.3) circle (0.05cm);
\fill[black] (0,0.6) circle (0.05cm);
\fill[black] (0,0.9) circle (0.05cm);
\draw (0,0) -- (0.1,0) arc (-90:90:0.45cm) -- (0,0.9);
\draw (0,0.6) -- (0.1,0.6) arc (90:-90:0.15cm) -- (0,0.3);
\draw (0,0) -- (-0.4,0) arc (-90:-270:0.15cm) -- (0,0.3);
\draw (0,0.6) -- (-0.4,0.6) arc (-90:-270:0.15cm) -- (0,0.9);
\end{tikzpicture}} 
\end{array}
=\ d\,, \qquad \langle e_1|e_1\rangle \ = \ \begin{array}{c}\scalebox{0.6}{
\begin{tikzpicture}[very thick]
\fill[black] (0,0.0) circle (0.05cm);
\fill[black] (0,0.3) circle (0.05cm);
\fill[black] (0,0.6) circle (0.05cm);
\fill[black] (0,0.9) circle (0.05cm);
\draw (0,0) -- (0.1,0) arc (-90:90:0.45cm) -- (0,0.9);
\draw (0,0.6) -- (0.1,0.6) arc (90:-90:0.15cm) -- (0,0.3);
\draw (0,0) -- (-0.1,0) arc (-90:-270:0.45cm) -- (0,0.9);
\draw (0,0.6) -- (-0.1,0.6) arc (90:270:0.15cm) -- (0,0.3);
\end{tikzpicture}} 
\end{array}
= \ d^2\,.
\ee
In other words, the basis is not orthogonal. It is straightforward to construct an orthonormal basis using Gram-Schmidt procedure. For example, the following linear combination of the basis vectors gives two normalized orhogonal vectors:
\be
\label{4pobasis}
|0\rangle 
\equiv \frac{1}{d}\begin{array}{c}\scalebox{0.6}{
\begin{tikzpicture}[very thick]
\fill[black] (0,0.0) circle (0.05cm);
\fill[black] (0,0.3) circle (0.05cm);
\fill[black] (0,0.6) circle (0.05cm);
\fill[black] (0,0.9) circle (0.05cm);
\draw (0,0) -- (0.4,0) arc (-90:90:0.15cm) -- (0,0.3);
\draw (0,0.6) -- (0.4,0.6) arc (-90:90:0.15cm) -- (0,0.9);
\end{tikzpicture}} 
\end{array}\,,\qquad |1\rangle 
\equiv \frac{1}{\sqrt{d^2-1}} \left(\begin{array}{c}\scalebox{0.6}{
\begin{tikzpicture}[very thick]
\fill[black] (0,0.0) circle (0.05cm);
\fill[black] (0,0.3) circle (0.05cm);
\fill[black] (0,0.6) circle (0.05cm);
\fill[black] (0,0.9) circle (0.05cm);
\draw (0,0) -- (0.1,0) arc (-90:90:0.45cm) -- (0,0.9);
\draw (0,0.6) -- (0.1,0.6) arc (90:-90:0.15cm) -- (0,0.3);
\end{tikzpicture}} 
\end{array}-\frac{1}{d}\begin{array}{c}\scalebox{0.6}{
\begin{tikzpicture}[very thick]
\fill[black] (0,0.0) circle (0.05cm);
\fill[black] (0,0.3) circle (0.05cm);
\fill[black] (0,0.6) circle (0.05cm);
\fill[black] (0,0.9) circle (0.05cm);
\draw (0,0) -- (0.4,0) arc (-90:90:0.15cm) -- (0,0.3);
\draw (0,0.6) -- (0.4,0.6) arc (-90:90:0.15cm) -- (0,0.9);
\end{tikzpicture}} 
\end{array}\right).
\ee

Finally, we implicitly assumed that the basis of Temperley-Lieb diagrams is nondegenerate, that is all the basis vectors are linearly independent. In the example of the 4-point Hilbert space the Gram matrix shows that it is not always the case. Namely,
\be
\det\langle e_i|e_j\rangle \ = \ d^2(d^2-1)
\ee
vanishes for a few special values, $d=0,\pm 1$. We will further discuss the degeneracy of the Temperley-Lieb basis in the next section. In general, apart from a finite number of special values of parameters $d$ or $A$, one can indeed assume the Temperley-Lieb diagrams to be linearly independent.

We now have a complete description of a quantum mechanics in terms of topological spaces -- sets of open curves. An infinite, but countable set of states in this model have a diagrammatic presentation. Generic states, however, are only linear combinations of the diagrammatic states. Kauffman's prescription for computing the bracket polynomial provides us with the technique to compute arbitrary overlaps of quantum states and, consequently, matrix elements of any operators represented topologically, such as one in equation~(\ref{knotasoverlap}).

In the next section we will give a three-dimensional definition of the same quantum mechanical model, which will allow us to gain additional intuition about the topological presentation. One subtlety, highlighted by the three-dimensional version, is the calculation of traces of the operators. In the one-dimensional case, calculation of trace is not intuitive: Closing a braid (an operator) in a tracelike manner, computes the Markov trace, rather then the ordinary trace. The regular trace should instead be computed as a sum of all diagonal elements. Embedding the construction in three dimensions clarifies the difference.  

%%%%%%%%%%%%%%%%%%%%%%%%%%%%%%%%%%%%%%%%%%%%%%%%%%%%%%%%%%%%%%%%%%%%

\subsection{Chern-Simons theory}
\label{sec:CS}

In this section we will briefly review how the model described in the previous section emerges from a three-dimensional quantum gauge theory. The review will be rather imprecise and its understanding is not necessary for discussion of the applications in the next section, but it might provide some additional intuition about the connection between topology, quantum mechanics gauge and conformal field theories (CFT). The main reference here is the work of Edward Witten on Chern-Simons theory and Jones polynomials~\cite{Witten:1988ze}. Further details can be found in lecture notes~\cite{Dunne:1998qy}.

Nonabelian Chern-Simons theory with gauge gauge group $G$ (which we will latter assume to be $SU(2)$) is given by the action
\be
\label{CSaction}
S_{\rm CS} \ = \ \frac{k}{4\pi}\int d^3x\ \epsilon^{\mu\nu\rho}\,\tr\!\left(A_\mu\partial_\nu A_\rho + \frac{2}{3}A_\mu A_\nu A_\rho\right),
\ee
where the trace is evaluated in the fundamental representation of $G$. Note that this covariant action is written without any use of the metric and therefore has vanishing stress-energy tensor, and consequently, Hamiltonian, according to~(\ref{energymomentum}).

The equations of motion of this theory state that the covariant field tensor is zero:
\be
F_{\mu\nu} \ = \ 0\,,
\ee
which implies that all the dynamical field configurations are pure gauge. The theory appears to be trivial, with no physical degrees of freedom.

To make this theory more interesting, one can couple the gauge field to some matter degrees of freedom. The simplest case is the one of coupling to nondynamical particles, which allows to keep the theory sufficiently simple, introducing some nontrivial observables at the same time. 

One way of doing it is to consider singular gauge transformations. Let us think that the Chern-Simons theory is defined in the space $I\times \Sigma$, where $I$ is a time interval and $\Sigma$ is a closed two-dimensional surface. In the gauge $A_0=0$, the equation of motion, corresponding to the variation with respect to $A_0$, becomes a constraint $F_{12}=0$, with indices $1$ and $2$ corresponding to the coordinates on $\Sigma$. There are two types of gauge transformations in the theory. The \emph{small} gauge transformations are the true ones -- they do not modify the constraint. The \emph{large} gauge transformations are singular -- they can introduce delta-function sources in the constraint:
\be
\label{GaussLaw}
\frac{k}{2\pi}F_{12}^a \ = \ \sum\limits_i T^a_{(i)}\delta^{(2)}(x-x_i)\,.
\ee
Here $x_i$ are the locations of the pointlike sources, which carry nonabelian charges characterized by the group generators $T^a$. 

This situation is similar to the Aharonov-Bohm effect. The field is zero everywhere except a finite number of points, where a finite flux of the field is localized. Since $\Sigma$ is closed the net flux should be zero. For the abelian case this would correspond to the condition of a zero net charge (on closed $\Sigma$ all the field lines should end somewhere). In the nonabelian case zero net charge is generalized to the condition of the trivial net representation of the group.

Equation~(\ref{GaussLaw}) can also be obtained from a theory modified by the explicit addition of the sources into the action:
\be
S_{\rm CS} \ \to \ S_{\rm CS} + \sum\limits_i\int_{\gamma_i} A_\mu\, dx^\mu\,.
\ee
(Singular gauge transformations essentially add such terms.)

One can canonically quantize Chern-Simons fields on $\Sigma$. The classical constraint~(\ref{GaussLaw}) needs to be imposed as an operator equation on the Hilbert space. The solution to this quantum constraint reduces the Hilbert space to a finite dimensional vector space labeled by classes of large gauge transformations, specifically, by the number of the sources on $\Sigma$ and by the associated representations. As above, the net representation of the sources is constrained to be trivial. 

It is also useful to discuss the path integral quantization of the Chern-Simons theory. Consider the path integral of a theory defined inside a three-dimensional space $M$ with boundary $\Sigma$. If, as before, the time direction is chosen orthogonally to $\Sigma$, then such a path integral can be thought of as producing a state in a Hilbert space:
\be
\label{PIstate}
 \Psi(\Sigma) \ = \ \int {\mathcal D}A\Big|_{A(\Sigma)=A_\Sigma} {\rm e}^{iS_{\rm CS}[M]}.
\ee
Here the path integral is calculated over field configurations in $M$ satisfying fixed boundary conditions on $\Sigma$. This is the standard state preparation by the Euclidean path integral, although Chern-Simons theory, being metric independent, does not distinguish between the Euclidean and Minkowski evolution.

Small gauge transformations leave the action, and the boundary conditions invariant. The integral over such configurations cancels in normalized vectors. Large gauge transformation produce sources and modify the boundary conditions according to the constraint~(\ref{GaussLaw}). Let us choose $\Sigma$ to be a 2-sphere $S^2$. We will also assume the gauge group to be $SU(2)$  and the representations of all sources to be fundamental. In the latter case the sources are nondynamical particles with spin $j=1/2$. 

Consider a boundary condition that prescribes four fundamental sources on $\Sigma$. Then space $M$ should be something like
\be
\label{CSstate1}
 \begin{array}{c}
     \includegraphics[scale=0.2]{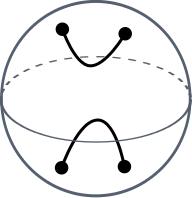} 
\end{array},
\ee
where the interior of the three-ball contains lines $\gamma_i$ connecting the boundary sources. However, some of the large gauge transformations may preserve the condition~(\ref{GaussLaw}) but modify the topology of lines $\gamma_i$ in a singular (nonsmooth) way. For example, one may also think of $M$ as
\be
\label{CSstate2}
\begin{array}{c}
     \includegraphics[scale=0.2]{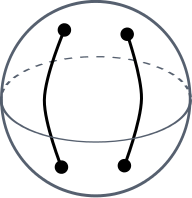} 
\end{array},
\ee
with an alternative way of connecting the sources. Indeed any modification of $\gamma$ in the coupling term
\be
\int_\gamma A_\mu dx^\mu
\ee
does not change the action unless another line or another part of $\gamma$ is crossed.

We conclude that topologically distinct ways of connecting the points on the sphere through its interior produce a priori distinct quantum states~(\ref{PIstate}). The connection with the one-dimensional TQFT of the previous section must be clear now. One can establish a map between the two models as follows:
\be
\begin{array}{c}
     \includegraphics[scale=0.2]{./figs/qubitbasis2.png} 
\end{array} \ \to \ \begin{array}{c}\scalebox{1}{
\begin{tikzpicture}[thick]
\draw[gray,opacity=0.5] (0,-0.2) -- (0,1.1);
\fill[black] (0,0.0) circle (0.05cm);
\fill[black] (0,0.3) circle (0.05cm);
\fill[black] (0,0.6) circle (0.05cm);
\fill[black] (0,0.9) circle (0.05cm);
\draw (0,0) -- (0.4,0) arc (-90:90:0.15cm) -- (0,0.3);
\draw (0,0.6) -- (0.4,0.6) arc (-90:90:0.15cm) -- (0,0.9);
\end{tikzpicture}} 
\end{array}\,,\qquad 
\begin{array}{c}
     \includegraphics[scale=0.2]{./figs/qubitbasis1.png} 
\end{array} \to \ \begin{array}{c}\scalebox{1}{
\begin{tikzpicture}[thick]
\draw[gray,opacity=0.5] (0,-0.2) -- (0,1.1);
\fill[black] (0,0.0) circle (0.05cm);
\fill[black] (0,0.3) circle (0.05cm);
\fill[black] (0,0.6) circle (0.05cm);
\fill[black] (0,0.9) circle (0.05cm);
\draw (0,0) -- (0.1,0) arc (-90:90:0.45cm) -- (0,0.9);
\draw (0,0.6) -- (0.1,0.6) arc (90:-90:0.15cm) -- (0,0.3);
\end{tikzpicture}} 
\end{array}.
\ee
This mapping is not unique and is fixed by ordering the sources on the sphere.

In order to construct a more precise map, and to see that states corresponding to the situation with four fundamental sources generate a two-dimensional vector space one has to provide more details about the quantization. It turns out that the action of the Chern-Simons theory with boundary $\Sigma$ can be reduced to a two-dimensional action on $\Sigma$. The resulting theory is called Wess-Zumino-Witten (WZW) theory -- a well-studied example of a CFT~\cite{Dunne:1998qy,Wess:1971yu,Witten:1983tw,Witten:1983ar,Gawedzki:1999bq}. 

For $G=SU(2)$ the spectrum of the WZW theory consists of operators labeled by \emph{integrable} representations of an associated $su(2)_k$ Kac-Moody algebra. This is equivalent to saying that the theory on $\Sigma$ only has particles with spin $j$ taking half-integer values between $0$ and $k/2$. Moreover, the particles are subject to the constraint of net zero spin.

Operators in the CFT satisfy the so-called fusion algebra, which is a particular form of the operator product expansion expressing the product of a pair of operators as an expansion in an appropriate operator basis:
\be
\Phi_k\cdot \Phi_l\ = \ \oplus_j N_{kl}^j \Phi_j\,. 
\ee
Here $\Phi_j$ represent conformal classes of the operators (operators modulo conformal transformations). In the WZW theory the basis is provided by the integrable representations, hence the fusion algebra is similar to the expansion of the tensor product of representations in the basis of the irreducible ones. The difference is that in the WZW theory only representations with $j\leq k/2$ are allowed. Numbers $N_{kl}^j$ are the multiplicities of the representations with spin $j$. Note that the irreducible representations that appear multiple times correspond to linearly independent operators.

Because of the net zero spin constraint, the product of all representations on $\Sigma$ must be projected on the subspace spanned by the trivial representations. For four spin $1/2$ this subspace has dimension two, in accord with the dimension of the space spanned by diagrams~(\ref{4pbasis}). Therefore any three vectors obtained from a path integral on $M$ with $\Sigma=S^2$ are linearly dependent. WZW data permit calculating the coefficients, cf.~\cite{Witten:1988ze}:\footnote{This is a $SU(2)$ result. The situation is more complicated in more general cases. See a recent discussion in~\cite{Anokhina:2024lbn,Anokhina:2024uso}.}
\be
-q\begin{array}{c}
     \includegraphics[scale=0.3]{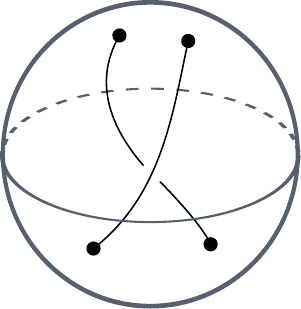} 
\end{array} 
+
q\begin{array}{c}
     \includegraphics[scale=0.3]{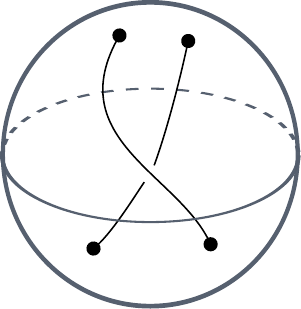} 
\end{array} 
+
(q^{1/2}-q^{-1/2})\begin{array}{c}
     \includegraphics[scale=0.3]{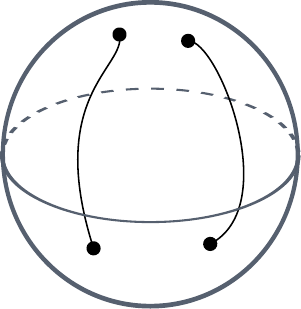} 
\end{array} 
=0\,,
\ee
where $q$ is determined in terms of the parameter $k$:
\be
\label{q}
q \ = \ \exp\left({\frac{2\pi i}{k+2}}\right)\,.
\ee
This is a slightly different version of skein relation~(\ref{skein}), or rather~(\ref{skein2}), which directly computes the Jones polynomial, without taking into account the framing factor. In order to derive~(\ref{skein2}) one needs to compensate for the framing factor rescaling the crossings, namely, in~(\ref{skein}) and~(\ref{iskein}) replace
\be
\begin{array}{c}
     \begin{tikzpicture}[thick]
     \draw (0,0.5) -- (0.3,0);
     \draw[line width=3.5,white] (0,0) -- (0.3,0.5);
     \draw (0,0) -- (0.3,0.5);
     \end{tikzpicture}
\end{array}
\longrightarrow
-A^3 \begin{array}{c}
     \begin{tikzpicture}[thick]
     \draw (0,0.5) -- (0.3,0);
     \draw[line width=3.5,white] (0,0) -- (0.3,0.5);
     \draw (0,0) -- (0.3,0.5);
     \end{tikzpicture}
\end{array}\,,
\qquad
\begin{array}{c}
     \begin{tikzpicture}[thick]
     \draw (0.3,0.5) -- (0.0,0);
     \draw[line width=3.5,white] (0.3,0) -- (0.0,0.5);
     \draw (0.3,0) -- (0.0,0.5);
     \end{tikzpicture}
\end{array}
\longrightarrow
-A^{-3}\begin{array}{c}
     \begin{tikzpicture}[thick]
     \draw (0.3,0.5) -- (0.0,0);
     \draw[line width=3.5,white] (0.3,0) -- (0.0,0.5);
     \draw (0.3,0) -- (0.0,0.5);
     \end{tikzpicture}
\end{array}\,,
\ee
and matching the parameter as follows:
\be
\label{CSparameters}
A \ = \ q^{1/4}\,, \qquad  d \ = \ -2\cos\left(\frac{\pi}{k+2}\right)\,.
\ee

Note that diagrams~(\ref{4pbasis}) can be interpreted as two independent ways of fusing spin $1/2$ particles into singlets. When viewed as evolution, the lines of four spin $1/2$ particles coalesce and disappear, which is interpreted as forming a singlet. An ``orthogonal'' possibility for the diagram of state $|e_0\rangle\sim |0\rangle$ is 
\be
\label{confblock}
|1\rangle \ \sim \ \begin{array}{c}\scalebox{1}{
\begin{tikzpicture}[thick]
\draw[gray,opacity=0.5] (0,-0.2) -- (0,1.1);
\fill[black] (0,0.0) circle (0.05cm);
\fill[black] (0,0.3) circle (0.05cm);
\fill[black] (0,0.6) circle (0.05cm);
\fill[black] (0,0.9) circle (0.05cm);
\draw (0,0) -- (0.4,0) arc (-90:90:0.15cm) -- (0,0.3);
\draw (0,0.6) -- (0.4,0.6) arc (-90:90:0.15cm) -- (0,0.9);
\draw[dashed] (0.55,0.15) -- (0.95,0.15) arc (-90:90:0.3cm) -- (0.55,0.75);
\end{tikzpicture}} 
\end{array}.
\ee
In the latter, two pairs of particles form a spin $1$ state instead of the singlet, and the pair of spin $1$ states (dashed line) later annihilate. State $|1\rangle$ can be expressed in terms of a pair of non-orthogonal diagrams according to~(\ref{4pobasis}). 

When analyzing the Gram matrix of states~(\ref{4pbasis}) we observed that it is degenerate if $d=-1$. For this value state $|1\rangle$ becomes null (has zero norm). On the other hand this value of $d$ corresponds to $k=1$, according to~(\ref{CSparameters}). The WZW theory does not have an integrable representation of spin $j=1$ in this case. 

For the Temperley-Lieb basis of $2n$ points the Gram matrix will be degenerate for values of $k<n$, in consistence with the integrable representations of the WZW theory. In this case the actual dimension of the Hilbert space associated with path integral~(\ref{PIstate}) is smaller than the naive dimension given by the Catalan number $C_n$. Conversely, if $k$ is kept sufficiently large, one can use $C_n$ as the correct value of the dimension. One can see that the simple one-dimensional TQFT model of section~\ref{sec:TQM} reproduces the basic features of the WZW CFT.

Diagrams similar to that of state $|0\rangle$ in~(\ref{4pobasis}) and state $|1\rangle$ in~(\ref{confblock}) are sometimes referred as conformal blocks of a relevant CFT. A more conventional way of drawing them is
\be
\label{conformalblock}
\begin{array}{c}
     \begin{tikzpicture}[thick]
         \draw (0,0) -- (0.75,0) node[anchor=south] {$j=0,1$} -- (1.5,0);
         \draw (-0.5,0.5) node[anchor=east] {$1/2$} -- (0,0) -- (-0.5,-0.5) node[anchor=east] {$1/2$};
         \draw (2,0.5) node[anchor=west] {$1/2$} -- (1.5,0) -- (2,-0.5) node[anchor=west] {$1/2$};
     \end{tikzpicture} 
\end{array}.
\ee
More precisely, conformal blocks are basis elements in the expansion of a correlation function (here 4-point) in a CFT, labeled by conformal classes (primary fields) in the intermediate channels of tree diagrams.

Finally, path integral representation also provides an explicit realization of a three-dimensional TQFT functor. Equation~(\ref{PIstate}) is implicitly a map from $M$ to a vector in a linear space. It also fixes the map between $\Sigma$ and a Hilbert space, as we explained. A scalar product of two states can be computed as a product of two functional integrals integrated over the boundary condition $A_\Sigma$:
\be
\langle\Phi|\Psi\rangle \ \sim \ \int\Dc A_\Sigma \int {\mathcal D}A\Big|_{A(\Sigma)=A_\Sigma}\int {\mathcal D}\bar{A}\Big|_{\bar{A}(\Sigma)=A_\Sigma}{\rm e}^{iS_{\rm CS}[M_1\cup M_2]}\,.
\ee
This is equivalent to a functional integral over a closed three-dimensional space, which produces a $\mathbb{C}$-number, as expected.

As an example, consider an overlap of two Chern-Simons states~$(\ref{CSstate1})$ and~$(\ref{CSstate2})$:
\be
\left\langle\begin{array}{c}
     \includegraphics[scale=0.2]{./figs/qubitbasis2.png} 
\end{array}\Bigg|\begin{array}{c}
     \includegraphics[scale=0.2]{./figs/qubitbasis1.png} 
\end{array}\right\rangle \ = \ \begin{array}{c}
     \includegraphics[scale=0.25]{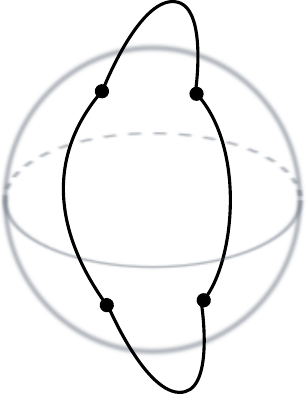} 
\end{array}. 
\ee
The result is a path integral with an insertion of a circular Wilson loop. The integral is computed over a theory defined in a closed 3-manifold obtained by gluing two 3-balls along an $S^2$ boundary. Such a 3-manifold is the sphere $S^3$. (It is a higher-dimensional generalization of gluing two discs along an $S^1$ to obtain $S^2$.) Hence, overlaps of states like~$(\ref{CSstate1})$ and~$(\ref{CSstate2})$ compute expectation values of Wilson loops in Chern-Simons theory.

It is the main result of~\cite{Witten:1988ze} that expectation values of knotted fundamental Wilson loops in the $SU(2)$ Chern-Simons theory compute Jones polynomials of the corresponding knots. For Wilson lines in other representations one obtains a generalization called \emph{colored Jones polynomials}, while for $G=SU(N)$ one obtains the HOMFLY-PT polynomials~\cite{Freyd:1985dx,Przytycki:2016inv}. 

One can also consider other examples of boundaries. A fundamental example is $\Sigma=T^2$ (two-dimensional torus). In this case a nondegenerate basis is provided by Wilson loops with all half-integer spins between $0$ and $k/2$ winding the noncontractible cycle of the torus. Quantum mechanics and quantum information applications based on this kind of states have been studied in many recent papers, e.g.~\cite{Salton:2016qpp,Balasubramanian:2016sro,Balasubramanian:2018por,Chun:2017hja,Dwivedi:2017rnj,Camilo:2019bbl,Buican:2019evc,Dwivedi:2020rlo,Fliss:2020yrd,Leigh:2021trp,Dwivedi:2021dix,Caputa:2024qkk,Ramirez-Valdez:2024rhf,Dwivedi:2024gzg,Balasubramanian:2025kaf}. This approach gives a direct association of a knot with a quantum state, but we will follow a different route in this review.

Another generalization is the case of multiple $S^2$ boundaries~\cite{Melnikov:2018zfn,Dwivedi:2019bzh,Melnikov:2022qyt,Melnikov:2023wwc,Chaves:2024rjs}. According to the TQFT axioms, these correspond to tensor products of Hilbert spaces discussed above. Examples of states in a tensor product of two 4-point spheres are given by the following diagrams:
\be
\begin{array}{c}
\includegraphics[height=0.1\linewidth]{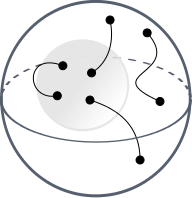}
\end{array}\,,
\qquad
\begin{array}{c}
     \includegraphics[height=0.1\linewidth]{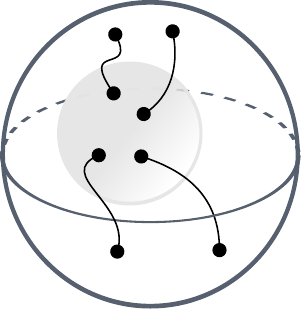} 
\end{array}\,.
\ee
The functional integral is computed over the filling of the space between two spheres. Wilson lines can connect to the same boundary $S^2$, or can connect different spheres. We can also map the above states to the one-dimensional model:
\be
\label{2qubitstate}
\begin{array}{c}
     \includegraphics[height=0.1\linewidth]{./figs/2qubit.pdf} 
\end{array} 
\quad \to \quad
\begin{array}{c}
     \begin{tikzpicture}
     \newcommand{\y}{1.8}
        \draw[draw=white,double=black,line width=1,rounded corners=2] (-0.45,0) -- (-0.45,-0.7) -- (0.45+\y,-0.7) -- (0.45+\y,0);
        \draw[draw=white,double=black,line width=1,rounded corners=2] (-0.15,0) -- (-0.15,-0.6) -- (0.15+\y,-0.6) -- (0.15+\y,0);
        \draw[draw=white,double=black,line width=1,rounded corners=2] (0.15,0) -- (0.15,-0.5) -- (-0.15+\y,-0.5) -- (-0.15+\y,0);
         \draw[draw=white,double=black,line width=1,rounded corners=2] (0.45,0) -- (0.45,-0.4) -- (-0.45+\y,-0.4) -- (-0.45+\y,0);
         \draw[gray,opacity=0.8,line width=2] (-0.75,0) -- (0.75,0);
         \foreach \x in {-0.45,-0.15,...,0.75}
           \fill[black] (\x,0) circle (0.075);
          \draw[gray,opacity=0.8,line width=2] (-0.75+\y,0) -- (0.75+\y,0); 
          \foreach \x in {-0.45,-0.15,...,0.75}
           \fill[black] (\x+\y,0) circle (0.075);
     \end{tikzpicture} 
\end{array} \ = \ |\Phi\rangle \,.
\ee

At the end of the previous section we mentioned a subtlety with evaluating traces of operators in the one-dimensional model. Note that the state shown in~(\ref{2qubitstate}) can also be viewed as an operator acting on a single copy of the 4-point Hilbert space. The trace of this operator is identical to the norm squared of the state. In the one-dimensional model the trace or the norm squared may naively appear as a tracelike closure of the diagram. Computing it in the three-dimensional theory gives
\be
\left\langle\begin{array}{c}
     \includegraphics[scale=0.25]{./figs/2qubit.pdf} 
\end{array}\Bigg|\begin{array}{c}
     \includegraphics[scale=0.25]{./figs/2qubit.pdf} 
\end{array}\right\rangle \ = \ \begin{array}{c}
     \includegraphics[scale=0.2]{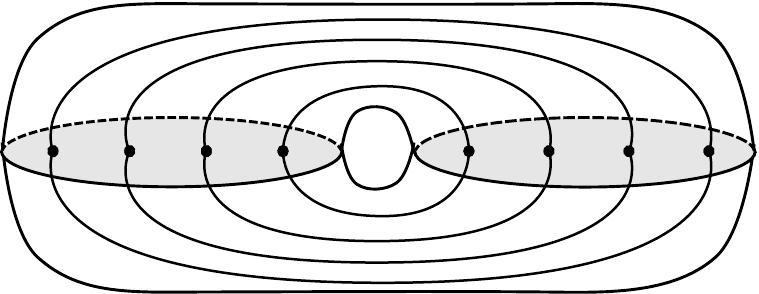} 
\end{array} \,.
\ee
Identifying the interior and the exterior spheres gives a 3-manifold different from $S^3$. It is not hard to see that this manifold is $S^2\times S^1$. It is not possible to draw the latter, so we rather showed it as a kind of a solid torus. Unlike the solid torus $S^2\times S^1$ does not have a boundary. It is topological invariants in $S^2\times S^1$ that compute the normal trace of an operator, while the invariants in $S^3$ (and the naive tracelike closures) compute the Markov traces.

%%%%%%%%%%%%%%%%%%%%%%%%%%%%%%%%%%%%%%%%%%%%%%%%%%%%%%%%%%%%%%%%%%%%
\subsection{TQFT in the Semiclassical Limit}
\label{sec:classlimit}

In Chern-Simons theory, the coupling constant $k$ (also called the \emph{level} of Chern-Simons theory) appears as a factor multiplying the Lagrangian, e.g.~(\ref{CSaction}). Hence, it has a role similar to the inverse Planck constant. In particular, the limit $k\to \infty$ localizes the path integral on classical configurations, computing the semiclassical wavefunction~(\ref{PIstate}). It is instructive to learn what happens with the topological model in this limit.

First, equations~(\ref{q}) and~(\ref{CSparameters}) instruct us that the classical limit corresponds to taking $A\to1$. In this limit the skein relation takes a symmetric form:
\be
\label{classskein}
\begin{array}{c}
     \begin{tikzpicture}[thick]
     \draw (0,0.5) -- (0.3,0);
     \draw[line width=3.5,white] (0,0) -- (0.3,0.5);
     \draw (0,0) -- (0.3,0.5);
     \end{tikzpicture}
\end{array}
\ = \ 
\begin{array}{c}
     \begin{tikzpicture}[thick]
     \draw (0,0) -- (0,0.5);
     \draw (0.3,0) -- (0.3,0.5);
     \end{tikzpicture}
\end{array}
+ 
\begin{array}{c}
     \begin{tikzpicture}[thick]
     \draw (0,0) -- (0,0.05) arc (180:0:0.15) -- (0.3,0);
     \draw (0,0.5) -- (0,0.45) arc (180:360:0.15) -- (0.3,0.5);
     \end{tikzpicture}
\end{array}
\ = \ 
\begin{array}{c}
     \begin{tikzpicture}[thick]
     \draw (0.3,0.5) -- (0.0,0);
     \draw[line width=3.5,white] (0.3,0) -- (0.0,0.5);
     \draw (0.3,0) -- (0.0,0.5);
     \end{tikzpicture}
\end{array},
\ee
that is, there is no difference between crossing types and the lines can pass through each other at no cost. Braid generators~(\ref{R}) take the form of  permutation generators. In other words, no knots can be formed in the classical limit.

This has an important consequence for the type of states, for which there exists a topological presentation. For states like the one of equation~(\ref{stateXX}), what really matters in the classical limit is the information about which point is connected to which, while the mutual relations of different connections are unimportant. Many states that are different for finite $k$ become indistinguishable in the classical limit -- a sort of a classical decoherence of the quantum state. The states split into a set of infinite equivalence classes generated by relation~(\ref{classskein}). The situation can be compared with neuroscience, where one builds maps of a brain, indicating which neuron is connected to which. Such maps are called \emph{connectomes}. Here we can call connectomes similar topological equivalence classes of quantum states. 

In the classical limit one does not need to worry about possible constraints imposed by the dependence of the calculus of $k$, such as degeneracy of the basis state like~(\ref{4pobasis}) and corrections to formula~(\ref{Catalan}) for the dimension of the Hilbert space. It is also interesting to consider another limit of this formula, for very large number of points, $n\to\infty$. One finds
\be
C_{n/2} \ \sim \ \frac{2^{n}}{(n/2)^{3/2}\sqrt{\pi}}\,, \qquad n\ \to \ \infty\,. 
\ee
The leading behavior of the dimension is that of the Hilbert space of $n$ qubits, or spins $j=1/2$. In the classical limit and for large number $n$ one can think of a line as the evolution of a single $j=1/2$ particle, as in the matrix representation of section~\ref{sec:matrix}. The logarithm of the dimension becomes an almost linear function of the number of spins.

%%%%%%%%%%%%%%%%%%%%%%%%%%%%%%%%%%%%%%%%%%%%%%%%%%%%%%%%%%%%%%%%%%%%
\section{Quantum Entanglement}
\label{sec:entanglement}

In the previous section we described an approach in which some quantum states can be understood as topological configurations (spaces). In this section we will show how this approach can be useful to understand and interpret fundamental concepts of quantum mechanics. The main subject of the discussion will be quantum entanglement.

The idea that entanglement can be thought as of topological tangling has been in the market for a while. Intuitive, but not rigorous, diagrammatic depiction of systems entangled in different ways is common in quantum information literature, cf.~\cite{Coecke:2017pic}. Proposal of Aravind~\cite{Aravind:1997bor}, comparing features of the topology of the Borromean rings~(\ref{Borromean}) with the Greenberger-Horne-Zeilinger (GHZ) state is often cited in this relation. (Observe that removing any of the rings in the Borromean link makes the other two disconnected, while measuring any of the qubits of the GHZ state in the computational basis, makes the other two separable.) Some early mathematical approach to the subject, which can be heuristically called ``entanglement=topology'', is reviewed in~\cite{Kauffman:2013bh}. More recently, a specific program to study different types of entanglement using Chern-Simons theory states has been carried out in~\cite{Salton:2016qpp,Balasubramanian:2016sro,Balasubramanian:2018por,Chun:2017hja,Dwivedi:2017rnj,Buican:2019evc,Dwivedi:2020rlo,Dwivedi:2021dix,Ramirez-Valdez:2024rhf,Melnikov:2018zfn,Dwivedi:2019bzh,Melnikov:2022qyt,Melnikov:2023wwc,Dwivedi:2024gzg,Balasubramanian:2025kaf} developing upon the ideas explored in~\cite{Dong:2008ft}.

%%%%%%%%%%%%%%%%%%%%%%%%%%%%%%%%%%%%%%%%%%%%%%%%%%%%%%%%%%%%%%%%%%%%
\subsection{Topological Tangling and Quantum Entanglement}
\label{sec:tangling}

Quantum entanglement is a property of quantum mechanical systems quantified by correlations. A canonical example of quantum entanglement is the correlation in the Einstein-Podolsky-Rosen (EPR) or Bell pair~\cite{Bohm:1989book}. An EPR-Bell pair can be represented by a pair of spins in the state
\be
\label{Bellstate}
|\Phi^+\rangle \ := \ \frac{1}{\sqrt{2}}\left(|0\rangle\otimes |0\rangle + |1\rangle\otimes |1\rangle\right).
\ee
Either of the spins can be measured in the state $|0\rangle$ or $|1\rangle$ with a 50 percent probability. The correlation is exhibited in the property that if one spin is measured in either of the two states, the state of the other, upon the first measurement, will be known with a 100 percent probability. 

As a first take let us discuss how an EPR-Bell pair can be described in a topological setup. For this let us explicitly compute the coefficients of state~(\ref{2qubitstate}) in the standard spin-orientation (computational) basis using methods of section~\ref{sec:TQM}.

In order to compute the coefficient of the basis element $|0\rangle\otimes |0\rangle$ we need to glue two $|0\rangle$ states from basis~(\ref{4pobasis}) to the open ends of state~(\ref{2qubitstate}).\footnote{From now on we will use simplified notations for the tensor products, e.g. $|00\rangle\equiv |0\rangle\otimes|0\rangle$.} The result is
\be
\label{coefs1}
\langle 00|\Phi\rangle \ =
    \frac{1}{d^2}\begin{array}{c}\scalebox{0.8}{
        \begin{tikzpicture}[thick]
            \newcommand{\y}{2}
         \newcommand{\z}{0}
                  \draw[rounded corners=2] (-0.45,0-\z) -- (-0.45,-0.7-\z) -- (0.45+\y,-0.7-\z) -- (0.45+\y,0-\z);
         \draw[rounded corners=2] (-0.15,0-\z) -- (-0.15,-0.6-\z) -- (0.15+\y,-0.6-\z) -- (0.15+\y,0-\z);
         \draw[rounded corners=2] (0.15,-\z) -- (0.15,-0.5-\z) -- (-0.15+\y,-0.5-\z) -- (-0.15+\y,0-\z);
         \draw[rounded corners=2] (0.45,0-\z) -- (0.45,-0.4-\z) -- (-0.45+\y,-0.4-\z) -- (-0.45+\y,0-\z);
        
         \draw[gray,opacity=0.3,line width=2] (-0.75,0) -- (0.75,0);
         \foreach \x in {-0.45,-0.15,...,0.75}
         \fill[black] (\x,-\z) circle (0.075);
         
         \draw[gray,opacity=0.3,line width=2] (-0.75+\y,0) -- (0.75+\y,0);
         \foreach \x in {-0.45,-0.15,...,0.75}
         \fill[black] (\x+\y,-\z) circle (0.075);
         
         \draw[rounded corners=2] (-0.45,-\z) -- (-0.45,0.6-\z) -- (-0.15,0.6-\z) -- (-0.15,-\z);
         \draw[rounded corners=2] (0.45,-\z) -- (0.45,0.6-\z) -- (0.15,0.6-\z) -- (0.15,-\z);
         
         \draw[rounded corners=2] (-0.45+\y,-\z) -- (-0.45+\y,0.6-\z) -- (-0.15+\y,0.6-\z) -- (-0.15+\y,-\z);
         \draw[rounded corners=2] (0.15+\y,-\z) -- (0.15+\y,0.6-\z) -- (0.45+\y,0.6-\z) -- (0.45+\y,-\z);
        \end{tikzpicture}} 
    \end{array} \ = \ \frac{d^2}{d^2} \ =  \ 1\,.
\ee
It is not much harder to compute the coefficient of $|01\rangle$. The only difference is that state $|1\rangle$ is given by a linear combination, so there will be two terms:
\be
\label{coefs2}
\langle 01|\Phi\rangle  = 
    \frac{1}{\sqrt{d^2-1}} 
    \left(\frac{1}{d}\!\begin{array}{c}
        \scalebox{0.6}{
        \begin{tikzpicture}[thick]
            \newcommand{\y}{2}
         \newcommand{\z}{0}     
         \draw[rounded corners=2] (-0.45,0-\z) -- (-0.45,-0.7-\z) -- (0.45+\y,-0.7-\z) -- (0.45+\y,0-\z);
         \draw[rounded corners=2] (-0.15,0-\z) -- (-0.15,-0.6-\z) -- (0.15+\y,-0.6-\z) -- (0.15+\y,0-\z);
         \draw[rounded corners=2] (0.15,-\z) -- (0.15,-0.5-\z) -- (-0.15+\y,-0.5-\z) -- (-0.15+\y,0-\z);
         \draw[rounded corners=2] (0.45,0-\z) -- (0.45,-0.4-\z) -- (-0.45+\y,-0.4-\z) -- (-0.45+\y,0-\z);
        
         \draw[gray,opacity=0.3,line width=2] (-0.75,0) -- (0.75,0);
         \foreach \x in {-0.45,-0.15,...,0.75}
         \fill[black] (\x,-\z) circle (0.075);
         
         \draw[gray,opacity=0.3,line width=2] (-0.75+\y,0) -- (0.75+\y,0);
         \foreach \x in {-0.45,-0.15,...,0.75}
         \fill[black] (\x+\y,-\z) circle (0.075);

         \draw[rounded corners=2] (-0.45,-\z) -- (-0.45,0.6-\z) -- (-0.15,0.6-\z) -- (-0.15,-\z);
         \draw[rounded corners=2] (0.45,-\z) -- (0.45,0.6-\z) -- (0.15,0.6-\z) -- (0.15,-\z);
         
         \draw[rounded corners=2] (-0.45+\y,-\z) -- (-0.45+\y,0.6-\z) -- (0.45+\y,0.6-\z) -- (0.45+\y,-\z);
         \draw[rounded corners=2] (-0.15+\y,-\z) -- (-0.15+\y,0.5-\z) -- (0.15+\y,0.5-\z) -- (0.15+\y,-\z);
        \end{tikzpicture}} 
    \end{array} \! - \
    \frac{1}{d^2}\!\begin{array}{c}\scalebox{0.6}{
        \begin{tikzpicture}[thick]
            \newcommand{\y}{2}
         \newcommand{\z}{0}

         \draw[rounded corners=2] (-0.45,-\z) -- (-0.45,0.6-\z) -- (-0.15,0.6-\z) -- (-0.15,-\z);
         \draw[rounded corners=2] (0.45,-\z) -- (0.45,0.6-\z) -- (0.15,0.6-\z) -- (0.15,-\z);
         
         \draw[rounded corners=2] (-0.45+\y,-\z) -- (-0.45+\y,0.6-\z) -- (-0.15+\y,0.6-\z) -- (-0.15+\y,-\z);
         \draw[rounded corners=2] (0.15+\y,-\z) -- (0.15+\y,0.6-\z) -- (0.45+\y,0.6-\z) -- (0.45+\y,-\z);
         
         \draw[rounded corners=2] (-0.45,0-\z) -- (-0.45,-0.7-\z) -- (0.45+\y,-0.7-\z) -- (0.45+\y,0-\z);
         \draw[rounded corners=2] (-0.15,0-\z) -- (-0.15,-0.6-\z) -- (0.15+\y,-0.6-\z) -- (0.15+\y,0-\z);
         \draw[rounded corners=2] (0.15,-\z) -- (0.15,-0.5-\z) -- (-0.15+\y,-0.5-\z) -- (-0.15+\y,0-\z);
         \draw[rounded corners=2] (0.45,0-\z) -- (0.45,-0.4-\z) -- (-0.45+\y,-0.4-\z) -- (-0.45+\y,0-\z);
        
         \draw[gray,opacity=0.3,line width=2] (-0.75,0) -- (0.75,0);
         \foreach \x in {-0.45,-0.15,...,0.75}
         \fill[black] (\x,-\z) circle (0.075);
         
         \draw[gray,opacity=0.3,line width=2] (-0.75+\y,0) -- (0.75+\y,0);
         \foreach \x in {-0.45,-0.15,...,0.75}
         \fill[black] (\x+\y,-\z) circle (0.075);

          \draw[rounded corners=2] (-0.45,-\z) -- (-0.45,0.6-\z) -- (-0.15,0.6-\z) -- (-0.15,-\z);
          \draw[rounded corners=2] (0.45,-\z) -- (0.45,0.6-\z) -- (0.15,0.6-\z) -- (0.15,-\z);
         
         \draw[rounded corners=2] (-0.45+\y,-\z) -- (-0.45+\y,0.6-\z) -- (-0.15+\y,0.6-\z) -- (-0.15+\y,-\z);
         \draw[rounded corners=2] (0.15+\y,-\z) -- (0.15+\y,0.6-\z) -- (0.45+\y,0.6-\z) -- (0.45+\y,-\z);   
        \end{tikzpicture}}     
    \end{array}\right) = \ 0\,.
\ee

Coefficient of $|10\rangle$ and $|11\rangle$ basis vectors are computed in a similar fashion. Altogether, we find
\be
\label{TopBellState}
|\Phi\rangle \ = \ |00\rangle + |11\rangle\,,
\ee
that is $|\Phi\rangle$ is the EPR-Bell pair $|\Phi^+\rangle$ up to a normalization.

Another way to see that~(\ref{2qubitstate}) is the state $|\Phi^+\rangle$ is to notice that by slightly bending the diagram we can transform the state into an operator
\be
\label{4pIdentity}
\begin{array}{c}\scalebox{0.8}{
     \begin{tikzpicture}[thick]
     \newcommand{\y}{1.8}
        \foreach \x in {-0.45,-0.15,...,0.75}
           \draw[black] (0,\x) -- (\y,\x);
        \draw[gray,opacity=0.8,line width=2] (0,-0.75) -- (0,0.75);
         \foreach \x in {-0.45,-0.15,...,0.75}
           \fill[black] (0,\x) circle (0.075);
          \draw[gray,opacity=0.8,line width=2] (0+\y,-0.75) -- (0+\y,0.75); 
          \foreach \x in {-0.45,-0.15,...,0.75}
           \fill[black] (0+\y,\x) circle (0.075);
     \end{tikzpicture}} 
\end{array}.
\ee
By the TQFT axioms, this must be the identity operator on the 4-point Hilbert space, which has dimension two. Therefore, it is
\be
|0\rangle\langle 0| + |1\rangle\langle 1|\,.
\ee
Conjugating the bra vector in this expression, that is transforming the operator into a state in a tensor product of two copies of the the Hilbert space, produces state~(\ref{TopBellState}).

Let us consider another example -- a state in a product of two 4-point Hilbert spaces given by the diagram
\be
\label{separablestate}
\begin{array}{c}\scalebox{0.8}{
        \begin{tikzpicture}[thick]
            \newcommand{\y}{2}
         \newcommand{\z}{0}  
         \draw[rounded corners=2] (-0.45,-\z) -- (-0.45,-0.6-\z) -- (-0.15,-0.6-\z) -- (-0.15,-\z);
         \draw[rounded corners=2] (0.45,-\z) -- (0.45,-0.6-\z) -- (0.15,-0.6-\z) -- (0.15,-\z);
         
         \draw[rounded corners=2] (-0.45+\y,-\z) -- (-0.45+\y,-0.6-\z) -- (-0.15+\y,-0.6-\z) -- (-0.15+\y,-\z);
         \draw[rounded corners=2] (0.15+\y,-\z) -- (0.15+\y,-0.6-\z) -- (0.45+\y,-0.6-\z) -- (0.45+\y,-\z);
         
         \draw[gray,opacity=0.3,line width=2] (-0.75,0) -- (0.75,0);
         \foreach \x in {-0.45,-0.15,...,0.75}
         \fill[black] (\x,-\z) circle (0.075);
         
         \draw[gray,opacity=0.3,line width=2] (-0.75+\y,0) -- (0.75+\y,0);
         \foreach \x in {-0.45,-0.15,...,0.75}
         \fill[black] (\x+\y,-\z) circle (0.075);
        \end{tikzpicture}} 
    \end{array}.
\ee
It is clear that such state should be simply state $|00\rangle$ up to normalization. This state is a product state with no correlations between spins. In any state of the form $|\psi_1\rangle\otimes |\psi_2\rangle$ measurements of the two parts are completely independent.

The two examples above illustrate a few properties of the topological description: First, correlations are expressed by lines (spaces) connecting parts of the system. Second, lack of connection means absence of correlations, with disconnected topological spaces representing product states. These observations can be generalized in a more general TQFT setting. Consider a pair of states in a tensor product of two Hilbert spaces corresponding to two manifolds $\Sigma_A$ and $\Sigma_B$:
\be
|\Psi_1\rangle  =  
  \begin{array}{c}\includegraphics[scale=0.25]{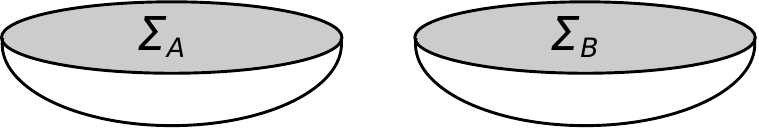}\end{array}\,,
\qquad
 |\Psi_2\rangle = 
  \begin{array}{c}\includegraphics[scale=0.25]{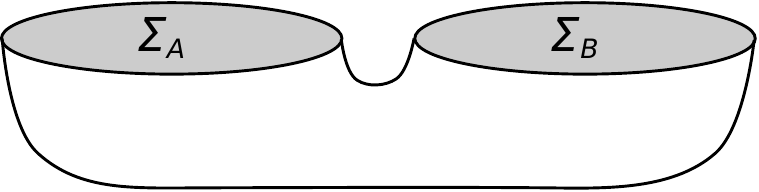}\end{array}\,.
\ee
In the first case the interiors of $\Sigma_A$ and $\Sigma_B$ are filled in such a way that the two boundaries remain disconnected. In the second, a cobordism fills the space between the two. 

States $|\Psi_1\rangle$ and $|\Psi_2\rangle$ illustrate two classes of states in quantum theory: separable and entangled. Separable state can be cast as a tensor product $|\Psi_1\rangle=|\psi_A\rangle\otimes |\psi_B\rangle$, with the two factors identifiable from the above diagram. An entangled state is the state in which such a factorization is not possible. In the next section we will give a formal proof of entanglement of $|\Psi_1\rangle$ and $|\Psi_2\rangle$ based on the derivation of von Neumann entropy.

The picture, in which entanglement emerged as space connecting subsystems is a recurring theme in different areas of theoretical physics in recent years. One example is the ER=EPR proposal by Juan Maldacena and Leonard Susskind~\cite{Maldacena:2013xja}, in which an EPR pair is compared with the Einstein-Rosen (ER) bridge -- a wormhole solution of general relativity~\cite{Israel:1976ur,Maldacena:2001kr}. A wormhole is a state of the Universe similar to~(\ref{2qubitstate}): the existence of a spatial bridge between two subuniverses implies the existence of quantum correlations between them. A similar idea is the ``emergence of space from entanglement'' in the context of AdS/CFT correspondence~\cite{VanRaamsdonk:2010pw}. The latter states a duality between a nongravitational theory in some space $\Sigma$ and a gravitational theory in some $M$ such that $\Sigma=\partial M$, is the boundary of $M$. In this context entanglement between different parts of $\Sigma$ corresponds to specific geometries of the spacetime manifold $M$.

%%%%%%%%%%%%%%%%%%%%%%%%%%%%%%%%%%%%%%%%%%%%%%%%%%%%%%%%%%%%%%%%%%%%
\subsection{SLOCC Classification}
\label{sec:SLOCC}

In the previous section we made an observation that disconnected topologies correspond to separable states in quantum theory, while connected ones are avatars of entangled states. In most cases we would like to have more information about the system. In particular, one may be interested in quantifying the amount of the correlation, or detail its structure. It is straightforward to elaborate on the previous idea and suggest that different kind of topologies encode different types and amounts of entanglement. The general problem, however, is to make such a classification useful for quantum tasks. By default, topology suggests an infinite number of different types of entanglement. Quantum Resource Theory, which views entanglement as a resource that can be used for different quantum tasks, prefers a finite classification.

One very common example of a classification is the one based on Stochastic Local Operations and Classical Communication (SLOCC)~\cite{Dur:2000zz}. It is established from the observation that local unitary operations do not change the amount of entanglement.\footnote{In this section we will keep the amount of entanglement, or correlation, as some abstract notion, since it depends, in general, on a specific measure used as a quantifier. In the next section we will focus on entanglement entropy as such a measure, which will make this discussion more precise.} In other words, if there is some amount of entanglement between parts $A$ and $B$ of the system, then unitary operators acting only on $A$ or only on $B$ will not change this amount. Conversely local nonunitary, but invertible (stochastic), or invertible nonlocal operations change the amount of entanglement.

It is somewhat less useful, however, to classify the entanglement through the action of unitaries only. If one considers a pair of qubits, for example, the action of local unitary operations, realized by $SU(2)$ matrices, is parameterized by only six real parameters, while one needs seven real parameters for a generic two-qubit state. This means that there will be a continuous family of orbits of quantum states, instead of discrete classes.  

One can extend the class of local operations to include all invertible ones. (Noninvertible ones are projectors. Projectors can reduce an entangled state to a separable one.) For two qubits, a pair of $SL(2)$ matrices is parameterized by twelve real numbers, sufficient to parameterize any two-qubit state. Therefore, one may suggest to classify entanglement types by finding the number of all nonequivalent states modulo action of local invertible operations. If the Hilbert space has the structure $\Hc = \Hc_{N_1}\otimes \cdots\otimes \Hc_{N_n}$, with $n$ subsystems with dimensions $N_k$, $k=1,\ldots,n$, the nonequivalent quantum states are described by the coset
\be
\frac{S^{2(N_1+N_2+\ldots + N_n)-1}}{SL(N_1)\otimes \cdots \otimes SL(N_n)}\,,
\ee
where $S^d$ is a $d$-dimensional sphere. For a pair of qudits, $n=2$, $N_1=N_2=d$, local invertible operations have $4d^2-4$ real parameters for the necessary $2d^2-1$ needed to define a two-qudit state. Consequently, the entanglement in all bipartite systems can be characterized as belonging to one of a finite number of SLOCC classes~\cite{Dur:2000zz}. We will discuss the case of bipartition in detail, while challenges of the multipartite entanglement classification will be mentioned in section~\ref{sec:multipartite}. 

A state in a tensor product of two finite-dimensional Hilbert spaces $\Hc_A\otimes \Hc_B$ can be characterized by the coefficient matrix $c_{ij}$:
\be
\label{statetensor}
|\Psi\rangle \ = \ \sum_{i,j}c_{ij}|ij\rangle\,,
\ee
where the expansion is in orthonormal bases $|i\rangle$ and $|j\rangle$ of choice in $\Hc_A$ and $\Hc_B$ respectively. Local operators $O_A$ and $O_B$ act on matrix $c_{ij}$ appropriately. Note that the rank of the matrix is an invariant if $O_A$ and $O_B$ are invertible. Hence, the SLOCC classes can be labeled by the rank.

Schmidt decomposition provides a convenient presentation of the coefficient matrix,  useful for quantifying the entanglement:
\be
 \sum_{i,j}c_{ij}|ij\rangle \ = \ \sum_{k=1}^{\min(N_A,N_B)}\sqrt{\lambda_k}|\tilde{k}\tilde{k}\rangle\,, \qquad \lambda_{\min(N_A,N_B)} \geq \cdots \geq \lambda_2\geq \lambda_1\geq 0\,.
\ee
Here $|\tilde{k}\rangle$ are the Schmidt bases, in which matrix elements are ``diagonal''. The rank of the coefficient matrix is the number of strictly positive $\lambda_k$. This defines the SLOCC class the state belongs to. The magnitudes of $\lambda_k$ could be used to quantify the amount of entanglement in each class.

Let us discuss how a similar structure can be built in the topological interpretation of bipartite entanglement. This structure will also provide an intuitive notion of its ``strength''. First, we will give examples of three types of operations appearing in the SLOCC classification, realized in terms of topologies.

Let us begin with state~(\ref{2qubitstate}), for which we will also use the alternative diagram~(\ref{4pIdentity}). Braiding is a natural topological operation that can be applied to either the left or right set of the boundary points, converting~(\ref{4pIdentity}) into a braid. In the construction of section~\ref{sec:TQM} braiding is a unitary operation: inverting the order and the type of the crossings produces the inverse of a braid group element. For example,
\be
\label{unitaryop}
\left(\begin{array}{c}
\scalebox{1}{\begin{tikzpicture}[thick]
\draw[gray,opacity=0.8,line width=2] (0,-0.25) -- (0,1.15);
\draw[gray,opacity=0.8,line width=2] (1.2,-0.25) -- (1.2,1.15);
\draw[rounded corners=2] (0,0) -- (1.2,0);
\draw[rounded corners=2] (0,0.6) -- (0.24,0.6) -- (0.32,0.7);
\draw[rounded corners=2] (0.4,0.8) -- (0.48,0.9) -- (1.2,0.9);
\draw[rounded corners=2] (0,0.3) -- (0.72,0.3) -- (0.96,0.6) -- (1.2,0.6);
\draw[rounded corners=2] (0,0.9) -- (0.24,0.9) -- (0.48,0.6) -- (0.72,0.6) -- (0.8,0.5);
\draw[rounded corners=2] (0.88,0.4) -- (0.96,0.3) -- (1.2,0.3);
\fill[black] (0,0.0) circle (0.05cm);
\fill[black] (0,0.3) circle (0.05cm);
\fill[black] (0,0.6) circle (0.05cm);
\fill[black] (0,0.9) circle (0.05cm);
\fill[black] (1.2,0.0) circle (0.05cm);
\fill[black] (1.2,0.3) circle (0.05cm);
\fill[black] (1.2,0.6) circle (0.05cm);
\fill[black] (1.2,0.9) circle (0.05cm);
\end{tikzpicture} }
\end{array}\right)^\dagger \ = \ \begin{array}{c}
\scalebox{1}{\begin{tikzpicture}[thick]
\draw[gray,opacity=0.8,line width=2] (0,-0.25) -- (0,1.15);
\draw[gray,opacity=0.8,line width=2] (-1.2,-0.25) -- (-1.2,1.15);
\draw[rounded corners=2] (0,0) -- (-1.2,0);
\draw[rounded corners=2] (0,0.6) -- (-0.24,0.6) -- (-0.32,0.7);
\draw[rounded corners=2] (-0.4,0.8) -- (-0.48,0.9) -- (-1.2,0.9);
\draw[rounded corners=2] (0,0.3) -- (-0.72,0.3) -- (-0.96,0.6) -- (-1.2,0.6);
\draw[rounded corners=2] (0,0.9) -- (-0.24,0.9) -- (-0.48,0.6) -- (-0.72,0.6) -- (-0.8,0.5);
\draw[rounded corners=2] (-0.88,0.4) -- (-0.96,0.3) -- (-1.2,0.3);
\fill[black] (0,0.0) circle (0.05cm);
\fill[black] (0,0.3) circle (0.05cm);
\fill[black] (0,0.6) circle (0.05cm);
\fill[black] (0,0.9) circle (0.05cm);
\fill[black] (-1.2,0.0) circle (0.05cm);
\fill[black] (-1.2,0.3) circle (0.05cm);
\fill[black] (-1.2,0.6) circle (0.05cm);
\fill[black] (-1.2,0.9) circle (0.05cm);
\end{tikzpicture} }
\end{array}\qquad \Longrightarrow \qquad 
\begin{array}{c}
\scalebox{1}{\begin{tikzpicture}[thick]
\draw[gray,opacity=0.3,line width=2] (0,-0.25) -- (0,1.15);
\draw[gray,opacity=0.8,line width=2] (1.2,-0.25) -- (1.2,1.15);
\draw[rounded corners=2] (0,0) -- (1.2,0);
\draw[rounded corners=2] (0,0.6) -- (0.24,0.6) -- (0.32,0.7);
\draw[rounded corners=2] (0.4,0.8) -- (0.48,0.9) -- (1.2,0.9);
\draw[rounded corners=2] (0,0.3) -- (0.72,0.3) -- (0.96,0.6) -- (1.2,0.6);
\draw[rounded corners=2] (0,0.9) -- (0.24,0.9) -- (0.48,0.6) -- (0.72,0.6) -- (0.8,0.5);
\draw[rounded corners=2] (0.88,0.4) -- (0.96,0.3) -- (1.2,0.3);
\fill[black] (0,0.0) circle (0.05cm);
\fill[black] (0,0.3) circle (0.05cm);
\fill[black] (0,0.6) circle (0.05cm);
\fill[black] (0,0.9) circle (0.05cm);
\fill[black] (1.2,0.0) circle (0.05cm);
\fill[black] (1.2,0.3) circle (0.05cm);
\fill[black] (1.2,0.6) circle (0.05cm);
\fill[black] (1.2,0.9) circle (0.05cm);
\draw[gray,opacity=0.8,line width=2] (-1.2,-0.25) -- (-1.2,1.15);
\draw[rounded corners=2] (0,0) -- (-1.2,0);
\draw[rounded corners=2] (0,0.6) -- (-0.24,0.6) -- (-0.32,0.7);
\draw[rounded corners=2] (-0.4,0.8) -- (-0.48,0.9) -- (-1.2,0.9);
\draw[rounded corners=2] (0,0.3) -- (-0.72,0.3) -- (-0.96,0.6) -- (-1.2,0.6);
\draw[rounded corners=2] (0,0.9) -- (-0.24,0.9) -- (-0.48,0.6) -- (-0.72,0.6) -- (-0.8,0.5);
\draw[rounded corners=2] (-0.88,0.4) -- (-0.96,0.3) -- (-1.2,0.3);
\fill[black] (0,0.0) circle (0.05cm);
\fill[black] (0,0.3) circle (0.05cm);
\fill[black] (0,0.6) circle (0.05cm);
\fill[black] (0,0.9) circle (0.05cm);
\fill[black] (-1.2,0.0) circle (0.05cm);
\fill[black] (-1.2,0.3) circle (0.05cm);
\fill[black] (-1.2,0.6) circle (0.05cm);
\fill[black] (-1.2,0.9) circle (0.05cm);
\end{tikzpicture} }
\end{array} \ = \ \begin{array}{c}
\scalebox{1}{\begin{tikzpicture}[thick]
\draw[gray,opacity=0.8,line width=2] (0,-0.25) -- (0,1.15);
\draw[gray,opacity=0.8,line width=2] (1.2,-0.25) -- (1.2,1.15);
\fill[black] (0,0.0) circle (0.05cm);
\fill[black] (0,0.3) circle (0.05cm);
\fill[black] (0,0.6) circle (0.05cm);
\fill[black] (0,0.9) circle (0.05cm);
\draw (0,0) -- (1.2,0);
\draw (0,0.6) -- (1.2,0.6);
\fill[black] (1.2,0.0) circle (0.05cm);
\fill[black] (1.2,0.3) circle (0.05cm);
\fill[black] (1.2,0.6) circle (0.05cm);
\fill[black] (1.2,0.9) circle (0.05cm);
\draw (1.2,0.3) -- (0,0.3);
\draw (1.2,0.9) -- (0,0.9);
\end{tikzpicture} }
\end{array}.
\ee

Natural examples of nonivertible operations are given by the Temperley-Lieb elements. By definition, Temperley-Lieb elements are projectors:
\be
\begin{array}{c} 
\begin{tikzpicture}[thick]
    \newcommand{\x}{1.2}
    \draw[gray,opacity=0.8,line width=2] (0,0) -- (0,1.3);
    \draw[rounded corners=2] (0,0.2) -- (0.4,0.2) -- (0.4,0.5) -- (0,0.5);
    \draw[rounded corners=2] (0,0.8) -- (0.4,0.8) -- (0.4,1.1) -- (0,1.1);
    \foreach \y in {0.2,0.5,...,1.1}
           \fill[black] (0,\y) circle (0.05cm);
    \draw[gray,opacity=0.3,line width=2] (\x-0,0) -- (\x-0,1.3);
    \draw[rounded corners=2] (\x-0,0.2) -- (\x-0.4,0.2) -- (\x-0.4,0.5) -- (\x-0,0.5);
    \draw[rounded corners=2] (\x-0,0.8) -- (\x-0.4,0.8) -- (\x-0.4,1.1) -- (\x-0,1.1);
    \foreach \y in {0.2,0.5,...,1.1}
           \fill[black] (\x-0,\y) circle (0.05cm);
    \draw[rounded corners=2] (\x+0,0.2) -- (\x+0.4,0.2) -- (\x+0.4,0.5) -- (\x+0,0.5);
    \draw[rounded corners=2] (\x+0,0.8) -- (\x+0.4,0.8) -- (\x+0.4,1.1) -- (\x+0,1.1);
     \draw[gray,opacity=0.8,line width=2] (2*\x-0,0) -- (2*\x-0,1.3);
    \draw[rounded corners=2] (2*\x-0,0.2) -- (2*\x-0.4,0.2) -- (2*\x-0.4,0.5) -- (2*\x-0,0.5);
    \draw[rounded corners=2] (2*\x-0,0.8) -- (2*\x-0.4,0.8) -- (2*\x-0.4,1.1) -- (2*\x-0,1.1);
    \foreach \y in {0.2,0.5,...,1.1}
           \fill[black] (2*\x-0,\y) circle (0.05cm);
\end{tikzpicture}
\end{array}
\ = \ d^2 \begin{array}{c} 
\begin{tikzpicture}[thick]
    \newcommand{\x}{1.2}
    \draw[gray,opacity=0.8,line width=2] (0,0) -- (0,1.3);
    \draw[rounded corners=2] (0,0.2) -- (0.4,0.2) -- (0.4,0.5) -- (0,0.5);
    \draw[rounded corners=2] (0,0.8) -- (0.4,0.8) -- (0.4,1.1) -- (0,1.1);
    \foreach \y in {0.2,0.5,...,1.1}
           \fill[black] (0,\y) circle (0.05cm);
    \draw[gray,opacity=0.8,line width=2] (\x-0,0) -- (\x-0,1.3);
    \draw[rounded corners=2] (\x-0,0.2) -- (\x-0.4,0.2) -- (\x-0.4,0.5) -- (\x-0,0.5);
    \draw[rounded corners=2] (\x-0,0.8) -- (\x-0.4,0.8) -- (\x-0.4,1.1) -- (\x-0,1.1);
    \foreach \y in {0.2,0.5,...,1.1}
           \fill[black] (\x-0,\y) circle (0.05cm);
\end{tikzpicture}
\end{array}.
\ee
It is clear that the above projector applied to the left or to the right of~(\ref{4pIdentity}) will transform the entangled state into separable state~(\ref{separablestate}). One of the main results can already be anticipated: The amount of entanglement depends on the number of lines that connect left and right.

A natural example of invertible but nonunitary operations also exists. It should be obvious that the following state does not correspond to a unitary matrix acting on~(\ref{4pIdentity}):
\be
\label{invertop}
\left(\begin{array}{c}
\scalebox{1}{\begin{tikzpicture}[thick]
\newcommand{\x}{0}
\draw[gray,opacity=0.8,line width=2] (0,-0.25) -- (0,1.15);
\draw[gray,opacity=0.8,line width=2] (1.2,-0.25) -- (1.2,1.15);
\draw (0,0.9) -- (0.5,0.9) arc (90:-30:0.15cm);
\draw (0.0,0.6) -- (0.5,0.6) arc (-90:-60:0.15cm);
\draw (1.2+\x,0.9) -- (0.7+\x,0.9) arc (90:120:0.15cm);
\draw (1.2+\x,0.6) -- (0.7+\x,0.6) arc (-90:-210:0.15cm);
\draw (0,0.3) -- (0.5,0.3) arc (90:-30:0.15cm);
\draw (0.0,0.) -- (0.5,0.) arc (-90:-60:0.15cm);
\draw (1.2+\x,0.3) -- (0.7+\x,0.3) arc (90:120:0.15cm);
\draw (1.2+\x,0.) -- (0.7+\x,0.) arc (-90:-210:0.15cm);
\fill[black] (1.2+\x,0.0) circle (0.05cm);
\fill[black] (1.2+\x,0.3) circle (0.05cm);
\fill[black] (1.2+\x,0.6) circle (0.05cm);
\fill[black] (1.2+\x,0.9) circle (0.05cm);
\fill[black] (0,0.0) circle (0.05cm);
\fill[black] (0,0.3) circle (0.05cm);
\fill[black] (0,0.6) circle (0.05cm);
\fill[black] (0,0.9) circle (0.05cm);
\end{tikzpicture}} 
\end{array}\right)^\dagger 
\ = \ 
\begin{array}{c}
\scalebox{1}{\begin{tikzpicture}[thick]
\newcommand{\x}{0}
\draw[gray,opacity=0.8,line width=2] (0,-0.25) -- (0,1.15);
\draw[gray,opacity=0.8,line width=2] (-1.2,-0.25) -- (-1.2,1.15);
\draw (0,0.9) -- (-0.5,0.9) arc (90:210:0.15cm);
\draw (0.0,0.6) -- (-0.5,0.6) arc (-90:-120:0.15cm);
\draw (-1.2-\x,0.9) -- (-0.7-\x,0.9) arc (90:60:0.15cm);
\draw (-1.2-\x,0.6) -- (-0.7-\x,0.6) arc (-90:30:0.15cm);
\draw (0,0.3) -- (-0.5,0.3) arc (90:210:0.15cm);
\draw (0.0,0.) -- (-0.5,0.) arc (-90:-120:0.15cm);
\draw (-1.2-\x,0.3) -- (-0.7-\x,0.3) arc (90:60:0.15cm);
\draw (-1.2-\x,0.) -- (-0.7-\x,0.) arc (-90:30:0.15cm);
\fill[black] (-1.2-\x,0.0) circle (0.05cm);
\fill[black] (-1.2-\x,0.3) circle (0.05cm);
\fill[black] (-1.2-\x,0.6) circle (0.05cm);
\fill[black] (-1.2-\x,0.9) circle (0.05cm);
\fill[black] (0,0.0) circle (0.05cm);
\fill[black] (0,0.3) circle (0.05cm);
\fill[black] (0,0.6) circle (0.05cm);
\fill[black] (0,0.9) circle (0.05cm);
\end{tikzpicture}} 
\end{array}, \qquad \text{but} \qquad \begin{array}{c}
\scalebox{1}{\begin{tikzpicture}[thick]
\newcommand{\x}{0}
\draw[gray,opacity=0.3,line width=2] (0,-0.25) -- (0,1.15);
\draw[gray,opacity=0.8,line width=2] (1.2,-0.25) -- (1.2,1.15);
\draw (0,0.9) -- (0.5,0.9) arc (90:-30:0.15cm);
\draw (0.0,0.6) -- (0.5,0.6) arc (-90:-60:0.15cm);
\draw (1.2+\x,0.9) -- (0.7+\x,0.9) arc (90:120:0.15cm);
\draw (1.2+\x,0.6) -- (0.7+\x,0.6) arc (-90:-210:0.15cm);
\draw (0,0.3) -- (0.5,0.3) arc (90:-30:0.15cm);
\draw (0.0,0.) -- (0.5,0.) arc (-90:-60:0.15cm);
\draw (1.2+\x,0.3) -- (0.7+\x,0.3) arc (90:120:0.15cm);
\draw (1.2+\x,0.) -- (0.7+\x,0.) arc (-90:-210:0.15cm);
\fill[black] (1.2+\x,0.0) circle (0.05cm);
\fill[black] (1.2+\x,0.3) circle (0.05cm);
\fill[black] (1.2+\x,0.6) circle (0.05cm);
\fill[black] (1.2+\x,0.9) circle (0.05cm);
\fill[black] (0,0.0) circle (0.05cm);
\fill[black] (0,0.3) circle (0.05cm);
\fill[black] (0,0.6) circle (0.05cm);
\fill[black] (0,0.9) circle (0.05cm);
\draw[gray,opacity=0.8,line width=2] (-1.2,-0.25) -- (-1.2,1.15);
\draw (0,0.9) -- (-0.5,0.9) arc (90:210:0.15cm);
\draw (0.0,0.6) -- (-0.5,0.6) arc (-90:-120:0.15cm);
\draw (-1.2-\x,0.9) -- (-0.7-\x,0.9) arc (90:60:0.15cm);
\draw (-1.2-\x,0.6) -- (-0.7-\x,0.6) arc (-90:30:0.15cm);
\draw (0,0.3) -- (-0.5,0.3) arc (90:210:0.15cm);
\draw (0.0,0.) -- (-0.5,0.) arc (-90:-120:0.15cm);
\draw (-1.2-\x,0.3) -- (-0.7-\x,0.3) arc (90:60:0.15cm);
\draw (-1.2-\x,0.) -- (-0.7-\x,0.) arc (-90:30:0.15cm);
\fill[black] (-1.2-\x,0.0) circle (0.05cm);
\fill[black] (-1.2-\x,0.3) circle (0.05cm);
\fill[black] (-1.2-\x,0.6) circle (0.05cm);
\fill[black] (-1.2-\x,0.9) circle (0.05cm);
\fill[black] (0,0.0) circle (0.05cm);
\fill[black] (0,0.3) circle (0.05cm);
\fill[black] (0,0.6) circle (0.05cm);
\fill[black] (0,0.9) circle (0.05cm);
\end{tikzpicture}} 
\end{array}\ \neq \ \begin{array}{c}
\scalebox{1}{\begin{tikzpicture}[thick]
\draw[gray,opacity=0.8,line width=2] (0,-0.25) -- (0,1.15);
\draw[gray,opacity=0.8,line width=2] (1.2,-0.25) -- (1.2,1.15);
\fill[black] (0,0.0) circle (0.05cm);
\fill[black] (0,0.3) circle (0.05cm);
\fill[black] (0,0.6) circle (0.05cm);
\fill[black] (0,0.9) circle (0.05cm);
\draw (0,0) -- (1.2,0);
\draw (0,0.6) -- (1.2,0.6);
\fill[black] (1.2,0.0) circle (0.05cm);
\fill[black] (1.2,0.3) circle (0.05cm);
\fill[black] (1.2,0.6) circle (0.05cm);
\fill[black] (1.2,0.9) circle (0.05cm);
\draw (1.2,0.3) -- (0,0.3);
\draw (1.2,0.9) -- (0,0.9);
\end{tikzpicture} }
\end{array}.
\ee
This example is not a braid, but a more general \emph{tangle}. We should expect this state to be invertible. Even though every point is connected by a line with a point on the same boundary, the line is entangled with the other boundary through a linking with other lines. The invertibility can be checked explicitly by computing the expansion of the above state in the computational basis, as we did in equations~(\ref{coefs1}) and~(\ref{coefs2}):
\be
\label{chainedstate}
\begin{array}{c}
\scalebox{1}{\begin{tikzpicture}[thick]
\newcommand{\x}{0}
\draw[gray,opacity=0.8,line width=2] (0,-0.25) -- (0,1.15);
\draw[gray,opacity=0.8,line width=2] (1.2,-0.25) -- (1.2,1.15);
\draw (0,0.9) -- (0.5,0.9) arc (90:-30:0.15cm);
\draw (0.0,0.6) -- (0.5,0.6) arc (-90:-60:0.15cm);
\draw (1.2+\x,0.9) -- (0.7+\x,0.9) arc (90:120:0.15cm);
\draw (1.2+\x,0.6) -- (0.7+\x,0.6) arc (-90:-210:0.15cm);
\draw (0,0.3) -- (0.5,0.3) arc (90:-30:0.15cm);
\draw (0.0,0.) -- (0.5,0.) arc (-90:-60:0.15cm);
\draw (1.2+\x,0.3) -- (0.7+\x,0.3) arc (90:120:0.15cm);
\draw (1.2+\x,0.) -- (0.7+\x,0.) arc (-90:-210:0.15cm);
\fill[black] (1.2+\x,0.0) circle (0.05cm);
\fill[black] (1.2+\x,0.3) circle (0.05cm);
\fill[black] (1.2+\x,0.6) circle (0.05cm);
\fill[black] (1.2+\x,0.9) circle (0.05cm);
\fill[black] (0,0.0) circle (0.05cm);
\fill[black] (0,0.3) circle (0.05cm);
\fill[black] (0,0.6) circle (0.05cm);
\fill[black] (0,0.9) circle (0.05cm);
\end{tikzpicture}} 
\end{array} \ = \ (A^4+A^{-4})^2|00\rangle + (1-A^{4})^2|11\rangle \,.
\ee
It is also useful to check the diagrammatic expansion of this diagram,
\be
\label{chainedstateexp}
{A^{-4}} 
\begin{array}{c}
\begin{tikzpicture}[thick]
\draw[gray,opacity=0.8,line width=2] (0,-0.25) -- (0,1.15);
\draw[gray,opacity=0.8,line width=2] (1.2,-0.25) -- (1.2,1.15);
\fill[black] (0,0.0) circle (0.05cm);
\fill[black] (0,0.3) circle (0.05cm);
\fill[black] (0,0.6) circle (0.05cm);
\fill[black] (0,0.9) circle (0.05cm);
\draw (0,0) -- (0.4,0) arc (-90:90:0.15cm) -- (0,0.3);
\draw (0,0.6) -- (0.4,0.6) arc (-90:90:0.15cm) -- (0,0.9);
\fill[black] (1.2,0.0) circle (0.05cm);
\fill[black] (1.2,0.3) circle (0.05cm);
\fill[black] (1.2,0.6) circle (0.05cm);
\fill[black] (1.2,0.9) circle (0.05cm);
\draw (1.2,0) -- (0.8,0) arc (-90:-270:0.15cm) -- (1.2,0.3);
\draw (1.2,0.6) -- (0.8,0.6) arc (-90:-270:0.15cm) -- (1.2,0.9);
\end{tikzpicture} 
\end{array}
 - (A^2-A^{-2}) \left(
\begin{array}{c}
\begin{tikzpicture}[thick]
\draw[gray,opacity=0.8,line width=2] (0,-0.25) -- (0,1.15);
\draw[gray,opacity=0.8,line width=2] (1.2,-0.25) -- (1.2,1.15);
\fill[black] (0,0.0) circle (0.05cm);
\fill[black] (0,0.3) circle (0.05cm);
\fill[black] (0,0.6) circle (0.05cm);
\fill[black] (0,0.9) circle (0.05cm);
\draw (0,0) -- (0.4,0) -- (1.2,0.0);
\draw (0,0.6) -- (0.4,0.6) arc (-90:90:0.15cm) -- (0,0.9);
\fill[black] (1.2,0.0) circle (0.05cm);
\fill[black] (1.2,0.3) circle (0.05cm);
\fill[black] (1.2,0.6) circle (0.05cm);
\fill[black] (1.2,0.9) circle (0.05cm);
\draw (1.2,0.3) -- (0,0.3);
\draw (1.2,0.6) -- (0.8,0.6) arc (-90:-270:0.15cm) -- (1.2,0.9);
\end{tikzpicture} 
\end{array}
 +  
\begin{array}{c}
\begin{tikzpicture}[thick]
\draw[gray,opacity=0.8,line width=2] (0,-0.25) -- (0,1.15);
\draw[gray,opacity=0.8,line width=2] (1.2,-0.25) -- (1.2,1.15);
\fill[black] (0,0.0) circle (0.05cm);
\fill[black] (0,0.3) circle (0.05cm);
\fill[black] (0,0.6) circle (0.05cm);
\fill[black] (0,0.9) circle (0.05cm);
\draw (0,0) -- (0.4,0) arc (-90:90:0.15cm) -- (0,0.3);
\draw (0,0.6) -- (1.2,0.6);
\fill[black] (1.2,0.0) circle (0.05cm);
\fill[black] (1.2,0.3) circle (0.05cm);
\fill[black] (1.2,0.6) circle (0.05cm);
\fill[black] (1.2,0.9) circle (0.05cm);
\draw (1.2,0) -- (0.8,0) arc (-90:-270:0.15cm) -- (1.2,0.3);
\draw (1.2,0.9) -- (0,0.9);
\end{tikzpicture} 
\end{array} \right)
 +  (1-A^{4})^2  
\begin{array}{c}
\begin{tikzpicture}[thick]
\draw[gray,opacity=0.8,line width=2] (0,-0.25) -- (0,1.15);
\draw[gray,opacity=0.8,line width=2] (1.2,-0.25) -- (1.2,1.15);
\fill[black] (0,0.0) circle (0.05cm);
\fill[black] (0,0.3) circle (0.05cm);
\fill[black] (0,0.6) circle (0.05cm);
\fill[black] (0,0.9) circle (0.05cm);
\draw (0,0) -- (1.2,0);
\draw (0,0.6) -- (1.2,0.6);
\fill[black] (1.2,0.0) circle (0.05cm);
\fill[black] (1.2,0.3) circle (0.05cm);
\fill[black] (1.2,0.6) circle (0.05cm);
\fill[black] (1.2,0.9) circle (0.05cm);
\draw (1.2,0.3) -- (0,0.3);
\draw (1.2,0.9) -- (0,0.9);
\end{tikzpicture} 
\end{array}.
\ee
The state is a linear combination of an entangled and a separable states, which implies that it should have more entanglement than the separable state and less entanglement than state~(\ref{4pIdentity}). The latter, being a Bell state, has the maximum possible amount of entanglement.

The difference between the unitary braids and invertible tangles is that the latter do not have local topological inverses, that is there are no diagrams that, when appended to either side, could untangle the tangle. This means that there is no way to transform state~(\ref{chainedstate}) into state~(\ref{4pIdentity}) via local topological operations, while the opposite conversion is straightforward. In Quantum Resource Theory this observation mirrors the statement that any entangled state can be obtained from the maximally entangled one with certainty, while the probability of the inverse transformation is finite~\cite{Dur:2000zz,Vidal:1999vh,Nielsen:1999zza}. 

It is also worth discussing the new diagrams appearing in expansion~(\ref{chainedstateexp}). One finds in the computational basis that
\be
\begin{array}{c}
\begin{tikzpicture}[thick]
\draw[gray,opacity=0.8,line width=2] (0,-0.25) -- (0,1.15);
\draw[gray,opacity=0.8,line width=2] (1.2,-0.25) -- (1.2,1.15);
\fill[black] (0,0.0) circle (0.05cm);
\fill[black] (0,0.3) circle (0.05cm);
\fill[black] (0,0.6) circle (0.05cm);
\fill[black] (0,0.9) circle (0.05cm);
\draw (0,0) -- (0.4,0) -- (1.2,0.0);
\draw (0,0.6) -- (0.4,0.6) arc (-90:90:0.15cm) -- (0,0.9);
\fill[black] (1.2,0.0) circle (0.05cm);
\fill[black] (1.2,0.3) circle (0.05cm);
\fill[black] (1.2,0.6) circle (0.05cm);
\fill[black] (1.2,0.9) circle (0.05cm);
\draw (1.2,0.3) -- (0,0.3);
\draw (1.2,0.6) -- (0.8,0.6) arc (-90:-270:0.15cm) -- (1.2,0.9);
\end{tikzpicture} 
\end{array} \ = \  \begin{array}{c}
\begin{tikzpicture}[thick]
\draw[gray,opacity=0.8,line width=2] (0,-0.25) -- (0,1.15);
\draw[gray,opacity=0.8,line width=2] (1.2,-0.25) -- (1.2,1.15);
\fill[black] (0,0.0) circle (0.05cm);
\fill[black] (0,0.3) circle (0.05cm);
\fill[black] (0,0.6) circle (0.05cm);
\fill[black] (0,0.9) circle (0.05cm);
\draw (0,0) -- (0.4,0) arc (-90:90:0.15cm) -- (0,0.3);
\draw (0,0.6) -- (1.2,0.6);
\fill[black] (1.2,0.0) circle (0.05cm);
\fill[black] (1.2,0.3) circle (0.05cm);
\fill[black] (1.2,0.6) circle (0.05cm);
\fill[black] (1.2,0.9) circle (0.05cm);
\draw (1.2,0) -- (0.8,0) arc (-90:-270:0.15cm) -- (1.2,0.3);
\draw (1.2,0.9) -- (0,0.9);
\end{tikzpicture} 
\end{array} \ = \ d |00\rangle\,.
\ee
In other words, both diagrams correspond to the separable state $|00\rangle$, as does the first diagram in~(\ref{chainedstateexp}). This observation goes against the intuition that the existence of connections between the boundaries implies correlations, and that the amount of correlations depends on the number of connections. However, what it really means is that this intuition (and the dependence of the entanglement of the number of connections) is nonlinear. In particular, a certain minimum amount of connections is needed to establish correlation, which is a feature of the specific topological quantum mechanics model used here.

Let us clarify the criteria for the minimal connection. If a pair of boundaries of a 1D TQFT are connected by only two lines than the boundaries are in fact disconnected. This was just seen in the example studied above:
\be
\label{surgery}
\begin{array}{c}
\begin{tikzpicture}[thick]
\draw[gray,opacity=0.8,line width=2] (0,-0.25) -- (0,1.15);
\draw[gray,opacity=0.8,line width=2] (1.2,-0.25) -- (1.2,1.15);
\fill[black] (0,0.0) circle (0.05cm);
\fill[black] (0,0.3) circle (0.05cm);
\fill[black] (0,0.6) circle (0.05cm);
\fill[black] (0,0.9) circle (0.05cm);
\draw (0,0) -- (0.4,0) -- (1.2,0.0);
\draw (0,0.6) -- (0.4,0.6) arc (-90:90:0.15cm) -- (0,0.9);
\fill[black] (1.2,0.0) circle (0.05cm);
\fill[black] (1.2,0.3) circle (0.05cm);
\fill[black] (1.2,0.6) circle (0.05cm);
\fill[black] (1.2,0.9) circle (0.05cm);
\draw (1.2,0.3) -- (0,0.3);
\draw (1.2,0.6) -- (0.8,0.6) arc (-90:-270:0.15cm) -- (1.2,0.9);
\end{tikzpicture} 
\end{array} \ = \ \frac{1}{d}\begin{array}{c}
\begin{tikzpicture}[thick]
\draw[gray,opacity=0.8,line width=2] (0,-0.25) -- (0,1.15);
\draw[gray,opacity=0.8,line width=2] (1.2,-0.25) -- (1.2,1.15);
\fill[black] (0,0.0) circle (0.05cm);
\fill[black] (0,0.3) circle (0.05cm);
\fill[black] (0,0.6) circle (0.05cm);
\fill[black] (0,0.9) circle (0.05cm);
\draw (0,0) -- (0.4,0) arc (-90:90:0.15cm) -- (0,0.3);
\draw (0,0.6) -- (0.4,0.6) arc (-90:90:0.15cm) -- (0,0.9);
\fill[black] (1.2,0.0) circle (0.05cm);
\fill[black] (1.2,0.3) circle (0.05cm);
\fill[black] (1.2,0.6) circle (0.05cm);
\fill[black] (1.2,0.9) circle (0.05cm);
\draw (1.2,0) -- (0.8,0) arc (-90:-270:0.15cm) -- (1.2,0.3);
\draw (1.2,0.6) -- (0.8,0.6) arc (-90:-270:0.15cm) -- (1.2,0.9);
\end{tikzpicture} 
\end{array}.
\ee
A similar cutting can be applied for any pair of boundaries connected by only a pair of lines. To see why this happens the diagram can be cut in the middle by inserting a completeness relation. Note that the completeness relation is formally
\be
\begin{array}{c}
\begin{tikzpicture}[thick]
\draw[gray,opacity=0.8,line width=2] (0,-0.25) -- (0,1.15);
\draw[gray,opacity=0.8,line width=2] (1.2,-0.25) -- (1.2,1.15);
\fill[black] (0,0.0) circle (0.05cm);
\fill[black] (0,0.3) circle (0.05cm);
\fill[black] (0,0.6) circle (0.05cm);
\fill[black] (0,0.9) circle (0.05cm);
\draw (0,0) -- (1.2,0);
\draw (0,0.6) -- (1.2,0.6);
\fill[black] (1.2,0.0) circle (0.05cm);
\fill[black] (1.2,0.3) circle (0.05cm);
\fill[black] (1.2,0.6) circle (0.05cm);
\fill[black] (1.2,0.9) circle (0.05cm);
\draw (1.2,0.3) -- (0,0.3);
\draw (1.2,0.9) -- (0,0.9);
\end{tikzpicture} 
\end{array} \ = \ \sum_i|i\rangle\langle i|\ = \ 
\begin{array}{c}
\begin{tikzpicture}[thick]
\draw[gray,opacity=0.8,line width=2] (0,-0.25) -- (0,1.15);
\draw[gray,opacity=0.8,line width=2] (1.2,-0.25) -- (1.2,1.15);
\fill[black] (0,0.0) circle (0.05cm);
\fill[black] (0,0.3) circle (0.05cm);
\fill[black] (0,0.6) circle (0.05cm);
\fill[black] (0,0.9) circle (0.05cm);
\draw (0,0) -- (1.2,0);
\draw (0,0.6) -- (1.2,0.6);
\fill[black] (1.2,0.0) circle (0.05cm);
\fill[black] (1.2,0.3) circle (0.05cm);
\fill[black] (1.2,0.6) circle (0.05cm);
\fill[black] (1.2,0.9) circle (0.05cm);
\draw (1.2,0.3) -- (0,0.3);
\draw (1.2,0.9) -- (0,0.9);
\end{tikzpicture} 
\end{array} 0\Big\rangle\Big\langle 0
\begin{array}{c}
\begin{tikzpicture}[thick]
\draw[gray,opacity=0.8,line width=2] (0,-0.25) -- (0,1.15);
\draw[gray,opacity=0.8,line width=2] (1.2,-0.25) -- (1.2,1.15);
\fill[black] (0,0.0) circle (0.05cm);
\fill[black] (0,0.3) circle (0.05cm);
\fill[black] (0,0.6) circle (0.05cm);
\fill[black] (0,0.9) circle (0.05cm);
\draw (0,0) -- (1.2,0);
\draw (0,0.6) -- (1.2,0.6);
\fill[black] (1.2,0.0) circle (0.05cm);
\fill[black] (1.2,0.3) circle (0.05cm);
\fill[black] (1.2,0.6) circle (0.05cm);
\fill[black] (1.2,0.9) circle (0.05cm);
\draw (1.2,0.3) -- (0,0.3);
\draw (1.2,0.9) -- (0,0.9);
\end{tikzpicture} 
\end{array}
+
\begin{array}{c}
\begin{tikzpicture}[thick]
\draw[gray,opacity=0.8,line width=2] (0,-0.25) -- (0,1.15);
\draw[gray,opacity=0.8,line width=2] (1.2,-0.25) -- (1.2,1.15);
\fill[black] (0,0.0) circle (0.05cm);
\fill[black] (0,0.3) circle (0.05cm);
\fill[black] (0,0.6) circle (0.05cm);
\fill[black] (0,0.9) circle (0.05cm);
\draw (0,0) -- (1.2,0);
\draw (0,0.6) -- (1.2,0.6);
\fill[black] (1.2,0.0) circle (0.05cm);
\fill[black] (1.2,0.3) circle (0.05cm);
\fill[black] (1.2,0.6) circle (0.05cm);
\fill[black] (1.2,0.9) circle (0.05cm);
\draw (1.2,0.3) -- (0,0.3);
\draw (1.2,0.9) -- (0,0.9);
\end{tikzpicture} 
\end{array} 1\Big\rangle\Big\langle 1
\begin{array}{c}
\begin{tikzpicture}[thick]
\draw[gray,opacity=0.8,line width=2] (0,-0.25) -- (0,1.15);
\draw[gray,opacity=0.8,line width=2] (1.2,-0.25) -- (1.2,1.15);
\fill[black] (0,0.0) circle (0.05cm);
\fill[black] (0,0.3) circle (0.05cm);
\fill[black] (0,0.6) circle (0.05cm);
\fill[black] (0,0.9) circle (0.05cm);
\draw (0,0) -- (1.2,0);
\draw (0,0.6) -- (1.2,0.6);
\fill[black] (1.2,0.0) circle (0.05cm);
\fill[black] (1.2,0.3) circle (0.05cm);
\fill[black] (1.2,0.6) circle (0.05cm);
\fill[black] (1.2,0.9) circle (0.05cm);
\draw (1.2,0.3) -- (0,0.3);
\draw (1.2,0.9) -- (0,0.9);
\end{tikzpicture} 
\end{array}
\,,
\ee
where the identity diagram is cut in the middle by the sum over separable products of basis elements. The dimension of the 4-point Hilbert space is two, so only two types of insertions appear in the sum. However, when only two lines are cut, there is only one  possible intermediate state:
\be
\begin{array}{c}
\begin{tikzpicture}[thick]
\draw[gray,opacity=0.8,line width=2] (0,-0.25) -- (0,0.55);
\draw[gray,opacity=0.8,line width=2] (1.2,-0.25) -- (1.2,0.55);
\fill[black] (0,0.0) circle (0.05cm);
\fill[black] (0,0.3) circle (0.05cm);
\draw (0,0) -- (1.2,0);
\fill[black] (1.2,0.0) circle (0.05cm);
\fill[black] (1.2,0.3) circle (0.05cm);
\draw (1.2,0.3) -- (0,0.3);
\end{tikzpicture} 
\end{array}\ = \ 
\frac{1}{d}\begin{array}{c}
\begin{tikzpicture}[thick]
\draw[gray,opacity=0.8,line width=2] (0,-0.25) -- (0,0.55);
\draw[gray,opacity=0.8,line width=2] (1.2,-0.25) -- (1.2,0.55);
\draw (0,0) -- (1.3,0) arc (-90:90:0.15) -- (0,0.3);
\fill[black] (0,0.0) circle (0.05cm);
\fill[black] (0,0.3) circle (0.05cm);
\fill[black] (1.2,0.0) circle (0.05cm);
\fill[black] (1.2,0.3) circle (0.05cm);
\newcommand{\x}{3.2}
\draw[gray,opacity=0.8,line width=2] (\x-0,-0.25) -- (\x-0,0.55);
\draw[gray,opacity=0.8,line width=2] (\x-1.2,-0.25) -- (\x-1.2,0.55);
\draw (\x-0,0) -- (\x-1.3,0) arc (270:90:0.15) -- (\x-0,0.3);
\fill[black] (\x-0,0.0) circle (0.05cm);
\fill[black] (\x-0,0.3) circle (0.05cm);
\fill[black] (\x-1.2,0.0) circle (0.05cm);
\fill[black] (\x-1.2,0.3) circle (0.05cm);
\end{tikzpicture} 
\end{array}\,, 
\ee
since the dimension of the 2-point Hilbert space is one (see a discussion of the same point in~\cite{Witten:1988ze}).

It is important to remember that the two lines can be cut only if these two lines are the only lines that connect two given boundaries. In terms of diagrams this means that one should be able to draw a closed contour over the diagram that is crossed by only these two lines and only once, as illustrated by the following example:
\be
\label{encircle}
\begin{array}{c}
\begin{tikzpicture}[thick]
\draw[gray,opacity=0.8,line width=2] (0,-0.25) -- (0,1.15);
\draw[gray,opacity=0.8,line width=2] (1.2,-0.25) -- (1.2,1.15);
\fill[black] (0,0.0) circle (0.05cm);
\fill[black] (0,0.3) circle (0.05cm);
\fill[black] (0,0.6) circle (0.05cm);
\fill[black] (0,0.9) circle (0.05cm);
\draw (0,0) -- (0.4,0) -- (1.2,0.0);
\draw (0,0.6) -- (0.2,0.6) arc (-90:90:0.15cm) -- (0,0.9);
\fill[black] (1.2,0.0) circle (0.05cm);
\fill[black] (1.2,0.3) circle (0.05cm);
\fill[black] (1.2,0.6) circle (0.05cm);
\fill[black] (1.2,0.9) circle (0.05cm);
\draw (1.2,0.3) -- (0,0.3);
\draw (1.2,0.6) -- (1.0,0.6) arc (-90:-270:0.15cm) -- (1.2,0.9);
\draw[gray,dashed,line width=2,opacity=0.5,rounded corners=3] (0.2,-0.35) -- (0.6,-0.35) -- (0.6,1.25) -- (-0.2,1.25) -- (-0.2,-0.35) -- (0.2,-0.35); 
\end{tikzpicture} 
\end{array} \ = \ \frac{1}{d}\begin{array}{c}
\begin{tikzpicture}[thick]
\draw[gray,opacity=0.8,line width=2] (0,-0.25) -- (0,1.15);
\draw[gray,opacity=0.8,line width=2] (1.2,-0.25) -- (1.2,1.15);
\fill[black] (0,0.0) circle (0.05cm);
\fill[black] (0,0.3) circle (0.05cm);
\fill[black] (0,0.6) circle (0.05cm);
\fill[black] (0,0.9) circle (0.05cm);
\draw (0,0) -- (0.4,0) arc (-90:90:0.15cm) -- (0,0.3);
\draw (0,0.6) -- (0.4,0.6) arc (-90:90:0.15cm) -- (0,0.9);
\fill[black] (1.2,0.0) circle (0.05cm);
\fill[black] (1.2,0.3) circle (0.05cm);
\fill[black] (1.2,0.6) circle (0.05cm);
\fill[black] (1.2,0.9) circle (0.05cm);
\draw (1.2,0) -- (0.8,0) arc (-90:-270:0.15cm) -- (1.2,0.3);
\draw (1.2,0.6) -- (0.8,0.6) arc (-90:-270:0.15cm) -- (1.2,0.9);
\end{tikzpicture} 
\end{array}.
\ee
In the 3D formulation of this topological quantum mechanics model this means that one should be able to draw a 2-sphere which cuts the space into two halves (interior and exterior) and is crossed by only two lines.

With some intuition gained we can start discussing the classification. We will first give characteristic representatives of entanglement classes and then explain how other states from each class can be obtained from the representatives. We will also show how to relate the classification with the SLOCC one and with the Schmidt decomposition.

In the two-qubit case we have two candidates:
\be
\label{2qubitSLOCC}
\begin{array}{c}
\begin{tikzpicture}[thick]
\draw[gray,opacity=0.8,line width=2] (0,-0.25) -- (0,1.15);
\draw[gray,opacity=0.8,line width=2] (1.2,-0.25) -- (1.2,1.15);
\fill[black] (0,0.0) circle (0.05cm);
\fill[black] (0,0.3) circle (0.05cm);
\fill[black] (0,0.6) circle (0.05cm);
\fill[black] (0,0.9) circle (0.05cm);
\draw (0,0) -- (0.4,0) -- (1.2,0.0);
\draw (0,0.6) -- (0.4,0.6) arc (-90:90:0.15cm) -- (0,0.9);
\fill[black] (1.2,0.0) circle (0.05cm);
\fill[black] (1.2,0.3) circle (0.05cm);
\fill[black] (1.2,0.6) circle (0.05cm);
\fill[black] (1.2,0.9) circle (0.05cm);
\draw (1.2,0.3) -- (0,0.3);
\draw (1.2,0.6) -- (0.8,0.6) arc (-90:-270:0.15cm) -- (1.2,0.9);
\end{tikzpicture} 
\end{array}
\qquad \text{and} \qquad
\begin{array}{c}
\begin{tikzpicture}[thick]
\draw[gray,opacity=0.8,line width=2] (0,-0.25) -- (0,1.15);
\draw[gray,opacity=0.8,line width=2] (1.2,-0.25) -- (1.2,1.15);
\fill[black] (0,0.0) circle (0.05cm);
\fill[black] (0,0.3) circle (0.05cm);
\fill[black] (0,0.6) circle (0.05cm);
\fill[black] (0,0.9) circle (0.05cm);
\draw (0,0) -- (1.2,0);
\draw (0,0.6) -- (1.2,0.6);
\fill[black] (1.2,0.0) circle (0.05cm);
\fill[black] (1.2,0.3) circle (0.05cm);
\fill[black] (1.2,0.6) circle (0.05cm);
\fill[black] (1.2,0.9) circle (0.05cm);
\draw (1.2,0.3) -- (0,0.3);
\draw (1.2,0.9) -- (0,0.9);
\end{tikzpicture} 
\end{array}\,,
\ee
which correspond to the separable and the Bell classes respectively (the same two classes that exist for two-qubit entanglement in the SLOCC classification). Any permutation of points on either side is a unitary braiding operation, which is merely a local change of basis, which should not affect nonlocal correlations. Therefore all diagrams obtained from the above by local braiding will belong to the corresponding classes. 

More generally, by applying invertible tangles to the diagrams in~(\ref{2qubitSLOCC}) one can obtain all possible diagrams which will fall into two classes
\be
\begin{array}{c}
\begin{tikzpicture}[thick]
\newcommand{\x}{2.4}
\draw[gray,opacity=0.8,line width=2] (0,-0.25) -- (0,1.15);
\draw[gray,opacity=0.8,line width=2] (\x,-0.25) -- (\x,1.15);
\draw (0,0) -- (\x,0.0);
\draw (0,0.3) -- (\x,0.3);
\draw (0,0.6) -- (1.0,0.6) arc (-90:90:0.15cm) -- (0,0.9);
\draw (\x,0.6) -- (1.4,0.6) arc (-90:-270:0.15cm) -- (\x,0.9);
\foreach \y in {0,0.3,...,1.0}
    \fill[black] (0,\y) circle (0.05cm);
\foreach \y in {0,0.3,...,1.0}
    \fill[black] (\x,\y) circle (0.05cm);
\fill[white,rounded corners=3] (0.6,-0.2) -- (0.9,-0.2) -- (0.9,1.1) -- (0.3,1.1) -- (0.3,-0.2) -- (0.6,-0.2);
\fill[gray,opacity=0.4,rounded corners=3] (0.6,-0.2) -- (0.9,-0.2) -- (0.9,1.1) -- (0.3,1.1) -- (0.3,-0.2) -- (0.6,-0.2);
\fill[white,rounded corners=3] (0.6+\x/2,-0.2) -- (0.9+\x/2,-0.2) -- (0.9+\x/2,1.1) -- (0.3+\x/2,1.1) -- (0.3+\x/2,-0.2) -- (0.6+\x/2,-0.2);
\fill[gray,opacity=0.4,rounded corners=3] (0.6+\x/2,-0.2) -- (0.9+\x/2,-0.2) -- (0.9+\x/2,1.1) -- (0.3+\x/2,1.1) -- (0.3+\x/2,-0.2) -- (0.6+\x/2,-0.2);
\node[rotate=90] at (0.6,0.45) {$O_A$};
\node[rotate=-90] at (0.6+\x/2,0.45) {$O_B$};
\end{tikzpicture} 
\end{array}
\qquad \text{and} \qquad
\begin{array}{c}
\begin{tikzpicture}[thick]
\draw[gray,opacity=0.8,line width=2] (0,-0.25) -- (0,1.15);
\draw[gray,opacity=0.8,line width=2] (1.2,-0.25) -- (1.2,1.15);
\fill[black] (0,0.0) circle (0.05cm);
\fill[black] (0,0.3) circle (0.05cm);
\fill[black] (0,0.6) circle (0.05cm);
\fill[black] (0,0.9) circle (0.05cm);
\draw (0,0) -- (1.2,0);
\draw (0,0.6) -- (1.2,0.6);
\fill[black] (1.2,0.0) circle (0.05cm);
\fill[black] (1.2,0.3) circle (0.05cm);
\fill[black] (1.2,0.6) circle (0.05cm);
\fill[black] (1.2,0.9) circle (0.05cm);
\draw (1.2,0.3) -- (0,0.3);
\draw (1.2,0.9) -- (0,0.9);
\fill[white,rounded corners=3] (0.6,-0.2) -- (0.9,-0.2) -- (0.9,1.1) -- (0.3,1.1) -- (0.3,-0.2) -- (0.6,-0.2);
\fill[gray,opacity=0.4,rounded corners=3] (0.6,-0.2) -- (0.9,-0.2) -- (0.9,1.1) -- (0.3,1.1) -- (0.3,-0.2) -- (0.6,-0.2);
\node[rotate=90] at (0.6,0.45) {2PI};
\end{tikzpicture} 
\end{array}\,,
\ee
where $O_A$ and $O_B$ are any local operators (tangles) and 2PI are all possible \emph{2-particle-irreducible} tangles (that is diagrams that cannot be cut into two disconnected pieces by cutting any two lines). Both types of states can be obtained by the action of local invertible operators $O_A\otimes O_B$ on the respective states~(\ref{2qubitSLOCC}). This action is class-preserving. 

Next, we would like to generalize this example to arbitrary bipartite systems. As we will see this generalization is not straightforward due to the mentioned nonlinearity of the TQFT model. One way the nonlinearity manifests itself is in the scaling of the dimension of the system, which grows exponentially with the number of points on the boundary, as prescribed by~(\ref{Catalan}). If we add a pair of points on the boundaries and consider the 6-point Hilbert space, the dimension jumps to five. The generalization of~(\ref{2qubitSLOCC}) are
\be
\begin{array}{c}
\begin{tikzpicture}[thick]
\draw[gray,opacity=0.8,line width=2] (0,-0.15) -- (0,1.65);
\draw[gray,opacity=0.8,line width=2] (1.2,-0.15) -- (1.2,1.65);
\draw (0,0) -- (0.4,0) arc (-90:90:0.15cm) -- (0,0.3);
\draw (0,0.6) -- (0.4,0.6) arc (-90:90:0.15cm) -- (0,0.9);
\draw (1.2,0) -- (0.8,0) arc (-90:-270:0.15cm) -- (1.2,0.3);
\draw (1.2,0.6) -- (0.8,0.6) arc (-90:-270:0.15cm) -- (1.2,0.9);
\draw (1.2,1.2) -- (0.8,1.2) arc (-90:-270:0.15cm) -- (1.2,1.5);
\draw (0,1.2) -- (0.4,1.2) arc (-90:90:0.15cm) -- (0,1.5);
\foreach \y in {0.0,0.3,...,1.6}
\fill[black] (0,\y) circle (0.05cm);
\foreach \y in {0.0,0.3,...,1.6}
\fill[black] (1.2,\y) circle (0.05cm);
\end{tikzpicture} 
\end{array}
\,, \qquad 
\begin{array}{c}
\begin{tikzpicture}[thick]
\draw[gray,opacity=0.8,line width=2] (0,-0.15) -- (0,1.65);
\draw[gray,opacity=0.8,line width=2] (1.2,-0.15) -- (1.2,1.65);
\draw (0,0) -- (0.4,0) -- (1.2,0.0);
\draw (0,0.6) -- (0.4,0.6) arc (-90:90:0.15cm) -- (0,0.9);
\draw (1.2,0.3) -- (0,0.3);
\draw (1.2,0.6) -- (0.8,0.6) arc (-90:-270:0.15cm) -- (1.2,0.9);
\draw (1.2,1.2) -- (0.8,1.2) arc (-90:-270:0.15cm) -- (1.2,1.5);
\draw (0,1.2) -- (0.4,1.2) arc (-90:90:0.15cm) -- (0,1.5);
\foreach \y in {0.0,0.3,...,1.6}
\fill[black] (0,\y) circle (0.05cm);
\foreach \y in {0.0,0.3,...,1.6}
\fill[black] (1.2,\y) circle (0.05cm);
\end{tikzpicture} 
\end{array}
\,, \qquad 
\begin{array}{c}
\begin{tikzpicture}[thick]
\draw[gray,opacity=0.8,line width=2] (0,-0.15) -- (0,1.65);
\draw[gray,opacity=0.8,line width=2] (1.2,-0.15) -- (1.2,1.65);
\draw (0,0) -- (0.4,0) -- (1.2,0.0);
\draw (0,1.2) -- (0.4,1.2) arc (-90:90:0.15cm) -- (0,1.5);
\draw (1.2,0.3) -- (0,0.3);
\draw (1.2,1.2) -- (0.8,1.2) arc (-90:-270:0.15cm) -- (1.2,1.5);
\draw (0,0.6) -- (1.2,0.6);
\draw (1.2,0.9) -- (0,0.9);
\foreach \y in {0.0,0.3,...,1.6}
\fill[black] (0,\y) circle (0.05cm);
\foreach \y in {0.0,0.3,...,1.6}
\fill[black] (1.2,\y) circle (0.05cm);
\end{tikzpicture} 
\end{array}
\,,\qquad 
\begin{array}{c}
\begin{tikzpicture}[thick]
\draw[gray,opacity=0.8,line width=2] (0,-0.15) -- (0,1.65);
\draw[gray,opacity=0.8,line width=2] (1.2,-0.15) -- (1.2,1.65);
\draw (0,0) -- (1.2,0);
\draw (0,0.6) -- (1.2,0.6);
\draw (1.2,0.3) -- (0,0.3);
\draw (1.2,0.9) -- (0,0.9);
\draw (0,1.2) -- (1.2,1.2);
\draw (1.2,1.5) -- (0,1.5);
\foreach \y in {0.0,0.3,...,1.6}
\fill[black] (0,\y) circle (0.05cm);
\foreach \y in {0.0,0.3,...,1.6}
\fill[black] (1.2,\y) circle (0.05cm);
\end{tikzpicture} 
\end{array}.
\ee
As follows from the above discussion, the first two diagrams are equivalent (they only differ in the normalization), so there are only three naive topological classes, against five in the SLOCC case. The three independent diagrams only capture matrices of rank one, two and five. This implies that no diagram can represent a rank three or rank four matrix. The latter can only be constructed via linear combinations of diagrams.

To circumvent this problem, we will make the model more ``linear'' using additional projectors~\cite{Melnikov:2022qyt}. Since the dimension of the $n$-point Hilbert space counts the number of possible singlets that can be formed by the representations of the points, we can note that four spin-one particles can form singlets in three different ways, four spin $3/2$ particles can make four singlets and four spin $j$ particles can form $2j+1$ singlets. Therefore to construct Hilbert spaces of arbitrary dimension we should rather consider representations other than fundamental. However, since the TQFT model was defined only for particles/Wilson lines in the fundamental representation we should find a way to express the calculations in terms of the fundamental particles, using tensor products of representations. A useful tool for this is the Jones-Wenzl (JW) projector, or symmetrizer~\cite{Jones:1985dw,Wenzl:1985seq,Kauffman:1994tem}, which we will now define.

Let us define a recursive series of diagrams, which belong to a Temperley-Lieb algebra: 
\be
\begin{tikzpicture}[baseline=2]
\fill[cyan] (0,0) rectangle (1,0.5);
\draw[thick] (0,0) rectangle (1,0.5);
\draw[ultra thick] (0.2,1) node[anchor=east] {$n$} -- (0.2,0.5);
\draw[ultra thick] (0.2,0) -- (0.2,-0.5);
\draw[thick] (0.5,-0.5) -- (0.5,0);
\draw[thick] (0.5,1) -- (0.5,0.5);
\draw[thick] (0.8,-0.5) -- (0.8,0);
\draw[thick] (0.8,1) -- (0.8,0.5);
\end{tikzpicture}\quad \ = \ \quad 
\begin{tikzpicture}[baseline=2]
\fill[cyan] (0,0) rectangle (1,0.5);
\draw[thick] (0,0) rectangle (1,0.5);
\draw[ultra thick] (0.3,-0.5) -- (0.3,0);
\draw[ultra thick] (0.3,1) node[anchor=east] {$n$} -- (0.3,0.5);
\draw[thick] (0.7,-0.5) -- (0.7,0);
\draw[thick] (0.7,1) -- (0.7,0.5);
\draw[thick] (1.2,-0.5) -- (1.2,1);
\end{tikzpicture}\quad \ - \ \frac{\Delta_n}{\Delta_{n+1}} \quad \begin{tikzpicture}[baseline=2]
\fill[cyan] (0,0.5) rectangle (1,0.8);
\draw[thick] (0,0.5) rectangle (1,0.8);
\draw[ultra thick] (0.3,-0.) -- (0.3,0.5);
\draw[ultra thick] (0.3,-0.5) -- (0.3,-0.3);
\draw[ultra thick] (0.3,1) node[anchor=east] {$n$} -- (0.3,0.8);
\draw[thick] (0.7,-0.)  arc (180:0:0.2) -- (1.1,-0.5);
\draw[thick] (0.7,1) -- (0.7,0.8);
\draw[thick] (0.7,-0.5) -- (0.7,-0.3);
\draw[thick] (1.1,1) -- (1.1,0.5) arc (0:-180:0.2);
\fill[cyan] (0,-0.3) rectangle (1,0.);
\draw[thick] (0,-0.3) rectangle (1,0.);
\end{tikzpicture}.
\ee
Here the thick line substitutes $n$ normal lines, so the diagrams correspond to an element of $TL_{n+2}$. The $(n+2)$-line symmetrizer is defined through an embedding of elements of $TL_{n+1}$. The coefficients of the recursion relations are 
\be
\Delta_{-1}\ = \ 0\,,\qquad \Delta_{0} \ = \ 1\,, \qquad \Delta_{n+1} \ = \ d\Delta_{n} - \Delta_{n-1}.
\ee
When appended to a set of $n+2$ lines, this projector singles out the representation of spin $j=(n+1)/2$.

Let us consider a few examples. The combination
\be
\label{P3}
\begin{tikzpicture}[baseline=2,thick]
\fill[cyan] (0,0) rectangle (1,0.5);
\draw[thick] (0,0) rectangle (1,0.5);
\draw[thick] (0.3,-0.3) -- (0.3,0.0);
\draw[thick] (0.3,0.8) -- (0.3,0.5);
\draw[thick] (0.7,-0.3) -- (0.7,0);
\draw[thick] (0.7,0.8) -- (0.7,0.5);
\end{tikzpicture}
\quad \ = \ \quad 
\begin{tikzpicture}[baseline=2]
\draw[thick] (0.3,-0.3) -- (0.3,0.8);
\draw[thick] (0.7,-0.3) -- (0.7,0.8);
\end{tikzpicture} \ - \ \frac{1}{d} \quad \begin{tikzpicture}[baseline=2]
\draw[thick] (0.7,-0.)  arc (180:0:0.2) -- (1.1,-0.3);
\draw[thick] (0.7,0.8) -- (0.7,0.5);
\draw[thick] (0.7,-0.3) -- (0.7,0);
\draw[thick] (1.1,0.8) -- (1.1,0.5) arc (0:-180:0.2);
\end{tikzpicture}
\ee
makes a $j=1$ particle from a pair of spins $j=1/2$. Note that the last diagram in the projector is a projector on a singlet. The above combination has already appeared in the construction of state $|1\rangle$~(\ref{4pobasis}). The fact that it is a projector orthogonal to the singlet projector makes state $|1\rangle$ orthogonal to state $|0\rangle$. The projector on the spin $3/2$ has the form
\be
\begin{array}{c}
       \begin{tikzpicture}[thick]
           \draw (0.1,0) -- (0.1,1);
           \draw (0.4,0) -- (0.4,1);
           \draw (0.7,0) -- (0.7,1);
           \fill[thick,cyan] (-0.1,0.25) rectangle (0.9,0.75);
\draw[thick] (-0.1,0.25) rectangle (0.9,0.75);
       \end{tikzpicture}
\end{array}
\quad \ = \ \quad 
\begin{array}{c}
       \begin{tikzpicture}[thick]
           \draw (0.1,0) -- (0.1,1);
           \draw (0.4,0) -- (0.4,1);
           \draw (0.7,0) -- (0.7,1);
       \end{tikzpicture}
\end{array}\quad \ - \ \frac{d}{d^2-1}\left(
\begin{array}{c}
       \begin{tikzpicture}[thick]
           \draw (0.7,0) -- (0.7,1);
           \draw[rounded corners=2] (0.1,0) -- (0.1,0.4) -- (0.4,0.4) -- (0.4,0);
           \draw[rounded corners=2] (0.1,1) -- (0.1,0.6) -- (0.4,0.6) -- (0.4,1);
       \end{tikzpicture}
\end{array}+\begin{array}{c}
       \begin{tikzpicture}[thick]
           \draw (0.1,0) -- (0.1,1);
           \draw[rounded corners=2] (0.4,0) -- (0.4,0.4) -- (0.7,0.4) -- (0.7,0);
           \draw[rounded corners=2] (0.4,1) -- (0.4,0.6) -- (0.7,0.6) -- (0.7,1);
       \end{tikzpicture}
\end{array}
\right)
\quad +
\quad \frac{1}{d^2-1} \left(
\begin{array}{c}
       \begin{tikzpicture}[thick]
           \draw[rounded corners=2] (0.1,0) -- (0.1,0.5) -- (0.7,0.5) -- (0.7,1);
           \draw[rounded corners=2] (0.4,0) -- (0.4,0.3) -- (0.7,0.3) -- (0.7,0);
           \draw[rounded corners=2] (0.1,1) -- (0.1,0.7) -- (0.4,0.7) -- (0.4,1);
       \end{tikzpicture}
\end{array},
+
\begin{array}{c}
       \begin{tikzpicture}[thick]
           \draw[rounded corners=2] (0.7,0) -- (0.7,0.5) -- (0.1,0.5) -- (0.1,1);
           \draw[rounded corners=2] (0.1,0) -- (0.1,0.3) -- (0.4,0.3) -- (0.4,0);
           \draw[rounded corners=2] (0.4,1) -- (0.4,0.7) -- (0.7,0.7) -- (0.7,1);
       \end{tikzpicture}
\end{array}
\right).
\label{P4}
\ee

JW projectors can be used to construct Hilbert spaces (and states) associated with arbitrary representations. The use of lines in the fundamental representations will give additional advantages for detailing the internal structure of the states. 

The space of four spin $j=1$ representations has dimension three and can be constructed using eight fundamental points and four JW projectors~(\ref{P3}). One choice of a basis in this space is
\be
|e_0\rangle \ = \ \begin{array}{c}
\begin{tikzpicture}[thick]
\fill[black] (0,0.0) circle (0.05cm);
\fill[black] (0,0.3) circle (0.05cm);
\fill[black] (0,0.6) circle (0.05cm);
\fill[black] (0,0.9) circle (0.05cm);
\draw (0,0) -- (0.3,0) arc (-90:90:0.45cm) -- (0,0.9);
\draw (0,0.6) -- (0.3,0.6) arc (90:-90:0.15cm) -- (0,0.3);
\fill[cyan] (0.1,-0.1) rectangle (0.3,0.4);
\draw (0.1,-0.1) rectangle (0.3,0.4);
\fill[cyan] (0.1,0.5) rectangle (0.3,1);
\draw (0.1,0.5) rectangle (0.3,1);
\newcommand\y{1.2};
\fill[black] (0,0.0+\y) circle (0.05cm);
\fill[black] (0,0.3+\y) circle (0.05cm);
\fill[black] (0,0.6+\y) circle (0.05cm);
\fill[black] (0,0.9+\y) circle (0.05cm);
\draw (0,0+\y) -- (0.3,0+\y) arc (-90:90:0.45cm) -- (0,0.9+\y);
\draw (0,0.6+\y) -- (0.3,0.6+\y) arc (90:-90:0.15cm) -- (0,0.3+\y);
\fill[cyan] (0.1,-0.1+\y) rectangle (0.3,0.4+\y);
\draw (0.1,-0.1+\y) rectangle (0.3,0.4+\y);
\fill[cyan] (0.1,0.5+\y) rectangle (0.3,1+\y);
\draw (0.1,0.5+\y) rectangle (0.3,1+\y);
\end{tikzpicture} 
\end{array}\,, \qquad
|e_1\rangle \ = \ \begin{array}{c}
\begin{tikzpicture}[thick]
\newcommand\y{1.2};
\fill[black] (0,0.0) circle (0.05cm);
\fill[black] (0,0.3) circle (0.05cm);
\fill[black] (0,0.6) circle (0.05cm);
\fill[black] (0,0.9) circle (0.05cm);
\draw (0,0) -- (0.3,0) arc (-90:90:1.05cm) -- (0,0.9+\y);
\draw (0,0.6) -- (0.3,0.6) arc (90:-90:0.15cm) -- (0,0.3);
\fill[black] (0,0.0+\y) circle (0.05cm);
\fill[black] (0,0.3+\y) circle (0.05cm);
\fill[black] (0,0.6+\y) circle (0.05cm);
\fill[black] (0,0.9+\y) circle (0.05cm);
\draw (0,0+\y) -- (0.9,0+\y) arc (90:-90:0.15cm) -- (0,0.9);
\draw (0,0.6+\y) -- (0.3,0.6+\y) arc (90:-90:0.15cm) -- (0,0.3+\y);
\fill[cyan] (0.1,-0.1) rectangle (0.3,0.4);
\draw (0.1,-0.1) rectangle (0.3,0.4);
\fill[cyan] (0.1,0.5) rectangle (0.3,1);
\draw (0.1,0.5) rectangle (0.3,1);
\fill[cyan] (0.1,-0.1+\y) rectangle (0.3,0.4+\y);
\draw (0.1,-0.1+\y) rectangle (0.3,0.4+\y);
\fill[cyan] (0.1,0.5+\y) rectangle (0.3,1+\y);
\draw (0.1,0.5+\y) rectangle (0.3,1+\y);
\end{tikzpicture} 
\end{array}\,, \qquad
|e_2\rangle \ = \ \begin{array}{c}
\begin{tikzpicture}[thick]
\newcommand\y{1.2};
\fill[black] (0,0.0) circle (0.05cm);
\fill[black] (0,0.3) circle (0.05cm);
\fill[black] (0,0.6) circle (0.05cm);
\fill[black] (0,0.9) circle (0.05cm);
\draw (0,0) -- (0.3,0) arc (-90:90:1.05cm) -- (0,0.9+\y);
\draw (0,0.6) -- (0.3,0.6) arc (-90:90:0.45cm) -- (0,0.3+\y);
\fill[black] (0,0.0+\y) circle (0.05cm);
\fill[black] (0,0.3+\y) circle (0.05cm);
\fill[black] (0,0.6+\y) circle (0.05cm);
\fill[black] (0,0.9+\y) circle (0.05cm);
\draw (0,0+\y) -- (0.3,0+\y) arc (90:-90:0.15cm) -- (0,0.9);
\draw (0,0.6+\y) -- (0.3,0.6+\y) arc (90:-90:0.75cm) -- (0,0.3);
\fill[cyan] (0.1,-0.1) rectangle (0.3,0.4);
\draw (0.1,-0.1) rectangle (0.3,0.4);
\fill[cyan] (0.1,0.5) rectangle (0.3,1);
\draw (0.1,0.5) rectangle (0.3,1);
\fill[cyan] (0.1,-0.1+\y) rectangle (0.3,0.4+\y);
\draw (0.1,-0.1+\y) rectangle (0.3,0.4+\y);
\fill[cyan] (0.1,0.5+\y) rectangle (0.3,1+\y);
\draw (0.1,0.5+\y) rectangle (0.3,1+\y);
\end{tikzpicture} 
\end{array}\,.
\label{qutritbasis}
\ee
Note that closing any pair of neighboring inputs or outputs in a JW projector gives zero,
\be
\label{platclosure}
\begin{tikzpicture}[baseline=2]
\draw[thick] (0,0) rectangle (1.5,0.5);
\draw[thick] (0.2,0.5) -- (0.2,0.7) arc (180:0:0.15) -- (0.5,0.5);
\draw[thick] (0.8,0.5) -- (0.8,0.85);
\draw[thick] (0.5,0) -- (0.5,-0.35);
\draw[thick] (0.2,0) -- (0.2,-0.35);
\draw[thick] (0.8,0) -- (0.8,-0.35);
\draw (1.2,0.7) node {\bf $\cdots$};
\draw (1.2,-0.2) node {\bf $\cdots$};
\end{tikzpicture}
\ = \  \begin{tikzpicture}[baseline=2]
\draw[thick] (0,0) rectangle (1.5,0.5);
\draw[thick] (0.5,0.5) -- (0.5,0.7) arc (180:0:0.15) -- (0.8,0.5);
\draw[thick] (0.2,0.5) -- (0.2,0.85);
\draw[thick] (0.5,0) -- (0.5,-0.35);
\draw[thick] (0.2,0) -- (0.2,-0.35);
\draw[thick] (0.8,0) -- (0.8,-0.35);
\draw (1.2,0.7) node {\bf $\cdots$};
\draw (1.2,-0.2) node {\bf $\cdots$};
\end{tikzpicture} \ = \ \begin{tikzpicture}[baseline=2]
\draw[thick] (0,0) rectangle (1.5,0.5);
\draw[thick] (0.2,0.5) -- (0.2,0.85);
\draw[thick] (0.8,0.5) -- (0.8,0.85);
\draw[thick] (0.5,0.5) -- (0.5,0.85);
\draw[thick] (0.2,0) -- (0.2,-0.2) arc (-180:0:0.15) -- (0.5,0);
\draw[thick] (0.8,0) -- (0.8,-0.35);
\draw (1.2,0.7) node {\bf $\cdots$};
\draw (1.2,-0.2) node {\bf $\cdots$};
\end{tikzpicture} \ = \ 0\,.
\ee
An easy way of seeing this is to realize that closing a pair of lines corresponding to projecting onto a singlet. Then the remaining lines are not enough to form a representation of the maximum spin. Therefore one can convince oneself that diagrams~(\ref{qutritbasis}) are the only possibility to close lines without crossings, so it is a generalization of the qubit Temperley-Lieb basis~(\ref{2qubitSLOCC}). As in the qubit case, one can construct an orthonormal qutrit basis. It is convenient to express it in terms of the projectors:
\be
\label{qutritobasis}
\Nc_0|0\rangle \ = \ \begin{array}{c}
\begin{tikzpicture}[thick]
\fill[black] (0,0.0) circle (0.05cm);
\fill[black] (0,0.3) circle (0.05cm);
\fill[black] (0,0.6) circle (0.05cm);
\fill[black] (0,0.9) circle (0.05cm);
\draw (0,0) -- (0.3,0) arc (-90:90:0.45cm) -- (0,0.9);
\draw (0,0.6) -- (0.3,0.6) arc (90:-90:0.15cm) -- (0,0.3);
\fill[cyan] (0.1,-0.1) rectangle (0.3,0.4);
\draw (0.1,-0.1) rectangle (0.3,0.4);
\fill[cyan] (0.1,0.5) rectangle (0.3,1);
\draw (0.1,0.5) rectangle (0.3,1);
\newcommand\y{1.2};
\fill[black] (0,0.0+\y) circle (0.05cm);
\fill[black] (0,0.3+\y) circle (0.05cm);
\fill[black] (0,0.6+\y) circle (0.05cm);
\fill[black] (0,0.9+\y) circle (0.05cm);
\draw (0,0+\y) -- (0.3,0+\y) arc (-90:90:0.45cm) -- (0,0.9+\y);
\draw (0,0.6+\y) -- (0.3,0.6+\y) arc (90:-90:0.15cm) -- (0,0.3+\y);
\fill[cyan] (0.1,-0.1+\y) rectangle (0.3,0.4+\y);
\draw (0.1,-0.1+\y) rectangle (0.3,0.4+\y);
\fill[cyan] (0.1,0.5+\y) rectangle (0.3,1+\y);
\draw (0.1,0.5+\y) rectangle (0.3,1+\y);
\end{tikzpicture} 
\end{array}\,, \qquad
\Nc_0|1\rangle \ = \ \begin{array}{c}
\begin{tikzpicture}[thick]
\newcommand\y{1.2};
\fill[black] (0,0.0) circle (0.05cm);
\fill[black] (0,0.3) circle (0.05cm);
\fill[black] (0,0.6) circle (0.05cm);
\fill[black] (0,0.9) circle (0.05cm);
\draw (0,0) -- (0.3,0) arc (-90:90:1.05cm) -- (0,0.9+\y);
\draw (0,0.6) -- (0.3,0.6) arc (90:-90:0.15cm) -- (0,0.3);
\fill[black] (0,0.0+\y) circle (0.05cm);
\fill[black] (0,0.3+\y) circle (0.05cm);
\fill[black] (0,0.6+\y) circle (0.05cm);
\fill[black] (0,0.9+\y) circle (0.05cm);
\draw (0,0+\y) -- (0.9,0+\y) arc (90:-90:0.15cm) -- (0,0.9);
\draw (0,0.6+\y) -- (0.3,0.6+\y) arc (90:-90:0.15cm) -- (0,0.3+\y);
\fill[cyan] (0.1,-0.1) rectangle (0.3,0.4);
\draw (0.1,-0.1) rectangle (0.3,0.4);
\fill[cyan] (0.1,0.5) rectangle (0.3,1);
\draw (0.1,0.5) rectangle (0.3,1);
\fill[cyan] (0.1,-0.1+\y) rectangle (0.3,0.4+\y);
\draw (0.1,-0.1+\y) rectangle (0.3,0.4+\y);
\fill[cyan] (0.1,0.5+\y) rectangle (0.3,1+\y);
\draw (0.1,0.5+\y) rectangle (0.3,1+\y);
\fill[cyan] (0.9,1.0) rectangle (1.5,1.1);
\draw (0.9,1.0) rectangle (1.5,1.1);
\end{tikzpicture} 
\end{array}\,, \qquad
\Nc_2|2\rangle \ = \ \begin{array}{c}
\begin{tikzpicture}[thick]
\newcommand\y{1.2};
\newcommand\x{0.2};
\fill[black] (0,0.0) circle (0.05cm);
\fill[black] (0,0.3) circle (0.05cm);
\fill[black] (0,0.6) circle (0.05cm);
\fill[black] (0,0.9) circle (0.05cm);
\draw (0,0) -- (0.3+\x,0) arc (-90:90:1.05cm) -- (0,0.9+\y);
\draw (0,0.6) -- (0.3+\x,0.6) arc (-90:90:0.45cm) -- (0,0.3+\y);
\fill[black] (0,0.0+\y) circle (0.05cm);
\fill[black] (0,0.3+\y) circle (0.05cm);
\fill[black] (0,0.6+\y) circle (0.05cm);
\fill[black] (0,0.9+\y) circle (0.05cm);
\draw (0,0+\y) -- (0.3+\x,0+\y) arc (90:-90:0.15cm) -- (0,0.9);
\draw (0,0.6+\y) -- (0.3+\x,0.6+\y) arc (90:-90:0.75cm) -- (0,0.3);
\fill[cyan] (0.1,-0.1) rectangle (0.3,0.4);
\draw (0.1,-0.1) rectangle (0.3,0.4);
\fill[cyan] (0.1,0.5) rectangle (0.3,1);
\draw (0.1,0.5) rectangle (0.3,1);
\fill[cyan] (0.1,-0.1+\y) rectangle (0.3,0.4+\y);
\draw (0.1,-0.1+\y) rectangle (0.3,0.4+\y);
\fill[cyan] (0.1,0.5+\y) rectangle (0.3,1+\y);
\draw (0.1,0.5+\y) rectangle (0.3,1+\y);
\fill[pink] (0.3+\x,1.0) rectangle (1.5+\x,1.1);
\draw (0.3+\x,1.0) rectangle (1.5+\x,1.1);
\end{tikzpicture} 
\end{array}\,,
\ee
with easily computable normalization factors $\Nc_{0,1,2}$. Here the blue boxes project on spin $j=1$ representation and the pink one -- on $j=2$, so that the three basis elements are labeled by spins $0$, $1$ and $2$ appearing in the intermediate channel. 

The qubit basis~(\ref{4pobasis}) can also be cast in a similar way using the JW projectors, with spins $0$ and $1$ appearing in the intermediate channel. Let us mention again the connection with conformal blocks, such as~(\ref{conformalblock}). The corresponding diagrams can be constructed explicitly with the help of the projectors, for example,
\be
\begin{array}{c}
\begin{tikzpicture}[thick]
\newcommand{\x}{2.5}
\newcommand{\y}{1}
\fill[black] (0,0.0) circle (0.05cm);
\fill[black] (0,0.1) circle (0.05cm);
\draw[rounded corners=2] (0,0) -- (0.4,0) -- (0.95,0.55) -- (0.4,1.1) -- (0,1.1);
\draw[rounded corners=2] (0,0.1) -- (0.4,0.1) -- (0.85,0.55) -- (0.4,1.0) -- (0,1.0);
\fill[black] (0,1.0) circle (0.05cm);
\fill[black] (0,1.1) circle (0.05cm);
\fill[cyan] (0.1,-0.1) rectangle (0.3,0.2);
\draw (0.1,-0.1) rectangle (0.3,0.2);
\fill[cyan] (0.1,-0.1+\y) rectangle (0.3,0.2+\y);
\draw (0.1,-0.1+\y) rectangle (0.3,0.2+\y);

\fill[black] (\x-0,0.0) circle (0.05cm);
\fill[black] (\x-0,0.1) circle (0.05cm);
\draw[rounded corners=2] (\x-0,0) -- (\x-0.4,0) -- (\x-0.95,0.55) -- (\x-0.4,1.1) -- (\x-0,1.1);
\draw[rounded corners=2] (\x-0,0.1) -- (\x-0.4,0.1) -- (\x-0.85,0.55) -- (\x-0.4,1.0) -- (\x-0,1.0);
\fill[black] (\x-0,1.0) circle (0.05cm);
\fill[black] (\x-0,1.1) circle (0.05cm);
\fill[cyan] (\x-0.1,-0.1) rectangle (\x-0.3,0.2);
\draw (\x-0.1,-0.1) rectangle (\x-0.3,0.2);
\fill[cyan] (\x-0.1,-0.1+\y) rectangle (\x-0.3,0.2+\y);
\draw (\x-0.1,-0.1+\y) rectangle (\x-0.3,0.2+\y);
\end{tikzpicture} 
\end{array}
\,,\qquad 
\begin{array}{c}
\begin{tikzpicture}[thick]
\newcommand{\x}{2.5}
\newcommand{\y}{1}
\fill[black] (0,0.0) circle (0.05cm);
\fill[black] (0,0.1) circle (0.05cm);
\draw[rounded corners=2] (0,0) -- (0.4,0) -- (0.9,0.5) -- (\x-0.9,0.5) -- (\x-0.4,0) -- (\x-0,0);
\draw[rounded corners=2] (0,0.1) -- (0.4,0.1) -- (0.85,0.55) -- (0.4,1.0) -- (0,1.0);
\fill[black] (0,1.0) circle (0.05cm);
\fill[black] (0,1.1) circle (0.05cm);

\fill[black] (\x-0,0.0) circle (0.05cm);
\fill[black] (\x-0,0.1) circle (0.05cm);
\draw[rounded corners=2] (0,\y+0.1) -- (0.4,\y+0.1) -- (0.9,\y/2+0.1) -- (\x-0.9,0.6) -- (\x-0.4,1.1) -- (\x-0,1.1);
\draw[rounded corners=2] (\x-0,0.1) -- (\x-0.4,0.1) -- (\x-0.85,0.55) -- (\x-0.4,1.0) -- (\x-0,1.0);
\fill[black] (\x-0,1.0) circle (0.05cm);
\fill[black] (\x-0,1.1) circle (0.05cm);
\fill[cyan] (\x-0.1,-0.1) rectangle (\x-0.3,0.2);
\draw (\x-0.1,-0.1) rectangle (\x-0.3,0.2);
\fill[cyan] (\x-0.1,-0.1+\y) rectangle (\x-0.3,0.2+\y);
\draw (\x-0.1,-0.1+\y) rectangle (\x-0.3,0.2+\y);
\fill[cyan] (0.1,-0.1) rectangle (0.3,0.2);
\draw (0.1,-0.1) rectangle (0.3,0.2);
\fill[cyan] (0.1,-0.1+\y) rectangle (0.3,0.2+\y);
\draw (0.1,-0.1+\y) rectangle (0.3,0.2+\y);
\fill[cyan] (\x/2-0.1,\y/2-0.1) rectangle (\x/2+0.1,\y/2+0.2);
\draw (\x/2-0.1,\y/2-0.1) rectangle (\x/2+0.1,\y/2+0.2);
\end{tikzpicture} 
\end{array}
\,,\qquad 
\begin{array}{c}
\begin{tikzpicture}[thick]
\newcommand{\x}{2.5}
\newcommand{\y}{1}
\fill[black] (0,0.0) circle (0.05cm);
\fill[black] (0,0.1) circle (0.05cm);
\draw[rounded corners=2] (0,0) -- (0.4,0) -- (0.8,0.4) -- (\x-0.8,0.4) -- (\x-0.4,0) -- (\x-0,0);
\draw[rounded corners=2] (0,0.1) -- (0.4,0.1) -- (0.8,0.5) -- (\x-0.8,0.5) -- (\x-0.4,0.1) -- (\x-0,0.1);
\fill[black] (0,1.0) circle (0.05cm);
\fill[black] (0,1.1) circle (0.05cm);

\fill[black] (\x-0,0.0) circle (0.05cm);
\fill[black] (\x-0,0.1) circle (0.05cm);
\draw[rounded corners=2] (0,\y+0.1) -- (0.4,\y+0.1) -- (0.8,\y/2+0.2) -- (\x-0.8,0.7) -- (\x-0.4,1.1) -- (\x-0,1.1);
\draw[rounded corners=2] (\x-0,1.0) -- (\x-0.4,1.0) -- (\x-0.8,0.6) -- (0.8,0.6) -- (0.4,1.0) -- (0,1.0);
\fill[black] (\x-0,1.0) circle (0.05cm);
\fill[black] (\x-0,1.1) circle (0.05cm);
\fill[cyan] (\x-0.1,-0.1) rectangle (\x-0.3,0.2);
\draw (\x-0.1,-0.1) rectangle (\x-0.3,0.2);
\fill[cyan] (\x-0.1,-0.1+\y) rectangle (\x-0.3,0.2+\y);
\draw (\x-0.1,-0.1+\y) rectangle (\x-0.3,0.2+\y);
\fill[cyan] (0.1,-0.1) rectangle (0.3,0.2);
\draw (0.1,-0.1) rectangle (0.3,0.2);
\fill[cyan] (0.1,-0.1+\y) rectangle (0.3,0.2+\y);
\draw (0.1,-0.1+\y) rectangle (0.3,0.2+\y);
\fill[pink] (\x/2-0.1,\y/2-0.2) rectangle (\x/2+0.1,\y/2+0.3);
\draw (\x/2-0.1,\y/2-0.2) rectangle (\x/2+0.1,\y/2+0.3);
\end{tikzpicture} 
\end{array}.
\ee
The diagrams constructed with lines with different spins also appear in the \emph{spin networks} proposed by Roger Penrose as a skeleton of spacetime~\cite{Penrose1971:ang} and later used in the construction of the states of (loop) quantum gravity~\cite{Rovelli:1995ac}. JW projectors take a proper care of the spin summation rules at the junctions. 

Using Hilbert spaces of representations of different spin, with the help of JW projectors, we can obtain another generalization of the entanglement classification~(\ref{2qubitSLOCC}). In the qutrit case the following diagrams will represent matrices of rank one, two and three, respectively~\cite{Melnikov:2022qyt}:
\be
\label{2qutritSLOCC}
\begin{array}{c}
\begin{tikzpicture}[thick]
\fill[pink] (-0.2,1.1) rectangle (0.1,1.6);
\fill[pink] (-0.2,-0.1) rectangle (0.1,0.4);
\fill[pink] (-0.2,0.5) rectangle (0.1,1.0);
\fill[pink] (-0.2,1.7) rectangle (0.1,2.2);
\fill[cyan] (1.1,1.1) rectangle (1.3,1.6);
\fill[cyan] (1.1,-0.1) rectangle (1.3,0.4);
\fill[cyan] (1.1,0.5) rectangle (1.3,1.0);
\fill[cyan] (1.1,1.7) rectangle (1.3,2.2);
\fill[black] (0,0.0) circle (0.05cm);
\fill[black] (0,0.3) circle (0.05cm);
\fill[black] (0,0.6) circle (0.05cm);
\fill[black] (0,0.9) circle (0.05cm);
\draw (0,0) -- (0.4,0) -- (1.2,0.0);
\draw (0,1.2) -- (0.1,1.2) arc (-90:90:0.45cm) -- (0,2.1);
\fill[black] (1.2,0.0) circle (0.05cm);
\fill[black] (1.2,0.3) circle (0.05cm);
\fill[black] (1.2,0.6) circle (0.05cm);
\fill[black] (1.2,0.9) circle (0.05cm);
\draw (1.2,0.3) -- (0,0.3);
\draw (1.2,1.2) -- (1.1,1.2) arc (-90:-270:0.45cm) -- (1.2,2.1);
\fill[black] (1.2,1.2) circle (0.05cm);
\fill[black] (1.2,1.5) circle (0.05cm);
\fill[black] (0,1.2) circle (0.05cm);
\fill[black] (0,1.5) circle (0.05cm);
\draw (0,0.6) -- (1.2,0.6);
\draw (1.2,0.9) -- (0,0.9);
\fill[black] (1.2,1.8) circle (0.05cm);
\fill[black] (1.2,2.1) circle (0.05cm);
\fill[black] (0,1.8) circle (0.05cm);
\fill[black] (0,2.1) circle (0.05cm);
\draw (0,1.8) -- (0.1,1.8) arc (90:-90:0.15) -- (0,1.5);
\draw (1.2,1.8) -- (1.1,1.8) arc (90:270:0.15) -- (1.2,1.5);
\end{tikzpicture} 
\end{array}
\,,\qquad \begin{array}{c}
\begin{tikzpicture}[thick]
\fill[pink] (-0.2,1.1) rectangle (0.1,1.6);
\fill[pink] (-0.2,-0.1) rectangle (0.1,0.4);
\fill[pink] (-0.2,0.5) rectangle (0.1,1.0);
\fill[pink] (-0.2,1.7) rectangle (0.1,2.2);
\fill[cyan] (1.1,1.1) rectangle (1.3,1.6);
\fill[cyan] (1.1,-0.1) rectangle (1.3,0.4);
\fill[cyan] (1.1,0.5) rectangle (1.3,1.0);
\fill[cyan] (1.1,1.7) rectangle (1.3,2.2);
\fill[black] (0,0.0) circle (0.05cm);
\fill[black] (0,0.3) circle (0.05cm);
\fill[black] (0,0.6) circle (0.05cm);
\fill[black] (0,0.9) circle (0.05cm);
\draw (0,0) -- (0.4,0) -- (1.2,0.0);
\draw (0,2.1) -- (1.2,2.1);
\fill[black] (1.2,0.0) circle (0.05cm);
\fill[black] (1.2,0.3) circle (0.05cm);
\fill[black] (1.2,0.6) circle (0.05cm);
\fill[black] (1.2,0.9) circle (0.05cm);
\draw (1.2,0.3) -- (0,0.3);
\draw (1.2,0.6) -- (0,0.6);
\fill[black] (1.2,1.2) circle (0.05cm);
\fill[black] (1.2,1.5) circle (0.05cm);
\fill[black] (0,1.2) circle (0.05cm);
\fill[black] (0,1.5) circle (0.05cm);
\draw (0,0.9) -- (1.2,0.9);
\draw (0,1.2) -- (1.2,1.2);
\fill[black] (1.2,1.8) circle (0.05cm);
\fill[black] (1.2,2.1) circle (0.05cm);
\fill[black] (0,1.8) circle (0.05cm);
\fill[black] (0,2.1) circle (0.05cm);
\draw (0,1.8) -- (0.1,1.8) arc (90:-90:0.15) -- (0,1.5);
\draw (1.2,1.8) -- (1.1,1.8) arc (90:270:0.15) -- (1.2,1.5);
\end{tikzpicture} 
\end{array}
\,, \qquad \text{and}\qquad  
\begin{array}{c}
\begin{tikzpicture}[thick]
\fill[pink] (-0.2,1.1) rectangle (0.1,1.6);
\fill[pink] (-0.2,-0.1) rectangle (0.1,0.4);
\fill[pink] (-0.2,0.5) rectangle (0.1,1.0);
\fill[pink] (-0.2,1.7) rectangle (0.1,2.2);
\fill[cyan] (1.1,1.1) rectangle (1.3,1.6);
\fill[cyan] (1.1,-0.1) rectangle (1.3,0.4);
\fill[cyan] (1.1,0.5) rectangle (1.3,1.0);
\fill[cyan] (1.1,1.7) rectangle (1.3,2.2);
\fill[black] (0,0.0) circle (0.05cm);
\fill[black] (0,0.3) circle (0.05cm);
\fill[black] (0,0.6) circle (0.05cm);
\fill[black] (0,0.9) circle (0.05cm);
\draw (0,0) -- (1.2,0);
\draw (0,0.6) -- (1.2,0.6);
\fill[black] (1.2,0.0) circle (0.05cm);
\fill[black] (1.2,0.3) circle (0.05cm);
\fill[black] (1.2,0.6) circle (0.05cm);
\fill[black] (1.2,0.9) circle (0.05cm);
\draw (1.2,0.3) -- (0,0.3);
\draw (1.2,0.9) -- (0,0.9);
\fill[black] (1.2,1.2) circle (0.05cm);
\fill[black] (1.2,1.5) circle (0.05cm);
\fill[black] (0,1.2) circle (0.05cm);
\fill[black] (0,1.5) circle (0.05cm);
\draw (0,1.2) -- (1.2,1.2);
\draw (1.2,1.5) -- (0,1.5);
\fill[black] (1.2,1.8) circle (0.05cm);
\fill[black] (1.2,2.1) circle (0.05cm);
\fill[black] (0,1.8) circle (0.05cm);
\fill[black] (0,2.1) circle (0.05cm);
\draw (0,1.8) -- (1.2,1.8);
\draw (1.2,2.1) -- (0,2.1);
\end{tikzpicture} 
\end{array}.
\ee
This statement can be checked using qutrit basis~(\ref{qutritobasis}). Moreover, breaking more connections between left and right will produce states equivalent to the separable state $|00\rangle$ expressed by the left diagram above, similarly to equivalence~(\ref{surgery}) for the two-qubit diagrams. Therefore for a pair of qutrits we find a classification analogous to the SLOCC one, based on the ranks of matrices. A generic 2-qutrit state will fall in one of the tree classes represented by irreducible, 4-particle-irreducible and reducible diagrams.  

This picture can be easily generalized to arbitrary 2-qudit situation. For a pair of Hilbert spaces of dimension $2j+1$ one considers the $(4j+2)$-point Hilbert space projected on the subspace of four spin-$j$ representations with four JW projectors. Considering all possible connections between the two sets of points by nonintersecting lines (planar diagrams), modulo permutations of points and subject to constraints~(\ref{platclosure}), as well as generalizations of~(\ref{surgery}), one obtains a set of diagrams labeling matrices of rank $r=1,2,\ldots,2j+1$.\footnote{Note that the equivalence classes of diagrams also correspond to the connectomes of the classical limit of Chern-Simons theory, discussed in section~\ref{sec:classlimit}, modulo permutations of the points in each Hilbert space factor.} Constraints~(\ref{surgery}) will relate all diagrams with at most $2j+1$ lines connecting left and right Hilbert spaces, so in order to save some drawing, only a half of each diagram needs to be drawn (the upper half in the conventions of this paper). Summarizing, for $\Hc=\mathbb{C}^n\otimes \mathbb{C}^n$ one draws
\begin{eqnarray*}
n=2\,: & & \begin{array}{c}
\begin{tikzpicture}[thick]
\fill[black] (0,0.6) circle (0.05cm);
\fill[black] (0,0.9) circle (0.05cm);
\draw (0,0.6) -- (0.4,0.6) arc (-90:90:0.15cm) -- (0,0.9);
\fill[black] (1.2,0.6) circle (0.05cm);
\fill[black] (1.2,0.9) circle (0.05cm);
\draw (1.2,0.6) -- (0.8,0.6) arc (-90:-270:0.15cm) -- (1.2,0.9);
\end{tikzpicture} 
\end{array}
\,,\qquad 
\begin{array}{c}
\begin{tikzpicture}[thick]
\fill[black] (0,0.6) circle (0.05cm);
\fill[black] (0,0.9) circle (0.05cm);
\draw (0,0.6) -- (1.2,0.6);
\fill[black] (1.2,0.6) circle (0.05cm);
\fill[black] (1.2,0.9) circle (0.05cm);
\draw (1.2,0.9) -- (0,0.9);
\end{tikzpicture} 
\end{array}\,, \\ && \\
n=3\,: & & \begin{array}{c}
\begin{tikzpicture}[thick]
\fill[black] (0,0.0) circle (0.05cm);
\fill[black] (0,0.3) circle (0.05cm);
\fill[black] (0,0.6) circle (0.05cm);
\fill[black] (0,0.9) circle (0.05cm);
\draw (0,0) -- (0.1,0) arc (-90:90:0.45) -- (0.,0.9);
\draw (0,0.6) -- (0.1,0.6) arc (90:-90:0.15cm) -- (0,0.3);
\fill[black] (1.2,0.0) circle (0.05cm);
\fill[black] (1.2,0.3) circle (0.05cm);
\fill[black] (1.2,0.6) circle (0.05cm);
\fill[black] (1.2,0.9) circle (0.05cm);
\draw (1.2,0.9) -- (1.1,0.9) arc (90:270:0.45) -- (1.2,0);
\draw (1.2,0.6) -- (1.1,0.6) arc (90:270:0.15cm) -- (1.2,0.3);
\end{tikzpicture} 
\end{array}
\,,\qquad  
\begin{array}{c}
\begin{tikzpicture}[thick]
\fill[black] (0,0.0) circle (0.05cm);
\fill[black] (0,0.3) circle (0.05cm);
\fill[black] (0,0.6) circle (0.05cm);
\fill[black] (0,0.9) circle (0.05cm);
\draw (0,0) -- (0.4,0) -- (1.2,0.0);
\draw (0,0.6) -- (0.1,0.6) arc (90:-90:0.15cm) -- (0,0.3);
\fill[black] (1.2,0.0) circle (0.05cm);
\fill[black] (1.2,0.3) circle (0.05cm);
\fill[black] (1.2,0.6) circle (0.05cm);
\fill[black] (1.2,0.9) circle (0.05cm);
\draw (1.2,0.9) -- (0,0.9);
\draw (1.2,0.6) -- (1.1,0.6) arc (90:270:0.15cm) -- (1.2,0.3);
\end{tikzpicture} 
\end{array}
\,,\qquad  
\begin{array}{c}
\begin{tikzpicture}[thick]
\fill[black] (0,0.0) circle (0.05cm);
\fill[black] (0,0.3) circle (0.05cm);
\fill[black] (0,0.6) circle (0.05cm);
\fill[black] (0,0.9) circle (0.05cm);
\draw (0,0) -- (1.2,0);
\draw (0,0.6) -- (1.2,0.6);
\fill[black] (1.2,0.0) circle (0.05cm);
\fill[black] (1.2,0.3) circle (0.05cm);
\fill[black] (1.2,0.6) circle (0.05cm);
\fill[black] (1.2,0.9) circle (0.05cm);
\draw (1.2,0.3) -- (0,0.3);
\draw (1.2,0.9) -- (0,0.9);
\end{tikzpicture} 
\end{array}\,,\\  && \\
n=4\,: & & \begin{array}{c}
\begin{tikzpicture}[thick]
\fill[black] (0,0.0) circle (0.05cm);
\fill[black] (0,0.3) circle (0.05cm);
\fill[black] (0,0.6) circle (0.05cm);
\fill[black] (0,0.9) circle (0.05cm);
\draw (0,0.0) -- (0.1,0.0) arc (-90:0:0.45cm) -- (0.55,1.05) arc (0:90:0.45) -- (0,1.5);
\draw (0,1.2) -- (0.1,1.2) arc (90:0:0.3cm) -- (0.4,0.6) arc (0:-90:0.3) -- (0,0.3);
\fill[black] (0,1.2) circle (0.05cm);
\fill[black] (0,1.5) circle (0.05cm);
\draw (0,0.6) -- (0.1,0.6) arc (-90:90:0.15cm) -- (0,0.9);
\fill[black] (1.2,0.0) circle (0.05cm);
\fill[black] (1.2,0.3) circle (0.05cm);
\fill[black] (1.2,0.6) circle (0.05cm);
\fill[black] (1.2,0.9) circle (0.05cm);
\fill[black] (1.2,1.2) circle (0.05cm);
\fill[black] (1.2,1.5) circle (0.05cm);
\draw (1.2,0.0) -- (1.1,0.0) arc (270:180:0.45cm) -- (0.65,1.05) arc (180:90:0.45) -- (1.2,1.5);
\draw (1.2,1.2) -- (1.1,1.2) arc (90:180:0.3cm) -- (0.8,0.6) arc (180:270:0.3) -- (1.2,0.3);
\draw (1.2,0.6) -- (1.1,0.6) arc (270:90:0.15cm) -- (1.2,0.9);
\end{tikzpicture} 
\end{array}
\,,\qquad \begin{array}{c}
\begin{tikzpicture}[thick]
\fill[black] (0,0.0) circle (0.05cm);
\fill[black] (0,0.3) circle (0.05cm);
\fill[black] (0,0.6) circle (0.05cm);
\fill[black] (0,0.9) circle (0.05cm);
\draw (0,0) -- (0.1,0) arc (-90:90:0.45) -- (0.,0.9);
\draw (0,0.6) -- (0.1,0.6) arc (90:-90:0.15cm) -- (0,0.3);
\fill[black] (1.2,0.0) circle (0.05cm);
\fill[black] (1.2,0.3) circle (0.05cm);
\fill[black] (1.2,0.6) circle (0.05cm);
\fill[black] (1.2,0.9) circle (0.05cm);
\draw (1.2,0.9) -- (1.1,0.9) arc (90:270:0.45) -- (1.2,0);
\draw (1.2,0.6) -- (1.1,0.6) arc (90:270:0.15cm) -- (1.2,0.3);
\fill[black] (0,-0.3) circle (0.05cm);
\fill[black] (0,1.2) circle (0.05cm);
\fill[black] (1.2,-0.3) circle (0.05cm);
\fill[black] (1.2,1.2) circle (0.05cm);
\draw (1.2,-0.3) -- (0,-0.3);
\draw (1.2,1.2) -- (0,1.2);
\end{tikzpicture} 
\end{array}
\,,\qquad  
\begin{array}{c}
\begin{tikzpicture}[thick]
\fill[black] (0,0.0) circle (0.05cm);
\fill[black] (0,0.3) circle (0.05cm);
\fill[black] (0,0.6) circle (0.05cm);
\fill[black] (0,0.9) circle (0.05cm);
\draw (0,0) -- (0.4,0) -- (1.2,0.0);
\draw (0,0.6) -- (0.1,0.6) arc (90:-90:0.15cm) -- (0,0.3);
\fill[black] (1.2,0.0) circle (0.05cm);
\fill[black] (1.2,0.3) circle (0.05cm);
\fill[black] (1.2,0.6) circle (0.05cm);
\fill[black] (1.2,0.9) circle (0.05cm);
\draw (1.2,0.9) -- (0,0.9);
\draw (1.2,0.6) -- (1.1,0.6) arc (90:270:0.15cm) -- (1.2,0.3);
\fill[black] (0,-0.3) circle (0.05cm);
\fill[black] (0,1.2) circle (0.05cm);
\fill[black] (1.2,-0.3) circle (0.05cm);
\fill[black] (1.2,1.2) circle (0.05cm);
\draw (1.2,-0.3) -- (0,-0.3);
\draw (1.2,1.2) -- (0,1.2);
\end{tikzpicture} 
\end{array}
\,,\qquad  
\begin{array}{c}
\begin{tikzpicture}[thick]
\fill[black] (0,0.0) circle (0.05cm);
\fill[black] (0,0.3) circle (0.05cm);
\fill[black] (0,0.6) circle (0.05cm);
\fill[black] (0,0.9) circle (0.05cm);
\draw (0,0) -- (1.2,0);
\draw (0,0.6) -- (1.2,0.6);
\fill[black] (1.2,0.0) circle (0.05cm);
\fill[black] (1.2,0.3) circle (0.05cm);
\fill[black] (1.2,0.6) circle (0.05cm);
\fill[black] (1.2,0.9) circle (0.05cm);
\draw (1.2,0.3) -- (0,0.3);
\draw (1.2,0.9) -- (0,0.9);
\fill[black] (0,-0.3) circle (0.05cm);
\fill[black] (0,1.2) circle (0.05cm);
\fill[black] (1.2,-0.3) circle (0.05cm);
\fill[black] (1.2,1.2) circle (0.05cm);
\draw (1.2,-0.3) -- (0,-0.3);
\draw (1.2,1.2) -- (0,1.2);
\end{tikzpicture} 
\end{array}\,.
\end{eqnarray*}
etc. These diagrams explicitly realize the idea that the amount of entanglement is quantified by the amount of connection that exists between two parties. Let us also note that the approach is directly generalized to pairs of Hilbert spaces with nonequal dimensions. In the latter case the number of possible connections will be limited by the the smallest of the two dimensions, as is the case of the matrix rank.

To close this section let us make an important disclaimer. The above analysis has completely ignored the dependence of the model of the parameter $A$ (or $d$, or $k$). As we have seen above, this parameter introduces additional restrictions on the structure of the TQFT Hilbert spaces. One manifestation of these restrictions is the dimension of the Hilbert space, which may be smaller than the naive count of singlet representations given by~(\ref{Catalan}). This happens because some of the vectors become null and disappear from the Hilbert space. As a result there are additional degeneracies between diagrammatic states, which make the analysis more subtle. In particular, some invertible operators might become projectors for specific values of $A$ at roots of unity. We will not enter in the details of such an analysis here. Instead, we will assume that either the parameter $k$ is not integer, or that it is a sufficiently large integer. (It is safe to assume that it is larger than the number of lines in all states of interest.) In both cases the mentioned complications do not occur.

Finally, we only considered the case of bipartite entanglement in this section. The situation with the multipartite entanglement will be mentioned in section~\ref{sec:multipartite}, together with some questions not answered in this review.

%%%%%%%%%%%%%%%%%%%%%%%%%%%%%%%%%%%%%%%%%%%%%%%%%%%%%%%%%%%%%%%%%%%%
\subsection{Entropy of Entanglement}
\label{sec:entropy}

So far we avoided working with a precise quantitative measure of entanglement. In this section we will discuss the von Neumann entropy, which provides one of the most common measures, and try to explain why it is natural for the topological formulation of quantum mechanics. 

The von Neumann, or entanglement entropy is defined in terms of density matrices. If a state is pure, that is we have complete access to the information about it, encapsulated in a vector $|\Psi\rangle$, then the density matrix associated with the state is given by the ``square'' of the state vector:
\be
\rho \ = \ |\Psi\rangle \langle \Psi|\,.
\ee

It is useful to already have a topological interpretation of the basic definitions. Pure state is a state isolated from the environment, that is lacking any correlations with anything that is not the state itself. In the TQFT language this means that, as a space, the state cannot have a component that is connected with something that is not the state (compactness). Consider an example of a bipartite state:
\be
\label{bipartitestate}
 |\Psi\rangle = 
  \begin{array}{c}\includegraphics[scale=0.25]{./figs/entangled}\end{array}\,.
\ee
The density matrix (a square of the state vector) can be drawn as two copies of the same state:
\be
\label{densitymatrix}
{\rho} = \begin{array}{c}\includegraphics[scale=0.25]{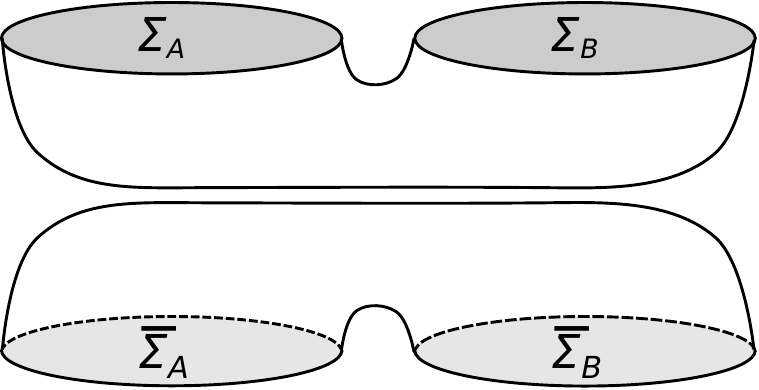}\end{array}.
\ee
Let us think of the diagram as being a generic topological space, with possible nontrivial features inside. One can draw the second copy oriented in a different way remembering that one of the copies should be conjugated.

State~(\ref{bipartitestate}) belongs to a product of subspaces $\Hc_{\Sigma_A}$ and $\Hc_{\Sigma_B}$. One can define the \emph{reduced density matrix} which is an effective description of the original state only in terms of one of the subsystems. For subsystem $A$ the reduced density matrix is defined as
\be
\rho_A \ : = \ \Tr_{\Hc_{\Sigma_B}}\rho\,,
\ee
where the trace is calculated with respect to the subsystem $B$. This \emph{partial trace} operation is equivalent to the matrix multiplication of two state tensors, cf.~(\ref{statetensor}):
\be
(\rho_A)_{i_Aj_A} \ =\ \sum_{i_B} c_{i_Ai_B}c^\dagger_{i_Bj_A}\,.
\ee
This operation can be interpreted as gluing two copies of state $|\Psi\rangle$ in~(\ref{densitymatrix}) along the boundary manifold $\Sigma_B$. The result is a space connecting two boundaries $\Sigma_A$ (cf.~\cite{Melnikov:2018zfn,Dong:2008ft}):
\be
\label{reduceddm}
\rho_A \ = \  \begin{array}{c}\includegraphics[scale=0.25]{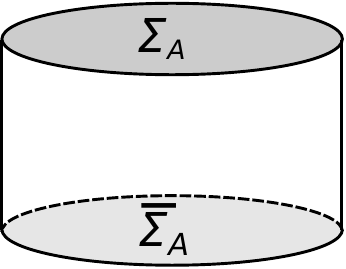}\end{array}.
\ee

Note that, given the reduced density matrix, recovering the state vector $|\psi_A\rangle$ is not generally possible. It is clear in the example above that there is no obvious $|\psi_A\rangle$ such that $\rho_A=|\psi_A\rangle\langle\psi_A|$. In other words, $\rho_A$ describes not a pure, but a mixed state. In the mixed state subsystem $A$ has correlations with subsystem $B$, but the degrees of freedom of $B$ are hidden in the description provided by the density matrix $\rho_A$. We can also see that if the original state $|\Psi\rangle$ is entangled with respect to some bipartition, as alleged by~(\ref{bipartitestate}), the reduced density matrix of either of the subsystems is mixed. 

Given a bipartition and a reduced density matrix, say $\rho_A$, the von Neumann entropy is given by
\be
S_{\rm vN} \ = \ - \Tr_{\Hc_{\Sigma_A}} \rho_A\log \rho_A\,.
\ee
To appreciate this formula let us formally compute the entropy for state~(\ref{bipartitestate}) and another state
\be
 |\Phi\rangle \  =  \
  \begin{array}{c}\includegraphics[scale=0.25]{./figs/separable}\end{array}.
\ee

For state $|\Phi\rangle$ we can introduce the density matrix and the reduced density matrix in the same way as for state $|\Psi\rangle$ above. In particular, 
\be
\label{reduceddm2}
\rho_A(\Phi) \ = \ \left(\begin{array}{c}\includegraphics[scale=0.2]{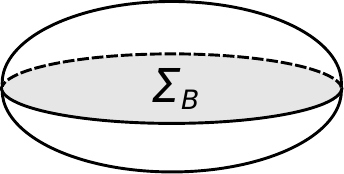}\end{array}\right)\times \begin{array}{c}\includegraphics[scale=0.25]{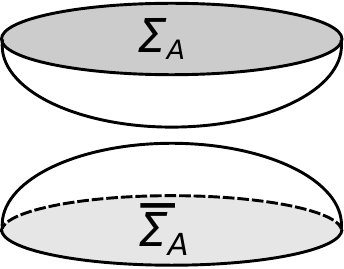}\end{array},
\ee
where the prefactor is a closed manifold ($\mathbb{C}$-number) corresponding to the trace of $\rho(\Phi)$ over $\Hc_{\Sigma_B}$. Density matrices~(\ref{reduceddm}) and~(\ref{reduceddm2}) need to be normalized, $\Tr_{\Hc_{\Sigma_A}}\rho_A=1$. It is not hard to see that the normalized density matrices have the form
\be
 \bar{\rho}_A(\Phi)  =  \left[\begin{array}{c}\includegraphics[scale=0.2]{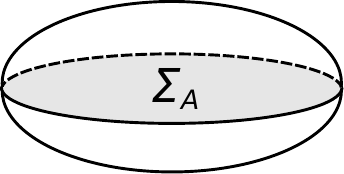}\end{array}\right]^{-1} \begin{array}{c}\includegraphics[scale=0.25]{./figs/densitym}\end{array},
 \qquad 
\bar{\rho}_A(\Psi) =  \left[\begin{array}{c}\includegraphics[scale=0.1,angle=0]{./figs/donut}\end{array}\right]^{-1} \begin{array}{c}\includegraphics[scale=0.25]{./figs/densityme}\end{array}.
\ee
The bars over $\rho$ now indicate that the matrices are normalized. The prefactor of the second density matrix is the result of the gluing the top and the bottom boundaries of~(\ref{reduceddm}). It is also a closed manifold representing a $\mathbb{C}$-number.

To apply the von Neumann formula to matrices in the diagrammatic form, it is convenient to apply the technique called \emph{replica trick}~\cite{Dong:2008ft}. It is based on the following relation:
\be
\Tr\rho\log\rho \ = \ \lim_{n\to1}\frac{d}{dn}\Tr\rho^n\,.
\ee
The procedure involves computing the analytic dependence of the trace of the $n$-th power of $\rho$ and taking the derivative, with a subsequent limit $n\to 1$. However, thinking in terms of the topological presentation, it is easy to concatenate spaces and compute $\Tr\rho^n$ when $n$ is integer, but not cutting them for fractional $n$. In other words, in general, one would need to assume some analytical continuation in $n$ before computing the derivative and the limit. In many situations that we will consider, it will be easy to do this, because $n$-dependence will only be present in the prefactors, and not in the matrices.

Let us assume that the cylinder representing state $|\Psi\rangle$ has translational invariance along its axis. In fact, for the following, one assumes a weaker condition that $\rho_A^2(\Psi)\propto \rho_A(\Psi)$ that is the condition of a general projector. (We will not assume anything about $|\Phi\rangle$.) It is a simple exercise to show that 
\be
\bar\rho_A^n(\Phi)  \ = \   \left[\begin{array}{c}\includegraphics[scale=0.2]{./figs/melonA}\end{array}\right]^{-1} \begin{array}{c}\includegraphics[scale=0.25]{./figs/densitym}\end{array}  \,,
\qquad  \bar\rho_A^n(\Psi)  \ = \ \left[\begin{array}{c}\includegraphics[scale=0.1,angle=0]{./figs/donut}\end{array}\right]^{-n} \begin{array}{c}\includegraphics[scale=0.25]{./figs/densityme}\end{array}.  
\ee
Importantly, the matrix structure appearing in the above expressions does not depend on $n$: the same matrices appear on the right hand side. For $|\Phi\rangle$, even the prefactor is $n$-independent. Consequently,
\be
\Tr\bar\rho_A^n(\Phi)  \ = \     1\,,
\qquad  \Tr\bar\rho_A^n(\Psi)  \ = \     \left[
\begin{array}{c}\includegraphics[scale=0.2]{./figs/donut}\end{array}
\right]^{1-n}.
\ee

Analytical continuation is straightforward for these expressions and one obtains
\be
\label{replicaentropy}
S_{\rm vN}(\Phi) \ = \ 0\,, \qquad S_{\rm vN}(\Psi) \ = \ \log \left[
\begin{array}{c}\includegraphics[scale=0.2]{./figs/donut}\end{array}
\right].
\ee
Let us discuss these results.

First, as any good measure of entanglement should do, $S_{\rm vN}$ vanishes on separable states. It is easy to verify this for any pure states also without invoking topology, but a lesson to learn here is that the von Neumann formula is constructed in such a way that it vanishes on everything that is factorized, or disconnected in terms of topology. This observation can assist in constructing other interesting quantities assessing entanglement, e.g.~\cite{Melnikov:2022qyt,Melnikov:2023nzn}.

Second, if the state is not separable, replica trick allows us to calculate the expression for the entropy easily if the space has translation symmetry along the ``evolution'' axis. In this case, the number corresponding to the closed manifold in the argument of log in~(\ref{replicaentropy}) is simply the trace of the identity operator on the Hilbert space $\Hc_{\Sigma_A}$, that is
\be
S_{\rm vN}(\Psi) \ = \ \log\dim\Hc_{\Sigma_A}\,.
\ee
In particular, for state~(\ref{2qubitstate}) $S_{\rm vN}(\Psi)=\log 2$, as it should be for a Bell state.

Let us slightly generalize this result for the class of reduced density matrices that are projectors. In terms of the one-dimensional TQFT of section~\ref{sec:TQM}, we will be interested in states of the form~\cite{Melnikov:2023wwc}
\be
\label{genconnectome}
\begin{array}{c}
\begin{tikzpicture}[thick]
\newcommand{\x}{1.2}
\fill[cyan] (-0.5,0.55) rectangle (0.2,1.2);
\fill[cyan, opacity=0.4] (-0.5,-0.1) rectangle (0.2,0.55);
\fill[pink] (\x-0.2,0.55) rectangle (\x+0.5,1.2);
\fill[pink,opacity=0.4] (\x-0.2,-0.1) rectangle (\x+0.5,0.55);
\foreach \y in {0.6,0.8,...,1.0}
            \draw (0,\y) -- (0.4,\y) arc (-90:90:0.05cm) -- (0,\y+0.1);
\foreach \y in {0.6,0.8,...,1.0}
            \draw (\x,\y) -- (\x-0.4,\y) arc (270:90:0.05cm) -- (\x,\y+0.1);
\foreach \y in {0,0.1,...,0.6}
            \draw (0,\y) -- (\x,\y);
\foreach \y in {0,0.1,...,1.2}
           \fill[black] (0,\y) circle (0.04cm);
\foreach \y in {0,0.1,...,1.2}
           \fill[black] (\x,\y) circle (0.04cm);
\draw (-0.25,0.25) node {$m$};
\draw (-0.25,0.9) node {$l$};
\end{tikzpicture} 
\end{array}.
\ee
This diagram shows a state in a Hilbert space, which is a tensor product of two copies of the Hilbert space of $l+m$ points. The state corresponds to a space with $l$ lines connecting points belonging to the same boundary, and $m$ lines connecting different boundaries, and is a characteristic example of a connectome class. Color shading is used to indicate different Hilbert spaces and subspaces within them.

The reduced density matrix, say, of the left subspace in this connectome state is expressed by the following diagram:
\be
\rho_L \ =\ 
\begin{array}{c}
\begin{tikzpicture}[thick]
\newcommand{\x}{1.2}
\newcommand{\z}{0}
\fill[cyan] (-0.5,0.55) rectangle (0.2,1.2);
\fill[cyan,opacity=0.4] (-0.5,-0.1) rectangle (0.2,0.55);
\fill[pink] (\x-0.2,0.55) rectangle (\x+0.2,1.2);
\fill[pink,opacity=0.4] (\x-0.2,-0.1) rectangle (\x+0.2,0.55);
\foreach \y in {0.6,0.8,...,1.0}
            \draw (0+\z,\y) -- (0.4+\z,\y) arc (-90:90:0.05cm) -- (0+\z,\y+0.1);
\foreach \y in {0.6,0.8,...,1.0}
            \draw (\x+\z,\y) -- (\x+\z-0.4,\y) arc (270:90:0.05cm) -- (\x+\z,\y+0.1);
\foreach \y in {0,0.1,...,0.6}
            \draw (0+\z,\y) -- (\x+\z,\y);
\foreach \y in {0,0.1,...,1.2}
           \fill[black] (0+\z,\y) circle (0.04cm);
\foreach \y in {0,0.1,...,1.2}
           \fill[black] (\x+\z,\y) circle (0.04cm);
\renewcommand{\z}{1.2}
\fill[cyan] (\x+\z-0.2,0.55) rectangle (\x+\z+0.5,1.2);
\fill[cyan,opacity=0.4] (\x+\z-0.2,-0.1) rectangle (\x+\z+0.5,0.55);
\foreach \y in {0.6,0.8,...,1.0}
            \draw (0+\z,\y) -- (0.4+\z,\y) arc (-90:90:0.05cm) -- (0+\z,\y+0.1);
\foreach \y in {0.6,0.8,...,1.0}
            \draw (\x+\z,\y) -- (\x+\z-0.4,\y) arc (270:90:0.05cm) -- (\x+\z,\y+0.1);
\foreach \y in {0,0.1,...,0.6}
            \draw (0+\z,\y) -- (\x+\z,\y);
\foreach \y in {0,0.1,...,1.2}
           \fill[black] (\x+\z,\y) circle (0.04cm);
\end{tikzpicture} 
\end{array}
\ = \ d^{l/2} \begin{array}{c}
\begin{tikzpicture}[thick]
\newcommand{\x}{1.2}
\newcommand{\z}{0}
\fill[cyan] (-0.5,0.55) rectangle (0.2,1.2);
\fill[cyan,opacity=0.4] (-0.5,-0.1) rectangle (0.2,0.55);
\fill[cyan] (\x-0.2,0.55) rectangle (\x+0.5,1.2);
\fill[cyan,opacity=0.4] (\x-0.2,-0.1) rectangle (\x+0.5,0.55);
\foreach \y in {0.6,0.8,...,1.0}
            \draw (0+\z,\y) -- (0.4+\z,\y) arc (-90:90:0.05cm) -- (0+\z,\y+0.1);
\foreach \y in {0.6,0.8,...,1.0}
            \draw (\x+\z,\y) -- (\x+\z-0.4,\y) arc (270:90:0.05cm) -- (\x+\z,\y+0.1);
\foreach \y in {0,0.1,...,0.6}
            \draw (0+\z,\y) -- (\x+\z,\y);
\foreach \y in {0,0.1,...,1.2}
           \fill[black] (0+\z,\y) circle (0.04cm);
\foreach \y in {0,0.1,...,1.2}
           \fill[black] (\x+\z,\y) circle (0.04cm);
\end{tikzpicture} 
\end{array}\,.
\ee
As a result of partial-tracing, shown above, some of the lines turned closed loops and were replaced by numerical factors according to the rules spelled in section~\ref{sec:TQM}. The reduced density matrix looks similar to the original state, but the left and the right boundaries are now copies of the Hilbert space $\Hc_L$ (or rather $\Hc_L$ and its dual $\Hc_L^\ast$).

We can normalize the reduced density matrix:
\be
\bar\rho_L \ = \ \frac{1}{D_m d^{l/2}} \begin{array}{c}
\begin{tikzpicture}[thick]
\newcommand{\x}{1.2}
\newcommand{\z}{0}
\fill[cyan] (-0.5,0.55) rectangle (0.2,1.2);
\fill[cyan,opacity=0.4] (-0.5,-0.1) rectangle (0.2,0.55);
\fill[cyan] (\x-0.2,0.55) rectangle (\x+0.5,1.2);
\fill[cyan,opacity=0.4] (\x-0.2,-0.1) rectangle (\x+0.5,0.55);
\foreach \y in {0.6,0.8,...,1.0}
            \draw (0+\z,\y) -- (0.4+\z,\y) arc (-90:90:0.05cm) -- (0+\z,\y+0.1);
\foreach \y in {0.6,0.8,...,1.0}
            \draw (\x+\z,\y) -- (\x+\z-0.4,\y) arc (270:90:0.05cm) -- (\x+\z,\y+0.1);
\foreach \y in {0,0.1,...,0.6}
            \draw (0+\z,\y) -- (\x+\z,\y);
\foreach \y in {0,0.1,...,1.2}
           \fill[black] (0+\z,\y) circle (0.04cm);
\foreach \y in {0,0.1,...,1.2}
           \fill[black] (\x+\z,\y) circle (0.04cm);
\draw (-0.25,0.25) node {$m$};
\draw (-0.25,0.9) node {$l$};
\end{tikzpicture} 
\end{array}\,.
\ee
Note that normalizing it, we get again $l/2$ trivially closed loops (replaced by numerical factors) and $m$ loops winding a nontrivial loop that appears when one computes the trace of the diagram (see comments at the end of sections~\ref{sec:TQM} and~\ref{sec:CS} about tracelike closure of diagrams in the topological quantum mechanical model). To properly evaluate the traces here one takes Markov trace for the subspace of $l$ points and normal trace for the subspace of $m$ points. For the latter, the result will be a trace of the identity on the subspace of $m$ points, that is the dimension $D_m$ of that space. 

Next, we compute the $n$-th power of this matrix:
\be
\bar\rho^n_L \ = \ \frac{1}{D_m^n d^{l/2}} \begin{array}{c}
\begin{tikzpicture}[thick]
\newcommand{\x}{1.2}
\newcommand{\z}{0}
\fill[cyan] (-0.5,0.55) rectangle (0.2,1.2);
\fill[cyan,opacity=0.4] (-0.5,-0.1) rectangle (0.2,0.55);
\fill[cyan] (\x-0.2,0.55) rectangle (\x+0.5,1.2);
\fill[cyan,opacity=0.4] (\x-0.2,-0.1) rectangle (\x+0.5,0.55);
\foreach \y in {0.6,0.8,...,1.0}
            \draw (0+\z,\y) -- (0.4+\z,\y) arc (-90:90:0.05cm) -- (0+\z,\y+0.1);
\foreach \y in {0.6,0.8,...,1.0}
            \draw (\x+\z,\y) -- (\x+\z-0.4,\y) arc (270:90:0.05cm) -- (\x+\z,\y+0.1);
\foreach \y in {0,0.1,...,0.6}
            \draw (0+\z,\y) -- (\x+\z,\y);
\foreach \y in {0,0.1,...,1.2}
           \fill[black] (0+\z,\y) circle (0.04cm);
\foreach \y in {0,0.1,...,1.2}
           \fill[black] (\x+\z,\y) circle (0.04cm);
\draw (-0.25,0.25) node {$m$};
\draw (-0.25,0.9) node {$l$};
\end{tikzpicture} 
\end{array}\,.
\ee
Any power of $\bar\rho_L$ has the same matrix structure because it is proportional to a projector. The only thing that distinguishes $\rho_L^n$ from $\rho_L$ is the normalization coefficient. Then
\be
\Tr \bar\rho_L^n \ = \ \frac{1}{D_m^n d^{l/2}}\ \Tr\begin{array}{c}
\begin{tikzpicture}[thick]
\newcommand{\x}{1.2}
\newcommand{\z}{0}
\fill[cyan] (-0.5,0.55) rectangle (0.2,1.2);
\fill[cyan,opacity=0.4] (-0.5,-0.1) rectangle (0.2,0.55);
\fill[cyan] (\x-0.2,0.55) rectangle (\x+0.5,1.2);
\fill[cyan,opacity=0.4] (\x-0.2,-0.1) rectangle (\x+0.5,0.55);
\foreach \y in {0.6,0.8,...,1.0}
            \draw (0+\z,\y) -- (0.4+\z,\y) arc (-90:90:0.05cm) -- (0+\z,\y+0.1);
\foreach \y in {0.6,0.8,...,1.0}
            \draw (\x+\z,\y) -- (\x+\z-0.4,\y) arc (270:90:0.05cm) -- (\x+\z,\y+0.1);
\foreach \y in {0,0.1,...,0.6}
            \draw (0+\z,\y) -- (\x+\z,\y);
\foreach \y in {0,0.1,...,1.2}
           \fill[black] (0+\z,\y) circle (0.04cm);
\foreach \y in {0,0.1,...,1.2}
           \fill[black] (\x+\z,\y) circle (0.04cm);
\draw (-0.25,0.25) node {$m$};
\draw (-0.25,0.9) node {$l$};
\end{tikzpicture} 
\end{array} \ = \ \frac{1}{D_m^n}\ \Tr\begin{array}{c}
\begin{tikzpicture}[thick]
\newcommand{\x}{1.2}
\newcommand{\z}{0}
\fill[cyan,opacity=0.4] (-0.5,-0.1) rectangle (0.2,0.55);
\fill[cyan,opacity=0.4] (\x-0.2,-0.1) rectangle (\x+0.5,0.55);
\foreach \y in {0,0.1,...,0.6}
            \draw (0+\z,\y) -- (\x+\z,\y);
\foreach \y in {0,0.1,...,0.6}
           \fill[black] (0+\z,\y) circle (0.04cm);
\foreach \y in {0,0.1,...,0.6}
           \fill[black] (\x+\z,\y) circle (0.04cm);
\draw (-0.25,0.25) node {$m$};
\end{tikzpicture} 
\end{array} \ = \ \frac{1}{D^{n-1}_m}\,.
\ee
Computing the entropy using the replica formula yields
\be
S_{\rm vN} \ = \ \log D_m\,.
\ee
That is the entropy is given by the logarithm of the dimension of the subspace of the Hilbert space that, in the given state, is ``correlated" with the other part of the system. Put differently, the entropy is a function of the number of connections between the left and the right subsystems.

The function expressing the entropy of a state like~(\ref{genconnectome}) takes a particularly simple form in the limit considered in section~\ref{sec:classlimit}. Take $k\to \infty$ and then $l\to \infty$ and $m\to\infty$. In this limit logarithm of $D_m$ is
\be
\label{ClassEntropy}
\log D_m \ \simeq \ m\log 2\,,
\ee
that is the entropy is indeed counting the number of connections between the two parties.

The main observation of this section is that the von Neumann entropy can be expressed as the amount of connectivity that exists between parties. Let us give another version of this relation. In the context of one-dimensional TQFT and a state expressed by lines connecting left and right subsystems let us draw an arbitrary curve that encircles one of the subsystems, as in~(\ref{encircle}). We can choose any curve, but let us take the one, which crosses the minimal number of lines. In the limit considered above the entropy is given by $\log2$ times the number of crossed lines. In the three-dimensional TQFT one should respectively find a surface that encloses one of the spheres and cuts the minimal number of lines.

The above definition only works for states of the connectome type~(\ref{genconnectome}) and more importantly, in the limit of large $k$ and large number of lines. The reason to introduce such a definition will be explained in section~\ref{sec:gravity} when we will mention connections with quantum gravity and Ryu-Takayanagi formula~\cite{Ryu:2006bv}.

%%%%%%%%%%%%%%%%%%%%%%%%%%%%%%%%%%%%%%%%%%%%%%%%%%%%%%%%%%%%%%%%%%%%
\subsection{Inequalities for the Entanglement entropy}
\label{sec:inequalities}

In the previous parts of section~\ref{sec:entanglement} we discussed expression of quantum entanglement in terms of topology and showed how a specific measure of entanglement -- the von Neumann entropy -- gets a topological expression. In this part we will discuss the entropy a little further and explain some of its well-known properties in topological terms. Specifically, we will consider inequalities satisfied by the entanglement entropy. The discussion becomes the most simple if we restrict ourselves to the limit of large $k$ and of large Hilbert spaces (that is Hilbert spaces of very large numbers of points) as introduced in section~\ref{sec:classlimit}.  We will refer to this as the connectome limit.

First, let us discuss the property of entanglement entropy called \emph{subadditivity}. Let us assume that in a given quantum system one can separate a subsystem $A$ and another subsystem $B$. Then the subadditivity of the entropy is the statement that the entropy of the union of $A$ and $B$ is not greater than the sum of entropies of $A$ and $B$:
\be
\label{subadditivity}
S_{\rm vN}(A+B)\leq S_{\rm vN}(A) + S_{\rm vN}(B)\,. 
\ee
Instead of trying to interpret this inequality right away, let us first produce its topological analog.

\begin{figure}
\begin{minipage}{0.45\linewidth}
\centering
\scalebox{1.5}{\begin{tikzpicture}[thick]
\draw (0,0.2) -- (1.5,0.2) -- (1.5,0) -- (0,0) -- (0,-0.2) -- (1.5,-0.2);
\draw (-0.2,0) -- (0.55,-1) -- (0.95,-1) -- (1.7,0);
\fill[white] (0,0) circle (0.3);
\draw (0,0) circle (0.3);
\draw (0,0) node {$A$};
\fill[white] (1.5,0) circle (0.3);
\draw (1.5,0) circle (0.3);
\draw (1.5,0) node {$B$};
\fill[white] (0.75,-1) ellipse (0.6 and 0.3);
\draw (0.75,-1) ellipse (0.6 and 0.3);
\draw (0.75,-1) node {$C$};
\fill[white] (0.75,0) circle (0.21);
\draw[blue] (0.75,0) node {\tiny $\ell_{AB}$};
\draw[blue,dashed,very thick] (0,0.4) arc (90:-120:0.4);
\draw[blue] (-0.15,0.45) node {\tiny $N_A$};
\draw[blue,dashed,very thick] (1.5,0.4) arc (90:300:0.4);
\draw[blue] (1.7,0.45) node {\tiny $N_B$};
\draw[blue,dashed,very thick] (0,-0.8) arc (120:60:1.5);
\draw[blue] (0.75,-0.42) node {\tiny $N_{C}$};
\end{tikzpicture}}
\end{minipage}
\hfill{
\begin{minipage}{0.45\linewidth}
\centering
\begin{tikzpicture}[thick]
\draw (0,0.2) -- (2.5,0.2) -- (2.5,0) -- (0,0) -- (0,-0.2) -- (2.5,-0.2);
\draw (-0.2,0) -- (1.25,-1.8) -- (2.5,-0.2) -- (2.5,0) -- (1.25,-1.6) -- (0,0);
\draw (1.25,-1.4) -- (1.25,-0.7) -- (2.75,0.);
\draw (1.25,-0.7) -- (-0.25,0.);
\fill[white] (0,0) circle (0.3);
\draw (0,0) circle (0.3);
\draw (0,0) node {$A$};
\fill[white] (2.5,0) circle (0.3);
\draw (2.5,0) circle (0.3);
\draw (2.5,0) node {$C$};
\fill[white] (1.25,-1.6) circle (0.3);
\draw (1.25,-1.6) circle (0.3);
\draw (1.25,-1.6) node {$B$};
\fill[white] (1.25,-0.6) ellipse (0.6 and 0.3);
\draw (1.25,-0.6) ellipse (0.6 and 0.3);
\draw (1.25,-0.6) node {$D$};
\end{tikzpicture}
\end{minipage}
}
    \caption{Topological presentation of correlations can be used to illustrate subadditivity (left) and strong subadditivity (right).}
    \label{fig:subadditivity}
\end{figure}
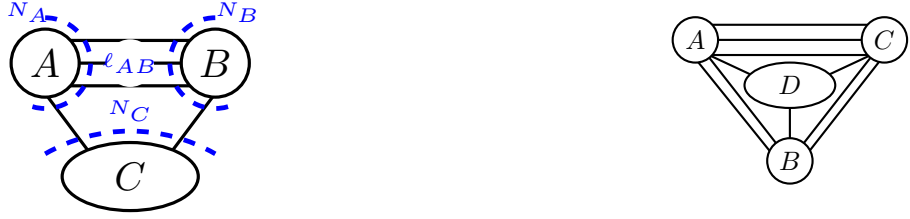

Consider a generic state in the connectome limit. Note that the union $A+B$ does not necessarily represent the whole system, so there is also a complement $\overline{A+B}=C$. Thus, a generic state will have the form shown in figure~\ref{fig:subadditivity} (left), where lines between subsystems $A$, $B$ and $C$ illustrate generic correlations between them. In the limit that we are applying, the correlations are quantified by the number of connections. Let $N_A$, $N_B$ and $N_C$ be the numbers of connections of the respective subsystems to the remaining parts. Let $\ell_{AB}$ be the number of connections exclusively between $A$ and $B$. Then a simple relation holds:
\be
N_A+N_B \ = \ N_C + 2\ell_{AB} \geq N_C \,.
\ee
Since the entanglement entropy is proportional to the number of connections, with $\log2$ coefficient, the subadditivity property~(\ref{subadditivity}) follows.

With the connectivity quantification of entanglement it is not hard to give an interpretation to the subadditivity. The correlation of $A$ and $B$ with the rest of the system cannot exceed the amount of correlations in $A$ and $B$ taken together, simply because there might exist correlations between $A$ and $B$. The amount of correlation between $A$ and $B$, in this case, is
\be
\ell_{AB} \ = \ (N_A + N_B - N_{C})/2\,.
\ee
This is the topological version of what is known in Information Theory as \emph{mutual information} $I(A,B)$,
\be
I(A,B) \ :=\ \Svn(A)+\Svn(B) - \Svn(A+B)\,.
\ee
If the system $A+B$ does not have a complement (isolated) then $\Svn(A+B)=0$, $\Svn(A)=\Svn(B)$ and  mutual information is essentially the same as the entanglement entropy.

Another very-well known inequality is \emph{strong subadditivity}. It is formulated for three subsystems $A$, $B$ and $C$. The inequality states
\be
\label{ssubadditivity}
\Svn(A+B+C) + \Svn(B) \ \leq \ \Svn(A+B) + \Svn(B+C)\,.
\ee
In terms of connectivity this inequality can be proven using the diagram shown in figure~\ref{fig:subadditivity} (right), where $D=\overline{A+B+C}$. Denoting the number of connections as in the previous case one finds that
\be
N_{D} \ = \  N_{\overline{A+B}} + N_{\overline{B+C}} - 2\ell_{AC} - N_B \leq  N_{\overline{A+B}} + N_{\overline{B+C}} - N_B\,,
\ee
where $\overline{A+B}$ and $\overline{B+C}$ denote the complements of $A+B$ and $B+C$ respectively. Again, in the connectome limit, this directly converts into the strong subadditivity~(\ref{ssubadditivity}). We see that in this case the ``obstacle'' is the existence of correlations between $A$ and $C$.

Strong subadditivity is notoriously difficult to prove in general~\cite{Lieb:1973cp}, but for the special class of states that we consider here it comes almost for free. Many more known inequalities can be tested in this way and some new ones can be designed~\cite{Melnikov:2023nzn,Melnikov:2023wwc}. It is important to keep in mind that in generic quantum states the relations derived in the connectome limit might not hold. Since it is an open question whether and how much the connectome relations may be violated, we will discuss some possibilities in section~\ref{sec:outlook}.

%%%%%%%%%%%%%%%%%%%%%%%%%%%%%%%%%%%%%%%%%%%%%%%%%%%%%%%%%%%%%%%%%%%%
\section{Quantum Algorithms}
\label{sec:algorithms}

As another set of applications of the topological visualization of quantum mechanics let us consider a pair of basic quantum algorithms. Quantum algorithms exploiting quantum correlations should be the most natural target for such a study. The example of the EPR pair puzzles us with the apparent instantaneity of interactions between causally separated particles. Quantum teleportation is a basic algorithm that capitalizes on this effect making an arbitrary state ``instantaneously'' available in a distant lab. Superdense coding is another example. We will discuss topological realization of both algorithms elucidating their inherent similarity.

%%%%%%%%%%%%%%%%%%%%%%%%%%%%%%%%%%%%%%%%%%%%%%%%%%%%%%%%%%%%%%%%%%%%
\subsection{Quantum Teleportation}
\label{sec:teleport}

\begin{figure}
    \begin{minipage}{0.45\linewidth}
    \centering
    \begin{tikzpicture}{thick}
        \fill (1,1.5) circle (0.1cm);
        \fill[white] (1,0.75) circle (0.15cm);
        \draw (0,0) node[anchor=east] {$\Phi^+_B$} -- (5,0) node[anchor=west] {$\psi$};
        \draw (0,0.75) node[anchor=east] {$\Phi^+_A$} -- (2,0.75);
        \draw (0,1.5) node[anchor=east] {$\psi$} -- (3,1.5);
        \draw (1,0.6) -- (1,1.5);
        \draw[double,line width=0.8] (2,0.75) -- (3,0.75) node[anchor=west] {$b$} -- (3,0);
        \draw[double,line width=0.8] (3,1.5) -- (4,1.5) node[anchor=west] {$a$} -- (4,0);
        \draw (1,0.75) circle (0.15cm);
        \fill[white] (1.7,1.25) rectangle (2.3,1.75);
        \draw (1.7,1.25) rectangle (2.3,1.75);
        \draw (2,1.5) node {$H$};
         \fill[white] (2.7,1.25) rectangle (3.3,1.75);
        \draw (2.7,1.25) rectangle (3.3,1.75);
        \draw (2.8,1.5) arc (180:0:0.2);
        \fill (3,1.45) circle (0.05cm);
        \draw (2.95,1.4) -- (3.2,1.65);
         \fill[white] (1.7,0.5) rectangle (2.3,1.);
        \draw (1.7,0.5) rectangle (2.3,1.);
        \draw (1.8,0.75) arc (180:0:0.2);
        \fill (2,0.7) circle (0.05cm);
        \draw (1.95,0.65) -- (2.2,0.9);
        \fill[white] (2.7,-0.25) rectangle (3.3,0.25);
        \draw (2.7,-0.25) rectangle (3.3,0.25);
        \draw (3,0) node {$X^b$};
        \fill[white] (3.7,-0.25) rectangle (4.3,0.25);
        \draw (3.7,-0.25) rectangle (4.3,0.25);
        \draw (4,0) node {$Z^a$};
    \end{tikzpicture}
    \end{minipage}
    \hfill{
    \begin{minipage}{0.45\linewidth}
    \centering
    \begin{tikzpicture}{thick}
        \fill (2.5,1.5) circle (0.1cm);
        \draw (0,0.75) node[anchor=east] {$\Phi^+_A$} -- (4.25,0.75);
        \draw (0,1.5) node[anchor=east] {$\Phi^+_B$} -- (4.25,1.5);
        \draw (2.5,0.6) -- (2.5,1.5);
        \draw[double,line width=0.8] (0.75,1.5) -- (0.75,2.25) node[anchor=east] {$b$};
        \draw[double,line width=0.8] (1.75,1.5) -- (1.75,2.25) node[anchor=east] {$a$};
        \draw[double,line width=0.8] (4.25,0.75) -- (5,0.75) node[anchor=west] {$b$};
        \draw[double,line width=0.8] (4.25,1.5) -- (5,1.5) node[anchor=west] {$a$};
        \draw (2.5,0.75) circle (0.15cm);
        \fill[white] (2.95,1.25) rectangle (3.55,1.75);
        \draw (2.95,1.25) rectangle (3.55,1.75);
        \draw (3.25,1.5) node {$H$};
         \fill[white] (3.95,1.25) rectangle (4.55,1.75);
        \draw (3.95,1.25) rectangle (4.55,1.75);
        \draw (4.05,1.5) arc (180:0:0.2);
        \fill (4.25,1.45) circle (0.05cm);
        \draw (4.2,1.4) -- (4.45,1.65);
         \fill[white] (3.95,0.5) rectangle (4.55,1.);
        \draw (3.95,0.5) rectangle (4.55,1.);
        \draw (4.05,0.75) arc (180:0:0.2);
        \fill (4.25,0.7) circle (0.05cm);
        \draw (4.2,0.65) -- (4.45,0.9);
        \fill[white] (0.45,1.25) rectangle (1.05,1.75);
        \draw (0.45,1.25) rectangle (1.05,1.75);
        \draw (0.75,1.5) node {$X^b$};
        \fill[white] (1.45,1.25) rectangle (2.05,1.75);
        \draw (1.45,1.25) rectangle (2.05,1.75);
        \draw (1.75,1.5) node {$Z^a$};
    \end{tikzpicture}
    \end{minipage}
    }
    \caption{The quantum circuit diagrams of quantum teleportation protocol (left) and dense coding (right) algorithms.}
    \label{fig:teleport}
\end{figure}
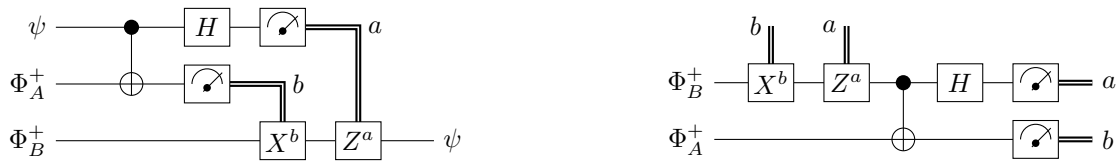

A textbook example of quantum teleportation, initially proposed by Charles Bennett, Gilles Brassard, Claude Crépeau, Richard Jozsa, Asher Peres, and William Wootters in~\cite{Bennett:1992tv}, consists of the following. Alice possesses an unspecified qubit state $|\psi\rangle$ that she wants to pass to Bob. Alice and Bob share a pair of entangled qubits -- the Bell state $|\Phi^+\rangle$~(\ref{Bellstate}). Alice can implement the protocol shown in figure~\ref{fig:teleport} (left) where the tree horizontal lines are evolution lines of three qubits: $\psi$, Alice's part of the entangled pair $\Phi^+_A$ and the Bob's part $\Phi^+_B$. Consequently, Alice applies an entangling CNOT gate on the two qubits in her possession and an Hadarmard gate on the first qubit. After that she measures both qubits in the computational basis and obtains $a=0$ or $a=1$ for the first qubit and $b=0$ or $b=1$ for the second. Next, Alice uses a classical communication channel to report the results of the measurements to Bob. In order to convert Bob's qubit to $|\psi\rangle$ all is needed is to apply a sequence of $X$ and $Z$ gates, specifically $Z^aX^b$ on the qubit. 

For our discussion we will use an even shorter version of the teleportation protocol. In this version Alice won't need to apply any gates on the qubits, but instead make the measurements in a different orthogonal basis: the Bell basis. The Bell basis consists of states
\be
\label{BellBasis}
|\Phi^\pm\rangle \ = \ \frac{|00\rangle\pm|11\rangle}{\sqrt{2}}\,, \qquad |\Psi^\pm\rangle \ = \ \frac{|01\rangle\pm|10\rangle}{\sqrt{2}}\,.
\ee
Again, Alice has to communicate the results of the measurement to Bob. To reconstruct precisely state $|\psi\rangle$ Bob needs to apply the sequence $Z^aX^b$ according to the table:
\begin{table}[h]
    \centering
    \begin{tabular}{||c|c|c||c|c|c||}
       \hline State & $a$ & $b$ & State & $a$ & $b$ \\
        \hline $\Phi^+$  & 0 & 0 & $\Psi^+$  & 0 & 1 \\
         \hline $\Phi^-$  & 1 & 0 & $\Psi^-$  & 1 & 1 \\
        \hline
    \end{tabular}
    \caption{The exponents of the operators in the sequence $Z^aX^b$ for different measurement outcomes.}
    \label{tab:teleport}
\end{table}

The second version of the protocol has a simple topological interpretation, e.g.~\cite{Kauffman:2019,Melnikov:2022vij,Kauffman:2005tel}. Entangled pair $|\Phi^+\rangle$, shared by Alice and Bob is given by the diagram of~(\ref{2qubitstate}). Let us assume that Alice has measured her pair of qubits also in state $|\Phi^+\rangle$. Since the measurement is a projection on the respective state, this outcome can be presented by the following diagram:
\be
\label{measureBell}
\begin{array}{c}
     \begin{tikzpicture}
     \newcommand{\y}{1.8}
     \newcommand{\z}{0.3}
        \draw[draw=white,double=black,line width=1,rounded corners=2] (-0.45,0) -- (-0.45,-0.7) -- (0.45+\y,-0.7) -- (0.45+\y,0);
        \draw[draw=white,double=black,line width=1,rounded corners=2] (-0.15,0) -- (-0.15,-0.6) -- (0.15+\y,-0.6) -- (0.15+\y,0);
        \draw[draw=white,double=black,line width=1,rounded corners=2] (0.15,0) -- (0.15,-0.5) -- (-0.15+\y,-0.5) -- (-0.15+\y,0);
         \draw[draw=white,double=black,line width=1,rounded corners=2] (0.45,0) -- (0.45,-0.4) -- (-0.45+\y,-0.4) -- (-0.45+\y,0);
         \draw[gray,opacity=0.8,line width=2] (-0.75,0) -- (0.75,0);
         \foreach \x in {-0.45,-0.15,...,0.75}
           \fill[black] (\x,0) circle (0.075);
          \draw[orange,opacity=0.8,line width=2] (-0.75+\y,0) -- (0.75+\y,0); 
          \foreach \x in {-0.45,-0.15,...,0.75}
           \fill[black] (\x+\y,0) circle (0.075);
           \draw[gray,opacity=0.8,line width=2] (-0.75-\y,0) -- (0.75-\y,0);
           \foreach \x in {-0.45,-0.15,...,0.75}
           \draw[draw=white,double=black,line width=1,rounded corners=2] (\x-\y,0) -- (\x-\y,-0.55);
           \foreach \x in {-0.45,-0.15,...,0.75}
           \fill[black] (\x-\y,0) circle (0.075);
           \fill[white,rounded corners=2](-0.6-\y,-0.3) rectangle (0.6-\y,-0.7);
           \draw[rounded corners=2]  (-0.6-\y,-0.3) rectangle (0.6-\y,-0.7);
           \draw (0-\y,-0.5) node {$\psi$};
           \draw[draw=white,double=black,line width=1,rounded corners=2] (-0.45-\y,0+\z) -- (-0.45-\y,0.7+\z) -- (0.45,0.7+\z) -- (0.45,0+\z);
        \draw[draw=white,double=black,line width=1,rounded corners=2] (-0.15-\y,0+\z) -- (-0.15-\y,0.6+\z) -- (0.15,0.6+\z) -- (0.15,0+\z);
        \draw[draw=white,double=black,line width=1,rounded corners=2] (0.15-\y,0+\z) -- (0.15-\y,0.5+\z) -- (-0.15,0.5+\z) -- (-0.15,0+\z);
         \draw[draw=white,double=black,line width=1,rounded corners=2] (0.45-\y,0+\z) -- (0.45-\y,0.4+\z) -- (-0.45,0.4+\z) -- (-0.45,0+\z);
           \draw[gray,opacity=0.8,line width=2] (-0.75,0+\z) -- (0.75,0+\z);
         \foreach \x in {-0.45,-0.15,...,0.75}
           \fill[black] (0+\x,0+\z) circle (0.075);
           \draw[gray,opacity=0.8,line width=2] (-0.75-\y,0+\z) -- (0.75-\y,0+\z);
         \foreach \x in {-0.45,-0.15,...,0.75}
           \fill[black] (0+\x-\y,0+\z) circle (0.075);
     \end{tikzpicture} 
\end{array}
\qquad \longrightarrow \qquad
\begin{array}{c}
   \begin{tikzpicture}[thick]
   \newcommand{\y}{0}
       \draw[orange,opacity=0.8,line width=2] (-0.75-\y,0) -- (0.75-\y,0);
           \foreach \x in {-0.45,-0.15,...,0.75}
           \draw[draw=white,double=black,line width=1,rounded corners=2] (\x-\y,0) -- (\x-\y,-0.55);
           \foreach \x in {-0.45,-0.15,...,0.75}
           \fill[black] (\x-\y,0) circle (0.075);
           \fill[white,rounded corners=2](-0.6-\y,-0.3) rectangle (0.6-\y,-0.7);
           \draw[rounded corners=2]  (-0.6-\y,-0.3) rectangle (0.6-\y,-0.7);
           \draw (0-\y,-0.5) node {$\psi$};
   \end{tikzpicture} 
\end{array}.
\ee
Here Bob's qubit is the one on the right, whose space is colored orange. It is obvious from the diagram that after the measurement, this qubit becomes $|\psi\rangle$ without any additional action required from Bob.

Now assume Alice measured her qubits in a different state, generated by the diagram appearing in~(\ref{unitaryop}), with some braiding involved:
\be
\label{measureBraid}
\begin{array}{c}
     \begin{tikzpicture}
     \newcommand{\y}{1.8}
     \newcommand{\z}{0.3}
     \newcommand{\xb}{-\y/2-0.6}
     \newcommand{\yb}{0.7}
        \draw[draw=white,double=black,line width=1,rounded corners=2] (-0.45,0) -- (-0.45,-0.7) -- (0.45+\y,-0.7) -- (0.45+\y,0);
        \draw[draw=white,double=black,line width=1,rounded corners=2] (-0.15,0) -- (-0.15,-0.6) -- (0.15+\y,-0.6) -- (0.15+\y,0);
        \draw[draw=white,double=black,line width=1,rounded corners=2] (0.15,0) -- (0.15,-0.5) -- (-0.15+\y,-0.5) -- (-0.15+\y,0);
         \draw[draw=white,double=black,line width=1,rounded corners=2] (0.45,0) -- (0.45,-0.4) -- (-0.45+\y,-0.4) -- (-0.45+\y,0);
         \draw[gray,opacity=0.8,line width=2] (-0.75,0) -- (0.75,0);
         \foreach \x in {-0.45,-0.15,...,0.75}
           \fill[black] (\x,0) circle (0.075);
          \draw[orange,opacity=0.8,line width=2] (-0.75+\y,0) -- (0.75+\y,0); 
          \foreach \x in {-0.45,-0.15,...,0.75}
           \fill[black] (\x+\y,0) circle (0.075);
           \draw[gray,opacity=0.8,line width=2] (-0.75-\y,0) -- (0.75-\y,0);
           \foreach \x in {-0.45,-0.15,...,0.75}
           \draw[draw=white,double=black,line width=1,rounded corners=2] (\x-\y,0) -- (\x-\y,-0.55);
           \foreach \x in {-0.45,-0.15,...,0.75}
           \fill[black] (\x-\y,0) circle (0.075);
           \fill[white,rounded corners=2](-0.6-\y,-0.3) rectangle (0.6-\y,-0.7);
           \draw[rounded corners=2]  (-0.6-\y,-0.3) rectangle (0.6-\y,-0.7);
           \draw (0-\y,-0.5) node {$\psi$};
           \draw[draw=white,double=black,line width=1,rounded corners=2] (-0.45-\y,0+\z) -- (-0.45-\y,1.3+\z) -- (0.45,1.3+\z) -- (0.45,0+\z);
        \draw[draw=white,double=black,line width=1,rounded corners=2] (-0.15-\y,0+\z) -- (-0.15-\y,1+\z) -- (0.15,1+\z) -- (0.15,0+\z);
        \draw[draw=white,double=black,line width=1,rounded corners=2] (0.15-\y,0+\z) -- (0.15-\y,0.7+\z) -- (-0.15,0.7+\z) -- (-0.15,0+\z);
         \draw[draw=white,double=black,line width=1,rounded corners=2] (0.45-\y,0+\z) -- (0.45-\y,0.4+\z) -- (-0.45,0.4+\z) -- (-0.45,0+\z);
\fill[white] (0+\xb,.2+\yb) rectangle (1.2+\xb,1.0+\yb);
\draw[rounded corners=2] (0+\xb,0.6+\yb) -- (0.24+\xb,0.6+\yb) -- (0.32+\xb,0.7+\yb);
\draw[rounded corners=2] (0.4+\xb,0.8+\yb) -- (0.48+\xb,0.9+\yb) -- (1.2+\xb,0.9+\yb);
\draw[rounded corners=2] (0+\xb,0.3+\yb) -- (0.72+\xb,0.3+\yb) -- (0.96+\xb,0.6+\yb) -- (1.2+\xb,0.6+\yb);
\draw[rounded corners=2] (0+\xb,0.9+\yb) -- (0.24+\xb,0.9+\yb) -- (0.48+\xb,0.6+\yb) -- (0.72+\xb,0.6+\yb) -- (0.8+\xb,0.5+\yb);
\draw[rounded corners=2] (0.88+\xb,0.4+\yb) -- (0.96+\xb,0.3+\yb) -- (1.2+\xb,0.3+\yb);
           \draw[gray,opacity=0.8,line width=2] (-0.75,0+\z) -- (0.75,0+\z);
         \foreach \x in {-0.45,-0.15,...,0.75}
           \fill[black] (0+\x,0+\z) circle (0.075);
           \draw[gray,opacity=0.8,line width=2] (-0.75-\y,0+\z) -- (0.75-\y,0+\z);
         \foreach \x in {-0.45,-0.15,...,0.75}
           \fill[black] (0+\x-\y,0+\z) circle (0.075);
     \end{tikzpicture} 
\end{array}
\qquad \longrightarrow \qquad
\begin{array}{c}
   \begin{tikzpicture}[thick]
   \newcommand{\y}{0}
   \newcommand{\z}{-1.2}
   \newcommand{\xb}{0.45}
   \newcommand{\yb}{-1.45}
       \draw[orange,opacity=0.8,line width=2] (-0.75-\y,0) -- (0.75-\y,0);
           \foreach \x in {-0.45,-0.15,...,0.75}
           \draw[draw=white,double=black,line width=1,rounded corners=2] (\x-\y,0) -- (\x-\y,-0.55+\z);
           \foreach \x in {-0.45,-0.15,...,0.75}
           \fill[black] (\x-\y,0) circle (0.075);
           \fill[white,rounded corners=2](-0.6-\y,-0.3+\z) rectangle (0.6-\y,-0.7+\z);
           \draw[rounded corners=2]  (-0.6-\y,-0.3+\z) rectangle (0.6-\y,-0.7+\z);
           \draw (0-\y,-0.5+\z) node {$\psi$};
\fill[white] (0.1+\xb,0+\yb) rectangle (-1+\xb,1.2+\yb);
\draw[rounded corners=2] (0+\xb,0+\yb) -- (0+\xb,1.2+\yb);
\draw[rounded corners=2] (-0.6+\xb,0+\yb) -- (-0.6+\xb,0.24+\yb) -- (-0.7+\xb,0.32+\yb);
\draw[rounded corners=2] (-0.8+\xb,0.4+\yb) -- (-0.9+\xb,0.48+\yb) -- (-0.9+\xb,1.2+\yb);
\draw[rounded corners=2] (-0.3+\xb,0+\yb) -- (-0.3+\xb,0.72+\yb) -- (-0.6+\xb,0.96+\yb) -- (-0.6+\xb,1.2+\yb);
\draw[rounded corners=2] (-0.9+\xb,0+\yb) -- (-0.9+\xb,0.24+\yb) -- (-0.6+\xb,0.48+\yb) -- (-0.6+\xb,0.72+\yb) -- (-0.5+\xb,0.8+\yb);
\draw[rounded corners=2] (-0.4+\xb,0.88+\yb) -- (-0.3+\xb,0.96+\yb) -- (-0.3+\xb,1.2+\yb);
   \end{tikzpicture} 
\end{array}.
\ee
In such a case, in order to obtain precisely state $|\psi\rangle$, Bob needs to undo the braiding induced by the measurement. Since braiding is a unitary operation, according to the discussion in section~\ref{sec:SLOCC} there exists a corresponding unitary. 

Note also the duality between the state measured by Alice and the operator needed to undo the braiding. The state can be thought of as created by the inverse braid operator acting on either of the two qubits of Bell state~(\ref{2qubitstate}). This highlights the importance of use of the Bell states in the teleportation protocol: the pair of qubits shared by Alice and Bob must be in a maximally entangled state. Any maximally entangled state can be generated by action of a local unitary (which acts on either Alice's or Bob's qubit) on state~(\ref{2qubitstate}) ensuring that its action can be undone in the protocol. For comparison, try to think of what would happen if the shared pair were obtained from the diagram in~(\ref{invertop}), corresponding to an invertible, but nonunitary operator.

For the same reason Alice must do measurements in a basis of maximally entangled states. Computational basis $|00\rangle$, $|01\rangle$, $|10\rangle$ and $|11\rangle$ is not suitable for this purpose. That is why, in the first version of the protocol, Alice needs to act by additional operators to prepare the measurement appropriately. One can try to do the same in the topological version and prepare Alice's qubits for measurements in the computational basis.\footnote{This was done in~\cite{Melnikov:2023nzn}, although with some important details missing. First, the measurement was assumed in a nonortogonal basis~(\ref{4pbasis}). Second, the entangling operator was not unitary (see the discussion at the end of section~\ref{sec:densecoding}).}

The TQFT approach developed in these notes also has a drawback. Measurements in the examples~(\ref{measureBell}) and~(\ref{measureBraid}) are not performed with respect to orthogonal states. The point is that it is not possible to define a set of orthogonal states in terms of tangles, in general. A linear combinations of diagrams must be considered instead, as in the case of basis~(\ref{4pobasis}). This is an inconvenience that does not allow us to construct an analog of table~\ref{tab:teleport} inserting basic tangle diagrams as measured states. Instead we will summarize the protocol by the following generalization:
\be
\begin{array}{c}
     \begin{tikzpicture}
     \newcommand{\y}{1.8}
     \newcommand{\z}{0.3}
        \draw[draw=white,double=black,line width=1,rounded corners=2] (-0.45,0) -- (-0.45,-0.7) -- (0.45+\y,-0.7) -- (0.45+\y,0);
        \draw[draw=white,double=black,line width=1,rounded corners=2] (-0.15,0) -- (-0.15,-0.6) -- (0.15+\y,-0.6) -- (0.15+\y,0);
        \draw[draw=white,double=black,line width=1,rounded corners=2] (0.15,0) -- (0.15,-0.5) -- (-0.15+\y,-0.5) -- (-0.15+\y,0);
         \draw[draw=white,double=black,line width=1,rounded corners=2] (0.45,0) -- (0.45,-0.4) -- (-0.45+\y,-0.4) -- (-0.45+\y,0);
         \draw[gray,opacity=0.8,line width=2] (-0.75,0) -- (0.75,0);
         \foreach \x in {-0.45,-0.15,...,0.75}
           \fill[black] (\x,0) circle (0.075);
          \draw[orange,opacity=0.8,line width=2] (-0.75+\y,0) -- (0.75+\y,0); 
          \foreach \x in {-0.45,-0.15,...,0.75}
           \fill[black] (\x+\y,0) circle (0.075);
           \draw[gray,opacity=0.8,line width=2] (-0.75-\y,0) -- (0.75-\y,0);
           \foreach \x in {-0.45,-0.15,...,0.75}
           \draw[draw=white,double=black,line width=1,rounded corners=2] (\x-\y,0) -- (\x-\y,-0.55);
           \foreach \x in {-0.45,-0.15,...,0.75}
           \fill[black] (\x-\y,0) circle (0.075);
           \fill[white,rounded corners=2](-0.6-\y,-0.3) rectangle (0.6-\y,-0.7);
           \draw[rounded corners=2]  (-0.6-\y,-0.3) rectangle (0.6-\y,-0.7);
           \draw (0-\y,-0.5) node {$\psi$};
           \draw[draw=white,double=black,line width=1,rounded corners=2] (-0.45-\y,0+\z) -- (-0.45-\y,0.7+\z) -- (0.45,0.7+\z) -- (0.45,0+\z);
        \draw[draw=white,double=black,line width=1,rounded corners=2] (-0.15-\y,0+\z) -- (-0.15-\y,0.6+\z) -- (0.15,0.6+\z) -- (0.15,0+\z);
        \draw[draw=white,double=black,line width=1,rounded corners=2] (0.15-\y,0+\z) -- (0.15-\y,0.5+\z) -- (-0.15,0.5+\z) -- (-0.15,0+\z);
         \draw[draw=white,double=black,line width=1,rounded corners=2] (0.45-\y,0+\z) -- (0.45-\y,0.4+\z) -- (-0.45,0.4+\z) -- (-0.45,0+\z);
           \draw[gray,opacity=0.8,line width=2] (-0.75,0+\z) -- (0.75,0+\z);
         \foreach \x in {-0.45,-0.15,...,0.75}
           \fill[black] (0+\x,0+\z) circle (0.075);
           \draw[gray,opacity=0.8,line width=2] (-0.75-\y,0+\z) -- (0.75-\y,0+\z);
         \foreach \x in {-0.45,-0.15,...,0.75}
           \fill[black] (0+\x-\y,0+\z) circle (0.075);
           \fill[white] (-0.2-\y/2,0.3+\z) rectangle (0.2-\y/2,0.8+\z);
           \draw (-0.2-\y/2,0.3+\z) rectangle (0.2-\y/2,0.8+\z);
           \draw (0-\y/2,0.55+\z) node {${\cal B}$};
     \end{tikzpicture} 
\end{array}
\qquad \longrightarrow \qquad
\begin{array}{c}
   \begin{tikzpicture}[thick]
   \newcommand{\y}{0}
   \newcommand{\z}{-1.2}
   \newcommand{\xb}{0.45}
   \newcommand{\yb}{-0.3}
       \draw[orange,opacity=0.8,line width=2] (-0.75-\y,0) -- (0.75-\y,0);
           \foreach \x in {-0.45,-0.15,...,0.75}
           \draw[draw=white,double=black,line width=1,rounded corners=2] (\x-\y,0) -- (\x-\y,-0.55+\z);
           \foreach \x in {-0.45,-0.15,...,0.75}
           \fill[black] (\x-\y,0) circle (0.075);
           \fill[white,rounded corners=2](-0.6-\y,-0.3+\z) rectangle (0.6-\y,-0.7+\z);
           \draw[rounded corners=2]  (-0.6-\y,-0.3+\z) rectangle (0.6-\y,-0.7+\z);
           \draw (0-\y,-0.5+\z) node {$\psi$};
           \fill[white](-0.6,-0.3+\yb) rectangle (0.6,-0.7+\yb);
           \draw  (-0.6,-0.3+\yb) rectangle (0.6,-0.7+\yb);
           \draw (0,-0.5+\yb) node {${\cal B}$};
   \end{tikzpicture} 
\end{array}.
\ee
Here ${\cal{B}}$ corresponds to a choice of orthogonal basis labeled by four unitary matrices, which generate this basis by a local action on Bell state $|\Phi^+\rangle$. In order to construct state $|\psi\rangle$ Bob needs to act on his qubit by an inverse of the corresponding unitary matrix.

According to~\cite{Coecke:2017pic} the diagrammatic interpretation of this sort was introduced independently in~\cite{Kauffman:2005tel}, using the TQFT methods, and in~\cite{Coecke:2004log,Abramsky:2004doh}, in the discussion of a more general categorical approach.

%%%%%%%%%%%%%%%%%%%%%%%%%%%%%%%%%%%%%%%%%%%%%%%%%%%%%%%%%%%%%%%%%%%%
\subsection{Dense Coding}
\label{sec:densecoding}

Superdense, or, simply, dense coding is a protocol closely related to quantum teleportation. It uses entanglement properties in a very similar way, allowing to use a quantum channel to simultaneously transmit more than a single bit of classical information. In the textbook version Alice and Bob share an entangled pair of qubits, so that Bob can encode two classical bits of information, sending only one qubit to Alice. The circuit diagram of the corresponding protocol is shown in figure~\ref{fig:teleport} (right).

Initially, Bob and Alice share an entangled pair of qubits $|\Phi^+\rangle$, where we label $\Phi_A^+$ and $\Phi_B^+$ Alice's and Bob's qubits respectively. Given two classical bits $a$ and $b$ Bob performs the encoding, applying the string $Z^aX^b$ on his qubit. He then sends his qubit to Alice. Alice applies a CNOT on the two qubits she now possesses and also an Hadamard on the Bob's qubit. If she now measures both qubits she will find hers in state $|a\rangle$ and Bob's in state $|b\rangle$.

Let us note that the CNOT and $H$ gate effectively replace the measurement basis by that of the Bell states. Consequently, there is a simpler (and a bit tautological) version of the protocol, consisting of application of the $Z^aX^b$ string and measuring in the Bell basis~(\ref{BellBasis}). Then the classical values encoded by Bob can be read off table~\ref{tab:teleport}, provided a specific Bell state as the result of the measurement. The topological version of this can be expressed by the diagram
\be
\label{densecode}
\begin{array}{c}
     \begin{tikzpicture}[thick]
        \newcommand{\y}{0}
          \draw[gray,opacity=0.8,line width=2] (0,-0.5+\y) -- (0,0.5+\y);
         \foreach \x in {-0.3,-0.1,...,0.3}
           \fill[black] (0,\x+\y) circle (0.065);
            \draw[gray,opacity=0.8,line width=2] (-1.5,-0.5+\y) -- (-1.5,0.5+\y);
         \foreach \x in {-0.3,-0.1,...,0.3}
           \fill[black] (-1.5,\x+\y) circle (0.065);
        \renewcommand{\y}{1.2}
        \foreach \x in {-0.3,-0.1,...,0.3}
        \draw[rounded corners=2] (0,\x) -- (-2.25+\x,\x) -- (-2.25+\x,\y-\x) -- (0,\y-\x);
        \draw[gray,opacity=0.8,line width=2] (0,-0.5+\y) -- (0,0.5+\y);
         \foreach \x in {-0.3,-0.1,...,0.3}
           \fill[black] (0,\x+\y) circle (0.065);
            \draw[gray,opacity=0.8,line width=2] (-1.5,-0.5+\y) -- (-1.5,0.5+\y);
         \foreach \x in {-0.3,-0.1,...,0.3}
           \fill[black] (-1.5,\x+\y) circle (0.065);
           \fill[white] (-1.,-0.45+\y) rectangle (-0.5,0.45+\y);
           \draw (-1.0,-0.45+\y) rectangle (-0.5,0.45+\y);
           \draw (-0.75,\y) node {${\cal B}$};
           \newcommand{\z}{0.3}
           \renewcommand{\y}{0}
            \draw[gray,opacity=0.8,line width=2] (0+\z,-0.5+\y) -- (0+\z,0.5+\y);
         \foreach \x in {-0.3,-0.1,...,0.3}
           \fill[black] (0+\z,\x+\y) circle (0.065);
            \renewcommand{\y}{1.2}
            \draw[gray,opacity=0.8,line width=2] (0+\z,-0.5+\y) -- (0+\z,0.5+\y);
         \foreach \x in {-0.3,-0.1,...,0.3}
           \fill[black] (0+\z,\x+\y) circle (0.065);
           \foreach \x in {-0.3,-0.1,...,0.3}
        \draw[rounded corners=2] (0+\z,\x) -- (0.8-\x+\z,\x) -- (0.8-\x+\z,\y-\x) -- (0+\z,\y-\x);
        \fill[white] (0.35+\z,0.4) rectangle (1.25+\z,-0.4+\y);
        \draw (0.35+\z,0.4) rectangle (1.25+\z,-0.4+\y);
        \draw (0.8+\z,\y/2) node {${\cal B}$};
           \end{tikzpicture}
\end{array}.
\ee
This diagram has three vertical sections. The left section is the initial state $|\Phi^+\rangle$ shared between Alice and Bob. In the central section Bob applies ${\cal B}[a,b]=Z^aX^b$, which, when appended to $|\Phi^+\rangle$, transforms the latter into one of the Bell basis states according to table~\ref{tab:teleport}. The right section is the measurement in the same Bell basis. Therefore, the outcome of the measurement will indeed tell which classical pair $(a,b)$ was encoded by Bob.

If one finds disappointing the triviality of diagram~(\ref{densecode}) one can also build a more neat example, similar to the circuit version shown in figure~\ref{fig:teleport}. Let us assume that Bob encodes classical bits using four braids, as shown in table~\ref{tab:densecode}, cf.~\cite{Melnikov:2023nzn}.
\begin{table}[h]
    \centering
    \begin{tabular}{||c|c|c||c|c|c||}
        \hline  $a$ & $b$ & Braid & $a$ & $b$ & Braid  \\
        \hline 0 & 0 &  \scalebox{0.8}{$\begin{array}{c} 
                \begin{tikzpicture}[thick]
                \newcommand{\y}{0}
          \draw[gray,opacity=0.8,line width=2] (0,-0.5+\y) -- (0,0.5+\y);
         \foreach \x in {-0.3,-0.1,...,0.3}
           \fill[black] (0,\x+\y) circle (0.065);
            \draw[gray,opacity=0.8,line width=2] (-1.5,-0.5+\y) -- (-1.5,0.5+\y);
         \foreach \x in {-0.3,-0.1,...,0.3}
           \fill[black] (-1.5,\x+\y) circle (0.065);
           \draw (-1.5,-0.3+\y) -- (0,-0.3+\y);
           \draw (-1.5,-0.1+\y) -- (0,-0.1+\y);
           \draw[teal] (-1.5,0.1+\y) -- (0,0.1+\y);
           \draw[teal] (-1.5,0.3+\y) -- (0,0.3+\y);
                \end{tikzpicture} 
           \end{array}$} & 1 & 0 & \scalebox{0.8}{$\begin{array}{c} 
                \begin{tikzpicture}[thick]
                \newcommand{\y}{0}
            \draw (-1.5,-0.3+\y) -- (0,-0.3+\y);
           \draw (-1.5,-0.1+\y) -- (0,-0.1+\y);
           \draw[teal,rounded corners=2] (-1.5,0.3+\y) -- (-0.9,0.3+\y) -- (-0.6,0.1+\y) -- (0,0.1+\y);
           \draw[draw=white,double=teal,line width=2,rounded corners=2] (-1.5,0.1+\y) -- (-0.9,0.1+\y) -- (-0.6,0.3+\y) -- (0,0.3+\y);
          \draw[gray,opacity=0.8,line width=2] (0,-0.5+\y) -- (0,0.5+\y);
         \foreach \x in {-0.3,-0.1,...,0.3}
           \fill[black] (0,\x+\y) circle (0.065);
            \draw[gray,opacity=0.8,line width=2] (-1.5,-0.5+\y) -- (-1.5,0.5+\y);
         \foreach \x in {-0.3,-0.1,...,0.3}
           \fill[black] (-1.5,\x+\y) circle (0.065);
                \end{tikzpicture} 
           \end{array}$} \\
        0 & 1 & \scalebox{0.8}{$\begin{array}{c} 
                \begin{tikzpicture}[thick]
                \newcommand{\y}{0}
                \draw[rounded corners=2] (-1.5,-0.3+\y) -- (-0.9,-0.3+\y) -- (-0.6,-0.1+\y) -- (0,-0.1+\y);
           \draw[draw=white,double=black,line width=2,rounded corners=2] (-1.5,-0.1+\y) -- (-0.9,-0.1+\y) -- (-0.6,-0.3+\y) -- (0,-0.3+\y);
           \draw[teal] (-1.5,0.1+\y) -- (0,0.1+\y);
           \draw[teal] (-1.5,0.3+\y) -- (0,0.3+\y);
          \draw[gray,opacity=0.8,line width=2] (0,-0.5+\y) -- (0,0.5+\y);
         \foreach \x in {-0.3,-0.1,...,0.3}
           \fill[black] (0,\x+\y) circle (0.065);
            \draw[gray,opacity=0.8,line width=2] (-1.5,-0.5+\y) -- (-1.5,0.5+\y);
         \foreach \x in {-0.3,-0.1,...,0.3}
           \fill[black] (-1.5,\x+\y) circle (0.065);
                \end{tikzpicture} 
           \end{array}$} & 1 & 1 & \scalebox{0.8}{$\begin{array}{c} 
                \begin{tikzpicture}[thick]
                \newcommand{\y}{0}
                \draw[teal,rounded corners=2] (-1.5,0.3+\y) -- (-0.9,0.3+\y) -- (-0.6,0.1+\y) -- (0,0.1+\y);
           \draw[draw=white,double=teal,line width=2,rounded corners=2] (-1.5,0.1+\y) -- (-0.9,0.1+\y) -- (-0.6,0.3+\y) -- (0,0.3+\y);
           \draw[rounded corners=2] (-1.5,-0.3+\y) -- (-0.9,-0.3+\y) -- (-0.6,-0.1+\y) -- (0,-0.1+\y);
           \draw[draw=white,double=black,line width=2,rounded corners=2] (-1.5,-0.1+\y) -- (-0.9,-0.1+\y) -- (-0.6,-0.3+\y) -- (0,-0.3+\y);
          \draw[gray,opacity=0.8,line width=2] (0,-0.5+\y) -- (0,0.5+\y);
         \foreach \x in {-0.3,-0.1,...,0.3}
           \fill[black] (0,\x+\y) circle (0.065);
            \draw[gray,opacity=0.8,line width=2] (-1.5,-0.5+\y) -- (-1.5,0.5+\y);
         \foreach \x in {-0.3,-0.1,...,0.3}
           \fill[black] (-1.5,\x+\y) circle (0.065);
                \end{tikzpicture} 
           \end{array}$}\\
         \hline
    \end{tabular}
    \caption{Braids used for encoding classical bits.}
    \label{tab:densecode}
\end{table}

The CNOT and $H$ gates will be simulated by some nonlocal braiding operation. Let us first assume that $a=1$ and $b=0$ are encoded. The protocol can be organized according to the following diagram:
\be
\begin{array}{c}
     \begin{tikzpicture}[thick]
        \newcommand{\y}{0}
        \renewcommand{\y}{1.2}
         \draw (-1.5,-0.3+\y) -- (0,-0.3+\y);
           \draw (-1.5,-0.1+\y) -- (0,-0.1+\y);
           \draw[teal,rounded corners=2] (-1.5,0.3+\y) -- (-0.9,0.3+\y) -- (-0.6,0.1+\y) -- (0,0.1+\y);
           \draw[draw=white,double=teal,line width=2,rounded corners=2] (-1.5,0.1+\y) -- (-0.9,0.1+\y) -- (-0.6,0.3+\y) -- (0,0.3+\y);
        \foreach \x in {-0.3,-0.1,...,-0.1}
        \draw[teal,rounded corners=2] (0,\x) -- (-2.25+\x,\x) -- (-2.25+\x,\y-\x) -- (-1.5,\y-\x);
        \foreach \x in {0.1,0.3,...,0.3}
        \draw[rounded corners=2] (0,\x) -- (-2.25+\x,\x) -- (-2.25+\x,\y-\x) -- (-1.5,\y-\x);
            \draw[gray,opacity=0.8,line width=2] (-1.5,-0.5) -- (-1.5,0.5+\y);
         \foreach \x in {-0.3,-0.1,...,0.3}
           \fill[black] (-1.5,\x+\y) circle (0.065);
           \foreach \x in {-0.3,-0.1,...,0.3}
           \fill[black] (-1.5,\x) circle (0.065);
           \newcommand{\z}{1.5}
           \renewcommand{\y}{0}
           \draw[teal,rounded corners=2] (-1.5+\z,-0.1+\y) -- (-3*1.5/4-0.1+\z,-0.1+\y) -- (-3*1.5/4+0.1+\z,0.1+\y) -- (-1.5/2-0.1+\z,0.1+\y) -- (-1.5/2+0.1+\z,0.3+\y) -- (0+\z,0.3+\y);
           \draw[teal,rounded corners=2] (-1.5+\z,-0.3+\y) -- (-1.5/2-0.1+\z,-0.3+\y) -- (-1.5/2+0.1+\z,-0.1+\y) -- (-1.5/4-0.1+\z,-0.1+\y) -- (-1.5/4+0.1+\z,0.1+\y) -- (0+\z,0.1+\y);
           \draw[draw=white,double=black,line width=2,rounded corners=2] (-1.5+\z,0.1+\y) -- (-3*1.5/4-0.1+\z,0.1+\y) -- (-3*1.5/4+0.1+\z,-0.1+\y) -- (-1.5/2-0.1+\z,-0.1+\y) -- (-1.5/2+0.1+\z,-0.3+\y) -- (0+\z,-0.3+\y);
           \draw[draw=white,double=black,line width=2,rounded corners=2] (-1.5+\z,0.3+\y) -- (-1.5/2-0.1+\z,0.3+\y) -- (-1.5/2+0.1+\z,0.1+\y) -- (-1.5/4-0.1+\z,0.1+\y) -- (-1.5/4+0.1+\z,-0.1+\y) -- (0+\z,-0.1+\y);
           \renewcommand{\y}{1.2}
           \newcommand{\w}{1.5}
           \foreach \x in {0.3,0.1,...,0.1}
        \draw[teal,rounded corners=2] (0,\x+\y) -- (1.5,\x+\y);
        \foreach \x in {-0.3,-0.1,...,-0.1}
        \draw[rounded corners=2] (0,\x+\y) -- (1.5,\x+\y);
        \draw[teal,rounded corners=2] (\w,0.3) -- (5*\w/4-0.1,0.3) -- (5*\w/4+0.1,-0.3+\y) -- (6*\w/4-0.1,-0.3+\y) -- (6*\w/4+0.1,-0.1+\y) -- (7*\w/4-0.1,-0.1+\y) -- (7*\w/4+0.1,0.1+\y) -- (2*\w,0.1+\y);
        \draw[teal,rounded corners=2] (\w,0.1) -- (6*\w/4-0.1,0.1) -- (6*\w/4+0.1,0.3) -- (7*\w/4-0.1,0.3) -- (7*\w/4+0.1,-0.3+\y) -- (2*\w,-0.3+\y);
        \draw[teal,rounded corners=2] (\w,0.1+\y) -- (6*\w/4-0.1,0.1+\y) -- (6*\w/4+0.1,0.3+\y) -- (2*\w,0.3+\y);
        \draw[draw=white,double=teal,line width=2,rounded corners=2] (\w,0.3+\y) -- (6*\w/4-0.1,0.3+\y) -- (6*\w/4+0.1,0.1+\y) -- (7*\w/4-0.1,0.1+\y) -- (7*\w/4+0.1,-0.1+\y) -- (2*\w,-0.1+\y);
        \draw[draw=white,double=black,line width=2,rounded corners=2] (\w,-0.3+\y) -- (5*\w/4-0.1,-0.3+\y) -- (5*\w/4+0.1,0.3) -- (6*\w/4-0.1,0.3) -- (6*\w/4+0.1,0.1) -- (7*\w/4-0.1,0.1) -- (7*\w/4+0.1,-0.1) -- (2*\w,-0.1);
        \draw[draw=white,double=black,line width=2,rounded corners=2] (\w,-0.1+\y) -- (6*\w/4-0.1,-0.1+\y) -- (6*\w/4+0.1,-0.3+\y) -- (7*\w/4-0.1,-0.3+\y) -- (7*\w/4+0.1,0.3) -- (2*\w,0.3);
        \draw[rounded corners=2] (\w,-0.1) -- (6*\w/4-0.1,-0.1) -- (6*\w/4+0.1,-0.3) -- (2*\w,-0.3);
        \draw[draw=white,double=black,line width=2,rounded corners=2] (\w,-0.3) -- (6*\w/4-0.1,-0.3) -- (6*\w/4+0.1,-0.1) -- (7*\w/4-0.1,-0.1) -- (7*\w/4+0.1,0.1) -- (2*\w,0.1);
           \draw[gray,opacity=0.8,line width=2] (0,-0.5) -- (0,0.5+\y);
         \foreach \x in {-0.3,-0.1,...,0.3}
           \fill[black] (0,\x+\y) circle (0.065);
           \foreach \x in {-0.3,-0.1,...,0.3}
            \fill[black] (0,\x) circle (0.065);
            \draw[gray,opacity=0.8,line width=2] (1.5,-0.5) -- (1.5,0.5+\y);
         \foreach \x in {-0.3,-0.1,...,0.3}
           \fill[black] (1.5,\x+\y) circle (0.065);
         \foreach \x in {-0.3,-0.1,...,0.3}
           \fill[black] (1.5,\x) circle (0.065);
            \draw[gray,opacity=0.8,line width=2] (2*\w,-0.5) -- (2*\w,0.5+\y);
         \foreach \x in {-0.3,-0.1,...,0.3}
           \fill[black] (2*\w,\x+\y) circle (0.065);
         \foreach \x in {-0.3,-0.1,...,0.3}
           \fill[black] (2*\w,\x) circle (0.065);
           \renewcommand{\z}{2*\w+0.3}
        \foreach \x in {-0.3,0.1,...,0.1}
        \draw[rounded corners=2] (0+\z,\x) -- (0.5+\z,\x) -- (0.5+\z,\x+0.2) -- (0+\z,\x+0.2);
        \draw[teal,rounded corners=2] (0+\z,\y-0.3) -- (0.6+\z,\y-0.3) -- (0.6+\z,0.3+\y) -- (0+\z,0.3+\y);
        \draw[teal,rounded corners=2] (0+\z,-0.1+\y) -- (0.4+\z,-0.1+\y) -- (0.4+\z,0.1+\y) -- (0+\z,0.1+\y);
            \draw[gray,opacity=0.8,line width=2] (0+\z,-0.5) -- (0+\z,0.5);
         \foreach \x in {-0.3,-0.1,...,0.3}
           \fill[black] (0+\z,\x) circle (0.065);
            \draw[gray,opacity=0.8,line width=2] (0+\z,-0.5+\y) -- (0+\z,0.5+\y);
         \foreach \x in {-0.3,-0.1,...,0.3}
           \fill[black] (0+\z,\x+\y) circle (0.065);
           \end{tikzpicture}
\end{array}.
\ee
This protocol is split into five sections. The first section, from the left, is state $|\Phi^+\rangle$, shared by Alice and Bob. The second section is Bob encoding the classical bits. In the third section Alice applies a permutation on her qubit, which is an analog of the Hadamard gate. In the forth section a nonlocal permutation gate is applied, again by Alice, who now has both qubits. This is the ``CNOT''. The fifth section shows the measurement. The diagram tells us that the result in this case will be state $|\hat{1}\rangle$ on the top qubit and state $|\hat{0}\rangle$ on the bottom, where the labels correspond to the nonorthonormal basis~(\ref{4pbasis}). 

Now let us encode the pair $a=0$, $b=1$. Using the same protocol we draw
\be
\begin{array}{c}
     \begin{tikzpicture}[thick]
        \newcommand{\y}{0}
        \renewcommand{\y}{1.2}
        \draw[rounded corners=2] (-1.5,-0.3+\y) -- (-0.9,-0.3+\y) -- (-0.6,-0.1+\y) -- (0,-0.1+\y);
           \draw[draw=white,double=black,line width=2,rounded corners=2] (-1.5,-0.1+\y) -- (-0.9,-0.1+\y) -- (-0.6,-0.3+\y) -- (0,-0.3+\y);
           \draw[teal] (-1.5,0.1+\y) -- (0,0.1+\y);
           \draw[teal] (-1.5,0.3+\y) -- (0,0.3+\y);
        \foreach \x in {-0.3,-0.1,...,-0.1}
        \draw[teal,rounded corners=2] (0,\x) -- (-2.25+\x,\x) -- (-2.25+\x,\y-\x) -- (-1.5,\y-\x);
        \foreach \x in {0.1,0.3,...,0.3}
        \draw[rounded corners=2] (0,\x) -- (-2.25+\x,\x) -- (-2.25+\x,\y-\x) -- (-1.5,\y-\x);
            \draw[gray,opacity=0.8,line width=2] (-1.5,-0.5) -- (-1.5,0.5+\y);
         \foreach \x in {-0.3,-0.1,...,0.3}
           \fill[black] (-1.5,\x+\y) circle (0.065);
           \foreach \x in {-0.3,-0.1,...,0.3}
           \fill[black] (-1.5,\x) circle (0.065);
           \newcommand{\z}{1.5}
           \renewcommand{\y}{0}
           \draw[teal,rounded corners=2] (-1.5+\z,-0.1+\y) -- (-3*1.5/4-0.1+\z,-0.1+\y) -- (-3*1.5/4+0.1+\z,0.1+\y) -- (-1.5/2-0.1+\z,0.1+\y) -- (-1.5/2+0.1+\z,0.3+\y) -- (0+\z,0.3+\y);
           \draw[teal,rounded corners=2] (-1.5+\z,-0.3+\y) -- (-1.5/2-0.1+\z,-0.3+\y) -- (-1.5/2+0.1+\z,-0.1+\y) -- (-1.5/4-0.1+\z,-0.1+\y) -- (-1.5/4+0.1+\z,0.1+\y) -- (0+\z,0.1+\y);
           \draw[draw=white,double=black,line width=2,rounded corners=2] (-1.5+\z,0.1+\y) -- (-3*1.5/4-0.1+\z,0.1+\y) -- (-3*1.5/4+0.1+\z,-0.1+\y) -- (-1.5/2-0.1+\z,-0.1+\y) -- (-1.5/2+0.1+\z,-0.3+\y) -- (0+\z,-0.3+\y);
           \draw[draw=white,double=black,line width=2,rounded corners=2] (-1.5+\z,0.3+\y) -- (-1.5/2-0.1+\z,0.3+\y) -- (-1.5/2+0.1+\z,0.1+\y) -- (-1.5/4-0.1+\z,0.1+\y) -- (-1.5/4+0.1+\z,-0.1+\y) -- (0+\z,-0.1+\y);
           \renewcommand{\y}{1.2}
           \newcommand{\w}{1.5}
           \foreach \x in {0.3,0.1,...,0.1}
        \draw[teal,rounded corners=2] (0,\x+\y) -- (1.5,\x+\y);
        \foreach \x in {-0.3,-0.1,...,-0.1}
        \draw[rounded corners=2] (0,\x+\y) -- (1.5,\x+\y);
        \draw[teal,rounded corners=2] (\w,0.3) -- (5*\w/4-0.1,0.3) -- (5*\w/4+0.1,-0.3+\y) -- (6*\w/4-0.1,-0.3+\y) -- (6*\w/4+0.1,-0.1+\y) -- (7*\w/4-0.1,-0.1+\y) -- (7*\w/4+0.1,0.1+\y) -- (2*\w,0.1+\y);
        \draw[teal,rounded corners=2] (\w,0.1) -- (6*\w/4-0.1,0.1) -- (6*\w/4+0.1,0.3) -- (7*\w/4-0.1,0.3) -- (7*\w/4+0.1,-0.3+\y) -- (2*\w,-0.3+\y);
        \draw[teal,rounded corners=2] (\w,0.1+\y) -- (6*\w/4-0.1,0.1+\y) -- (6*\w/4+0.1,0.3+\y) -- (2*\w,0.3+\y);
        \draw[draw=white,double=teal,line width=2,rounded corners=2] (\w,0.3+\y) -- (6*\w/4-0.1,0.3+\y) -- (6*\w/4+0.1,0.1+\y) -- (7*\w/4-0.1,0.1+\y) -- (7*\w/4+0.1,-0.1+\y) -- (2*\w,-0.1+\y);
        \draw[draw=white,double=black,line width=2,rounded corners=2] (\w,-0.3+\y) -- (5*\w/4-0.1,-0.3+\y) -- (5*\w/4+0.1,0.3) -- (6*\w/4-0.1,0.3) -- (6*\w/4+0.1,0.1) -- (7*\w/4-0.1,0.1) -- (7*\w/4+0.1,-0.1) -- (2*\w,-0.1);
        \draw[draw=white,double=black,line width=2,rounded corners=2] (\w,-0.1+\y) -- (6*\w/4-0.1,-0.1+\y) -- (6*\w/4+0.1,-0.3+\y) -- (7*\w/4-0.1,-0.3+\y) -- (7*\w/4+0.1,0.3) -- (2*\w,0.3);
        \draw[rounded corners=2] (\w,-0.1) -- (6*\w/4-0.1,-0.1) -- (6*\w/4+0.1,-0.3) -- (2*\w,-0.3);
        \draw[draw=white,double=black,line width=2,rounded corners=2] (\w,-0.3) -- (6*\w/4-0.1,-0.3) -- (6*\w/4+0.1,-0.1) -- (7*\w/4-0.1,-0.1) -- (7*\w/4+0.1,0.1) -- (2*\w,0.1);
           \draw[gray,opacity=0.8,line width=2] (0,-0.5) -- (0,0.5+\y);
         \foreach \x in {-0.3,-0.1,...,0.3}
           \fill[black] (0,\x+\y) circle (0.065);
           \foreach \x in {-0.3,-0.1,...,0.3}
           \fill[black] (0,\x) circle (0.065);
            \draw[gray,opacity=0.8,line width=2] (1.5,-0.5) -- (1.5,0.5+\y);
         \foreach \x in {-0.3,-0.1,...,0.3}
           \fill[black] (1.5,\x+\y) circle (0.065);
         \foreach \x in {-0.3,-0.1,...,0.3}
           \fill[black] (1.5,\x) circle (0.065);
            \draw[gray,opacity=0.8,line width=2] (2*\w,-0.5) -- (2*\w,0.5+\y);
         \foreach \x in {-0.3,-0.1,...,0.3}
           \fill[black] (2*\w,\x+\y) circle (0.065);
         \foreach \x in {-0.3,-0.1,...,0.3}
           \fill[black] (2*\w,\x) circle (0.065);
           \renewcommand{\z}{2*\w+0.3}
        \foreach \x in {-0.3,0.1,...,0.1}
        \draw[teal,rounded corners=2] (0+\z,\x+\y) -- (0.5+\z,\x+\y) -- (0.5+\z,\x+0.2+\y) -- (0+\z,\x+0.2+\y);
        \draw[rounded corners=2] (0+\z,-0.3) -- (0.6+\z,-0.3) -- (0.6+\z,0.3) -- (0+\z,0.3);
        \draw[rounded corners=2] (0+\z,-0.1) -- (0.4+\z,-0.1) -- (0.4+\z,0.1) -- (0+\z,0.1);
            \draw[gray,opacity=0.8,line width=2] (0+\z,-0.5) -- (0+\z,0.5);
         \foreach \x in {-0.3,-0.1,...,0.3}
           \fill[black] (0+\z,\x) circle (0.065);
            \draw[gray,opacity=0.8,line width=2] (0+\z,-0.5+\y) -- (0+\z,0.5+\y);
         \foreach \x in {-0.3,-0.1,...,0.3}
           \fill[black] (0+\z,\x+\y) circle (0.065);
           \end{tikzpicture}
\end{array}.
\ee
By inspecting the part of the diagram before the measurement one should be able to see that it represents the state shown in the measurement part. In other words, as a result of the protocol the top qubit is found in the $|\hat{0}\rangle$ state and the bottom -- in $|\hat{1}\rangle$. Now it should be straightforward  to conclude that the remaining two cases $a=b=0$ and $a=b=1$ will produce pairs $|\hat{0}\hat{0}\rangle$ and $|\hat{1}\hat{1}\rangle$ respectively.

Using the basis $|\hat{a}\hat{b}\rangle$ is not suitable for a physical experiment, so one should pass to an orthogonal basis instead. As already mentioned, the flaw of the diagrammatic basis used in the present approach is in the absence of orthogonal diagrams. Therefore one will need to consider linear combinations of the diagrams. To measure states in the computational basis $|ab\rangle$ the encoding transformations should be appropriate linear combinations of diagrams in table~(\ref{tab:densecode}). The generalized protocol is then
\be
\label{densecoding}
\begin{array}{c}
     \begin{tikzpicture}[thick]
        \newcommand{\y}{0}
        \renewcommand{\y}{1.2}
        \draw[rounded corners=2] (-1.5,-0.3+\y) -- (-0.9,-0.3+\y) -- (-0.6,-0.1+\y) -- (0,-0.1+\y);
           \draw[draw=white,double=black,line width=2,rounded corners=2] (-1.5,-0.1+\y) -- (-0.9,-0.1+\y) -- (-0.6,-0.3+\y) -- (0,-0.3+\y);
           \draw[teal] (-1.5,0.1+\y) -- (0,0.1+\y);
           \draw[teal] (-1.5,0.3+\y) -- (0,0.3+\y);
           \fill[white] (-1.,-0.4+\y) rectangle (-0.5,0.4+\y);
           \draw (-1,-0.4+\y) rectangle (-0.5,0.4+\y);
           \node[rotate=-90] at (-0.75,\y) {\small ${\cal B}_{[a,b]}$};
        \foreach \x in {-0.3,-0.1,...,-0.1}
        \draw[teal,rounded corners=2] (0,\x) -- (-2.25+\x,\x) -- (-2.25+\x,\y-\x) -- (-1.5,\y-\x);
        \foreach \x in {0.1,0.3,...,0.3}
        \draw[rounded corners=2] (0,\x) -- (-2.25+\x,\x) -- (-2.25+\x,\y-\x) -- (-1.5,\y-\x);
            \draw[gray,opacity=0.8,line width=2] (-1.5,-0.5) -- (-1.5,0.5+\y);
         \foreach \x in {-0.3,-0.1,...,0.3}
           \fill[black] (-1.5,\x+\y) circle (0.065);
           \foreach \x in {-0.3,-0.1,...,0.3}
           \fill[black] (-1.5,\x) circle (0.065);
           \newcommand{\z}{1.5}
           \renewcommand{\y}{0}
           \draw[teal,rounded corners=2] (-1.5+\z,-0.1+\y) -- (-3*1.5/4-0.1+\z,-0.1+\y) -- (-3*1.5/4+0.1+\z,0.1+\y) -- (-1.5/2-0.1+\z,0.1+\y) -- (-1.5/2+0.1+\z,0.3+\y) -- (0+\z,0.3+\y);
           \draw[teal,rounded corners=2] (-1.5+\z,-0.3+\y) -- (-1.5/2-0.1+\z,-0.3+\y) -- (-1.5/2+0.1+\z,-0.1+\y) -- (-1.5/4-0.1+\z,-0.1+\y) -- (-1.5/4+0.1+\z,0.1+\y) -- (0+\z,0.1+\y);
           \draw[draw=white,double=black,line width=2,rounded corners=2] (-1.5+\z,0.1+\y) -- (-3*1.5/4-0.1+\z,0.1+\y) -- (-3*1.5/4+0.1+\z,-0.1+\y) -- (-1.5/2-0.1+\z,-0.1+\y) -- (-1.5/2+0.1+\z,-0.3+\y) -- (0+\z,-0.3+\y);
           \draw[draw=white,double=black,line width=2,rounded corners=2] (-1.5+\z,0.3+\y) -- (-1.5/2-0.1+\z,0.3+\y) -- (-1.5/2+0.1+\z,0.1+\y) -- (-1.5/4-0.1+\z,0.1+\y) -- (-1.5/4+0.1+\z,-0.1+\y) -- (0+\z,-0.1+\y);
           \renewcommand{\y}{1.2}
           \newcommand{\w}{1.5}
           \foreach \x in {0.3,0.1,...,0.1}
        \draw[teal,rounded corners=2] (0,\x+\y) -- (1.5,\x+\y);
        \foreach \x in {-0.3,-0.1,...,-0.1}
        \draw[rounded corners=2] (0,\x+\y) -- (1.5,\x+\y);
        \draw[teal,rounded corners=2] (\w,0.3) -- (5*\w/4-0.1,0.3) -- (5*\w/4+0.1,-0.3+\y) -- (6*\w/4-0.1,-0.3+\y) -- (6*\w/4+0.1,-0.1+\y) -- (7*\w/4-0.1,-0.1+\y) -- (7*\w/4+0.1,0.1+\y) -- (2*\w,0.1+\y);
        \draw[teal,rounded corners=2] (\w,0.1) -- (6*\w/4-0.1,0.1) -- (6*\w/4+0.1,0.3) -- (7*\w/4-0.1,0.3) -- (7*\w/4+0.1,-0.3+\y) -- (2*\w,-0.3+\y);
        \draw[teal,rounded corners=2] (\w,0.1+\y) -- (6*\w/4-0.1,0.1+\y) -- (6*\w/4+0.1,0.3+\y) -- (2*\w,0.3+\y);
        \draw[draw=white,double=teal,line width=2,rounded corners=2] (\w,0.3+\y) -- (6*\w/4-0.1,0.3+\y) -- (6*\w/4+0.1,0.1+\y) -- (7*\w/4-0.1,0.1+\y) -- (7*\w/4+0.1,-0.1+\y) -- (2*\w,-0.1+\y);
        \draw[draw=white,double=black,line width=2,rounded corners=2] (\w,-0.3+\y) -- (5*\w/4-0.1,-0.3+\y) -- (5*\w/4+0.1,0.3) -- (6*\w/4-0.1,0.3) -- (6*\w/4+0.1,0.1) -- (7*\w/4-0.1,0.1) -- (7*\w/4+0.1,-0.1) -- (2*\w,-0.1);
        \draw[draw=white,double=black,line width=2,rounded corners=2] (\w,-0.1+\y) -- (6*\w/4-0.1,-0.1+\y) -- (6*\w/4+0.1,-0.3+\y) -- (7*\w/4-0.1,-0.3+\y) -- (7*\w/4+0.1,0.3) -- (2*\w,0.3);
        \draw[rounded corners=2] (\w,-0.1) -- (6*\w/4-0.1,-0.1) -- (6*\w/4+0.1,-0.3) -- (2*\w,-0.3);
        \draw[draw=white,double=black,line width=2,rounded corners=2] (\w,-0.3) -- (6*\w/4-0.1,-0.3) -- (6*\w/4+0.1,-0.1) -- (7*\w/4-0.1,-0.1) -- (7*\w/4+0.1,0.1) -- (2*\w,0.1);
           \draw[gray,opacity=0.8,line width=2] (0,-0.5) -- (0,0.5+\y);
         \foreach \x in {-0.3,-0.1,...,0.3}
           \fill[black] (0,\x+\y) circle (0.065);
           \foreach \x in {-0.3,-0.1,...,0.3}
           \fill[black] (0,\x) circle (0.065);
            \draw[gray,opacity=0.8,line width=2] (1.5,-0.5) -- (1.5,0.5+\y);
         \foreach \x in {-0.3,-0.1,...,0.3}
           \fill[black] (1.5,\x+\y) circle (0.065);
         \foreach \x in {-0.3,-0.1,...,0.3}
           \fill[black] (1.5,\x) circle (0.065);
            \draw[gray,opacity=0.8,line width=2] (2*\w,-0.5) -- (2*\w,0.5+\y);
         \foreach \x in {-0.3,-0.1,...,0.3}
           \fill[black] (2*\w,\x+\y) circle (0.065);
         \foreach \x in {-0.3,-0.1,...,0.3}
           \fill[black] (2*\w,\x) circle (0.065);
           \renewcommand{\z}{2*\w+0.3}
        \foreach \x in {-0.3,0.1,...,0.1}
        \draw[teal,rounded corners=2] (0+\z,\x+\y) -- (0.5+\z,\x+\y) -- (0.5+\z,\x+0.2+\y) -- (0+\z,\x+0.2+\y);
        \draw[rounded corners=2] (0+\z,-0.3) -- (0.6+\z,-0.3) -- (0.6+\z,0.3) -- (0+\z,0.3);
        \draw[rounded corners=2] (0+\z,-0.1) -- (0.4+\z,-0.1) -- (0.4+\z,0.1) -- (0+\z,0.1);
            \draw[gray,opacity=0.8,line width=2] (0+\z,-0.5) -- (0+\z,0.5);
         \foreach \x in {-0.3,-0.1,...,0.3}
           \fill[black] (0+\z,\x) circle (0.065);
            \draw[gray,opacity=0.8,line width=2] (0+\z,-0.5+\y) -- (0+\z,0.5+\y);
         \foreach \x in {-0.3,-0.1,...,0.3}
           \fill[black] (0+\z,\x+\y) circle (0.065);
           \fill[white,rounded corners=2] (0.3+\z,-0.4+\y) rectangle (0.7+\z,0.4+\y);
           \draw[rounded corners=2] (0.3+\z,-0.4+\y) rectangle (0.7+\z,0.4+\y);
           \node[rotate=90] at (0.5+\z,\y) {$a$};
           \fill[white,rounded corners=2] (0.3+\z,-0.4) rectangle (0.7+\z,0.4);
           \draw[rounded corners=2] (0.3+\z,-0.4) rectangle (0.7+\z,0.4);
           \node[rotate=90] at (0.5+\z,0) {$b$};
           \end{tikzpicture}
\end{array}.
\ee
It is straightforward to compute the expansion coefficients of ${\cal B}[a,b]$ in terms of the braids in table~\ref{tab:densecode}. For example,  
\be
\begin{array}{c} 
                \begin{tikzpicture}[thick]
                \newcommand{\y}{0}
            \draw (-1.5,-0.3+\y) -- (0,-0.3+\y);
           \draw (-1.5,-0.1+\y) -- (0,-0.1+\y);
           \draw[teal,rounded corners=2] (-1.5,0.3+\y) -- (-0.9,0.3+\y) -- (-0.6,0.1+\y) -- (0,0.1+\y);
           \draw[draw=white,double=teal,line width=2,rounded corners=2] (-1.5,0.1+\y) -- (-0.9,0.1+\y) -- (-0.6,0.3+\y) -- (0,0.3+\y);
          \draw[gray,opacity=0.8,line width=2] (0,-0.5+\y) -- (0,0.5+\y);
         \foreach \x in {-0.3,-0.1,...,0.3}
           \fill[black] (0,\x+\y) circle (0.065);
            \draw[gray,opacity=0.8,line width=2] (-1.5,-0.5+\y) -- (-1.5,0.5+\y);
         \foreach \x in {-0.3,-0.1,...,0.3}
           \fill[black] (-1.5,\x+\y) circle (0.065);
            \fill[white] (-1.,-0.4+\y) rectangle (-0.5,0.4+\y);
           \draw (-1,-0.4+\y) rectangle (-0.5,0.4+\y);
           \node[rotate=-90] at (-0.75,\y) {\small ${\cal B}_{[1,0]}$};
                \end{tikzpicture} 
           \end{array}
\ = \ \frac{1}{d\sqrt{d^2-1}}\left(\begin{array}{c} 
                \begin{tikzpicture}[thick]
                \newcommand{\y}{0}
            \draw (-1.5,-0.3+\y) -- (0,-0.3+\y);
           \draw (-1.5,-0.1+\y) -- (0,-0.1+\y);
           \draw[teal,rounded corners=2] (-1.5,0.3+\y) -- (-0.9,0.3+\y) -- (-0.6,0.1+\y) -- (0,0.1+\y);
           \draw[draw=white,double=teal,line width=2,rounded corners=2] (-1.5,0.1+\y) -- (-0.9,0.1+\y) -- (-0.6,0.3+\y) -- (0,0.3+\y);
          \draw[gray,opacity=0.8,line width=2] (0,-0.5+\y) -- (0,0.5+\y);
         \foreach \x in {-0.3,-0.1,...,0.3}
           \fill[black] (0,\x+\y) circle (0.065);
            \draw[gray,opacity=0.8,line width=2] (-1.5,-0.5+\y) -- (-1.5,0.5+\y);
         \foreach \x in {-0.3,-0.1,...,0.3}
           \fill[black] (-1.5,\x+\y) circle (0.065);
                \end{tikzpicture} 
           \end{array} - \frac{1}{d}\begin{array}{c} 
                \begin{tikzpicture}[thick]
                \newcommand{\y}{0}
          \draw[gray,opacity=0.8,line width=2] (0,-0.5+\y) -- (0,0.5+\y);
         \foreach \x in {-0.3,-0.1,...,0.3}
           \fill[black] (0,\x+\y) circle (0.065);
            \draw[gray,opacity=0.8,line width=2] (-1.5,-0.5+\y) -- (-1.5,0.5+\y);
         \foreach \x in {-0.3,-0.1,...,0.3}
           \fill[black] (-1.5,\x+\y) circle (0.065);
           \draw (-1.5,-0.3+\y) -- (0,-0.3+\y);
           \draw (-1.5,-0.1+\y) -- (0,-0.1+\y);
           \draw[teal] (-1.5,0.1+\y) -- (0,0.1+\y);
           \draw[teal] (-1.5,0.3+\y) -- (0,0.3+\y);
                \end{tikzpicture} 
           \end{array}\right). 
\ee
Note also that the protocol has some cosmetic differences with the protocol in figure~\ref{fig:teleport} (right). In particular, in~(\ref{densecoding}), the analog of $H$ is applied before the analog of CNOT. Also the assignments of $a/b$ to Alice's/Bob's qubits are swapped. The reader is invited to redesign the protocol to better match the one in figure~\ref{fig:teleport}.

Let us also mention another subtlety highlighted by protocol~(\ref{densecoding}). Note that sections of the diagram (vertical gray lines) involve a Hilbert space of eight points, rather than a tensor product of two four-point Hilbert spaces. This means that the initial two-qubit state is embedded into a large Hilbert space (in general, of dimension 14). This is not a problem as long as the result of the protocol is restricted to belong to the four-dimensional two-qubit subspace. It is instructive to check what would happen if a tensor product of two four-point spaces were used.

\begin{figure} 
    $$
    \begin{array}{c} 
    \includegraphics[width=0.2\linewidth,angle=-90]{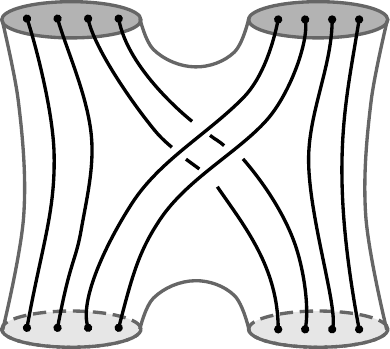}\end{array} \qquad \Longrightarrow \qquad \begin{array}{c}\includegraphics[width=0.2\linewidth,angle=-90]{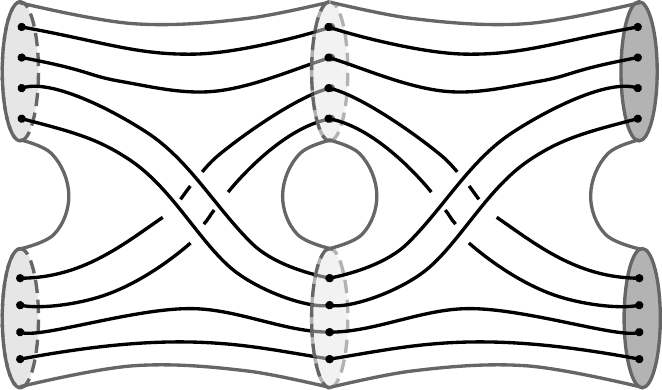}\end{array} \quad \neq \quad 
    \begin{array}{c}\includegraphics[width=0.2\linewidth,angle=-90]{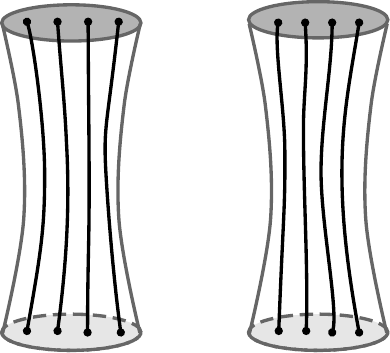}\end{array}
    $$
    \caption{Topological operators acting on tensor products of two Hilbert spaces are in general nonunitary. Contracting an operator with its unitary conjugate creates a topological defect, preventing the result to be an identity matrix.}
    \label{fig:nonunitop}
\end{figure}

First, application of a nonlocal, two-qubit, operation would result in a topological defect -- a hole in the middle:
\be
\label{holecreation}
\begin{array}{c}
     \includegraphics[scale=0.2]{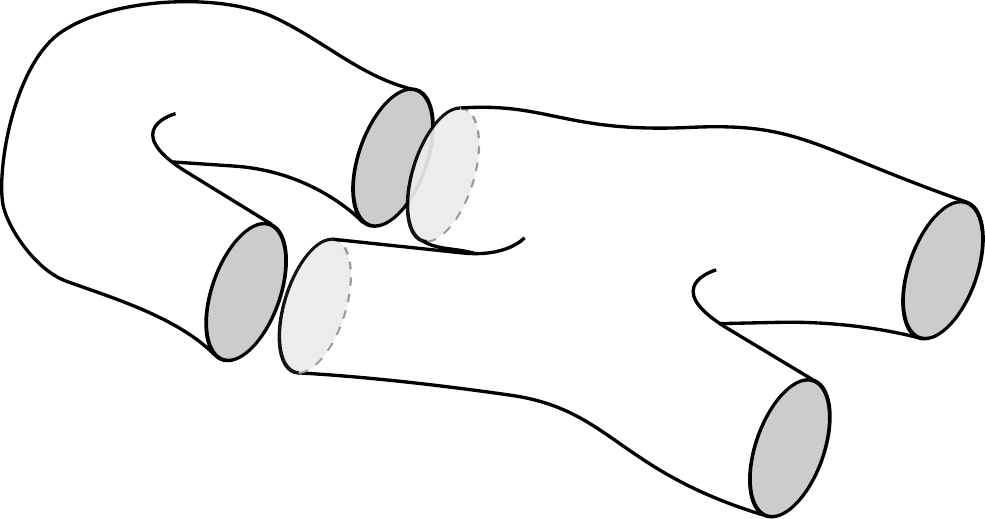} 
\end{array}
\qquad \longrightarrow \qquad
\begin{array}{c}
     \includegraphics[scale=0.2]{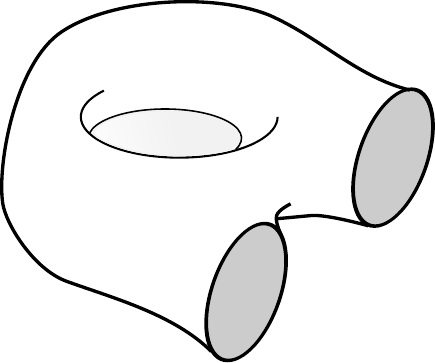} 
\end{array}.
\ee
Such a whole would obstruct realization of the necessary final states: One would not be able to freely move the lines around. Second, the operator, of the type appearing in~(\ref{holecreation}) would not in general be unitary, for a similar reason. Figure~\ref{fig:nonunitop} illustrates the fact that concatenating the operator and its Hermitian conjugate will not in general produce the identity due to a topological defect.

These comments complement a more naive version of this topological protocol considered in~\cite{Melnikov:2023nzn}.

%%%%%%%%%%%%%%%%%%%%%%%%%%%%%%%%%%%%%%%%%%%%%%%%%%%%%%%%%%%%%%%%%%%%
\section{Outlook}
\label{sec:outlook}

Hopefully the present notes convincingly demonstrate the potential and functionality of topology and, in particular, knot theory in the discussion of quantum mechanics. We have shown that states and operators can be thought as of topological spaces, equipped with a set of basic operations, such as braiding. The formalism of topological quantum field theories, or path integral formalism, gave precise mathematical meaning to such an association. The approach of the notes was to use the association directly to derive properties of quantum mechanics from topology, before ascertaining them in the standard matrix description. In the meantime, the precise mathematical description was available at every step of the discussion through a specific chosen TQFT model. In other words, the heuristic diagrammatic intuition (the picturalism of~\cite{Coecke:2017pic}) can always be tested by explicit precise computations.

We have also encountered a few complications related with the topological approach, which do not seem to be intractable, but rather call for a better diagrammatic approach or a better choice of the TQFT. One of these complications is the specific choice of the basis of the Temperley-Lieb diagrams, e.g.~(\ref{4pbasis}). Despite a simpler diagrammatic technique and connections to various interesting models, the basis is not orthogonal, which haunted us several times in this review. The basis of conformal blocks, more commonly used in topological quantum computation, for example, provides a set of orthogonal diagrams, but obscures the connection with knots and arguably complicates intuition. 

The choice of spheres $S^2$ as surfaces encoding TQFT Hilbert spaces comes with strings attached: the dimension of the Hilbert space is a complicated nonlinear function of the number of punctures. In this sense, choosing tori, as in the link complement states of~\cite{Balasubramanian:2016sro} gives a simpler model, although complicating the topology at the same time. An important complication arises in the attempts to work with multiboundary surfaces and tensor products of Hilbert spaces. As mentioned at the end of section~\ref{sec:densecoding} naive braiding operators in such spaces are not unitary. Instead, one seems to be forced to embed tensor products in spaces with connected boundaries and associated large Hilbert spaces, e.g.~\cite{Melnikov:2022vij,Chaves:2024rjs}.

The above are some obvious technical challenges of the present approach. In the following two sections a more general set of questions will be outlined, in application to which the topological paradigm can be further explored.

%%%%%%%%%%%%%%%%%%%%%%%%%%%%%%%%%%%%%%%%%%%%%%%%%%%%%%%%%%%%%%%%%%%%
\subsection{Multipartite Entanglement}
\label{sec:multipartite}

In section~\ref{sec:SLOCC} we reviewed classification of bipartite entanglement and showed that a classification similar to SLOCC is possible in terms of the number of connections (Wilson lines) held by the two subsystems. How does this classification work for multipartite systems?

For a 3-qubit system the SLOCC classification expects four distinct classes, two of which are biseparable (either all three qubits are separable, or two of them in the Bell-type entanglement). The other two classes correspond to genuinely tripartite entanglement: the GHZ class and the W class~\cite{Dur:2000zz}. We can try to use the same idea as in section~\ref{sec:SLOCC} and draw all possible ``connectome diagrams'' of three $S^2$ boundaries with four punctures~\cite{Melnikov:2022qyt}:
\be
\newcommand{\x}{0.6}
\begin{array}{c}
\scalebox{\x}{\begin{tikzpicture}[thick]
\fill[black] (0,0.0) circle (0.05cm);
\fill[black] (0,0.2) circle (0.05cm);
\fill[black] (0,0.4) circle (0.05cm);
\fill[black] (0,0.6) circle (0.05cm);
\draw (0,0) -- (0.2,0) arc (-90:90:0.1) -- (0,0.2);
\draw (0,0.4) -- (0.2,0.4) arc (-90:90:0.1) -- (0,0.6);
\fill[black] (1.8,0.0) circle (0.05cm);
\fill[black] (1.8,0.2) circle (0.05cm);
\fill[black] (1.8,0.4) circle (0.05cm);
\fill[black] (1.8,0.6) circle (0.05cm);
\draw (1.8,0.4) -- (1.6,0.4) arc (270:90:0.1) -- (1.8,0.6);
\draw (1.8,0.) -- (1.6,0.) arc (270:90:0.1) -- (1.8,0.2);
\fill[black] (0.6,1.2) circle (0.05cm);
\fill[black] (0.8,1.2) circle (0.05cm);
\fill[black] (1.,1.2) circle (0.05cm);
\fill[black] (1.2,1.2) circle (0.05cm);
\draw (0.6,1.2) -- (0.6,1.0) arc (-180:0:0.1) -- (0.8,1.2);
\draw (1.0,1.2) -- (1.0,1.0) arc (-180:0:0.1) -- (1.2,1.2);
\end{tikzpicture}} 
\end{array}
\quad
\begin{array}{c}
\scalebox{\x}{\begin{tikzpicture}[thick]
\fill[black] (0,0.0) circle (0.05cm);
\fill[black] (0,0.2) circle (0.05cm);
\fill[black] (0,0.4) circle (0.05cm);
\fill[black] (0,0.6) circle (0.05cm);
\draw (0,0.4) -- (0.2,0.4) arc (-90:90:0.1) -- (0,0.6);
\draw (0,0) -- (1.8,0.0);
\fill[black] (1.8,0.0) circle (0.05cm);
\fill[black] (1.8,0.2) circle (0.05cm);
\fill[black] (1.8,0.4) circle (0.05cm);
\fill[black] (1.8,0.6) circle (0.05cm);
\draw (1.8,0.2) -- (0,0.2);
\draw (1.8,0.4) -- (1.6,0.4) arc (270:90:0.1) -- (1.8,0.6);
\fill[black] (0.6,1.2) circle (0.05cm);
\fill[black] (0.8,1.2) circle (0.05cm);
\fill[black] (1.,1.2) circle (0.05cm);
\fill[black] (1.2,1.2) circle (0.05cm);
\draw (0.6,1.2) -- (0.6,1.0) arc (-180:0:0.1) -- (0.8,1.2);
\draw (1.0,1.2) -- (1.0,1.0) arc (-180:0:0.1) -- (1.2,1.2);
\end{tikzpicture}} 
\end{array}
\quad
\begin{array}{c}
\scalebox{\x}{\begin{tikzpicture}[thick]
\fill[black] (0,0.0) circle (0.05cm);
\fill[black] (0,0.2) circle (0.05cm);
\fill[black] (0,0.4) circle (0.05cm);
\fill[black] (0,0.6) circle (0.05cm);
\fill[black] (1.8,0.0) circle (0.05cm);
\fill[black] (1.8,0.2) circle (0.05cm);
\fill[black] (1.8,0.4) circle (0.05cm);
\fill[black] (1.8,0.6) circle (0.05cm);
\fill[black] (0.6,1.2) circle (0.05cm);
\fill[black] (0.8,1.2) circle (0.05cm);
\fill[black] (1.,1.2) circle (0.05cm);
\fill[black] (1.2,1.2) circle (0.05cm);
\draw (0.6,1.2) -- (0.6,1.0) arc (0:-90:0.4) -- (0.,0.6);
\draw (0.8,1.2) -- (0.8,1.0) arc (0:-90:0.6) -- (0.,0.4);
\draw (1.0,1.2) -- (1.0,1.0) arc (180:270:0.6) -- (1.8,0.4);
\draw (1.2,1.2) -- (1.2,1.0) arc (180:270:0.4) -- (1.8,0.6);
\draw (0,0) -- (0.2,0) arc (-90:90:0.1) -- (0,0.2);
\draw (1.8,0.) -- (1.6,0.) arc (270:90:0.1) -- (1.8,0.2);
\end{tikzpicture}}
\end{array}
\quad
\begin{array}{c}
\scalebox{\x}{\begin{tikzpicture}[thick]
\fill[black] (0,0.0) circle (0.05cm);
\fill[black] (0,0.2) circle (0.05cm);
\fill[black] (0,0.4) circle (0.05cm);
\fill[black] (0,0.6) circle (0.05cm);
\draw (0,0.4) -- (0.2,0.4) arc (90:-90:0.1) -- (0,0.2);
\draw (0,0) -- (1.8,0.0);
\fill[black] (1.8,0.0) circle (0.05cm);
\fill[black] (1.8,0.2) circle (0.05cm);
\fill[black] (1.8,0.4) circle (0.05cm);
\fill[black] (1.8,0.6) circle (0.05cm);
\draw (1.8,0.4) -- (1.6,0.4) arc (90:270:0.1) -- (1.8,0.2);
\fill[black] (0.6,1.2) circle (0.05cm);
\fill[black] (0.8,1.2) circle (0.05cm);
\fill[black] (1.,1.2) circle (0.05cm);
\fill[black] (1.2,1.2) circle (0.05cm);
\draw (0.6,1.2) -- (0.6,1.0) arc (0:-90:0.4) -- (0.,0.6);
\draw (0.8,1.2) -- (0.8,1.0) arc (-180:0:0.1) -- (1.0,1.2);
\draw (1.2,1.2) -- (1.2,1.0) arc (180:270:0.4) -- (1.8,0.6);
\end{tikzpicture}}
\end{array}
\quad
\begin{array}{c}
\scalebox{\x}{\begin{tikzpicture}[thick]
\fill[black] (0,0.0) circle (0.05cm);
\fill[black] (0,0.2) circle (0.05cm);
\fill[black] (0,0.4) circle (0.05cm);
\fill[black] (0,0.6) circle (0.05cm);
\fill[black] (1.8,0.0) circle (0.05cm);
\fill[black] (1.8,0.2) circle (0.05cm);
\fill[black] (1.8,0.4) circle (0.05cm);
\fill[black] (1.8,0.6) circle (0.05cm);
\draw (1.8,0.2) -- (0,0.2);
\draw (1.8,0.) -- (0,0.);
\draw (1.8,0.6) -- (0,0.6);
\draw (1.8,0.4) -- (0,0.4);
\fill[black] (0.6,1.2) circle (0.05cm);
\fill[black] (0.8,1.2) circle (0.05cm);
\fill[black] (1.,1.2) circle (0.05cm);
\fill[black] (1.2,1.2) circle (0.05cm);
\draw (0.6,1.2) -- (0.6,1.0) arc (-180:0:0.1) -- (0.8,1.2);
\draw (1.0,1.2) -- (1.0,1.0) arc (-180:0:0.1) -- (1.2,1.2);
\end{tikzpicture}}
\end{array}
\quad
\begin{array}{c}
\scalebox{\x}{\begin{tikzpicture}[thick]
\fill[black] (0,0.0) circle (0.05cm);
\fill[black] (0,0.2) circle (0.05cm);
\fill[black] (0,0.4) circle (0.05cm);
\fill[black] (0,0.6) circle (0.05cm);
\draw (0,0.4) -- (1.8,0.4);
\draw (0,0) -- (1.8,0.0);
\fill[black] (1.8,0.0) circle (0.05cm);
\fill[black] (1.8,0.2) circle (0.05cm);
\fill[black] (1.8,0.4) circle (0.05cm);
\fill[black] (1.8,0.6) circle (0.05cm);
\draw (0,0.2) -- (1.8,0.2);
\fill[black] (0.6,1.2) circle (0.05cm);
\fill[black] (0.8,1.2) circle (0.05cm);
\fill[black] (1.,1.2) circle (0.05cm);
\fill[black] (1.2,1.2) circle (0.05cm);
\draw (0.6,1.2) -- (0.6,1.0) arc (0:-90:0.4) -- (0.,0.6);
\draw (0.8,1.2) -- (0.8,1.0) arc (-180:0:0.1) -- (1.0,1.2);
\draw (1.2,1.2) -- (1.2,1.0) arc (180:270:0.4) -- (1.8,0.6);
\end{tikzpicture}}
\end{array}
\quad
\begin{array}{c}
\scalebox{\x}{\begin{tikzpicture}[thick]
\fill[black] (0,0.0) circle (0.05cm);
\fill[black] (0,0.2) circle (0.05cm);
\fill[black] (0,0.4) circle (0.05cm);
\fill[black] (0,0.6) circle (0.05cm);
\draw (0,0) -- (1.8,0.0);
\fill[black] (1.8,0.0) circle (0.05cm);
\fill[black] (1.8,0.2) circle (0.05cm);
\fill[black] (1.8,0.4) circle (0.05cm);
\fill[black] (1.8,0.6) circle (0.05cm);
\draw (0,0.2) -- (1.8,0.2);
\fill[black] (0.6,1.2) circle (0.05cm);
\fill[black] (0.8,1.2) circle (0.05cm);
\fill[black] (1.,1.2) circle (0.05cm);
\fill[black] (1.2,1.2) circle (0.05cm);
\draw (0.6,1.2) -- (0.6,1.0) arc (0:-90:0.4) -- (0.,0.6);
\draw (0.8,1.2) -- (0.8,1.0) arc (0:-90:0.6) -- (0.,0.4);
\draw (1.0,1.2) -- (1.0,1.0) arc (180:270:0.6) -- (1.8,0.4);
\draw (1.2,1.2) -- (1.2,1.0) arc (180:270:0.4) -- (1.8,0.6);
\end{tikzpicture}} 
\end{array}.
\ee
The first diagram clearly depicts a completely separable state of three qubits, but also diagrams number two, three and four, since in those diagrams parties are connected with other parties by at most two lines. In such a case the lines can be cut, as explained in section~\ref{sec:SLOCC}. Similarly, diagrams number five and six are biseparable, corresponding to a Bell pair and a separable qubit. The last diagram expresses the state
\be
\begin{array}{c}
\begin{tikzpicture}[thick]
\fill[black] (0,0.0) circle (0.05cm);
\fill[black] (0,0.2) circle (0.05cm);
\fill[black] (0,0.4) circle (0.05cm);
\fill[black] (0,0.6) circle (0.05cm);
\draw (0,0) -- (1.8,0.0);
\fill[black] (1.8,0.0) circle (0.05cm);
\fill[black] (1.8,0.2) circle (0.05cm);
\fill[black] (1.8,0.4) circle (0.05cm);
\fill[black] (1.8,0.6) circle (0.05cm);
\draw (0,0.2) -- (1.8,0.2);
\fill[black] (0.6,1.2) circle (0.05cm);
\fill[black] (0.8,1.2) circle (0.05cm);
\fill[black] (1.,1.2) circle (0.05cm);
\fill[black] (1.2,1.2) circle (0.05cm);
\draw (0.6,1.2) -- (0.6,1.0) arc (0:-90:0.4) -- (0.,0.6);
\draw (0.8,1.2) -- (0.8,1.0) arc (0:-90:0.6) -- (0.,0.4);
\draw (1.0,1.2) -- (1.0,1.0) arc (180:270:0.6) -- (1.8,0.4);
\draw (1.2,1.2) -- (1.2,1.0) arc (180:270:0.4) -- (1.8,0.6);
\end{tikzpicture} 
\end{array} \ =\  |000\rangle + \frac{1}{\sqrt{d^2 - 1}} |111\rangle\,,
\ee
which is not exactly the GHZ state, but a state in the GHZ class. No connectome diagram exists to represent the W class.

The SLOCC classification points to existence of some properties of entanglement that are not captured by the ``classical'' connectome diagrams. (Recall that in the classical limit the connectomes label large classes of diagrams.) Perhaps, a more refined topological classification exists, and there is a diagram representing a genuine W state. See some suggestions in~\cite{Balasubramanian:2018por}, although those apply to higher dimensional Hilbert spaces, where SLOCC does not work. Note that W state certainly exists as a linear combination of diagrams, but there is no single diagram that would reduce to it in the classical limit.

Classical limit was also essential in the derivation of entanglement inequalities in section~\ref{sec:inequalities}. The above observation reminds that one should not take them for granted in the full quantum regime. The study of connectome inequalities in~\cite{Melnikov:2023nzn} showed that many known inequalities for entanglement entropy can be reproduced by the simple counting of the connectome method, although in some cases the latter produces identities, which saturate the inequalities. This is the case of monogamy of mutual information, which reads
\be
\label{MImonogamy}
S(A+B) + S(B+C) + S(A+C) \ \geq \ S(A+B+C) + S(A) + S(B) + S(C)\,.
\ee
and a family of similar relations studied in~\cite{Bao:2015bfa}, for example. It is not hard to check, that counting connections indeed saturates this inequality.

Are there quantum states that violate the connectome inequalities? As in the example of the W state, inaccessible to connectomes, such state would have to be ``more quantum'' then the semiclassical states satisfying the connectome inequalities.

%%%%%%%%%%%%%%%%%%%%%%%%%%%%%%%%%%%%%%%%%%%%%%%%%%%%%%%%%%%%%%%%%%%%
\subsection{Quantum Gravity}
\label{sec:gravity}

Knot theory explicitly appears in the construction of states of loop quantum gravity~\cite{Rovelli:1995ac,Gambini:2001book}. Quantum states in this theory are similar to the Chern-Simons states used in this review, although they are based on closed lines, or closed networks, which are analogs of conformal blocks mentioned here, but with no open ends. Consequently, some questions addressed in this review, such as properties of multipartite entanglement, can be studied in the context of loop quantum gravity, using similar techniques. See~\cite{Cepollaro:2023bqe} for a recent example.

In the string theory approach to quantum gravity the relevant framework appears in the context of the AdS/CFT (or holographic) correspondence~\cite{Maldacena:1997re}. The AdS/CFT correspondence postulates the equivalence, or \emph{duality}, of two descriptions of the same quantum theory, valid for a large class of theories: one is a gravitational theory in the bulk of some space and another is a theory without gravity defined on the boundary of that space (see also~\cite{Gubser:1998bc,Witten:1998qj}). The general idea of such a duality can be explained in terms of a TQFT, and, in particular, Chern-Simons theory, which is a prototypical example of the holographic correspondence. 

Recall that in the Atiyah's axioms, the boundary of the topological space encodes the Hilbert space. The bulk cobordisms serve as one way of discriminating between different states in this Hilbert space, but not a unique way. In the 3D Chern-Simons case the boundary theory can be described by a 2D Wess-Zumino-Witten model, which is then a CFT-dual description of the 3D gauge theory. More abstractly, the theory and the specific states of the boundary theory can be constructed in terms of representations of the braid group action on the punctures on the boundary. One can think of the construction in section~\ref{sec:matrix}
as of an example of such a representation. In this case, braids themselves are emerging bulk representations of the action of given operators in the Hilbert space.

The analogy between Chern-Simons states and gravity can be made even more precise if one observes that three-dimensional gravity in anti de Sitter (AdS) space has a description in terms of a Chern-Simons theory with a noncompact gauge group~\cite{Achucarro:1986uwr,Witten:1988hc}. Due to the noncompactness, the treatment of this theory is more difficult than the $SU(2)$ case considered here (see for example~\cite{Maloney:2007ud,Maloney:2020nni} for the discussion of the main issues), but in principle, questions similar to the ones discussed in this review can be addressed in this framework too.

We will close this discussion by giving the interpretation of the classical limit of the entanglement entropy formula in terms of the famous result of Shinsei Ryu and Tadashi Takayanagi in holography~\cite{Ryu:2006bv}. The Ryu-Takayanagi formula expresses the entropy of a region of the boundary (subsystem $A$) in terms of gravitational data in the bulk of the dual description. Namely, the entropy is proportional to the area of the minimal-area surface in the bulk anchored at the boundary of the boundary region $A$: 
\be
S_{\rm vN} \ = \ \min_{\{\gamma\}}\frac{{\rm Area}[\gamma]}{4G_{\rm N}}\,.
\ee
Here $\gamma$ is a surface in the bulk, whose area must be minimized, and $G_{\rm N}$ is the Newton constant.

In section~\ref{sec:entropy} we observed that for the connectome states, which are characteristic states of Chern-Simons in the classical limit, the entropy is computed by the number of lines connecting the two parties. In the limit of large number of Wilson lines the result is given by~(\ref{ClassEntropy}). We noted that the number of connections is counted by a surface that encircles one of the subsystems and crosses the minimal number of connections. If one assumes that each line comes with a flux proportional to the inverse Newton constant then the Ryu-Takayanagi formula follows~\cite{Melnikov:2023wwc}. Similar counting emerges in the tensor network construction of holography~\cite{Pastawski:2015qua} and in the ``bit thread'' models~\cite{Freedman:2016zud,Headrick:2022nbe}.

Chern-Simons, or TQFT states provide at least a heuristic picture of many concepts and results discussed in the AdS/CFT over the past years. Besides the correspondence itself, the emergence of space from entanglement, the Ryu-Takayanagi formula, entropy inequalities, discussed here, one can also mention Witten diagrams~\cite{Witten:1998qj}, bulk reconstruction~\cite{Harlow:2018fse}, quantum error correcting codes~\cite{Almheiri:2014lwa}, tensor networks~\cite{Jahn:2021uqr}, replica wormholes and black hole information paradox~\cite{Penington:2019npb,Almheiri:2019psf,Penington:2019kki,Almheiri:2019qdq} among the topics, where the present ``categorical'' or ``pictorial'' approach may be of certain use. It should also be useful for understanding quantum gravity itself.

\paragraph{Acknowledgments.} I would like to thank Julio Fabris and the Federal University of Espirito Santo in Vitoria for hosting the IV Patricio Letelier School on Mathematical Physics, for which the first version of these notes was prepared. I would also like to thank João Aires, Rafael Chaves, Andrei Mironov, Sergey Mironov, Alexei Morozov, Andrey Morozov, Marcos Neves, Jefferson Oliveira, Luigy Pinto, Davide Poderini and Vivek Singh for useful discussions and collaboration on related projects. I am grateful to Marcia Tenser for reading the manuscript and suggesting important improvements. This work was supported by the Simons Foundation award number 1023171-RC, grants of the Brazilian National Council for Scientific and Technological Development (CNPq) number 308580/2022-2 and 404274/2023-4, and grant 1699/24 IIF-FINEP of the Brazilian Ministry of Science of Technology.

\providecommand{\href}[2]{#2}\begingroup\raggedright\endgroup

%\bibliographystyle{JHEP}
%\bibliography{refs}

\end{document}